\documentclass[rmp,twocolumn]{revtex4-1}
\usepackage{graphicx}
\usepackage{amssymb,amsmath,enumerate}
\usepackage[colorlinks,allcolors=blue]{hyperref}
\usepackage{color}

\newcommand{\be}{\begin{eqnarray}}
\newcommand{\ee}{\end{eqnarray}}
\newcommand{\D}{\mathrm{d}}
\newcommand{\s}{\sigma}

\begin{document}

\title{Edwards Statistical Mechanics for Jammed Granular Matter}

\author{Adrian Baule$^1$, Flaviano Morone$^2$, 
Hans J. Herrmann$^3$, 
and Hern\'an A. Makse$^{2}\footnote{Correspondence to: hmakse@lev.ccny.cuny.edu}$ }

\affiliation{ $^1$School of Mathematical Sciences, Queen Mary
  University of London, London E1 4NS, UK\\ $^2$Levich
  Institute and Physics Department, City College of New York, New
  York, New York 10031, USA\\ 
$^3$ETH Z\"urich, Computational Physics for Engineering Materials,
Institute for Building Materials, Wolfgang-Pauli-Str.~27, HIT, CH-8093 Z\"urich, Switzerland}

\begin{abstract}
In 1989, Sir Sam Edwards made the visionary proposition to treat
jammed granular materials using a volume ensemble of equiprobable
jammed states in analogy to thermal equilibrium statistical mechanics,
despite their inherent athermal features. Since then, the statistical
mechanics approach for jammed matter -- one of the very few
generalizations of Gibbs-Boltzmann statistical mechanics to out of
equilibrium matter -- has garnered an extraordinary amount of
attention by both theorists and experimentalists. Its importance stems
from the fact that jammed states of matter are ubiquitous in nature
appearing in a broad range of granular and soft materials such as
colloids, emulsions, glasses, and biomatter. Indeed, despite being one
of the simplest states of matter -- primarily governed by the steric
interactions between the constitutive particles -- a theoretical
understanding based on first principles has proved exceedingly
challenging. Here, we review a systematic approach to jammed matter
based on the Edwards statistical mechanical ensemble. We discuss the
construction of microcanonical and canonical ensembles based on the
volume function, which replaces the Hamiltonian in jammed systems. The
importance of approximation schemes at various levels is emphasized
leading to quantitative predictions for ensemble averaged quantities
such as packing fractions and contact force distributions. An overview
of the phenomenology of jammed states and experiments, simulations,
and theoretical models scrutinizing the strong assumptions underlying
Edwards' approach is given including recent results suggesting the
validity of Edwards ergodic hypothesis for jammed states. A
theoretical framework for packings whose constitutive particles range
from spherical to non-spherical shapes like dimers, polymers,
ellipsoids, spherocylinders or tetrahedra, hard and soft, frictional,
frictionless and adhesive, monodisperse and polydisperse particles in
any dimensions is discussed providing insight into an unifying phase
diagram for all jammed matter. Furthermore, the connection between the
Edwards' ensemble of metastable jammed states and metastability in
spin-glasses is established. This highlights that the packing problem
can be understood as a constraint satisfaction problem for excluded
volume and force and torque balance leading to a unifying framework
between the Edwards ensemble of equiprobable jammed states and
out-of-equilibrium spin-glasses.

\end{abstract}

\date{\today}

\maketitle

\tableofcontents

\section{Introduction}

\begin{figure*}
\begin{center}
{\bf (a)}
\includegraphics[width=0.8\columnwidth]{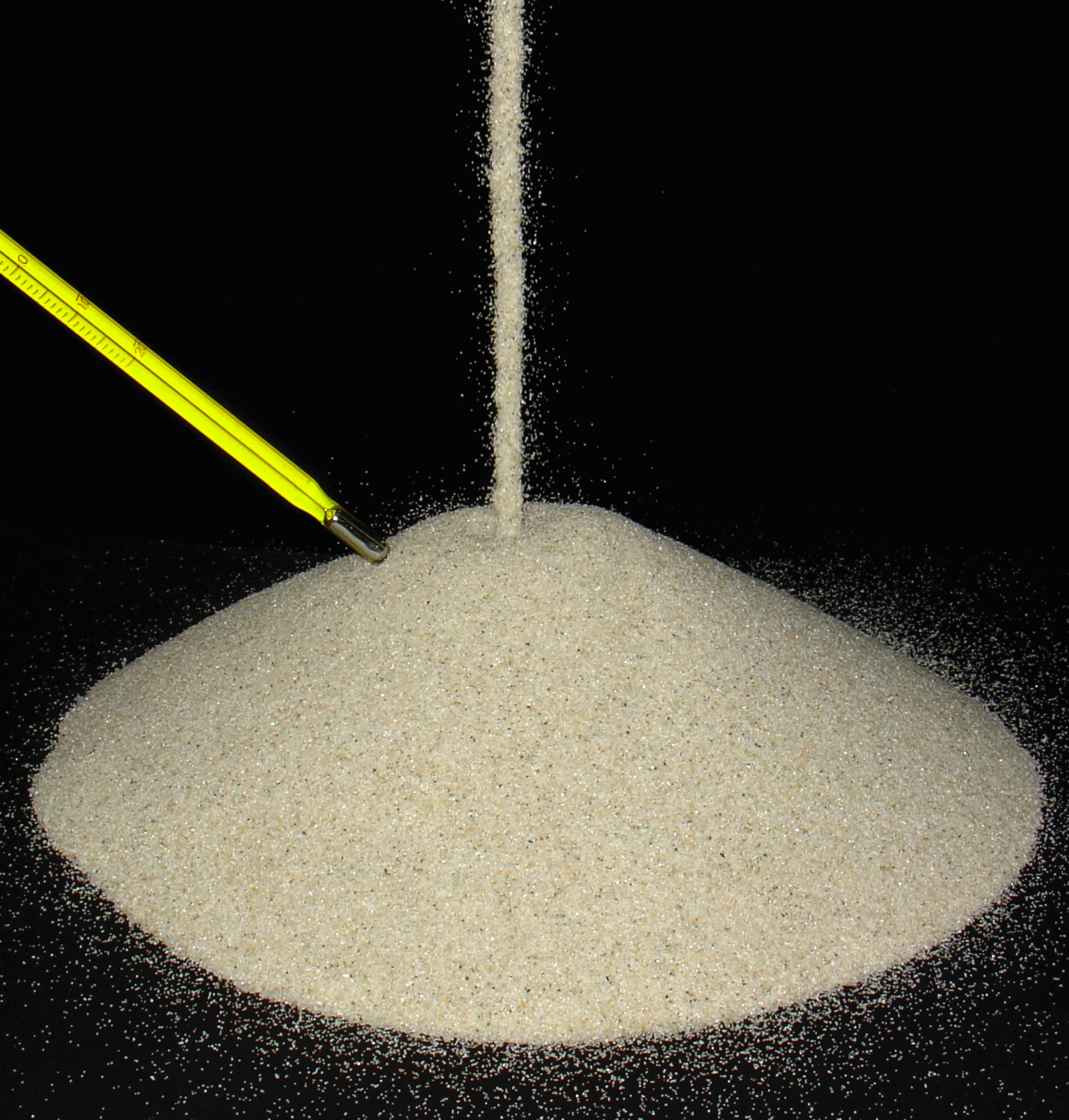}
\hspace{2cm} {\bf (b)}
\includegraphics[height=7.2cm]{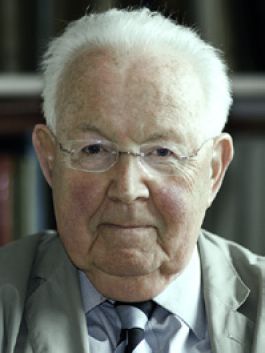}
\caption{{\bf (a)} (Color online) Pouring grains into a sandpile is
  the simplest example of a jamming transition from a flowing state to
  a mechanically stable jammed state. However, this simplicity can be
  deceiving. In this review we show that building sandpiles is at the
  core of one of the most profound problems in disordered media.  From
  the glass transition to novel phases in anisotropic colloidal
  systems, pouring grains in a pile is the emblematic system to master
  with tremendous implications on all sort of soft materials, from
  glasses, colloids, foams and emulsions to biomatter. Edwards'
  endeavour to tame granular matter is condensed in the attempt of
  measuring the `temperature' of the sandpile.  {\bf (b)} Sir Sam F. Edwards (February 1st 1928 -- July 7th
    2015) \cite{Warner:2017aa}. S.~F. Edwards first introduced the
  intriguing idea that a far-from-equilibrium, jammed granular matter
  could be described using methods from equilibrium statistical
  mechanics. In the Edwards' ensemble, macroscopic quantities are
  computed as flat averages over force- and torque-balanced
  configurations, which leads to a natural definition of a
  configurational `granular' temperature known as the compactivity.}
\label{sirsam} 
\end{center}
\end{figure*}

Materials composed of macroscopic grains such as sand, sugar, and ball
bearings are ubiquitous in our everyday experience. Nevertheless, a
fundamental description of both static and dynamic properties of
granular matter has proven exceedingly challenging. Take for example
the pouring of sand into a sandpile, Fig.~\ref{sirsam}a. This process
can be considered as a simple example of a fluid-to-solid phase
transition of a multi-particle system. However, it is not clear
whether this transition is governed by a variational principle of an
associated thermodynamic quantity like the free energy in equilibrium
systems. Granular materials do not explore different configurations in
the absence of external driving because thermal fluctuations induce
negligible particle motion at room temperature and inter-grain
dissipation and friction quickly drain the kinetic energy from the
system. On the other hand, the {\it jammed} state of granular matter
bears a remarkable resemblance with an amorphous solid in thermal equilibrium:
both are able to sustain a non-zero shear stress; the phase transition
from liquid to solid states and the analogous {\it jamming transition}
in grains are both governed by one or a few macroscopic control
parameters; and, when using certain packing-generation protocols,
macroscopic observables, such as the packing fraction, are largely
reproducible.

Jamming transitions not only occur in granular media, but also in soft
materials such as colloidal suspensions which may asymptotically reach
jamming under centrifugation, compressed emulsions, foams, glasses and
spin-glasses below their glass transition temperature and biological
materials such as cells, DNA and protein packing.  Even more broadly,
the jamming transition pertains to a larger family of computational
problems named Constraint Satisfaction Problems (CSP)
\cite{Krzakala:2007aa}. These problems involve finding the values of a
set of variables satisfying simultaneously all the constraints imposed
on those variables and maximizing (or minimizing) an objective
function. For example, in the problem of sphere packings, the goal is
to minimize the volume occupied by the packing subject to the
geometrical constraint of non-overlapping particles and the mechanical
constraints of force and torque balance at mechanical equilibrium. In
general, packing problems play a central role in various fields of
science in addition to physics, such as discrete mathematics, number
theory and information theory.  An example of practical interest is
the problem of efficient data transmission through error-correcting
codes, which is deeply related to the optimal packing of (Hamming)
spheres in a high-dimensional space \cite{Conway:1999aa}. The common
feature of all packing problems is the existence of a phase
transition, the jamming transition, separating the phase where the
constraints are satisfiable from a phase where they are unsatisfiable.

The existence of constraints in physical systems causes, in general, a
significant metastability. Metastability is the phenomenon by which
the system remains confined for a relatively long time in suboptimal
regions of the phase space. It is related to the rough energy (or free
energy) landscape characterized by the presence of many non-trivially
related minima as a function of the microscopic configurations (or the
macroscopic states). Metastability is, indeed, the {\it leitmotiv} in
most complex physical systems, whatever its origin. For example, in
granular materials metastability arises from geometrical and
mechanical constraints, but it is found also in spin glasses, which
are magnetic systems with competing ferromagnetic and
antiferromagnetic exchange interactions. In spin glasses, the
emergence of metastability is due to frustration, which is the
inability of the system to satisfy simultaneously all local ordering
requirements. Notwithstanding their differences, these two physical
systems, jammed grains and spin-glasses, exhibit a remarkably similar
organization of their metastable states, a fact that stimulates our
search for further analogies within these systems and common
explanations. It is, indeed, this analogue approach, as best
exemplified by the encompassing vision of Sir Sam Edwards
\cite{Goldbart:2005aa}, that may shed new light on the solution to
jamming problems otherwise doomed to remain obscure.

Due to their substantial metastability, these systems are
fundamentally out-of-equilibrium even in a macroscopically quiescent
state. Nevertheless, the commonalities with equilibrium many body systems
suggest that ideas from equilibrium statistical mechanics might be
useful. In this review, we consider theories for jammed matter based
on generalizations of equilibrium ensembles. These statistical
mechanics-based approaches were pioneered by Sir Sam F. Edwards in the
late 1980s (Fig.~\ref{sirsam}b).

Investigations of the structural properties of jammed packings are
much older. In fact, the related problem of identifying the densest
packing of objects has an illustrious history in the mathematical literature~\cite{Kepler:1611aa,Weaire:2008aa}.  Exact
mathematical proofs of the densest packings are extremely challenging
even for spherical particles. The Kepler conjecture of 1611 stating
that the densest arrangement of spheres in three spatial dimensions
(3d) is a face-centered-cubic (FCC) crystal with a packing volume
fraction $\phi_{\rm fcc}=\pi/(3\sqrt{2})\approx 0.74048...$ remained an
unsolved mathematical problem for almost four
centuries~\cite{Kepler:1611aa,Hales:2005aa}. Systematic experiments
on disordered hard-sphere packings began in the 1960s with the work by
Bernal~\cite{Bernal:1960aa,Bernal:1964aa}. These experiments are
conceptually simple, yet give fundamental insight into the structure
of dense liquids, glasses, and jammed systems. Equally sized spherical
particles were placed into a container and compactified by shaking or
tapping the system until no further volume reduction was detected. These experiments typically yielded configurations with
packing fraction $\phi_{\rm rcp} \approx 0.64$, which is historically
referred to as random close packing (RCP).

In order to apply a statistical mechanical framework to these jammed
systems, it is first necessary to identify the variables characterizing
the state of the system macroscopically. Clearly, the system energy is
not suitable, since it may either not be conserved (for frictional
dissipative particles) or not be relevant (for frictionless hard particles). On the
other hand, an obvious state variable is the system volume. In fact, unlike in equilibrium
systems, the volume in jammed systems is not an externally imposed
fixed variable, but rather depends on the microscopic configuration of
the grains. Edwards first extraordinary insight was to parametrize the
ensemble of jammed states by the volume function
$\mathcal{W}(\{\mathbf{r}_i,\mathbf{\hat{t}}_i\})$, as a function of
the $N$ particles' positions $\{\mathbf{r}_i\}$ and orientations
$\{\mathbf{\hat{t}}_i\}$, as a replacement for the Hamiltonian in the
equilibrium ensembles
\cite{Edwards:1989aa,Mehta:1990aa,Edwards:1991aa,Edwards:1994aa}.

A second crucial point in the development of the Edwards granular
statistical mechanics is a proper definition of the jammed state. It
is important to note that only jammed configurations
$\{\mathbf{r}_i,\mathbf{\hat{t}}_i\}$ are included in the ensemble. A
definition of what we mean by jammed state is not a trivial task and
will be treated rigorously in the next section. Assuming that an unambiguous definition of
metastable jammed state can be expressed analytically, then a
statistical mechanics approach to granular matter proceeds by analogy
with equilibrium systems. In this case, the volume function allows for
the definition of a granular entropy leading to both microcanonical
and canonical formulations of the volume ensembles. This implies, in
particular, the existence of an intensive parameter conjugate to the
volume. This temperature-like parameter was called {\it compactivity}
by Edwards.

The full Edwards ensemble is characterized by the macroscopic
volume and, further, by the stress of the packing. Since analytical treatments of the
full ensemble are challenging, one typically considers suitable
approximations. Neglecting correlations between the volume and the
stress leads to a volume ensemble under the condition of isostaticity \cite{Song:2008aa}. The
core of this review will be devoted to elaborate on a mean-field
formulation of the Edwards volume ensemble that can potentially lead
to a unifying phase diagram encompassing all jammed matter ranging
from systems made of spherical to non-spherical particles, with
friction or adhesion to frictionless particles, monodisperse and
polydisperse systems and in any dimension. Likewise, we describe frameworks for stress and force statistics alone, such as the stress
ensemble \cite{Henkes:2007aa,Chakraborty:2010aa}, force network
ensemble \cite{Snoeijer:2004ab,Tighe:2010aa,Bouchaud:2002aa}, and belief propagation
for force transmission \cite{Bo:2014aa}.

Edwards statistical mechanical ensemble rely on two assumptions: (i) Ergodicity and (ii) Equiprobability of microstates. These assumptions have been scrutinized in the literature, and the questions raised in this context will be reviewed here. Despite these critiques, the Edwards' approach has been used to
describe a wide range of jammed and glassy materials. Early works
adopted the concept of inherent structures from glasses
\cite{Coniglio:2000aa,Coniglio:2001aa,Coniglio:2002ab,Fierro:2002ab,Coniglio:1996aa}
and effective temperatures
\cite{Kurchan:2000aa,Kurchan:2001aa,Makse:2002aa,Ciamarra:2006aa,Ono:2002aa,OHern:2004aa,Cugliandolo:2011aa}
with applications to plasticity \cite{Lieou:2012aa}. More recent
approaches are based on replica theory for hard-sphere glasses
\cite{Parisi:2010aa,Charbonneau:2017aa}. Valuable insight is gained from models that
exhibit both jamming and glass transitions
\cite{Krzakala:2007aa,Mari:2009aa,Ikeda:2012aa}. In this review, we emphasize that the Edwards ensemble can be recast as a constraint
satisfaction problems, which allows for an unifying view of
hard-sphere glasses and spin-glasses through a synthesis applied at
the foundation of granular statistical mechanics.

This review is organized as follows. In Sec.~\ref{Sec:theories} we
discuss the foundations of the ensemble approach via the rigorous
definition of metastable jammed states, and the construction of
microcanonical and canonical ensembles based on the volume function
and stress-moment tensor, which play the role of the Hamiltonian in
jammed systems. In Sec.~\ref{Sec:tests} we collect empirical results on the phenomenology of jammed states. Moreover, we review results from experiments,
simulations, and theoretical models that test the ergodic and uniform
measure underlying the ensemble approach. In
Sec.~\ref{Sec:volume} we consider volume ensembles and their
mean-field description, which provides quantitative predictions for
ensemble averaged quantities such as the packing fraction of spherical
and non-spherical particles.  In Sec.~\ref{Sec:constraints} we discuss
a unification between the Edwards ensemble of jammed matter and
theories based on ideas from glass/spin glass theories under the CSP
paradigm. In Sec.~\ref{Sec:outlook} we finally close with a summary
and a collection of open questions for future work.

In recent years a number of reviews have appeared dealing with more
specific aspects of granular matter: \cite{Richard:2005aa} (granular
compaction), \cite{Makse:2005aa} (jammed emulsions),
\cite{Chakraborty:2010aa,Bi:2015aa} (stress ensembles),
\cite{Tighe:2010aa} (force network ensemble), \cite{Cugliandolo:2011aa,Qiong:2014aa} (effective temperatures). The present review is also complementary to other reviews on jammed granular
matter, which do not specifically discuss the Edwards thermodynamics:
\cite{Jaeger:1996aa,Alexander:1998aa,Kadanoff:1999aa,Parisi:2010aa,Torquato:2010aa,Hecke:2010aa,Liu:2010aa,Borzsonyi:2013aa,Charbonneau:2017aa}. Rather than replacing
these reviews, our work puts these topics into the general context of Edwards statistical mechanics and provides an overview of the immense amount
of literature related to Edwards ensemble approaches.

\section{Statistical Mechanics for jammed granular matter}

\label{Sec:theories}

In a jammed system all particle motion is prevented due to
the confinement by the neighbouring particles. The transition to a
jammed state is thus not controlled by the temperature as conventional
phase transitions in systems at thermal equilibrium, but by
geometrical and mechanical constraints imposed by all particles in the system. Therefore, jammed states can be
regarded as the set of solutions in the general class of Constraint
Satisfaction Problems (CSP), which we term {\it Jamming Satisfaction Problem} (JSP), where the constraints are fixed by the mechanical
stability of the blocked configurations of grains. From this
standpoint, the jamming problem has a wider scope than the pure
physical significance, encompassing the broader class of CSPs: the
unique feature of the packing problem in the large universe of CSPs is
that this system allows for a direct and relatively simple
experimental test of theoretical predictions.

\subsection{Definition of jammed states}
\label{Ch2:Sec:stability}

We consider an assembly of $N$ (for the sake of simplicity)
monodisperse particles described by the configurations of the
particles
$\{\mathbf{r}_1,\mathbf{\hat{t}}_1;...;\mathbf{r}_N,\mathbf{\hat{t}}_N\}$,
where $\mathbf{r}_i$ denotes the $i$th particle's position (of its
center of mass) and $\mathbf{\hat{t}}_i$ its orientation. The first
problem we address concerns the definition of a blocked configuration
of the particles, i.e, the jammed states. To be jammed the system has
to satisfy both excluded volume and mechanical
constraints. The excluded volume constraint enforces that particles do not overlap, and its mathematical implementation depends on the shape of the particles. For a system
of monodisperse hard-spheres, this constraint takes on the following form: \be
|\mathbf{r}_i-\mathbf{r}_j|\geq2R\ , \qquad \mbox{(equal-size hard
  spheres)}
\label{eq:hardcore}
\ee which means that the centers of any pair of particles $i$ and $j$
must be at a distance twice as large as their radius $R$.
The hard-core constraint in Eq.~\eqref{eq:hardcore} is valid only for monodisperse spheres, but it can be generalized to polydisperse and nonspherical particles.

\begin{center}
\begin{figure}
\includegraphics[width=0.6\columnwidth]{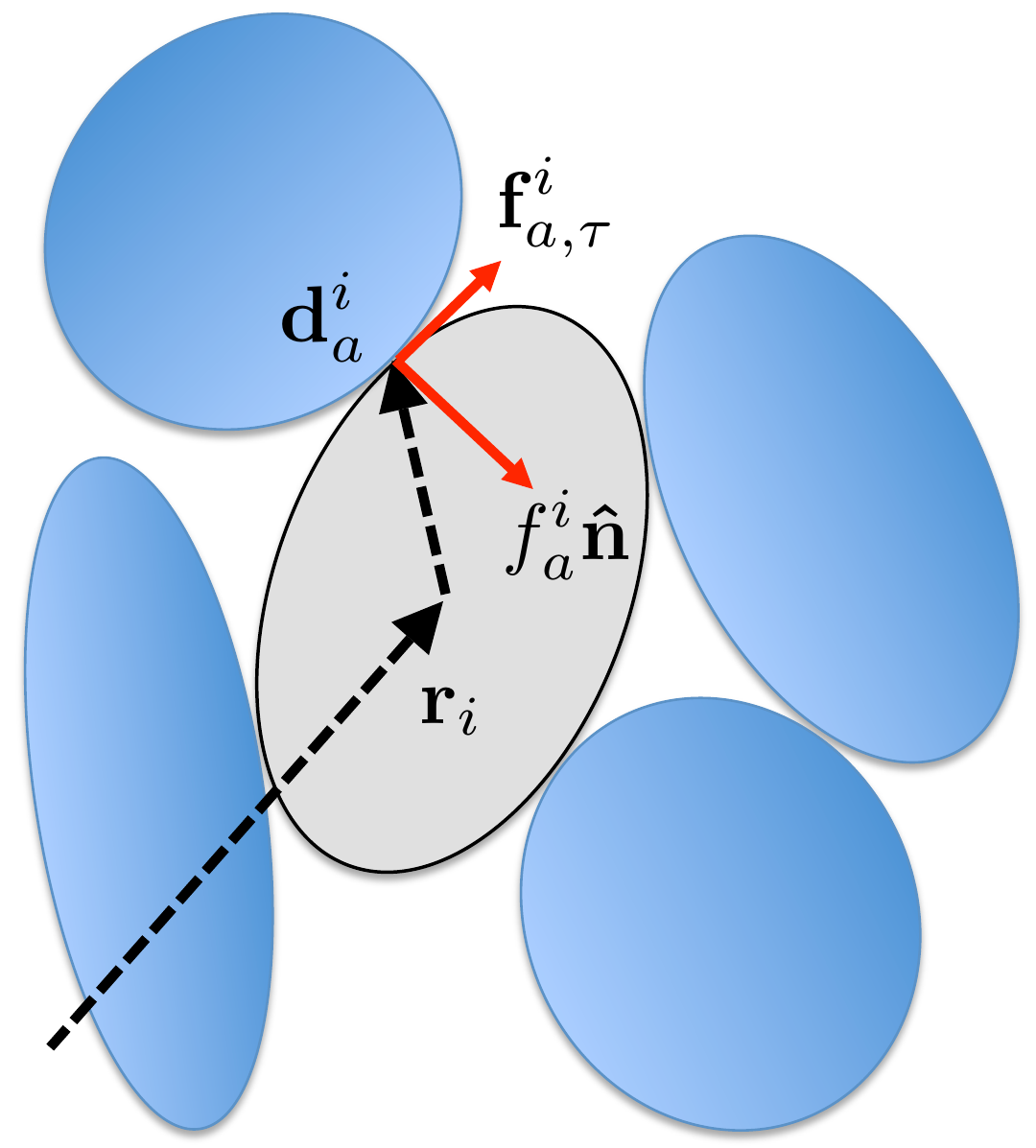}
\caption{Parametrization of a jammed configuration involving 5
  non-spherical grains. The tangential $\mathbf{f}^i_{a,\tau}$ and
  normal force vectors $f^i_{a,n} {\hat n}_a^i$ at contact $a$ on
  particle $i$ are shown. $d^i_a$ indicates the vector from the center
  of particle $i$ to the contact point $a$ between one of its
  neighbours. $r_i$ gives the location of the center of particle $i$.
  The grey-shaded particle is mechanically stable if all forces and
  torques generated at the four contact points cancel (see
  Eqs.~(\ref{Ch2:forceb},\ref{Ch2:torqueb})).}
\label{Fig:forcetorque}
\end{figure}
\end{center}

The excluded volume constraint is necessary but not sufficient by
itself to determine whether a configuration of particles is
jammed. Indeed, it has to be supplemented by a constraint enforcing
the mechanical stability of the system, requiring that particles
satisfy the force and torque balance conditions. We denote by
$\mathbf{d}_a^i$ the vector connecting $\mathbf{r}_i$ and the $a$th
contact on the $i$th particle. At this contact there is a
corresponding force vector $\mathbf{f}_a^i$ on particle $i$ arising
from the contacting particle. With this notation we can formulate the
conditions of force and torque balances for a particle of general
shape: \be
\label{Ch2:forceb}
\sum_{a\in\partial i}\mathbf{f}_a^i&=& 0,\qquad\qquad i=1,...,N\\
\label{Ch2:torqueb}
\sum_{a\in\partial i}\mathbf{d}_a^i\times\mathbf{f}_a^i&=&
0,\qquad\qquad i=1,...,N
\ee where the notation $\partial i$ denotes
the set of contacts of particle $i$. Equations~(\ref{Ch2:forceb}--\ref{Ch2:torqueb}) apply to both
frictional and frictionless particles. In the latter case there is
only one single force component in the normal direction
\be
\label{Ch2:normalf}
\mathbf{f}_a^i=-f_a^i\mathbf{\hat{n}}_a^i \hspace{.5cm} {\rm
  (frictionless)},
\ee where $\mathbf{\hat{n}}_a^i$ denotes the normal
unit vector at the contact point, which depends on the particle
shape. For frictional particles, we can decompose $\mathbf{f}_a^i$
into a normal component $f^i_{a,n}$ and a force vector in the tangent
plane $\mathbf{f}^i_{a,\tau}$ (see Fig.~\ref{Fig:forcetorque}). Coulomb's law with friction coefficient
$\mu$ is then expressed by the inequality \be
\label{Ch2:coulomb}
|\mathbf{f}^i_{a,\tau}|\le \mu f^i_{a,n} \hspace{.5cm} {\rm (frictional)}.  \ee 
If the interparticle forces are purely repulsive, as in most of the cases
treated in this review, we also have the condition:
\be 
\label{Ch2:repulsive}
\mathbf{d}_a^i\cdot\mathbf{f}_a^i<0.  \ee Finally, Newton's third
law implies, that two particles $i, j$ in contact at $a$ satisfy:
 \be
\label{Ch2:third}
\mathbf{f}_a^i=-\mathbf{f}_a^j.
\ee

\subsection{Metastability of jammed states}
\label{Sec:meta}

Having defined the necessary and sufficient conditions for a granular
system to be jammed, we now provide a finer description of jammed
states, based on the concept of metastability, i.e., their stability
with respect to particles displacements. A characterization similar
to the one proposed here appeared already in \cite{Torquato:2001aa}, where the authors defined the concept of
jamming categories for metastable packings. The similarities with the
classification of the jammed states in \cite{Torquato:2001aa} are
discussed in parallel with the classification presented next.

\begin{figure}[h!]
\begin{center}
\includegraphics[width=.8\columnwidth]{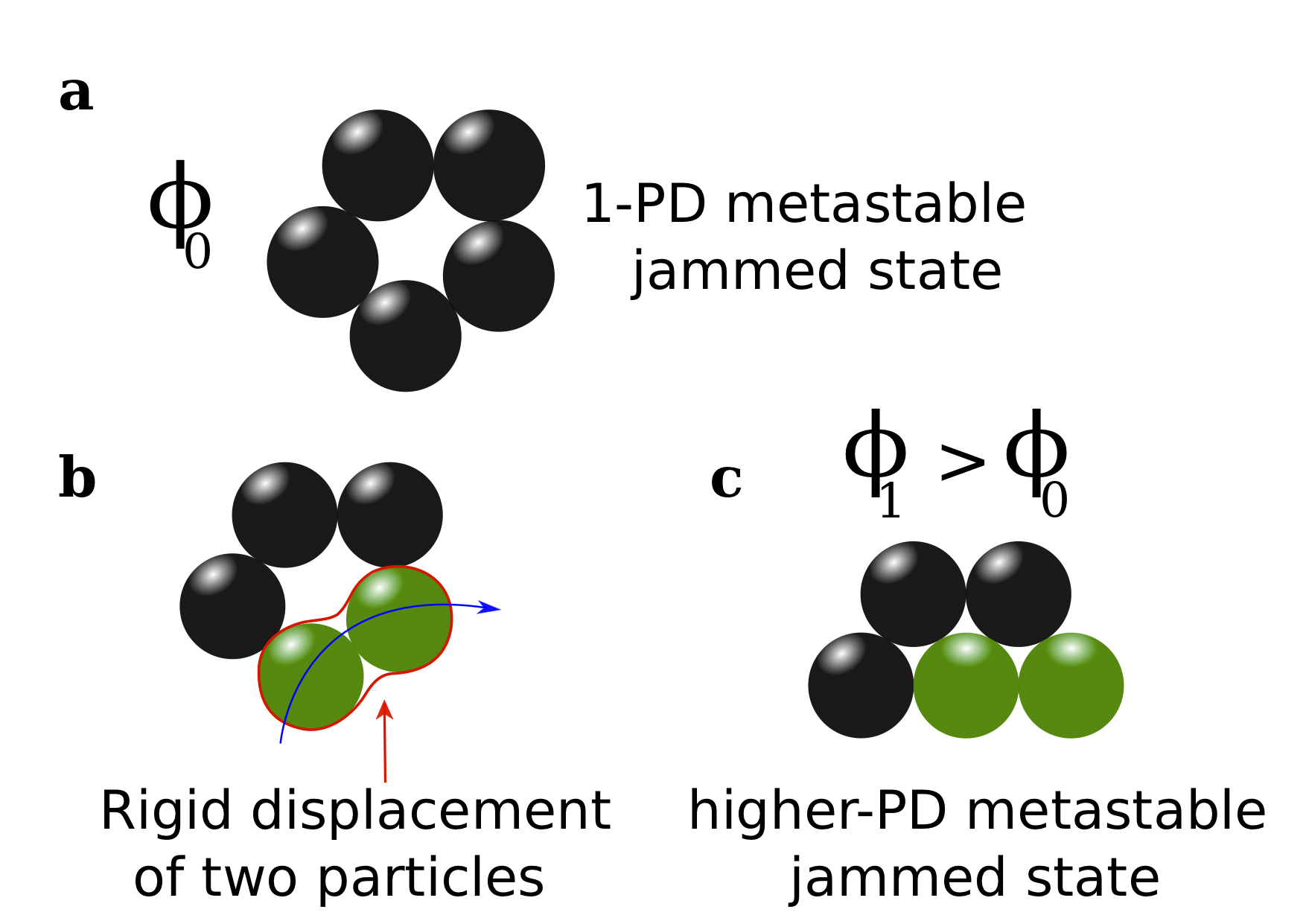}
\caption{ (a) Example of a 1-Particle-Displacement jammed state: no
  particle can increase the volume fraction by displacing itself while
  keeping the others fixed in their positions. It is assumed that a
  membrane is keeping the particles in place or that they are
  surrounded and kept in place by a rigid container. (b) The 1-PD metastable state in (a) is not stable under 2PD. Simultaneous displacement of two particles: to escape the 1-PD metastable trap,
  two contacting particles are displaced while keeping the others
  fixed in their positions. (c) Higher order metastable jammed state:
  after the move in (b), a new metastable jammed state is reached
  having higher stability than the original one in (a).}
\label{fig:1-PD}
\end{center}
\end{figure}

To define properly the metastable jammed states we need to specify
with respect to what type of displacements they are metastable. More
precisely, if we start from an initially jammed state satisfying
Eqs.~(\ref{eq:hardcore})--(\ref{Ch2:third})
and then displace a set of particles, how do we decide if the
initial state is stable under this move? A helpful discriminant is
the volume $V$ or equivalently the volume fraction of the packing
$\phi$ defined as the ratio of the volume occupied by the particles to
the total volume of the system and the number of particles involved in the displacement. Thus, consider an initially jammed
state, and assume you can displace only one particle at a time. If the
volume fraction of the packing is not increasing whatever particle you
move, then we may assert that the packing is stable against any single
particle displacement. We call this type of jammed state a
$1$-Particle-Displacement ($1$-PD) metastable jammed state, which is
defined as a configuration whose volume fraction cannot be increased
by the displacement of one single particle, see Fig.~\ref{fig:1-PD}a. However, $\phi$ may be increased by moving a set of two or more particles at the same time. The definition of $1$-PD metastable jammed states is the same as the definition of {\it local} jamming in \cite{Torquato:2001aa}, stating that in a locally jammed configuration no single particle can be displaced while keeping the positions of all other particles fixed.

We can now extend this definition to jammed states which are stable
with respect to the simultaneous displacement of multiple
particles. Specifically, we define a $k$-Particle-Displacement ($k$-PD)
metastable jammed state as a configuration whose volume fraction
cannot be increased by the simultaneous displacement of any contacting
subset of $1,2,\dots, k$ particles. Again, we find this definition
quite similar to the definition of the {\it collective} jamming category
in \cite{Torquato:2001aa}, which states that in collectively jammed
configurations no subset of particles can be simultaneously displaced
so that its members move out of contact with one another and with the
remaining set. Following the definitions given above a {\it ground
state} of the system is a configuration whose volume fraction cannot be
increased by the simultaneous displacement of any finite number of
particles. A ground state of jamming corresponds to the $k\to \infty$
limit of a $k$-PD metastable jammed state, the $\infty$-PD jammed
ground state.

\begin{figure*}
\begin{center}
{\bf (a)}
\includegraphics[height=5.8cm]{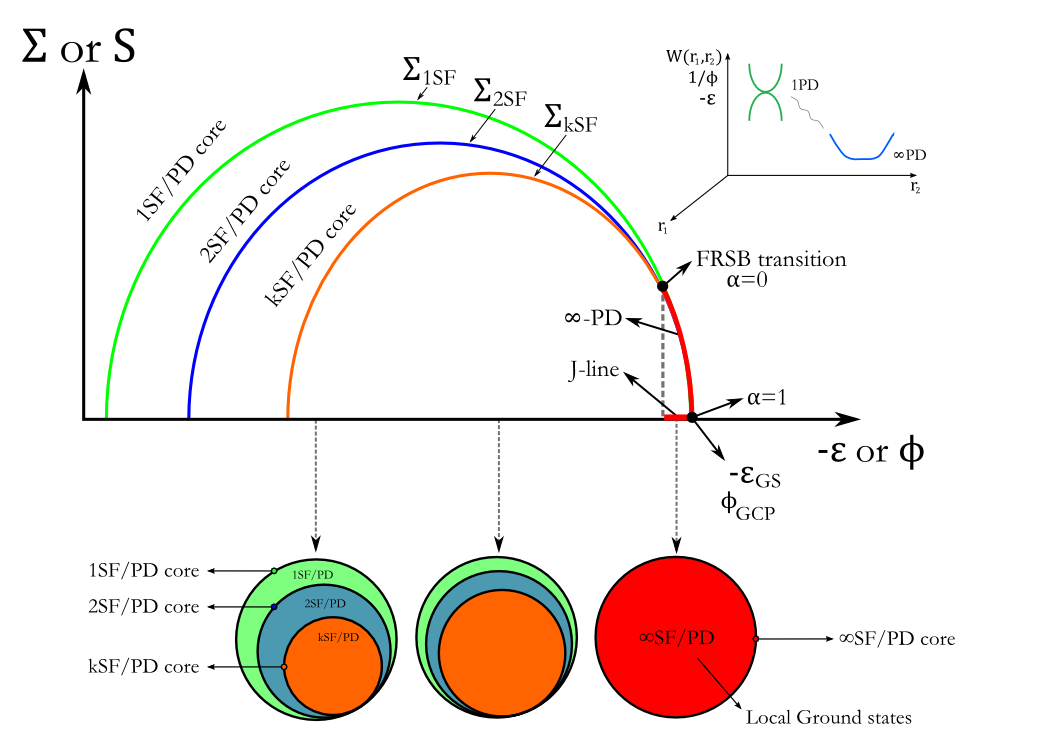}
\hspace{0.5cm} {\bf (b)}
\includegraphics[height=5.8cm]{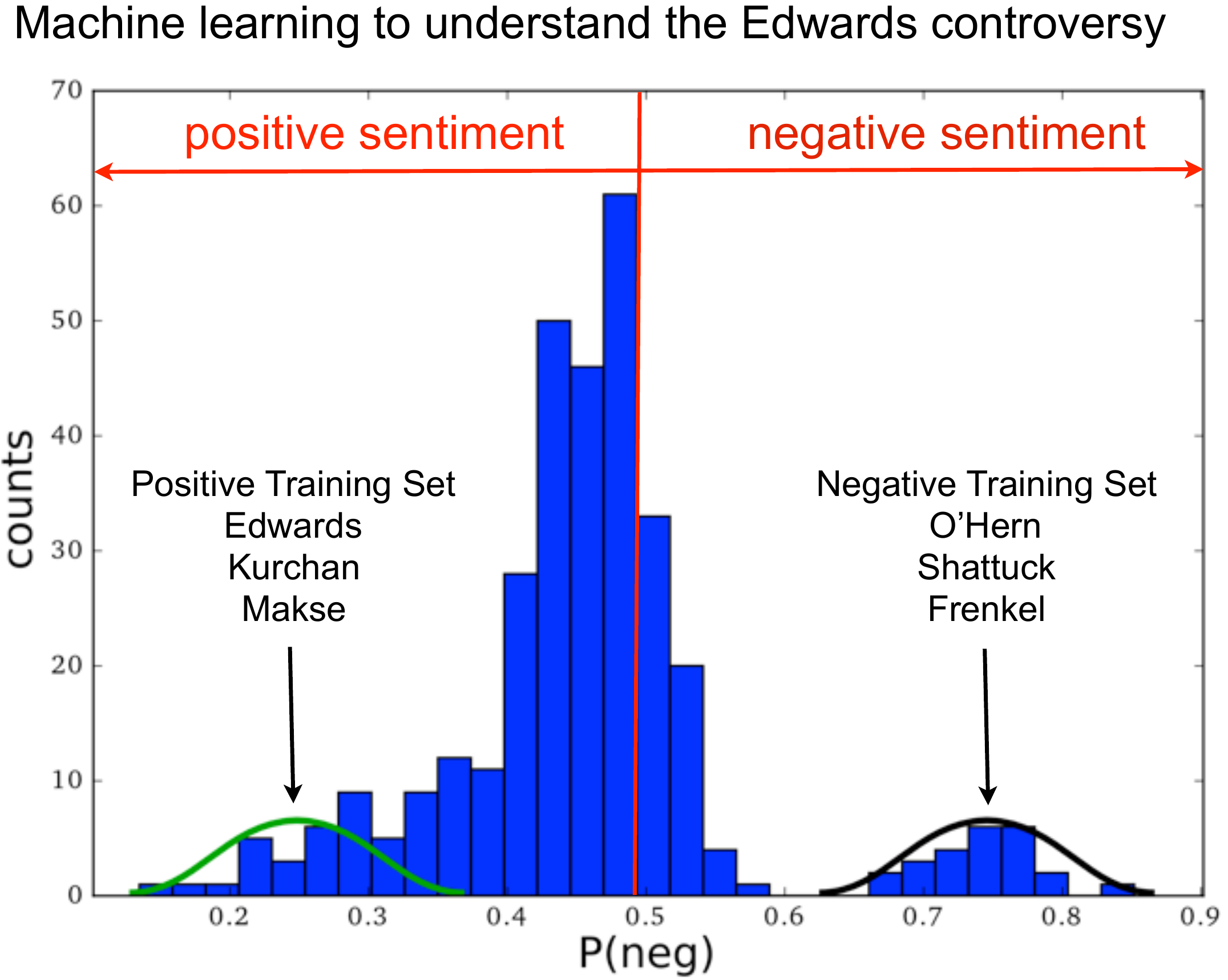}
\caption{\label{fig:PD2} (Colors online) (a) Unification between
  Edwards statistical mechanics of jammed matter and the mean field
  picture of spin glasses. Main panel: Edwards entropy $S(\phi)$
  versus the volume fraction, $\phi$, of the metastable states $k$-PD
  and ground state $\infty$-PD in the jamming model as well as the
  analogous complexity $\Sigma(\epsilon)$ versus the negative energy
  density, $-\epsilon$, in the spin glass models in terms of the
  equivalent $k$-SF metastable and $\infty$-SF ground state. The
  J-line corresponds to the ground states between $[\phi_{\rm
      th},\phi_{\rm GCP}]$ and are $\infty$-PD states with only
  positive eigenvalues for the Hessian. This line is obtained by
  changing $\alpha=k/N$ between $[0, 1]$ and $k \to \infty$ and $N\to
  \infty$ below the full replica symmetry breaking transition as
  indicated.  Top right panel: Schematic representation of the
  metastable states 1-PD and ground state $\infty$-PD of jamming in
  the volume landscape (the analogous $1$-SF metastable states and the
  $\infty$-SF ground state in the energy landscape of spin glasses is
  a function of the spin configuration $\sigma$ instead of $\bf{r}$).
  Lower panel: Organization of the $k$-PD metastable states into a
  hierarchy of successively nested $k$-PD cores ($k$-cores).  (b)
  Machine learning classifier applied to the abstracts of 581 papers
  citing the original Edwards paper~\cite{Edwards:1989aa} to classify
  the sentiment of the citing authors regarding the validity of the
  Edwards ergodic hypothesis. We construct two training sets of papers
  of the authors indicated in the figure based on the positive
  sentiment (showing results in agreement with Edwards and assigned
  $P($neg$)=0$) and negative sentiment (in disagreement with Edwards,
  assigned $P($neg$)=1$).  The training sets are then re-classified as
  indicated. In the negative training set we do not include the recent
  paper of the Frenkel group showing the validity of Edwards ensemble
  at the jamming transition \cite{Martiniani:2017aa} (we thank the
  authors for allowing us to use their papers as training sets).  We
  use the same machine learning methods from \cite{Bovet:2016aa} used
  to predict presidential elections from Twitter activity, see also
  \cite{Bohannon:2017aa}, \url{http://bit.ly/2nSjHuI}. While the
  majority of authors are neutral, the classifier also identifies two
  polarized groups; the positive sentiment being the largest one going
  gradually from neutral to extreme positive. A gap at $P($neg$) =
  0.6$ separates this group from the negative one. Recent work
  \cite{Martiniani:2017aa} seems to justify this ``wisdom of crowds''
  effect, see Section \ref{Sec_Edwardstest}.}
\end{center}
\end{figure*}

In the following section we will introduce the volume function
$\mathcal{W}({\bf r})$ to parametrize the system volume as a function
of the particles' positions.  It is useful then to classify the $k$-PD
metastable jammed states in terms of the minima of this function.
More precisely, we identify the $k$-PD metastable jammed states as
those states that satisfy the geometrical and mechanical constraints
and are local minima of $\mathcal{W}({\bf r})$. For example, $1$-PD
metastable states are those configurations ${\bf r}^*$ for which
$\mathcal{W}({\bf r})$ is convex around ${\bf r}^*$ under
1-Particle-Displacements, but non-convex under
$k$-Particle-Displacements with $k>1$, see Fig.~\ref{fig:PD2}a. Here,
convex means that all the eigenvalues of the Hessian of
$\mathcal{W}({\bf r})$ evaluated at the configurations ${\bf r}^*$ are
positive, while non-convex means that there exists at least one
negative eigenvalue in the spectrum of the Hessian.  Similarly, $k$-PD
metastable states are those configurations ${\bf r}^*$ for which
$\mathcal{W}({\bf r})$ is convex around ${\bf r}^*$ under any
$k'$-Particle-Displacements with $k'\le k$, and non-convex under any
$k'$-Particle-Displacements with $k'>k$.  A simple example of a $1$-PD
metastable jammed state is shown in Fig.~\ref{fig:1-PD}a.

Interestingly, in spin-glass systems the (energetically) metastable
states can be defined in a similar way, not with respect to volume but
with respect to energy. The analog of the $1$-PD metastable jammed
state is, for a spin glass, the $1$-spin-flip ($1$-SF) metastable
state, defined as a configuration whose energy cannot be lowered by
the flip of any single spin. Similarly the $k$-spin-flip ($k$-SF)
metastable state, akin to the $k$-PD metastable jammed state, is a
configuration whose energy cannot be lowered by the flip of any
cluster of $1,2,\dots, k$ spins. Moreover, for spin glasses, several
rigorous results on metastable states are known, including their
probabilities, basins of attraction, and how they are sampled by
various dynamics \cite{Newman:1999aa}. These results are explained in
detail in Section \ref{Sec:constraints} along with their granular
counterpart. The analogy between grains, hard-sphere glasses, and spin
glasses has been reviewed in \cite{Dauchot:2007aa} and is
described in Table~\ref{Ch2:table1} and Fig.~\ref{fig:PD2}a.

Protocols to generate jammed packings usually lead to a non-zero fraction of particles ($2-5\%$), which remain mobile even though all other particles are $\infty$-PD jammed. These particles are called {\it rattlers} and can be displaced within a cage without changing the volume function.

Now that we have a rigorous definition for the jammed states and their
metastable classification, we address the crucial problem of how to
describe their statistical mechanics. Consider a granular material
undergoing vertical tapping. After tapping, the system relaxes into a
jammed state. Subsequent tapping will allow the system to explore
other jammed states. An important question arises: how does the tapping
dynamics sample the jammed states, or what is the probability measure for
jammed states obtained from tapping? 

\begin{table*}[ht]
\centering
\begin{tabular}{| c | c | c | c |}
\hline 
 & Granular matter & Hard-Sphere Glasses & Spin-Glasses\\ 
\hline 
\hline 
& & &  \\
Thermodynamic descriptor\ & Volume function & Density functional & Hamiltonian\\
& $\mathcal{W}(\mathbf{q})$ & $\mathcal{S}[\rho({\bf r})]$ & $\mathcal{H}(\boldsymbol{\s})$\\
& & & \\
\hline
& & &  \\
Lagrange multiplier \ & Compactivity $X$ & Pressure $P$  & Temperature $T$ \\
& & & \\
\hline
& & &  \\
Entropy \ & Edwards entropy $S(V) $ & Configurational entropy $\Sigma$  & Complexity $\Sigma$ \\
& & & \\
\hline
\hline
& & &  \\
Metastable states\ &  Minima of $\mathcal{W}(\mathbf{q})$ &  Minima of $\mathcal{S}[\rho({\bf r})]$ &  Minima of $\mathcal{H}(\boldsymbol{\s})$ \\
& & & \\
 &  $+$ jamming constraint &   & at $T=0$  \\
& & &  \\
\hline
& & &  \\
Local metastable\ &  $1$-Particle-Displacement  &   &  $1$-Spin-Flip ($T=0$) \\
& & &  \\
Collective metastable\ &  $k$-Particle-Displacement  &   &  $k$-Spin-Flip ($T=0$)\\
& & &  \\
Global metastable\ &  $\infty$-Particle-Displacement  &  $\phi\in[\phi_{\rm th},\phi_{\rm GCP})$ &  $\infty$-Spin-Flip ($T=0$)\\
& $0\le \alpha<1$ & & $0\le \alpha<1$ \\
& & & \\
Ground state & $\infty$-Particle-Displacement & $\phi_{\rm GCP}$ & $\infty$-Spin-Flip ($T=0$)\\
& $\alpha=1$ & & $\alpha=1$\\
& & &  \\
\hline
\end{tabular}
\caption{\label{Ch2:table1}Synoptic view of unifying framework to
  understand the thermodynamics, relevant observables and
  classification of metastable states in granular matter, hard-sphere
  glasses and spin-glasses. The four categories of jamming are defined according
to their metastability: local metastable ($1$-PD/SF
stable); collective metastable ($k$-PD/SF stable with finite $1<k<\infty$);
globally metastable ($\infty$-PD/SF stable, but with $0\le \alpha<1$, where $\alpha=k/N$ for $k,N\to\infty$); and the
true global ground state ($\infty$-PD/SF stable and $\alpha=1$).}
\end{table*}

\subsection{Edwards statistical ensemble for granular matter}

\label{Ch2:Sec:edwardsens}

In 1989 Edwards made the remarkable proposal that the macroscopic
properties of static granular matter can be calculated as ensemble
averages over equiprobable jammed microstates controlled by the system
volume \cite{Edwards:1989aa}. The thermodynamics of powders was
created with this claim \cite{Edwards:1994aa}:

{\it ``We assume that
  when $N$ grains occupy a volume $V$ they do so in such a way that
  all configurations are equally weighted. We assume this; it is the
  analog of the ergodic hypothesis of conventional thermal physics}."

This idea is very suggestive because it turns a complicated dynamical
problem into a relatively simpler equilibrium problem. Such an
equilibrium sampling in a non-equilibrium system has been recently
also adopted by several authors in the glass community to study the
ground state of amorphous packings as the infinite pressure limit of
metastable glassy states described by equilibrium statistical
mechanics \cite{Parisi:2010aa,Charbonneau:2017aa}. Here, in the
so-called Monasson construction \cite{Monasson:1995aa}, a modified
equilibrium average over metastable states is taken, supplemented by
the additional assumption that such metastable states become jammed
states in the infinite-pressure limit. In such a limit, the states are
sampled flatly and the Monasson construction is exactly the Edwards
ensemble. Even more, it turns out that mean-field glass models
relaxing at zero temperature have exactly Edwards ergodicity property
\cite{Kurchan:2001aa}: at long times any nonequilibrium observable is
correctly given by the typical value it takes over all local energy
minima of the appropriate energy density. The original idea put
forward by Edwards is basically to take the flat average at the end,
i.e., in the jammed state.

Under the Edwards ergodic hypothesis, granular matter should be
amenable to an equilibrium statistical mechanical treatment, where the
role of energy is played by the volume, and all the jammed states at a
fixed volume are equally probable. In granular assemblies consisting
of dry particles in a size range above a few microns, the thermal
energy at room temperature can be neglected and neither equilibrium
entropy nor free energy can be used as thermodynamic potentials to
describe the system. Nevertheless, for large enough particle numbers,
statistical ideas seem relevant: Macroscopic observables such as the
packing fraction are robustly reproduced for a given protocol. If
operations manipulating individual particles are neglected, granular
assemblies are thus described by well defined macrostates that
correspond to many different microscopic configurations. Instead of
the energy, one can equivalently take the volume as the key variable
characterizing the macrostate of a static assembly. S.~F. Edwards
insight has suggested to consider the volume of a granular assembly
analogous to the energy of an equilibrium system: Unlike in typical
equilibrium systems, the volume is not an externally fixed parameter,
but depends on the microscopic configuration of the particles
including positions and orientations. This suggests to introduce a
volume function $\mathcal{W}(\{\mathbf{r}_i,\mathbf{\hat{t}}_i\})$
giving the system volume as a function of the particles' positions
$\mathbf{r}_i$ and orientations $\mathbf{\hat{t}}_i$ equivalent to the
Hamiltonian $\mathcal{H}(\{\mathbf{p}_i,\mathbf{r}_i\})$, $i=1,...,N$.

With this analogy, all concepts of equilibrium statistical mechanics
can be carried over into the realm of non-thermal static granular
systems opening the door for the use of thermal concepts for athermal
systems, i.e., there is a whole new statistical mechanics emerging
from the point which, in conventional, thermal, statistical mechanics corresponds to $T = 0$, $S = 0$ \cite{Edwards:2008aa}. For an in-depth treatment of
equilibrium statistical mechanics we refer to standard textbooks
\cite{Landau:1980aa,Pathria:2011aa,Huang:1987aa}. In particular, one
can introduce the concept of a granular entropy $S(V)$ as a measure of
the number of microstates $\Omega(V)$ for a given fixed volume $V$
\be \label{Ch2:grentropy}S(V)&=&\lambda \log
\Omega(V),\\ \Omega(V)&=&\int\D\mathbf{q}\,\delta(V-\mathcal{W}(\mathbf{q}))\Theta_{\rm
  jam}.
\label{Ch2:grMicrostates}
\ee Here, we use the shorthand notation
$\mathbf{q}=\{\mathbf{r}_i,\mathbf{\hat{t}}_i\}$ and $\int
\D\mathbf{q}=\prod_{i=1}^N\int\D\mathbf{r}_i\oint\D\mathbf{\hat{t}}_i$.
The parameter $\lambda$ ensures the correct dimension of $S$ as volume
(set to unity in the following).

The function $\Theta_{\rm jam}$ in Eq.~(\ref{Ch2:grMicrostates}) is
crucial. It is there to admit only microstates in the ensemble that
are jammed by enforcing the excluded volume and mechanical stability constraints in Eqs.~(\ref{eq:hardcore})--(\ref{Ch2:third}). Only these rigid states lead to a static assembly at fixed volume. While
this function has been treated lightly in earlier studies of Edwards
thermodynamics, it contains most of the interesting physics of the
problem and therefore will be treated carefully in the remaining of
this review. More precisely, $\Theta_{\rm jam}$ admits only the solutions of the Jamming Satisfaction Problem (JSP), which reads for monodisperse hard-spheres:
\be
\begin{aligned}
\Theta_{\rm jam} = &\ \ \prod_{i,j=1}^N\theta\Big(
|\mathbf{r}_i-\mathbf{r}_j|-2R\Big)
\hspace{.8cm} {\rm hard-core\ (spherical)}\\
&\times\prod_{i=1}^N \delta\left(\sum_{a\in\partial i}\mathbf{f}_a^i\right) 
\hspace{1.9cm} {\rm force\ balance}\\
&\times\prod_{i=1}^N \delta\left(\sum_{a\in\partial i}\mathbf{d}_a^i\times\mathbf{f}_a^i\right)
\hspace{1.05cm} {\rm torque\ balance}\\
&\times\prod_{i=1}^N\prod_{a\in\partial i}\theta\left(\mu f^i_{a,n} - |\mathbf{f}^i_{a,\tau}|\right)
\hspace{.4cm} {\rm Coulomb\ friction}\\
&\times\prod_{i=1}^N\prod_{a\in\partial i}\theta\left( -\mathbf{d}_a^i\cdot\mathbf{f}_a^i \right)
\hspace{1.2cm} {\rm repulsive\ forces}\\
&\times\prod_{\rm{all\ contacts\ {\it a}}}\delta(\mathbf{f}_a^i+\mathbf{f}_a^j)
\hspace{1.3cm} {\rm Newton\ 3^{rd}\ law}\ . 
\label{eq:thetajam}
\end{aligned}
\ee 

Implicit in this microcanonical description is again the underlying
assumption of equiprobability: The distribution of jammed
configurations $\mathbf{q}$ at a given volume is uniform: 
\be P_{\rm
  mic}(\mathbf{q})&=&\Omega(V)^{-1}\delta(V-\mathcal{W}(\mathbf{q}))\Theta_{\rm
  jam}.  \ee
The definition of $\Theta_{\rm jam}$ deserves a crucial clarification.
According to the classification of metastable jammed states given
previously, when constructing the volume ensemble we have to specify
what type of metastable jammed states we are considering at the fixed
volume V.  The crucial point is that $k$-PD jammed states are
fundamentally different for different values of $k$, and hence there
is no reason, in principle, to assign them the same statistical weight
across all the values of $k$. In other words, when we fix the volume
$V$, we consider as equiprobable only the jammed state corresponding
to the same metastable class, i.e., with the same $k$. This is
evident in the language of jammed categories: a locally jammed state
(=$1$-PD) is substantially different from a collectively jammed state
(=$k$-PD), and it cannot be claimed, {\it a priori}, that they are found with equal
probability in a tapping experiment, even if they may have the same
density. An identical situation applies to metastable states in
spin-glasses and disordered ferromagnets where the equiprobability of
the metastable states has been rigorously studied
\cite{Newman:1999aa}.

This clarification is very important, and indeed it is at the origin
of many headaches when trying to prove or disprove Edwards conjecture. In the absence of a first principle derivation of Edwards statistical mechanics, there has been a long standing controversy on
it validity, as illustrated in Fig.~\ref{fig:PD2}b. Even if this condition did not appear in the
original formulation by Edwards, it is nevertheless a quite obvious
requirement, especially in light of analogous exact results in
spin-glasses and hard-sphere glasses
\cite{Newman:1999aa,Parisi:2010aa}.  The reason to not make explicit
this further condition was presumably the feeling of Edwards that the
jammed states that only matter in granular media are the ones
corresponding to $k=\infty$, i.e. the ``ground states" (see however
\cite{Edwards:2004aa} for a more detailed discussion). Here, we extend
Edwards idea also to jammed states with $k<\infty$.  Summing it up,
the correct reading of the assumption about the probability measure
over jammed states must take into account the restriction to the
states within the same $k$-PD class, a condition that must be included
in the definition of $\Theta_{\rm jam}$ as an additional
constraint. In practice this can be done after having defined the
volume function of the system, which provides an unambiguous
definition of mechanically metastable states via its convexity, much
in the same way as for spin-glasses, the Hamiltonian allows one to
properly define the energetically metastable states, i.e. its local
minima \cite{Newman:1999aa}. This topic will be discussed in detail in
Section~\ref{Sec:constraints}.

\bigskip

In principle the Edwards conjecture can be correct or not, and a
case-by-case analysis is required to establish its validity. In granular systems, Liouville's theorem for the conservation of phase
space volume under time evolution (the cornerstone of conventional
equilibrium statistical mechanics) does not hold, leading to nonzero phase space
compressibility. The reason is the strongly dissipative nature of
granular assemblies, which are dominated by static frictional forces;
although an intuitive proof for the use of $\mathcal{W}$ in granular
thermodynamics has been sketched by the analogous proof of the
Boltzmann equation (H-theorem) \cite{Edwards:2004aa}.

In this ensemble, statistical averages of observables are assumed to
be equal to time averages over single trajectories, provided the
actual dynamics is ergodic. This can be induced by external
drive, such as infinitesimally small tapping or very slow
shearing. Since the drive induces fluctuations of the packing
configuration, and thus fluctuations of the volume, one can similarly
introduce a canonical picture (without change in particle number). The
analogue of temperature is called compactivity $X$, whose inverse is
the derivative of the granular entropy \be X^{-1}=\frac{\partial
  S(V)}{\partial V}.  \ee For a real granular system, the compactivity
can be thought of as a measure of how more compact the system can
possibly be. Large values of $X$ indicate a loose or ``fluffy" (but
mechanically stable) configuration, whose volume could be reduced
further under rearrangement.

The canonical distribution follows from the maximization of the Gibbs
entropy just as in thermal equilibrium under the constraint of a fixed
average volume \be V=\int\D\mathbf{q}\,\mathcal{W}(\mathbf{q})\,P_{\rm
  can}(\mathbf{q})\ , \ee and has the standard Gibbs form and
canonical partition function:\be
\label{Ch2:vcan}
P_{\rm
  can}(\mathbf{q})&=&\frac{1}{\mathcal{Z}}e^{-\mathcal{W}(\mathbf{q})/X}\Theta_{\rm jam},\\ \mathcal{Z}&=&\int\D\mathbf{q}\,e^{-\mathcal{W}(\mathbf{q})/X}\Theta_{\rm jam}.
  \label{Ch2:pfcan}
  \ee

If we follow the analogy with equilibrium thermodynamics, the concepts
of granular entropy and compactivity translate into postulated laws of
a granular thermodynamics \cite{Edwards:2004aa}:

\textit{Zeroth law.} A consistent picture of compactivity as a
temperature-like parameter requires the notion of equilibration: Two
systems in physical contact should equilibrate to the same
compactivity. The required ``volume'' transfer is achieved by the
external drive, but needs to avoid any mixing of the particles.

\textit{First law.} The analogy with granular matter is not clear as a
distinction between heat and work is not useful for jammed granular
materials.

\textit{Second law.} In any natural process, the granular entropy
always increases. The second law forms the basis of Edwards
statistical mechanics.
 
\textit{Third law.} Our qualitative discussion of compactivity
suggests that entropy should thus be a monotonically increasing
function of $X$: Loose packings at high $X$ can be realized in many
more configurations than dense packings at low $X$. In the limit $X\to
0$ we can thus postulate that $S(V)\to {\rm const}$. The limiting entropy will be finite for any disordered arrangement, while $S(V ) = 0$ is only achieved for a fully ordered non-degenerate crystal structure.

\bigskip

Up to now we have considered only the volume $V$ as the relevant
variable to characterize the jammed state of a granular
system. However, this is not the general case. Indeed, when the system is shaken the grains will fill a volume $V$
and exert a stress $\hat{\Sigma}$ on the boundary. Shaking after
shaking, the system explores presumably typical configurations in the
configuration phase space, which are subject to the constraint on $V$
and also on $\hat{\Sigma}$. Consequently, the entropy of the system
$S(V,\hat{\Sigma})$ must then be computed as a function of those
observables, which in the microcanonical ensemble can be defined as \be
S(V,\hat{\Sigma})&=&\log \int\D
\mathbf{q}\,\delta(V-\mathcal{W}(\mathbf{q}))\delta(V\hat{\Sigma}-\hat{\Phi}(\mathbf{q}))\Theta_{\rm jam}\,\nonumber\\
\label{Ch2:volstress}
\ee
where 
\be
\label{Ch2:hatsigmai}
\hat{\sigma}_i=\sum_{a\in\partial i}\mathbf{d}_a^i\otimes
\mathbf{f}_a^i
\ee
is the stress tensor associated with particle $i$ and the sum
\be
\label{Ch2:stressmoment}
\hat{\Phi}=\sum_{i=1}^N\hat{\sigma}_i=\sum_{i=1}^N\sum_{a\in\partial i}\mathbf{d}_a^i\otimes
\mathbf{f}_a^i 
\ee 
is the macroscopic force-moment tensor.

In analogy to the volume ensemble, there should thus exist a
temperature-like Lagrange multiplier associated with the stress. Since
$\hat{\Sigma}$ is a tensor, this quantity is also a tensor, which can
be defined as \be \hat{\Lambda}_{ij}=V\frac{\partial
  \hat{\Sigma}_{ij}}{\partial S}.  \ee The tensor $\hat{\Lambda}$ is referred
to as angoricity from the Greek word {\it ankhos} for stress
\cite{Blumenfeld:2009aa}.

A simplification occurs if the stress $\hat{\Sigma}$ is a simple
hydrostatic pressure $\hat{\Sigma}=p$. In this case the angoricity
degenerates to the scalar quantity $\Lambda=V\partial p/\partial S$.

Considerable progress in a theoretical description of granular matter
could be achieved from pure volume and stress/force ensembles, which
appear as limits of the full description
Eq.~(\ref{Ch2:volstress}). We discuss these in detail in the
following. On the other hand, it has been suggested that volume and stress
ensembles are necessarily interdependent, which would require more
sophisticated approaches to deal with their correlations
\cite{Pugnaloni:2010aa,Blumenfeld:2012aa}.

\subsection{Volume ensemble}

Pure volume ensembles neglect the force degrees of freedom. This
is reasonable, e.g., in isostatic systems, where all forces are
uniquely determined from the configurational degrees of freedom. In
this case, the statistical volume ensemble is fully specified by the
volume function Eq.~(\ref{Ch2:vcan}), which relies on a suitable space
tessellation.

\subsubsection{Conventions for space tessellation}

\label{Ch2:Sec:space}

In the case of a Hamiltonian
there is a unique way to define the energy as a function of the
particle configurations, typically in terms of a superposition of all
particles' individual kinetic and potential energy plus the energy
contribution due to interactions. Such a decomposition is not
straightforward in the case of the volume function. Nevertheless, it
is natural to express $\mathcal{W}$ in the form of a
superposition \be
\label{Ch2:volsum}
\mathcal{W}(\mathbf{q})=\sum_{i=1}^N\mathcal{W}_i(\mathbf{q}) \ee of
non-overlapping volume elements that tesselate the space occupied by
the packing. $\mathcal{W}_i$ is the volume associated with each of the
$N$ particles. Crucially, this volume is not a function of the
configuration of the $i$th particle only. Naively, one could imagine
that $\mathcal{W}_i$ depends solely on the configurations of particles
in the first coordination shell. However, such a restriction is mathematically not sufficient and does not apply in general, e.g., in
the Voronoi tesselation. The collective nature of the
systems' response to perturbations induces dependencies on particles
further away. Moreover, even if one considers only particles in the first
coordination shell as a first approximation, a precise definition of
$\mathcal{W}_i$ is not straightforward. The key problem is to
reference individual particles, so that their neighbours can be
defined. While this is easily achieved in a regular crystalline
packing, the difficulties originating from a disordered contact
network have been realized early on \cite{Edwards:1989aa,Mounfield:1994aa}. Below we review the different
definitions of $\mathcal{W}_i$ in historical order.

\paragraph{Tensorial formulation}

A first solution to the problem of defining $\mathcal{W}(\mathbf{q})$
was proposed in \cite{Edwards:2001aa}. Introducing the tensor
\cite{Edwards:1999aa,Edwards:1999ab} $\hat{F}_i=\sum_{j\in\partial
  i}\mathbf{r}_{ij}\otimes\mathbf{r}_{ij}$, where $\mathbf{r}_{ij}$
is the separation vector of particles $i$ and $j$, we can define the
volume associated with particle $i$ as $\mathcal{W}_i=2\sqrt{\det \hat{F}_i}$, which involves only contacting particles. The resulting total volume
$\mathcal{W}=\sum_{i=1}^N\mathcal{W}_i$ is thus only an approximation
of the exact volume occupied by all $N$ particles. Formal corrections
that allow for an exact definition of $\mathcal{W}$ have been
suggested, but the quantities specifying correlations of tensors
belonging to nearest neighbours are intractable for any practical
purposes \cite{Edwards:2001aa}.

\paragraph{Quadrons}

In 2d, a definition of $\mathcal{W}_i$, such that
Eq.~(\ref{Ch2:volsum}) is exact can be obtained by analysing planar packings
in terms of loops and voids \cite{Ball:2002aa,Blumenfeld:2003aa}, leading to area-tesselating quadrilateral
elements referred to as quadrons. In 2d one can show that the number of quadrons is identical to the
number of configurational degrees of freedoms
\cite{Blumenfeld:2003aa,Blumenfeld:2006aa}, motivating the use of the
quadrons as the elementary ``particles" of the system on which the
statistical mechanics is based. In 3d this coincidence is
no longer valid \cite{Blumenfeld:2006aa}, thus limiting the applicability of 
the quadrons to realistic systems. Even in 2d it has been noted that the exact tesselation is only valid in the absence of non-convex voids, which are actually present in a gravitational field \cite{Ciamarra:2007aa}.

\paragraph{Delaunay tessellation}

For a set of points specifying, e.g., the centres of spheres in a
packing, elementary Delaunay cells are simplexes with vertices at the
centres of neighbouring particles. In 2d the simplexes are triangles
defined such that no other point lies inside the circumcircle of a
given triangle. In 3d the simplexes are likewise tetrahedra defined
such that no other point lies inside the circumsphere of a given
tetrahedron. In both cases a space filling set of cells is obtained,
which, however, is not uniquely associated with a given set of
particles. Thus, it is not possible to cast this tesselation into the
form of Eq.~(\ref{Ch2:volsum}), reducing its applicability to realistic
systems. The Delaunay tessellation has been used to analyse the
volume statistics of disordered sphere packings
\cite{Finney:1970aa,Hiwatari:1984aa,Aste:2005ab,Aste:2006aa,Aste:2007aa,Klumov:2014aa}, and is the
cornerstone in Hales' proof of the Kepler conjecture.

\begin{figure}
\begin{center}
(a)\\
\includegraphics[width=0.8\columnwidth]{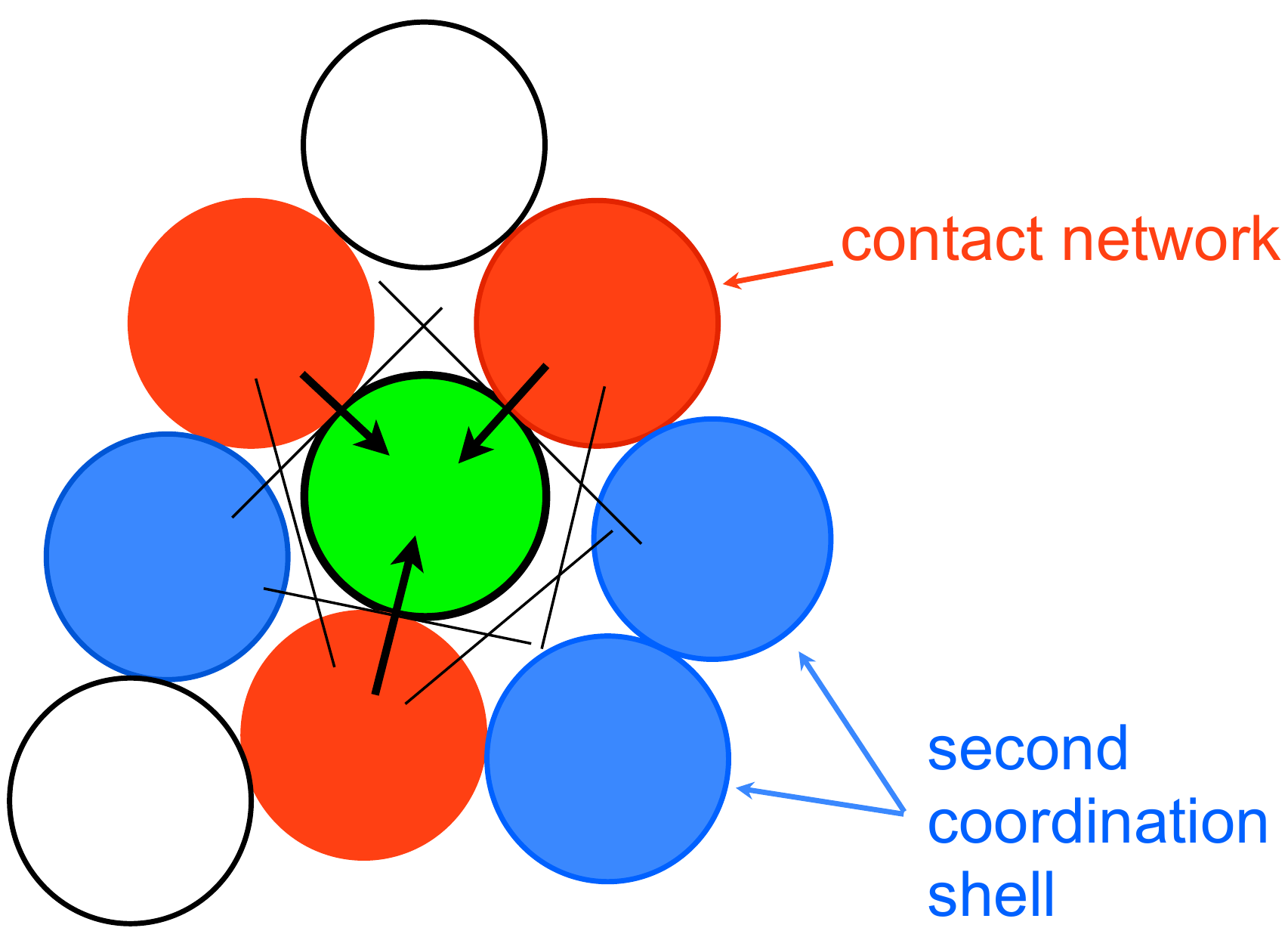}\\
\vspace{0.5cm}
(b)\\
\includegraphics[width=0.8\columnwidth]{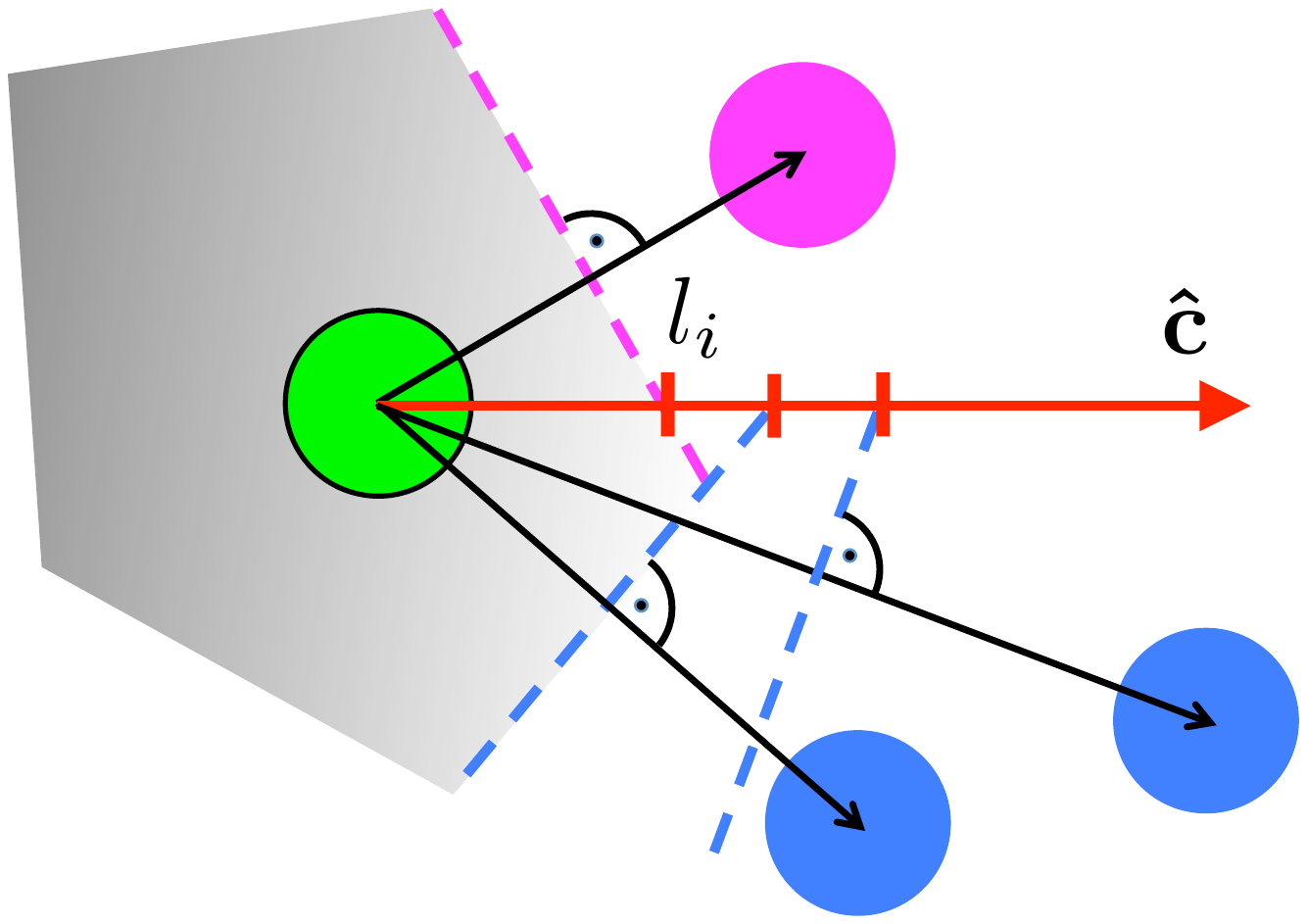}
\caption{\label{Fig:Ch2:voronoi} (Colors online) Illustration of the
  Voronoi tessellation in a packing of monodisperse disks. (a) In this
  case the Voronoi boundary (VB) between two particles is the plane
  perpendicular to the separation vector at half the distance (see
  Eq.~(\ref{Ch2:vbsphere})). The VBs of the reference particle (green)
  with the particles in the first and second coordination shell are
  indicated with thin black lines. (b) The volume of a Voronoi cell
  associated with a given particle is defined as the amount of space
  that is closer to the surface of that particle than to the surface
  of any other particle. The cell boundary
  $l_i(\mathbf{q},\mathbf{\hat{c}})$ in a given direction
  $\mathbf{\hat{c}}$ for a configuration $\mathbf{q}$ thus follows
  from the global minimization Eq.~(\ref{Ch2:globmin}) and the cell
  volume from the orientational integral
  Eq.~(\ref{Ch2:voronoidef}). In the figure the contributed VBs of all
  particles along $\mathbf{\hat{c}}$ are indicated. The pink particle
  contributes the smallest VB, which thus defines the boundary of the
  Voronoi cell (indicated in grey). We also refer to this particle as
  ``Voronoi particle" along the direction $\mathbf{\hat{c}}$.}
\end{center}
\end{figure}

\paragraph{Voronoi tessellation}

A straightforward way to tessellate the volume of a packing is to
associate that amount of space with particle $i$ that is closer to it
than to any other particle (Fig.~\ref{Fig:Ch2:voronoi}), thus making
full use of the form Eq.~(\ref{Ch2:volsum}). This defines the Voronoi
tesselation, first introduced by the Ukrainian mathematician
G. F. Voronoi in 1908, which is now widely used in mathematics and
many applied areas \cite{Aurenhammer:1991aa,Okabe:2000aa}. In the case of spheres or points, the
Voronoi tessellation is dual to the Delaunay decomposition: the
centres of the circumspheres are just the vertices of the Voronoi
graph. 

Before we define the volume $\mathcal{W}_i$, we first introduce the
Voronoi boundary (VB). The VB between two particles is defined as the
hypersurface that contains all the points that are equidistant to the
surfaces of both particles
\cite{Baule:2013aa,Portal:2013aa,Schaller:2013ab}. If we fix our
coordinate system at the centre of mass of particle $i$ (and also
assume its orientation fixed), we can parametrize the VB in terms of
the direction $\mathbf{\hat{c}}$ from particle $i$
(Fig.~\ref{Fig:Ch2:voronoi}b). A point on the VB is found at
$s\mathbf{\hat{c}}$, where $s$ depends on the relative position
$\mathbf{r}_{ij}$ and orientation $\mathbf{\hat{t}}_{ij}$ of the two
particles:
$s=s(\mathbf{r}_{ij},\mathbf{\hat{t}}_{ij};\mathbf{\hat{c}})$. The
value of $s$ is obtained from two conditions:
\begin{enumerate}
\item The point $s\mathbf{\hat{c}}$ has the minimal distance to the surfaces of each of the two objects along the direction $\mathbf{\hat{c}}$.
\item Both distances are the same.
\end{enumerate}
As an example, the VB between two spheres of equal radii is the same
as the VB between two points at the centres of the spheres. Therefore,
condition $1$ is trivially satisfied for every $s$ and condition $2$
translates into the equation
$(s\mathbf{\hat{c}})^2=(s\mathbf{\hat{c}}-\mathbf{r}_{ij})^2$, leading
to \be
\label{Ch2:vbsphere}
s=\frac{r_{ij}}{2\mathbf{\hat{c}}\cdot\mathbf{\hat{r}}_{ij}}, \ee i.e., the
VB is the plane perpendicular to the separation vector
$\mathbf{r}_{ij}$ at half the separation (see Fig.~\ref{Fig:Ch2:voronoi}a). Already for two spheres of
unequal radii, the VB is a curved surface. Taking into account the
different radii $R_i$ and $R_j$, the second condition becomes
$s-R_i=\sqrt{(s\mathbf{\hat{c}}-\mathbf{r}_{ij})^2}-R_j$, which has
the solution \cite{Danisch:2010aa}: \be
\label{Ch2:VBsphere}
s=\frac{1}{2}\frac{r_{ij}^2-(R_i-R_j)^2}{\mathbf{\hat{c}}\cdot\mathbf{\hat{r}}_{ij}-(R_i-R_j)}.
\ee Finding a solution for both conditions 1. and 2. for general
non-spherical objects is non-trivial \cite{Baule:2013aa,Portal:2013aa}
and will be discussed in Sec.~\ref{Ch4:Sec:vbnonsp}.

Having defined the VB, the exact mathematical formula for $\mathcal{W}_i(\mathbf{q})$ in $d$ dimensions is given by the orientational integral: \be
\label{Ch2:voronoidef}
\mathcal{W}_i(\mathbf{q})=\frac{1}{d}\oint\D \mathbf{\hat{c}}
\,l_i(\mathbf{q},\mathbf{\hat{c}})^d, \ee
where $l_i(\mathbf{q},\mathbf{\hat{c}})$ is the boundary of the Voronoi cell
in the direction $\mathbf{\hat{c}}$. This boundary depends on all $N$
particle configurations $\mathbf{q}$ in terms of a global
minimization: $l_i(\mathbf{q},\mathbf{\hat{c}})$ is the minimum among
all VBs in the direction $\mathbf{\hat{c}}$ between particle $i$ and
all other $N-1$ particles in the packing (see Fig.~\ref{Fig:Ch2:voronoi}b). Formally, \be
\label{Ch2:globmin}
l_i(\mathbf{q},\mathbf{\hat{c}})=
\min_{j:s>0}\;s(\mathbf{r}_{ij},\mathbf{\hat{t}}_{ij},\mathbf{\hat{c}}).
\ee Clearly, the global minimization over all particles $j$ defining
$\mathcal{W}_i$ in Eq.~(\ref{Ch2:globmin}) is highly difficult to
treat analytically. The Voronoi volume of a particle depends on the
position of all the other particles in the packing; clearly, a
many-body interaction. The precise knowledge of the microscopic
configurations of all particles is intractable in the thermodynamic
limit. Nevertheless, the Voronoi convention has been shown to be the
most useful way of defining the volume function, since it is well defined for any dimension and captures the effect of different particle shapes. The technical
challenges can be circumvented by: {\it (i)} decomposing non-spherical shapes into overlapping and
intersecting spheres leading to analytically tractable expressions for
the VB; {\it (ii)} coarse-graining the volume function over a mesoscopic
length-scale, which avoids the global minimization problem.

This approach
\cite{Song:2008aa,Baule:2013aa} turns the volume ensemble into a
predictive framework for packings, as discussed in detail in
Sec.~\ref{Sec:volume}. Interestingly, the Voronoi cell of a particle can be interpreted as
its available volume in the packing. This correspondence can be
demonstrated by considering a soft interparticle potential and
evaluating the free volume for a given potential energy before taking
the hard core limit \cite{Song:2010aa}. Analyzing the statistics of
the Voronoi cells also provides deeper insight into structural
features of packings, e.g., by quantifying the cell shape anisotropies
\cite{Medvedev:1987aa,Luchnikov:1999aa,Schroeder-Turk:2010aa,Schaller:2015aa}.

\subsubsection{Statistical mechanics of planar assemblies using quadrons}

The quadron convention of the volume function $\mathcal{W}$ has been
used in \cite{Blumenfeld:2003aa} to calculate the partition function
of the volume ensemble explicitly. If correlations
  between particle positions are neglected, analytical results can be
  obtained by introducing suitable approximations for $\Theta_{\rm
    jam}$. The partition function is then analytically tractable and
  leads likewise to predictions for the average quadron volume and
  fluctuations \cite{Blumenfeld:2003aa}. The quadron approach also
  allows to assess the effect of correlations. The lowest order
  correlations originate from intergranular loops and can thus be
  considered as background fluctuations. In the case of circular
  particles with three neighbours one finds that taking into account
  correlations only due to the intergranular loops reduces the packing
  density at high compactivity, but increases it at low
  compactivity. In addition, the difference in density due to
  correlations is shown to be relatively small at around 2--4\%, which
  suggests that correlation-free models might be sufficiently accurate
  to capture many packing properties \cite{Blumenfeld:2003aa}.

\subsubsection{$\Gamma$-distribution of volume cells}

The analysis of the statistics of volume cells in
  sphere packings reveals an interesting universality irrespective of
  packing protocols and volume conventions. In
  \cite{Aste:2006aa,Aste:2007aa} experimental packings of
  $\sim145,000$ spherical glass beads were prepared with fluidized bed
  techniques and structural features investigated with X-ray
  tomography. The PDFs of cell volumes in the Delaunay convention for
  18 different experiments show a surprising collapse onto a unique
  master curve. The master curve is the $\Gamma$-distribution $f(V,k)=\frac{(V-V_{\rm
     min})^{k-1}}{\Gamma(k)\chi^k}e^{-(V-V_{\rm min})/\chi}$, with
 shape parameter $k$ and scale parameter $\chi=(\left<V\right>-V_{\rm
   min})/k$. Such a $\Gamma$-distribution has been shown to capture
 well the volume statistics in a large variety of jammed systems
 \cite{Aste:2007aa,Aste:2008aa,Aste:2008ab,Frenkel:2008aa,Matsushima:2014aa,Lechenault:2006aa,Oquendo:2016aa}. Its
 possible universality has been motivated by statistical mechanical
 arguments applied to independent elementary volume cells
 \cite{Aste:2007aa,Aste:2008aa} assuming that the cells are
 uncorrelated. Even though the data collapse on a
 $\Gamma$-distribution is remarkable, it is not clear if it is indeed
 a signature of a jammed state. A Poisson point process, e.g., leads
 likewise to a distribution of Voronoi cell volumes that is well
 described by a $\Gamma$-distribution
 \cite{Kumar:1992aa,Ferenc:2007aa,Lazar:2013aa}.

\subsection{Stress and force ensemble}

\subsubsection{Force tilings}

\begin{figure}
\includegraphics[width=8cm]{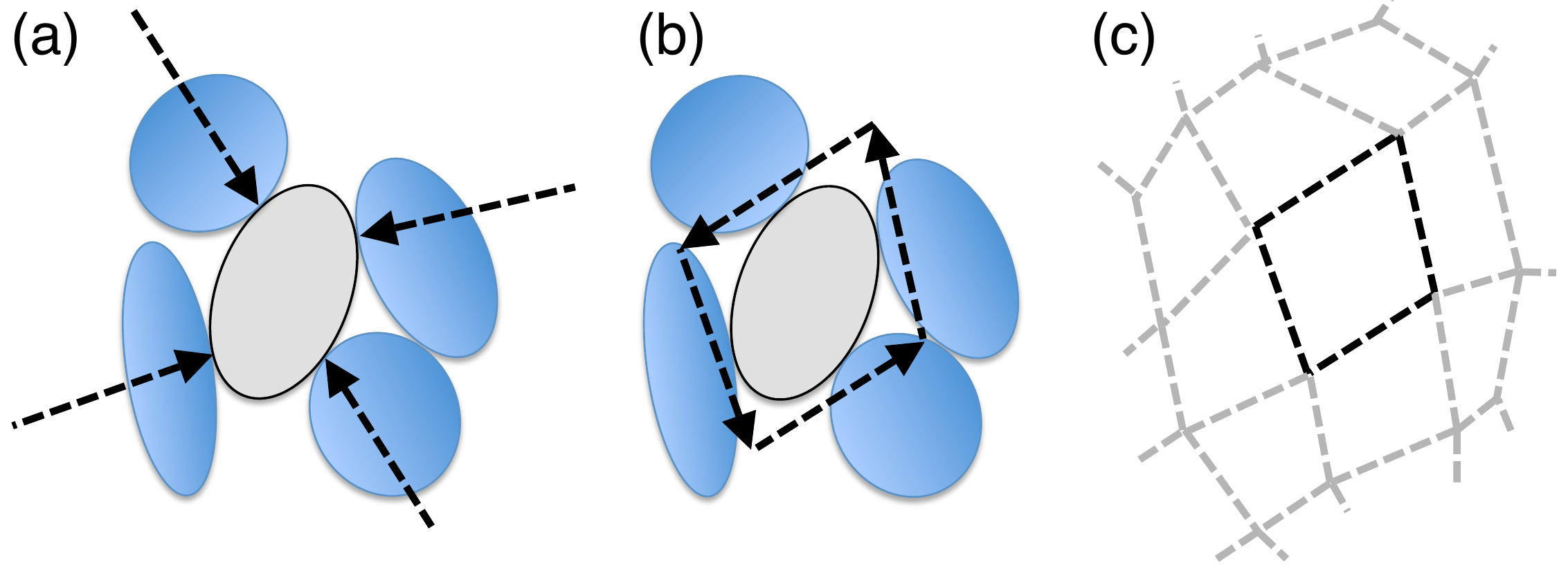}
\caption{\label{Fig:forcetiles}. Illustration of a Maxwell-Cremona tessellation. (a,b) Rotating the contact force vectors by $\pi/2$ and joining them tip to end leads to a tile that can be associated with an individual particle. (b,c) Due to force balance every tile is closed and the collection of tiles tesselates the plane. }
\end{figure}

It has already been noted in the mid 19th century that the contact
forces in a 2d packing can be mapped to a tessellation of the plane,
the so called Maxwell-Cremona tessellation
\cite{Maxwell:1864aa,Cremona:1890aa}. An individual tile in the tessellation arises
from the contact forces acting on a particle $i$: the boundary of the
tile is constructed by rotating all force vectors by $\pi/2$ and
joining them tip to end leading to a polygon (see
Fig.~\ref{Fig:forcetiles}a,b). If the forces on the particle all
balance the polygon is closed, because its boundary is the sum of all
contact forces. Moreover, due to Newton's third law the tiles of
contacting particles always have a side of equal length and
orientation, which, for a $N$ particle packing satisfying force
balance leads to a tessellation of the plane without any gaps
(Fig.~\ref{Fig:forcetiles}c). Note that the
condition of torque balance is not required to construct the
tiles. The Maxwell-Cremona tessellation underlies the mapping of
contact forces to auxiliary forces such as the void forces
\cite{Satake:1993aa}, loop forces \cite{Ball:2002aa}, and height
fields \cite{Henkes:2005aa} (see Sec.~\ref{Ch2:Sec:parastress}).

An important observation is that any rearrangement of forces changes
the area of individual tiles $A_i$, but leaves the overall area of the
tessellation invariant if force balance is maintained and boundary
forces are unchanged. This means that the total area is an invariant
under these force rearrangements
\cite{Tighe:2008aa,Tighe:2010ab,Tighe:2011aa} \be
\label{Ch2:area}
\sum_{i=1}^NA_i={\rm const},
\ee
where the sum runs over all tiles in the tessellation. Another manifestation is the conservation of the stress-moment tensor \cite{Ball:2002aa,Henkes:2005aa,Henkes:2007aa}. Eq.~(\ref{Ch2:area}) only holds for frictionless grains. In frictional systems, the force tiles are non-convex and self-intersecting polygons, which makes the tiling graph non planar and the individual tile areas do not sum up to the overall area \cite{Bi:2015aa}.

Maximum entropy methods in the spirit of E.~D. Jaynes information
theoretic approach to statistical mechanics \cite{Jaynes:1957aa,Jaynes:1957ab} have been applied to the
problem of force statistics in a number of works
\cite{Bagi:1997aa,Bagi:2003aa,Kruyt:2002aa,Rothenburg:2009aa,Goddard:2004aa,Radeke:2004aa,Ngan:2003aa,Ngan:2004aa,Metzger:2004aa,Metzger:2005aa}.

\subsubsection{Force network ensemble}

The force network ensemble (FNE)
\cite{Snoeijer:2004ab,Tighe:2008aa,Tighe:2010aa,Tighe:2010ab,Tighe:2011aa}
motivated by work of Bouchaud \cite{Bouchaud:2002aa} is based on a
separation of scales relevant for the particle configurations and
forces. In quantitative terms, one can introduce the parameter \be
\epsilon=\frac{\left<f_{ij}\right>}{\left<r_{ij}\right>}\left<\frac{\D
  f_{ij}}{\D r_{ij}}\right>^{-1}, \ee where $\left<...\right>$ denotes
an average over all particles in the packing and we introduce the
notation $f_{ij}$ for the normal force component $f_a^i$ of contact
$a$ on particle $i$ with particle $j$. For $\epsilon\ll 1$ variations
of the forces of order $\left<f\right>$ only result in vanishing
changes in the particle positions $\mathbf{r}_{ij}$. If the forces are
underdetermined, i.e., not uniquely fixed by the force and torque
balance equations, the forces are thus uncoupled from the
configurational degrees of freedom. The FNE considers a fixed contact
network (a fixed set of $\{\mathbf{r}_{ij}\}$) and constructs an
ensemble of contact forces $\{f_{ij}\}$ with the following properties:
{\it (i)} The forces are a priori uniformly distributed as in the
Edwards ensemble; {\it (ii)} Force and torque balance equations are
imposed as constraints; {\it (iii)} Forces are repulsive $\forall
f_{ij}\ge 0$ and satisfy the Coulomb condition
Eq.~(\ref{Ch2:coulomb}); {\it (iv)} A fixed external pressure
$\mathcal{P}$ sets an overall force scale. For a small number of
spheres the resulting force distribution can be derived exactly
\cite{Snoeijer:2004ab}. For larger packings maximum entropy arguments
can be used \cite{Tighe:2008aa,Tighe:2010ab,Tighe:2011aa}. The
underlying assumptions imply that the FNE is in principle applicable
to frictional hyperstatic systems, but is mathematically well defined
also for frictionless particles.

For an isostatic system at jamming the force network ensemble is not needed, since the contact geometry uniquely defines the contact forces \cite{Lerner:2013aa,Charbonneau:2015aa,Gendelman:2016aa}. In this case, an approximation of $P(f)$ can be calculated with the cavity method assuming a locally tree-like contact geometry corresponding to an assumption of replica symmetry (RS) \cite{Bo:2014aa}. We note that a correct determination of $P(f)$ requires one to take into account subtle correlations between particle positions that exist at
jamming beyond RS, and that are neglected in \cite{Bo:2014aa}, which in the end fails to
account for the non-trivial power-laws of $P(f)$ at jamming. A similar
situation appears in the approximative calculation at 1RSB using replicas, which also fails to predict the correct exponents \cite{Parisi:2010aa}. As discussed in Sec.~\ref{Ch5:Sec:cavity}, the correct
calculation needs to be performed at the full-RSB level since the jamming
line is deep in the Gardner phase of the model.

\subsubsection{Stress ensemble}
\label{Ch2:Sec:parastress}

A statistical ensemble based on the stress-moment tensor is
conveniently constructed by introducing auxiliary force variables
based on the voids surrounded by contacting particles in 2d
\cite{Ball:2002aa,Henkes:2005aa}. If we choose the centre of an
arbitrary void as the origin of a height field, we can construct the
height vectors $\mathbf{h}_\nu$ iteratively as \cite{Henkes:2005aa}
\be \mathbf{h}_\nu=\mathbf{f}^i_{a}+\mathbf{h}_\mu.  \ee Here, $\mu$,
$\nu$ label voids and $\mathbf{f}^i_{a}$ is the force vector at the contact that is crossed when going
from the centre of void $\mu$ to the centre of void $\nu$. Since the contact forces on a particle sum to zero due to force balance, the height vectors are well defined and represent a one-to-one mapping of the contact forces. The microscopic stress tensor of a single grain, Eq.~(\ref{Ch2:hatsigmai}),
$\hat{\sigma}_i$ can then be expressed in terms of the height fields \cite{Ball:2002aa} \be
\hat{\sigma}_i=\sum_{a\in\partial i}(\mathbf{r}_{a1}+\mathbf{r}_{a2})\otimes\mathbf{h}_\mu,
\ee where $\mathbf{r}_{a1}$ and $\mathbf{r}_{a2}$ denote the vectors connecting void $a$ with the contact points. The macroscopic force-moment tensor Eq.~(\ref{Ch2:stressmoment}) of a
macroscopic assembly of $N$ particles occupying area $A$ in the
quadron convention is thus \be
\hat{\Phi}=\sum_{i=1}^N\hat{\sigma}_i=\sum_{\mu\in \partial
  A}(\mathbf{r}_{\mu1}+\mathbf{r}_{\mu2})\otimes\mathbf{h}_\mu.  \ee
The sum in the last expressions runs only over all voids defining the
boundary of the area $A$, since all contributions from particles in
the bulk cancel. We see that $\hat{\Phi}$ is conserved under
rearrangement of the contact forces in the bulk that preserve force
balance, which is a manifestation of the area conservation
Eq.~(\ref{Ch2:area}). Therefore, packings with different values of
$\hat{\Phi}$ can not be transformed into each other by rearranging the
bulk forces. This allows us to define a granular entropy $S=\log
\Omega(A,\hat{\Phi},N)$ via the number of force configurations
$\Omega(A,\hat{\Phi},N)$ leading to a given $\hat{\Phi}$.

In order to obtain the canonical distribution, we divide the system
into a small partition of size $m$ and the remaining system $N-m$,
which acts as a reservoir. For frictionless isotropic systems the only independent part of
$\hat{\Phi}$ is the trace $\Gamma={\rm tr}\, \hat{\Phi}$, which
represents a simple hydrostatic pressure $p=\Gamma/A$. In this case,
the formalism simplifies and the canonical distribution is
\cite{Henkes:2007aa,Henkes:2009aa} \be
\label{Ch2:stresscan2}
P(\Gamma_m)=\frac{\Omega_m(\Gamma_m)}{\mathcal{Z}(\alpha)}e^{-\alpha
  \Gamma_m},\qquad \Gamma_m=\sum_{i,j}d_{ij}F_{ij}, \ee where $\alpha=\log \Omega_N(\Gamma)/\partial
\Gamma$ and the sum is taken over all contact vectors and forces in the $m$-particle
cluster.

Eq.~(\ref{Ch2:stresscan2}) leads to the following testable
predictions:
\begin{itemize}
\item All subregions in an equilibrated packing $k$ should have the
  same granular temperature $\alpha_k$. Thus measuring $P(\Gamma_m)$
  in two packings $k$ and $k'$ yields the ratio \cite{Henkes:2007aa}
  \be
\label{Ch2:scanpred1}
\log\left[\frac{P_k(\Gamma_m)P_{k'}(\Gamma'_m)}{P_k(\Gamma'_m)P_{k'}(\Gamma_m)}\right]=(\alpha_k-\alpha_{k'})(\Gamma_m-\Gamma'_m).
\ee
Moreover, the distribution $P_k(\Gamma_m)$ satisfies the scaling \cite{Henkes:2007aa}
\be
\label{Ch2:scanpred2}
P_k(\Gamma_m)=P_{k'}(\Gamma_m)e^{-(\alpha_k-\alpha_{k'})\Gamma_m}.  \ee
Eqs.~(\ref{Ch2:scanpred1},\ref{Ch2:scanpred1}) require that packings
$k$ and $k'$ are sufficiently close in density to neglect changes in
$\Omega$ due to different volumes.

\item At the isostatic point the partition sum $Z(\alpha)$ can be
  evaluated analytically by summing over all force degrees of freedom
  assuming a uniform distribution. In a monodisperse system of
  spheres, this yields the predictions \cite{Henkes:2009aa}:
  $\Omega(\Gamma_m)=\Gamma^{2m}$ for $m\gg 1$ and \be
\label{Ch2:scanpred3}
\alpha=\frac{Nz_{\rm iso}}{2\left<\Gamma\right>}, \ee where
$\left<\Gamma\right>=-\partial \log Z/\partial \alpha$. We also obtain
the exponential force distribution \be P(F)\propto e^{-\alpha r_0 F},
\ee where $r_0$ is the sphere radius.

\end{itemize}

Simulations of soft sphere systems have confirmed predictions
Eqs.~(\ref{Ch2:scanpred1},\ref{Ch2:scanpred2}) for different packing
densities \cite{Henkes:2007aa}. Eq.~(\ref{Ch2:scanpred3}) has also
been shown close to the $J$-point, but deviations are observed for
larger densities, where instead the relation
$\alpha=Na\left<z\right>/\Gamma_N$ is observed. Here, $a$ increases
monotonically from $a=2$ for $\left<z\right>>z_{\rm iso}$
\cite{Henkes:2009aa}.

\section{Phenomenology of jammed states and scrutinization of the Edwards ensemble}
\label{Sec:tests}

In this section we first describe the phenomenological results
characterizing the jammed states and then proceed to review work
dedicated to test the Edwards assumption of equiprobability of jammed
states.

\subsection{Jamming in soft and hard sphere systems}

\label{Ch2:Sec:softspheres}

Over the past two decades, considerable progress has been made in our understanding of jammed particles packings.  Here we summarize the main results of this work needed for the remainder of this review.  One can refer to several recent review articles for more details. 
\cite{Hecke:2010aa,Liu:2010aa,Bi:2015aa,Torquato:2010aa,Charbonneau:2017aa}

\subsubsection{Isostaticity in jammed packings}
\label{Sec:isostatic}

The average coordination number in packings is approximately estimated by
naive Maxwell counting arguments
\cite{Maxwell:1870aa,Alexander:1998aa} which consider the force
variables constrained only by force and torque balance
Eqs.~(\ref{Ch2:forceb},\ref{Ch2:torqueb}) and Newton's third law
Eq.~(\ref{Ch2:third}), but ignore the crucial constraints of Coulomb,
Eqs.~(\ref{Ch2:coulomb}), and repulsive forces,
Eq.~(\ref{Ch2:repulsive}). In particular, attractive forces are
allowed, contradicting the fact that the forces are purely repulsive,
Eq.~(\ref{Ch2:repulsive}). With these caveats in mind, one obtains an estimation of the
average coordination number $z$ assuming: {\it (i)} all degrees of freedom (dofs) in the packing are constrained by contacts (for periodic boundary conditions); {\it (ii)} the number of contacts will be minimal for a generic disordered packing. As a consequence, packings of
frictionless particles should satisfy (see appendix~\ref{App:isostatic}) \be
\label{Ch2:iso}
z=2d_{\rm f}.  \ee When Eq.~(\ref{Ch2:iso}) is satisfied the packing
is \textit{isostatic} under the naive Maxwell counting argument: the
number of force and torque balance equations exactly equals the number
of contact force components. Therefore, the configurational dofs fully
determine the force dofs and vice versa, which allows to construct
ensembles based on only configurational or force dofs. Since isostatic
packings have the minimal number of contacts for a geometrically rigid
packings they are also referred to as marginally stable
\cite{Muller:2015aa}. Packings with $z$ smaller or larger than the
isostatic value are referred to as hypostatic and hyperstatic,
respectively.

Equation~(\ref{Ch2:iso})
predicts that packings of frictionless spheres have $z=6$, while
rotationally symmetric shapes such as spheroids and spherocylinders
have $z=10$ and fully asymmetric shapes have $z=12$. The isostaticity
for spheres is indeed widely observed to hold very closely in experiments and simulations
for both soft and hard sphere systems. In fact it has been shown \cite{Moukarzel:1998ab} that non-cohesive sphere packings become exactly isostatic, when their stiffness goes to infinity. However, if we consider a small
deformation from the spherical shape to, e.g., a spheroid, the
isostatic condition would predict a discontinuous jump in the average
coordination number from $z=6$ to $z=10$.
Instead, one finds that packings of non-spherical shapes are in general
hypostatic with a smooth increase from the spherical isostatic $z$
value under deformation
\cite{Williams:2003aa,Donev:2004aa,Donev:2007aa,Wouterse:2009aa,Schreck:2012aa}. These
hypostatic packings are indeed mechanically stable if the effect
of the shape curvature at the contact point is taken into account \cite{Roux:2000aa,Donev:2007aa}. As a
consequence, one can construct configurations that are mechanically
stable even though there are fewer contacts than configurational dofs
per particle (see Sec.~\ref{Ch4:Sec:degenerate}). 
Interestingly, also for larger aspect ratios the
average coordination number generally stays below the isostatic value,
which is just slightly lower for spheroids and fully asymmetric
ellipsoids \cite{Donev:2004aa}, but exhibits a much stronger decrease
for spherocylinders \cite{Williams:2003aa,Wouterse:2009aa,Zhao:2012aa,Baule:2013aa}.

For polyhedral particles with flat faces and edges the above counting
arguments need to be modified, since, e.g., two touching faces
constrain more than a single configurational dof. In
\cite{Jaoshvili:2010aa} it has been suggested to associate every
contact with the number of configurational dofs that are constrained
by it: Contact of two faces $\to$ 3 constraints; face and edge contact
$\to$ 2 constraints; face and vertex, edge and edge contacts $\to$ 1
constraint. With these correspondences the isostaticity of disordered
jammed packings of tetrahedra and other Platonic solids could indeed
be demonstrated \cite{Jaoshvili:2010aa,Jiao:2011aa,Smith:2011aa}.

For frictional particles the contact counting argument provides the range of coordination numbers $4\le z \le 6$ for spheres and $4\le
z \le 12$ for general shapes (see appendix~\ref{App:isostatic}). For spheres it is generally observed
that $z\to 6$ for a friction coefficient $\mu\to 0$ (frictionless
limit) and $z\to 4$ for $\mu\to\infty$ (infinitely rough spheres) (see
Sec.~\ref{Ch2:Sec:softspheres}). For intermediate $\mu$ sphere packings
are thus generally hyperstatic. Hyperstaticity is also found for
frictional ellipsoids \cite{Schaller:2015ab} and frictional
tetrahedra, when the different types of contact are translated into
constraints on the configurational dofs \cite{Neudecker:2013aa}.

The Coulomb condition Eq.~(\ref{Ch2:coulomb}) restricts the possible force
configurations compared with the infinitely rough limit: A stable
force configuration with a certain $z(\mu)$ is also stable for all
larger $\mu$ values. Any determined value $z(\mu)$ is thus in
principle a lower bound on the possible combinations of $z$ and $\mu$,
although it might not be possible to generate these combinations in
practice. This highlights that $z(\mu)$ is not unique and
depends strongly on the history of the packing generation. It should
be stressed that the above isostatic conjectures are valid only under
the naive Maxwell counting argument ignoring the repulsive nature of
the interactions and the inequalities derived from Coulomb
conditions. A model generalizing Maxwell arguments to this more
realistic scenario was proposed in \cite{Bo:2014aa} suggesting the existence of a well defined lower bound on $z(\mu)$ (see Sec.~\ref{Ch5:Sec:cavity}).

\subsubsection{Packing of soft spheres}

\label{Sec:softsphere}

So far we have treated only hard spheres. A packing of
soft spheres with radius $R$ is modelled by repulsive
normal forces: \cite{Johnson:1985aa,Landau:1959aa}:
\begin{equation} f_{a,n}^i = k_n \xi ^{\alpha}, 
\label{fn}
\end{equation}
where the normal overlap is $\xi= (1/2)[2R -
  |\mathbf{r}_1-\mathbf{r}_2|]>0$, and $\mathbf{r}_{1,2}$ are the
positions of the grain centres. The normal force acts only in
compression, $f_{a,n}^i = 0$ when $\xi<0$. The effective stiffness
$k_n=\frac{8}{3} \mu_g R^{1/2} / (1-\pi_g)$ is defined in terms of the
shear modulus of the grains $\mu_g$ and the Poisson ratio $\pi_g$ of
the material from which the grains are made (typically $\mu_g=29$ GPa
and $\pi_g = 0.2$, for spherical glass beads).
The exponent $\alpha$ is typically chosen among two possibilities:
{\it (i)} $\alpha=1$ for simple harmonic springs, and {\it (ii)}
$\alpha=3/2$ for 3d spherical geometries at the contact
(Hertz forces). 

The situation in the presence of a tangential force, $\mathbf{f}_{a,\tau}^i$, is more
complicated. In the case of spheres under oblique loading, the
tangential contact force was calculated by Mindlin \cite{Mindlin:1949aa}. For
the special case where the partial increments do not involve microslip
at the contact surface (i.e., $|\Delta f_{a,\tau}^i| < \mu \Delta f_{a,n}^i$, where
$\mu$ is the static friction coefficient between the spheres,
typically $\mu=0.3$) Mindlin \cite{Mindlin:1949aa} showed that the 
incremental tangential force is
\begin{equation}
 \Delta f_{a,\tau}^i= k_t \xi^{1/2} \Delta s,
\label{ft} \end{equation} where $k_t = 8 \mu_g R^{1/2} / (2-\pi_g)$, and
the variable $s$ is defined such that the relative shear
displacement between the two grain centers is $2s$. This is
called the Mindlin ``no-slip'' solution.

Typical packing preparation protocols employ Molecular Dynamics compressing an initially loose gas
\cite{Makse:2004aa,Makse:2000aa,Makse:1999aa}. In 2d it is necessary
to use bidisperse mixtures in order to avoid crystallization. Other
protocols start from a random configuration corresponding to a large
``temperature'' $T=\infty$ initial state. Jammed packings at $T=0$
are generated by bringing the system to the closest energy minimum
using conjugate-gradient techniques to minimize the energy of the
system, which is well defined for frictionless systems
\cite{OHern:2002aa}. Another protocol for numerically constructing jammed states consists in putting particles at random positions above the packing at a certain height and letting particles settle under gravity \cite{Herrmann:1993ab}. Also sophisticated experimental realizations of this procedure have been developed \cite{Pouliquen:1997aa}.

In the $T=0$ limit or the mechanical equilibrium state assemblies of
these particles exhibit a transition to the jammed state. There exists in
particular a critical packing density $\phi_c$ characterizing the
onset of jamming at which the static shear moduli $G_\infty$ and the
pressure $p$ (and therefore, the static bulk modulus as well) become
zero simultaneously (under decompression) and the coordination number attains the isostatic value \cite{Makse:1999aa}. For
finite $N$ the precise value of $\phi_c$ depends on the initial $T$ state
and the protocol employed, but scaling behavior of $G_\infty$ and $p$
for each of the different $\alpha$ values is observed when using the
distance to jamming $\phi-\phi_c$ as a control parameter for packings near isostaticity. The critical
density $\phi_c$ in the $T=0$ limit and zero shear stress is referred
to as \textit{J-point} \cite{OHern:2002aa}.  For quenches starting at infinite
temperature, in the thermodynamic limit $N\to\infty$ the distribution
of $\phi_c$ values converges to a delta function at a value
$\phi^*=0.639\pm0.001$ for frictionless monodisperse spheres in
3d. The J-point thus obtained is close to values typically found for random close packings (RCP) of
hard spheres.

The following power-law scalings have been observed by many studies
and are independent of polydispersity or dimensionality \cite{Hecke:2010aa,Liu:2010aa,Makse:1999aa,Makse:2000aa,Makse:2004aa,OHern:2002aa,OHern:2003aa,Zhang:2005aa,Majmudar:2007aa}:
\begin{itemize}
\item Pressure:
\be
p\sim (\phi-\phi_c)^{\alpha}
\ee
\item Static bulk modulus:
\be
\label{Ch2:bulkmod}
B_\infty\sim (\phi-\phi_c)^{\alpha-1}
\ee
\item Static shear modulus:
\be
\label{Ch2:shearmod}
G_\infty\sim (\phi-\phi_c)^{\alpha-1/2}
\ee
\item Average coordination number:
\be
\label{Ch2:cscaling}
z-z_c\sim (\phi-\phi_c)^{1/2}, \ee where $z_c$, the critical
coordination number measured at $\phi_c$, agrees in fact with the
isostatic value $z=2d_{\rm f}$.
\end{itemize}
The square root scaling of $z-z_c$ is observed for all $\alpha$
values, which indicates that this scaling is only due to the packing
geometry independent of the interaction potential. The scaling of the
pressure can be interpreted as an affine response of the packing to
deformations. This argument, which is usually
referred as the Effective Medium Approximation in granular matter
\cite{Walton:1987aa,Digby:1981aa,Norris:1997aa,Makse:1999aa,Makse:2004aa,Jenkins:2005aa,Wyart:2010aa,During:2013aa,DeGiuli:2014aa,DeGiuli:2014ab,DeGiuli:2015aa},
also predicts an exponent $\alpha-1$ for the bulk modulus
Eq.~(\ref{Ch2:bulkmod}) (proportional to the second derivative of the
energy) as observed (although the scaling law has a different
prefactor as expected from affine deformations). However, the shear
modulus should then also scale with an exponent $\alpha-2$, which is
not observed in Eq.~(\ref{Ch2:shearmod}), highlighting the effects of non-affine motion under
shear \cite{Makse:1999aa,Makse:2004aa,Magnanimo:2008aa}.
The observed scaling of the shear modulus has been
reproduced in models of disordered solids by taking into account the
non-affine response within an approximate analytical scheme
\cite{Zaccone:2011aa}. Equation~(\ref{Ch2:cscaling}) has been shown to
be a bound for stability in \cite{Wyart:2005ab} based on physical
arguments and confirmed analytically in a replica calculation of the
perceptron model of jamming \cite{Franz:2015aa}. Lattice models that exhibit critical behavior related to Eqs.~(\ref{Ch2:bulkmod})--(\ref{Ch2:cscaling}) capture the jamming transition in terms of a percolation transition ($k$-core or bootstrap percolation) \cite{Toninelli:2006aa,Schwarz:2006aa}.

Anomalous behavior at point J is also indicated in the density of
normal mode frequencies
\cite{OHern:2003aa,Wyart:2005aa,Wyart:2005ab,Silbert:2005aa,Silbert:2009aa,DeGiuli:2014aa,Charbonneau:2015ab}. In a crystal the low frequency excitations are sound modes with a
vibrational density of states $\sim \omega^{d-1}$ (Debye scaling). In
a disordered packing theoretical arguments based on marginal stability predict instead \cite{DeGiuli:2014aa}
\be\label{Ch3:WyartDOS}
D(\omega) \sim
	\begin{cases}
	\omega^{d-1} & \omega \ll \omega_0 \\
	\omega^2/\omega^{*2}  & \omega_0 \ll \omega \ll \omega^* \\
	\text{constant} & \omega \gg \omega^*
	\end{cases},
\ee which is also exhibited by the perceptron model
\cite{Franz:2015aa} and found in simulations of jammed soft spheres in
dimensions 3--7 \cite{Charbonneau:2015ab,Lerner:2016aa,Mizuno:2017aa}. The $\omega^2/\omega^{*2}$ scaling has also been observed in emulsion experiments \cite{Lin:2016aa}. In
Eq.~(\ref{Ch3:WyartDOS}), $\omega^*$ is a characteristic frequency
that vanishes at jamming as \be
\label{omega}
\omega^*\sim z-z_c
\ee
and $\omega_0$ is a small threshold frequency.

At jamming the density of states thus stays non-zero for arbitrary
small frequencies. This highlights that at point J there is an excess
of low frequency modes compared with crystals. This anomaly is
sometimes seen analogous to the Boson peak observed in glassy
materials \cite{Franz:2015aa}. The vanishing crossover frequency
$\omega^*$ allows to identify a length scale $l^*$, which diverges
upon reaching point J as: $l^*\sim (z-z_c)^{-1}$
\cite{Wyart:2005aa}. Such a diverging length scale has been observed
numerically in the vibrational eigenmodes and in the response to point
perturbations
\cite{Silbert:2005aa,Ellenbroek:2006aa,Ellenbroek:2009ab}. However,
theoretical arguments predict for point responses
$l^*\sim(z-z_c)^{-1/2}$ \cite{Lerner:2014aa}. The length scale $l^*$
has been computed in \cite{Wyart:2010aa,During:2013aa}. Diverging length scales when
approaching point J from below have also been identified related to
velocity correlation functions \cite{Olsson:2007aa} and clusters of
moving particles \cite{Drocco:2005aa}. When approaching point J from
above finite point correlation functions are not sufficient to detect
such a length scale. Instead, point to set correlation functions are
necessary, which can provide a quantitative description of the
sensitivity of force propagation in granular materials to boundary
conditions \cite{Mailman:2011aa,Mailman:2012aa}.

The concept of frequency dependent complex-valued effective mass
$M_{\rm eff}(\omega)$ \cite{Hsu:2009aa} obtained as the packing is
subjected to a vertical acceleration at a given frequency is directly
related to the vibrational density of states \cite{Hu:2014ab}. Indeed,
the vibrational density of states can be accessed experimentally
through the measurement of $M_{\rm eff}(\omega)$ via a pole
decomposition of the normal modes of the system \cite{Hu:2014ab}. By
measuring the stress dependence of the effective mass, it was shown
that the scaling of the characteristic frequency $\omega^*$ deviates
from the mean field prediction Eq.~(\ref{omega}) in
real frictional packings \cite{Hu:2014ab}. Furthermore, the presence of dissipative
modes can be studied via the imaginary part of the complex
valued effective mass \cite{Hu:2014aa,Johnson:2015aa}. 

When friction is added, the observed packing densities and
coordination numbers at point J are generally smaller than RCP
\cite{Makse:2000aa,Silbert:2002ab,Kasahara:2004aa,Shundyak:2007aa,Silbert:2010aa,Papanikolaou:2013aa,Shen:2014aa}. As
a function of the friction coefficient $\mu$ the densities decrease
monotonically from $\phi\approx 0.64$ for frictionless spheres to
$\phi\approx 0.55$ in the limit of infinitely rough
spheres. Experiments find much lower packing fractions in the large
friction limit \cite{Farrell:2010aa}. The densities are also dependent
on the packing preparation for the same $\mu$ highlighting the history
dependence of frictional packings. An open question is whether there
is a well-defined lower bound on the packing density for a given
$\mu$, which could specify random loose packing (RLP) densities
\cite{Onoda:1990aa,Makse:2000aa}: the lowest density packings that are
mechanically stable.  Extremely low density mechanically stable
packings can be generated with additional attractive interactions,
e.g., due to adhesion. Adhesive packings of spheres are discussed in
Sec.~\ref{Ch4:Sec:adhesion}.

Likewise, the coordination number decreases monotonically for $\mu\ge
0$ from the isostatic frictionless value $2d_{\rm f}$, reaching the
frictional isostatic value $z_{\rm iso}^\mu=d+1$ in the limit
$\mu\to\infty$. Frictional packings are thus in general hyperstatic,
so that particle configurations do not uniquely determine the contact
forces. How this indeterminacy depends on the friction coefficient and affects the mechanical properties has been investigated in detail using contact dynamics by \cite{Unger:2005aa}. It was also found that the contacts with large indeterminacy are also those contacts that make up force chains \cite{McNamara:2004aa}.

The following scaling results at point J have been obtained in
simulations of frictional soft spheres with Hertz-Mindlin forces
\cite{Makse:2000aa,Zhang:2005aa,Shundyak:2007aa,Somfai:2007aa,Silbert:2010aa,Henkes:2010aa}. For
the coordination number one finds a scaling analogous to
Eq.~(\ref{Ch2:cscaling}) \be z-z_c\sim z_0(\mu)(\phi-\phi_c)^{1/2},
\ee where $z_c\approx 2d_{\rm f}$ is the frictionless isostatic value
at point J and $z_0(\mu)$ a weakly $\mu$-dependent prefactor. However,
other quantities like the critical frequency $\omega^*$ and the
bulk/shear modulus do not scale with $\phi-\phi_c$ contrary to the
frictionless case. One finds \be \omega^*\sim z-z_{\rm iso}^\mu,\qquad
G_\infty/B_\infty\sim z-z_{\rm iso}^\mu. \ee By comparison,
Eqs.~(\ref{Ch2:bulkmod},\ref{Ch2:shearmod},\ref{Ch2:cscaling}) predict
the scaling $G_\infty/B_\infty\sim z-z_c$. Therefore, one can conclude
that the critical observables generally scale with the distance to
isostaticity \cite{Wyart:2005ac}.

\subsubsection{Packing of hard spheres}

The structural properties of packings have been investigated in
considerable detail with computer simulations and experiments of hard
spheres satisfying constraints Eq.~(\ref{eq:hardcore}). Hard sphere
results should coincide with soft spheres at zero pressure. A widely
used simulation algorithm for jammed hard particles is the
Lubachevsky-Stillinger (LS) algorithm \cite{Lubachevsky:1990aa}. Here,
starting from a random initial configuration of spheres in a
volume with periodic boundary conditions generated, e.g., by random
sequential addition of spheres, the sphere radii are expanded
uniformly with a rate $\lambda$. Collisions occur due to the expansion
of the particles, which are resolved in an
event-driven manner. Forces can be calculated from the rate of exchange of momentum per unit time. Eventually, a jammed state is reached with
diverging collision rates at the contacts and typically 2-3\% of rattlers that remain unjammed. The
properties of the final state are then independent of the random
initial state, but depend on the expansion rate. For $\lambda\to 0$
the system is in equilibrium leading to crystallization, while for
small $\lambda>0$ the system is able to reach a quasiequilibrium
jammed state with a density $\phi(\lambda)$. These states have been
characterized as long-lived metastable glass states which in infinite
dimensions are described \cite{Parisi:2010aa} by the replica symmetry
breaking (RSB) theory adapted from the solution of the
Sherrington-Kirkpatrick (SK) model of spin-glasses
\cite{Sherrington:1975aa} (see Secs.~\ref{Ch2:Sec:rcp} and
\ref{Sec:constraints}).

An advanced numerical technique
that can deal with perfectly rigid particles and at the same time obtain the
contact forces precisely is Contact Dynamics (CD), as reviewed for
instance in \cite{Radjai:2009aa}. In fact, granular structures turn
out to be more stable under gravity when using CD than any other
numerical method \cite{McNamara:2004aa}. CD has been used extensively
to explore force networks, their fluctuations and their
indeterminacies in frictional packings, see e.g. \cite{Unger:2005aa}.

Experiments of hard sphere packings go back to the seminal work by
Bernal and Scott
\cite{Bernal:1960aa,Bernal:1960ab,Scott:1960aa,Scott:1962aa}.  Indeed,
in the old days Mason, a postgraduate student of Bernal, took on the
task of shaking glass balls in a sack and 'freezing' the resulting
configuration by pouring wax over the whole system.  He would then
carefully take the packing apart, ball by ball, noting the positions
of contacts for each particle.  Since this labor-intensive method
patented half a century ago, yet still used in recent studies
\cite{Donev:2004aa}, other groups have extracted data at the level of
the constituent particles using x-ray tomography
\cite{Richard:2003aa,Aste:2004ab,Aste:2005aa,Saadatfar:2012aa}. The most sophisticated experiment for granular matter to date has
resolved coordinates of up to 380000 spheres using X-ray tomography
\cite{Aste:2004ab,Aste:2005aa}. The packing densities achieved are in
general sensitive to the packing protocol, friction, and
polydispersity. The effect of boundary walls can be reduced by
focusing the analysis on bulk particles or preparing the walls with
randomly glued spheres. Mechanically stable disordered packings of
spheres are typically found in the range $\phi\approx 0.55$ --
$0.64$. Empirical studies have shown that one can identify different
density regions depending on variations in the protocol
\cite{Aste:2005ab}: {\it (i)} $\phi\approx 0.55$ -- $0.58$: packings
are only created by reducing the effect of gravity
\cite{Onoda:1990aa}; {\it (ii)} $\phi\approx 0.58$ -- $0.61$: packings
are unstable under tapping; {\it (iii)} $\phi\approx 0.61$ -- $0.64$:
packings are generated by tapping and compression
\cite{Knight:1995aa,Nowak:1997aa,Nowak:1998aa,Philippe:2002aa}. Packings
in the range $\phi\approx 0.64$ -- $0.74$, i.e., up to the FCC crystal
density are usually only generated by introducing local crystalline
order. This has been achieved experimentally by pouring spheres of equal size homogeneously over plate, that vibrates horizontally at a very low frequency \cite{Pouliquen:1997aa}. The attained density depends on the frequency. A similar range of densities is obtained by flux deposition of spheres into a container with a templated surface \cite{Panaitescu:2014aa}.

\begin{figure}
\begin{center}
\includegraphics[width=0.85\columnwidth]{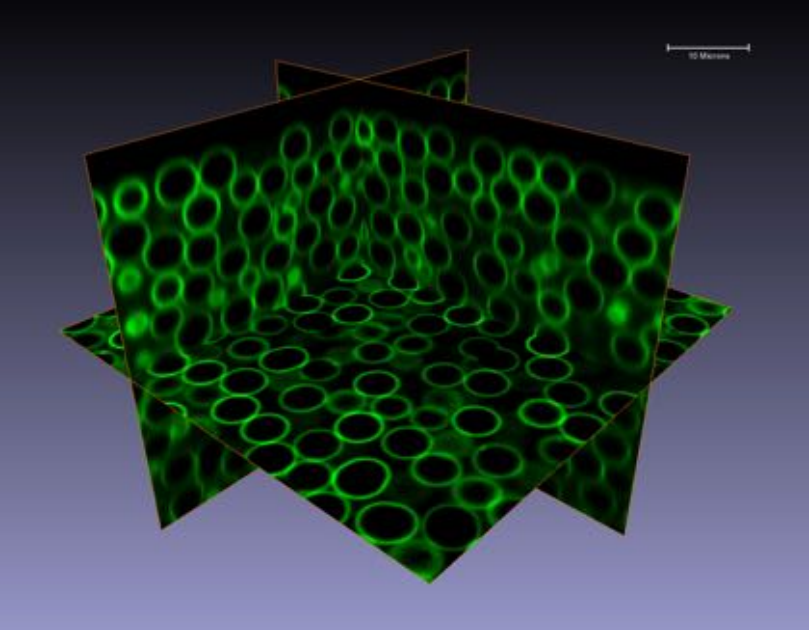}
\caption{\label{Fig:kyeyune} 3D confocal image of a colloidal packing
  showing green fluorescence on the particles' surface. The method of
  \cite{Kyeyune-Nyombi:2018aa} improves detection resolution of the
  particle contact network using fluorescent exclusion effects at the
  contact point. Structural properties of the colloidal packing near
  marginal stability that required high resolution contact detection
  thus become experimentally accessible (see
  Table~\ref{table:kyeyune}). From \cite{Kyeyune-Nyombi:2018aa}.}
\end{center}
\end{figure}

Establishing the number of contacting spheres in experiments is
somewhat challenging. The celebrated Bernal packings
\cite{Bernal:1960aa} find a coordination number close to $z=6$, while
compressed jammed emulsions near the jamming transition studied by
confocal microscopy \cite{Brujic:2007aa} finds an average coordination
$\langle z \rangle = 6.08$, close to the isostatic
conjecture. One generally finds that larger densities coincide with
larger values of $z$ exhibiting a monotonic increase over the range
$\phi\approx 0.55$ -- $0.64$ from $z\approx 4$ -- $7$
\cite{Aste:2004ab,Aste:2005aa,Aste:2005ab,Aste:2006ab} largely in
agreement with simulation results on frictional soft-sphere systems at
small pressure. A new method for contact detection in
  jammed colloids using fluorescent exclusion effects at the contact
  point has been developed in \cite{Kyeyune-Nyombi:2018aa}. The method
  improves detection resolution and allows precise determination of
  the small force distributions, coordination number, vibrational
  density of states, and pair correlations (see Fig.~\ref{Fig:kyeyune}).
  
The following consensus on the structural properties
  of the pair correlation function $g_2(r)$ of jammed hard-spheres has
  been reached from simulations and experiments for a variety of
  protocols:
\begin{itemize}
\item A delta function peak at $r=\sigma$ due to contacting particles,
  where $\sigma=2R$ is the contact radius. The area under the peak is
  the average coordination number, which has the isostatic value
  $z_{\rm iso}=2d_{\rm f}=6$ at jamming in frictionless systems.
\item A power-law divergence due to a large number of near-contacting
  particles
  \be
  \label{Ch2:gammaexp}
  g_2(r)\sim (r-\sigma)^{-\gamma}.  \ee The exponent
  $\gamma$ has been measured as $\gamma\approx 0.4$ in simulations of
  hard spheres \cite{Donev:2005aa,Skoge:2006aa,Lerner:2013aa,Charbonneau:2012aa} and $\gamma\approx
  0.5$ in simulations of stiff soft spheres 
  \cite{Silbert:2002aa,OHern:2003aa,Silbert:2006aa}. The value depends on whether rattlers are included or not in the numerical protocol. Theoretical arguments based on the marginal stability of jammed packings provide \cite{Muller:2015aa}
  \be
  \label{Ch2:gammatheta}
  \gamma= 1/(2+\theta),
  \ee
  where $\theta$ is the exponent of the force distribution: $P(f)\sim f^\theta$. Empirical studies find $\theta\approx 0.2-0.5$ (see Sec.~\ref{Ch2:Sec:forces}).

\item A split second peak at $r=\sqrt{3}\sigma$ and $r=2\sigma$ away
  from contact. The precise shapes of the two peaks have not been
  clearly established yet. Simulations show a strong asymmetry of the
  $r=2\sigma$ peak. The values $2\sigma$ and $\sqrt{3}\sigma$ have
  been related to the contact network: $2\sigma$ is the maximal
  distance between two particles sharing one neighbour, while
  $\sqrt{3}\sigma$ is the maximal distance between two particles
  sharing two \cite{Clarke:1993aa}. The split-second peak is
  indicative of structural order between the first and second
  coordination shells. However, no signs of crystalline order have
  been observed.
\item Long-range order $g_2(r)-1\sim -r^{-4}$ for $r\to \infty$
  \cite{Donev:2005ab}. This is equivalent to a non-analytic behavior
  of the structure factor $S(k)\sim|k|$ for $k\to 0$, which is
  typically only seen in systems with long-range interactions and is
  uncharacteristic for liquids. The fact that $S(0)=0$ is
  characteristic of a hyperuniform system
  \cite{Torquato:2003aa}. However, the validity of hyperuniformity at jamming has recently been questioned \cite{Ozawa:2017aa,Wu:2015aa,Ikeda:2015aa,Ikeda:2017aa}.
  \end{itemize}

\subsubsection{The nature of random close packing}

\label{Ch2:Sec:rcp}

\begin{figure*}
\begin{center}
{\bf (a)}
\includegraphics[width=0.4\textwidth]{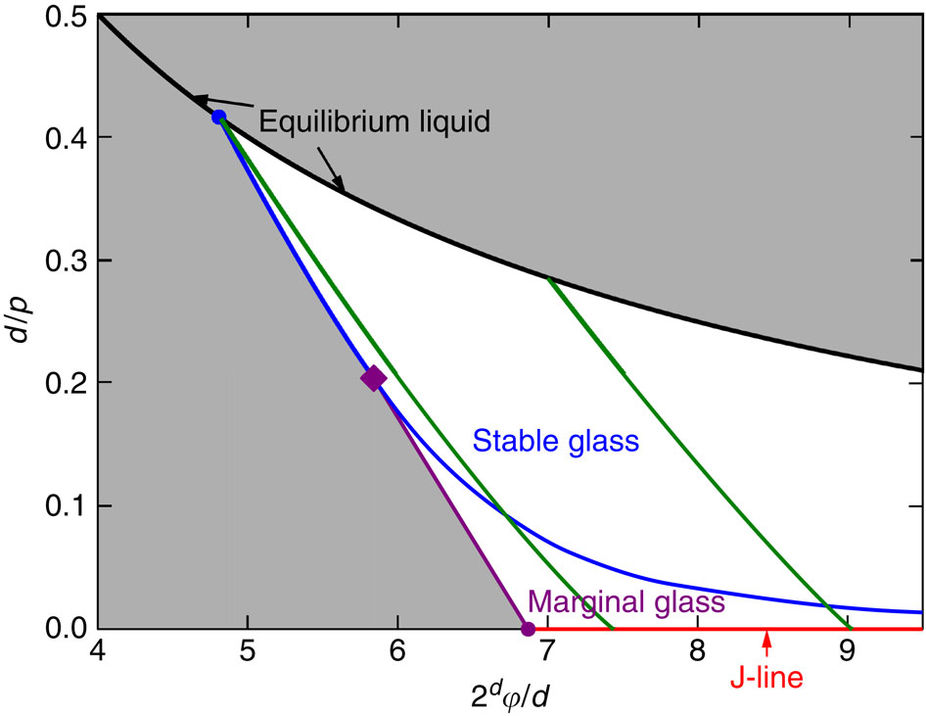}
\hspace{0.2cm} {\bf (b)}
\includegraphics[width=0.51\textwidth]{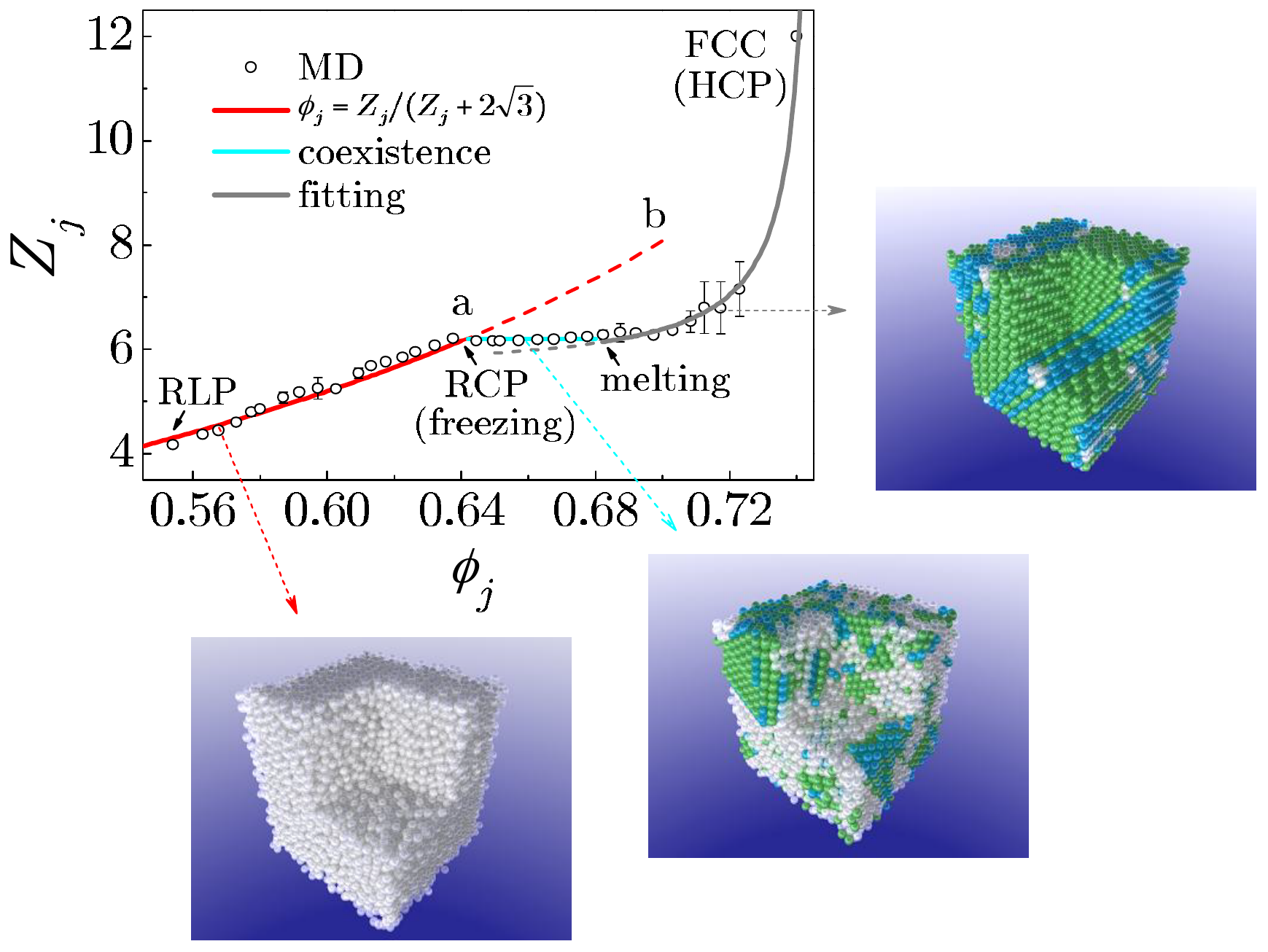}
\caption{\label{Fig:rcps} (Colors online) (a) Phase
    diagram in $d\to\infty$ obtained from the non-equilibrium sampling
    of glassy states \cite{Charbonneau:2017aa}. Glassy states exist in
    the white region between the continuation of the equilibrium
    equation of state (black) and the infinite pressure J-line. The
    blue line denotes the Gardner phase transition separating stable
    and marginally stable glass states. Glass states are possible for
    densities $>\phi_{\rm d}$ at which metastable states first appear in the liquid. Compressing
    the glass states to $p\to\infty$ yields jammed states on the
    J-line $\phi_j\in[\phi_{\rm th},\phi_{\rm GCP}]$. From
    \cite{Charbonneau:2014aa}. (b) Interpretation of RCP in a 3d
    system made of monodisperse spheres as a first order freezing
    transition between disordered and ordered phases. In low
    dimensional systems (3d and specially 2d) crystallization prevails
    around RCP and precludes the appearance of the J-line as discussed
    in \cite{Parisi:2010aa}.  The coordination number $z_j$ is plotted
    versus the volume fraction $\phi_j$ for each packing at
    jamming. One can identify: (i) a disordered branch which can be
    fitted by the equation of state~(\ref{Ch4:phiz}) derived in
    Sec.~\ref{Ch4:Sec:volume3d}; (ii) a coexistence region; and (iii)
    an ordered branch. White particles are random clusters, light blue are HCP
    and green are FCC clusters. The dashed line from $a\to b$ denotes
    the states beyond crystallization, which can be reached upon deformation of the particles (see Fig.~\ref{Ch4:Fig_phasediagram}). From \cite{Jin:2010aa}.}
\end{center}
\end{figure*}

\begin{table*}[ht]
\centering
\begin{tabular}{|c | c | c |}
\hline
$\qquad$Density$\qquad$  & Definition & $\quad$Value in $d=3\qquad$ \\ 
\hline \hline
$\phi_{\rm d}$ & The liquid state splits in an exponential number of states & $\approx 0.58$  \\
$\phi_{\rm K}$ & Ideal glass phase transition -- jump in compressibility & $\approx 0.62$ \\
$\phi_{\rm th}$& Divergence of the pressure of the less dense states & $\approx 0.64$ \\
$\phi_{\rm GCP}$& Divergence of the pressure of the ideal glass & $\approx 0.68$\\
\hline
\end{tabular}
\caption{\label{Ch3:table_phi1}Density values when compressing a liquid state until jamming avoiding crystallization \cite{Parisi:2010aa,Charbonneau:2017aa}.}
\end{table*}

\begin{table*}[ht]
\centering
\begin{tabular}{|c | c | c |}
\hline $\qquad$Density$\qquad$ & Definition & $\quad$Value in
$d=3\qquad$ \\ \hline \hline $\phi_{\rm rlp}$ & Random loose packing:
lowest density of a mechanically stable packing & $\frac{1}{1+\sqrt{3}/2}= 0.536...$ \cite{Song:2008aa}
\\ $\phi_{\rm rcp}$ & Random close packing &
$\frac{1}{1+1/\sqrt{3}} = 0.634...$ \cite{Song:2008aa}\\ $\phi_{\rm f}$ & Packing freezing
point of a 1st order transition & $\approx 0.64$ \cite{Jin:2010aa}\\ $\phi_{\rm m}$ &
Packing melting point of a 1st order transition & $\approx 0.68$ \cite{Jin:2010aa}\\ $\phi_{\rm
  fcc}$ & Density of the FCC crystal &
$\pi/(3\sqrt{2})=0.74048...$\\ \hline
\end{tabular}
\caption{\label{Ch3:table_phi2}Density values when crystallization is
  not suppressed. The values for $\phi_{\rm rlp}$ and $\phi_{\rm rcp}$
  are determined within the Edwards ensemble using a coarse-grained
  volume function \cite{Song:2008aa,Jin:2010aa} (see
  Sec.~\ref{Ch4:Sec:ensemble}).}
\end{table*}

The nature of RCP of frictionless hard spheres and whether it is
indeed a well-defined concept has been a long-standing issue. In
\cite{Torquato:2000aa} it has been argued that ``random" and
``close-packed" are at odds with each other, since inducing partial
order typically increases packing densities, such that both can not be
maximized simultaneously. As an alternative it has been suggested to
use a more quantitative approach based, e.g., on a metric detecting bond-orientational order \cite{Steinhardt:1983aa}. RCP can then be replaced by the concept of a ``maximally
random jammed" (MRJ) packing: The packing with the minimal order among
all jammed ones. In practice, all possible order
  metrics would need to be checked to identify a truly random state,
  which is of course not feasible. Nevertheless, many different
  packing protocols and algorithms seem to robustly achieve disordered
  packings with maximal densities around $\phi\approx 0.64$, which
  coincides with the densities of MRJ packings for many different
  order parameters \cite{Torquato:2010aa}. Despite early attempts to
  explain this reproducibility, e.g., based on maximum entropy
  arguments \cite{OHern:2002aa,OHern:2003aa} and liquid state theory
  \cite{Aste:2004aa,Kamien:2007aa}, there is now a general consensus
  that jamming densities can be obtained over a range of densities
  depending on the preparation protocol if crystallization is
  suppressed
  \cite{Skoge:2006aa,Ciamarra:2010aa,Chaudhuri:2010aa,Ozawa:2012aa,Charbonneau:2012aa,Hermes:2010aa}. This
  leads to the concept of a \textit{J-line}, which was first proposed
  theoretically in the context of a replica solution of hard sphere
  glasses at the mean-field level ($d\to\infty$) \cite{Parisi:2005aa}
  and other fully connected models \cite{Mari:2009aa}. In the presence
  of polydispersity in the particle size or in higher dimensions,
  crystallization is strongly suppressed and the physics of the glass
  transition is expected to dominate the corresponding jamming
  transition. If jamming is approached from the equilibrium fluid
  phase, the resulting jammed states are then essentially the infinite
  pressure limits of glassy states. A deep understanding of jamming in
  this scenario has been provided by exact solutions for $d\to\infty$
  using both dynamical mode-coupling type approaches
  \cite{Maimbourg:2016aa,Kurchan:2016aa} and static approaches adapted
  from the solution of the Sherrington-Kirkpatrick model of
  spin-glasses
  \cite{Rainone:2016aa,Parisi:2010aa,Charbonneau:2014aa,Charbonneau:2014ab,Franz:2015aa}. Remarkably,
  the full RSB $d\to\infty$ solution predicts scaling exponents for
  $g_2(r)$, Eq.~\eqref{Ch2:gammaexp}, and the force distribution
  $P(f)$, Eq.~\eqref{Eq:theta} (see next section), that are in
  agreement with finite dimensional measurements for a range of $d$
  values even in 3d \cite{Charbonneau:2014aa,Charbonneau:2014ab}. This
  remarkable agreement between an infinite dimensional mean-field
  theory and 3d simulations indicates that, at jamming, there is a
  strong suppression of fluctuations, first of all thermal
  fluctuations by definition, but, more importantly, sample to sample
  fluctuations which are known to be stronger than thermal
  fluctuations. Similar agreement between an infinite-dimensional
  result and finite dimensions is not observed for the
  finite-temperature glass transition. Thus, the critical properties
  of jamming related to marginal stability appear independent of
  dimensionality. For a recent review on the $d\to\infty$ solution of
  hard sphere glasses, we refer to \cite{Charbonneau:2017aa}. An overview of the different density values discussed in the following is given in Tables~\ref{Ch3:table_phi1},\ref{Ch3:table_phi2}.

Briefly, in this scenario a glass transition
  interrupts the continuation of the liquid equation of state
  considered in \cite{Aste:2004aa,Kamien:2007aa} at densities
  $\phi_j\in [\phi_{\rm d},\phi_{\rm K}]$, where $\phi_{\rm d}$
  signals the dynamical glass transition at the density at which many
  metastable states first appear in the liquid phase and $\phi_{\rm
    K}$ is the Kauzmann density of the ideal glass. Upon compression of the metastable states (taking some care in the preparation protocol \cite{Charbonneau:2017aa}) the
  pressure diverges at jamming densities $\phi_j\in [\phi_{\rm
      th},\phi_{\rm GCP}]$. The lower limit is the threshold density $\phi_{\rm
    th}\approx 0.64$ calculated in \cite{Parisi:2010aa}, although it should be noted that the values calculated with replica theory come with a large error bar due to the approximation of the liquid equation of state \cite{Mangeat:2016aa}. The
  maximal density is the glass close packing $\phi_{\rm GCP}\approx
  0.68$ corresponding to the infinite pressure limit of the
  ideal glass $\phi_{\rm K}$. Therefore, the ground state of jamming
  can be achieved in a whole range of densities along a J-line
  $\phi_j\in[\phi_{\rm th},\phi_{\rm GCP}]$ depending on the density
  of the metastable glass phase $\phi\in [\phi_{\rm d},\phi_{\rm K}]$
  that is compressed to jamming. Before jamming is reached the glass
  undergoes a transition to a Gardner phase, where the configuration
  space is fragmented into an infinite fractal hierarchy of
  disconnected regions, which, in turn, brings about isostaticity and
  marginal stability
  \cite{Charbonneau:2014aa,Charbonneau:2014ab}. Indeed the states on the $J$-line are all stable under all possible
particle rearrangements with $k\to\infty$ in the thermodynamic limit
$N\to\infty$, thus corresponding to the ground state of jamming, as discussed in Fig.~\ref{fig:PD2}a. On the other hand, they differ in the fraction $\alpha=k/N\sim$ const. of particle
rearrangements required for stability.

Such a viewpoint is motivated by analogy with the full RSB solution of the $p$-spin
glass \cite{Crisanti:2006aa}, which is the spin glass model corresponding to the full-RSB solution of infinite dimensional spheres underlying the J-line \cite{Charbonneau:2014ab}. By varying $\alpha$ one obtains states on the J-line: the value $\alpha=0$ corresponds to the states at the lower
density $\phi_{\rm th}$, while $\alpha=1$ corresponds to the true
global ground state of jamming at the largest density $\phi_{\rm
  GCP}$. Metastable $k$-PD states with finite $k$ are achieved with
lower packing fractions as depicted in Fig.~\ref{fig:PD2}a and in Table~\ref{Ch2:table1}.
  
We conclude that the truly global ground state is actually only one of
the possible $\infty$-PD stable states and corresponds to the point
$\alpha=1$, which is at $\phi_{\rm GCP}$. The other states along the
J-line, obtained by varying $0\le \alpha <1$, can be thought of as
globally metastable (in reality they also belong to the ground state
of the J-line). On the basis of this picture, we propose four categories of jamming according
to their metastability as explained in Table~\ref{Ch2:table1}: local metastable ($1$-PD
stable), collective metastable ($k$-PD stable with finite $1<k<\infty$),
globally metastable ($\infty$-PD stable but with $0\le \alpha<1$, and the
true global ground state ($\infty$-PD stable and $\alpha=1$). In
particular the J-line corresponds to globally metastable states
($\infty$-stable) while the ground state corresponds to $\phi_{\rm
  GCP}$.

Interestingly, the phase diagram that arises from the
  $d\to\infty$ solution, which corresponds to a particular packing
  protocol, can be reproduced by sampling over glassy states with a
  modified (non-equilibrium) measure
  \cite{Parisi:2010aa,Charbonneau:2014ab,Charbonneau:2017aa} (see
  Fig.~\ref{Fig:rcps}a). Possible glass states are then predicted in
  the white region of Fig.~\ref{Fig:rcps}a bounded by the metastable
  continuation of the equilibrium liquid and the J-line. In this
  approach the Gardner transition (blue line) separates stable and
  marginally stable states. Crucially, for infinite pressure this
  non-equilibrium sampling assigns equal probability to each jammed
  state at a given density, i.e., it agrees with Edwards uniform
  measure. Therefore, the non-equilibrium sampling of glassy states at
  the ground state is another generalization of the Edwards ensemble
  to finite pressures. Since the critical jamming exponents calculated in this approach are the same as those from the full RSB solution \cite{Rainone:2016aa}, we conclude that the observed phenomenology of jamming is at least consistent with Edwards assumption of equiprobability in the values of the exponents. Edwards statistical mechanics thus captures key features of the jamming phenomenology, a fact that is increasingly being recognized  \cite{Charbonneau:2017aa,Sharma:2016aa}. Highly sophisticated simulations have recently confirmed the validity of Edwards assumptions at the jamming transition as well \cite{Martiniani:2017aa} (see Sec.~\ref{Sec_Edwardstest}).

Furthermore, these results highlight the fact that packing problems,
and more generally CSPs, undergo a phase transition separating a
satisfiable (SAT) (hypostatic or under-constrained) regime from an
unsatisfiable (UNSAT) (hyperstatic or over-constrained) phase, as one
varies the ratio of constraints over variables. The jamming transition
is equivalent to this SAT-UNSAT phase transition in the broad class of
continuous CSPs, which are conjectured to belong to the same
"super-universality" class based on models displaying SAT/UNSAT like
the celebrated perceptron model \cite{Franz:2015aa,Franz:2015ab} which
admits a much simpler solution at the full RSB level than the
hard-sphere glass.

If crystallization is \textit{not} suppressed, compressing an equilibrium liquid of monodisperse spheres can lead to partial crystalline order \cite{Jin:2010aa,Francois:2013aa,Hanifpour:2014aa,Hanifpour:2015aa,Anikeenko:2007aa,Anikeenko:2008aa,Klumov:2011aa,Klumov:2014aa,Radin:2008aa,Kapfer:2012aa}. Using the granular entropy of Edwards statistical mechanics as treated in Sec.~\ref{Ch2:Sec:edwardsens}, then allows to identify the onset of crystalline order with the freezing point of a first order transition, which is found at $\phi_{\rm
  f}\approx 0.64$ \cite{Jin:2010aa}. Likewise, a melting point appears at $\phi_{\rm m} \approx 0.68$. Between these two densities a coexistence of
disordered and ordered states exists at the coordination number of
isostaticity $z=6$ (see Fig.~\ref{Fig:rcps}b). Defining RCP in this scenario as the freezing point, two branches then exist: a disordered branch from
the RLP at $\phi_{\rm rlp}\approx0.54$ up to the freezing point $\phi_{\rm
  f}\approx 0.64$ and an ordered branch from the melting point
$\phi_{\rm m} \approx 0.68$ to FCC at $\phi_{\rm fcc}=0.74...$. The
signature of this disorder-order transition is a discontinuity in the
entropy density of jammed configurations as a function of the
compactivity. This highlights the fact that beyond RCP, denser packing fractions of monodisperse spheres can only be reached by partial crystallization up to the homogeneous FCC crystal phase in agreement with the interpretation of RCP as a MRJ state \cite{Torquato:2000aa}. Indeed,
RCPs are known to display sharp structural changes
\cite{Anikeenko:2007aa,Anikeenko:2008aa,Klumov:2011aa,Klumov:2014aa,Aristoff:2009aa,Radin:2008aa,Kapfer:2012aa}
signalling the onset of crystallization \cite{Torquato:2010aa}. The first-order transition scenario observed numerically in
  \cite{Jin:2010aa} has been verified in a set of experiments of 3d hard
  sphere packings
  \cite{Francois:2013aa,Hanifpour:2014aa,Hanifpour:2015aa}. In
  \cite{Francois:2013aa} the onset of crystallization at the freezing
  point $\phi_{\rm f}\approx 0.64$ has been identified from the
  variance of the Voronoi volume fluctuations \cite{Jin:2010aa}, a
  ``granular specific heat" \cite{Aste:2008aa}, and the frequency of
  polytetrahedral structures. The coexistence line at isostaticity between
  $\phi_{\rm f}\approx 0.64$ and $\phi_{\rm m} \approx 0.68$ has been
  observed not only for frictionless packings but also for frictional
  ones, where high densities have been achieved by applying intense
  vibrations \cite{Hanifpour:2014aa,Hanifpour:2015aa}.

The existence of the first-order crystallization transition at RCP is
expected to be dominant in a finite dimensional 3d system of equal
size spheres and therefore excludes the appearance of the interesting
glassy phases discussed above unless crystallization is suppressed by heterogeneities like polydispersity. Interestingly, the values of the limiting densities $
[\phi_{\rm th},\phi_{\rm GCP}]$ coincide approximately with the
densities of the melting and freezing points in the first-order
transition obtained for monodisperse 3d systems
\cite{Jin:2010aa}. However, this coincidence is most likely coincidental since these states are unrelated. It should be
noted that the analysis of structure and order parameters is generally
supportive of the existence of a glass-crystal coexistence mixture in
the density region $0.64\le \phi\le 0.68$ in monodisperse sphere
packings where crystallization dominates over the glass phase.  All the
(maximally random) jammed states along the segment $[\phi_{\rm th},
  \phi_{\rm GCP}]$ can be made denser at the cost of introducing some
partial crystalline order.  Support for an order/disorder transition at
$\phi_{\rm f}$ is also obtained from the increase of polytetrahedral
substructures up to RCP and its consequent decrease upon
crystallization \cite{Anikeenko:2008aa}.

The connection of the replica approach with the Edwards ensemble for
jammed disordered states is summarized in Table \ref{Ch2:table1} and
Fig.~\ref{fig:PD2}a and will be discussed in detail in
Sec.~\ref{Sec:constraints}. The hierarchy of metastable jammed states
$k$-PD with $k\in[1,\infty)$ is analogous to $k$-SF with
  $k\in[1,\infty)$ metastable states in spin-glasses which in turn are
    related to the continuity of jammed states along the J-line. This
    is the picture emerging from a full RSB solution, at the
    mean-field level of fully connected systems, like the SK model of
    spin-glasses \cite{Sherrington:1975aa}. Thus, we expect that a
    continuous jamming line of states should emerge from the Edwards
    ensemble solution of the JSP, since it is another realization of a
    typical NP-hard CSP.

On the other hand, the mean field solution of the Edwards volume
ensemble \cite{Song:2008aa} reviewed in Sec.~\ref{Sec:volume}
predicts a single jamming point at RCP, Eq.~(\ref{Ch4:phircp}), $\phi_{\rm rcp} = \frac{1}{1+1/\sqrt{3}} \approx 0.634$ for $z=6$.  This
prediction corresponds to the ensemble average over a coarse-grained
Voronoi volume for a fixed coordination number.  Since an ensemble
average over all packings at a fixed coordination number is performed in
the coarse-graining of the volume function, the obtained volume
fractions $\phi_{\rm rcp}$ are in fact averaged over the J-line
predicted by the replica method.  Thus, $\phi_{\rm rcp}$ can be associated to the
state with the largest entropy (largest complexity) along $[\phi_{\rm
    th}, \phi_{\rm GCP}]$, expected to be near the highest entropic
state $\phi_{\rm th}$ in the replica theory picture. Indeed,
high-dimensional calculations performed in
Sec. \ref{Ch4:Sec:dimension} support this conjecture: the scaling
obtained with dimension $d$ of the Edwards prediction for RCP and
$\phi_{\rm th}$ agree within a prefactor, see Eqs. (\ref{highd}) and
(\ref{rsb-d}) below.

New possibilities to study densely packed states are opened up by including activity on the particle level (self-propulsion), which shifts the glass transition closer to random close packing \cite{Ni:2013aa}.

\subsubsection{Force statistics}
\label{Ch2:Sec:forces}

It has been realized early on that jammed granular aggregates exhibit
non-uniform stress fields due to arching effects
\cite{Jaeger:1996aa,Cates:1998aa}. More recent work has focused on the
interparticle contact force network. The key quantity is the force
distribution $P(f)$, which exhibits characteristic features at jamming
as observed in both experiments
\cite{Zhou:2006aa,Liu:1995aa,Mueth:1998aa,Lovoll:1999aa,Erikson:2002aa,Brujic:2003aa,Brujic:2003ab,Corwin:2005aa,Makse:2000aa,Kyeyune-Nyombi:2018aa}
and simulations
\cite{Radjai:1996aa,OHern:2001aa,Tkachenko:2000aa,Makse:2000aa}:
\begin{itemize}
\item $P(f)$ has a peak at small forces (approximately at the mean
  force $\left<f\right>$). This peak has been argued to represent a
  characteristic signature of jamming \cite{OHern:2001aa}.
\item For large forces, the decay of $P(f)$ has been generally
  measured as exponential. Although a faster than exponential decay
  has also been observed in experiments \cite{Majmudar:2005aa} and
  simulations \cite{Eerd:2007aa}.
 \end{itemize}
These properties are observed in both hard and soft sphere systems,
largely independent of the force law.
 
For $f\to 0^+$, $P(f)$ converges to a power-law \be P(f)\sim f^\theta, \,\,\,\,\,\,  f\to 0^+,
\label{Eq:theta} 
\ee with some uncertainty regarding the value of the exponent:
$\theta\approx 0.2 - 0.5$. The existence of this power-law has been
explained by the marginal stability of the packing which is controlled
by small forces \cite{Wyart:2012aa}. As a consequence, $\theta$ is
related to the exponent $\gamma$ of near contacting neighbours by Eq.~(\ref{Ch2:gammatheta}). A more detailed investigation of the
excitation modes related to the opening and closing of contacts
suggests that there are in fact two relevant exponents $\theta_{\rm
  e}$ and $\theta_{\rm l}$ \cite{Lerner:2013aa}: $\theta_{\rm e}$
corresponding to motions of particles extending through the entire
systems; and $\theta_{\rm l}$ corresponding to a local buckling of
particles. A marginal stability analysis provides $\gamma=(2+\theta_{\rm e})^{-1}=(1-\theta_{\rm
  l})/2$ \cite{Muller:2015aa}, which has also been demonstrated numerically \cite{Lerner:2013aa}. Asymptotically $\theta=\min(\theta_{\rm l},\theta_{\rm
  e})$ and thus $\theta=\theta_l\approx 0.2$ for $\gamma\approx 0.4$.

Theoretically, one step replica symmetry 1RSB theory for fully
connected hard sphere packings in infinite dimensions predicts
$\theta=0$ \cite{Parisi:2010aa}, while the full RSB calculation
provides a non-zero $\theta=0.42..$ and $\gamma=0.41..$
\cite{Charbonneau:2014aa,Charbonneau:2014ab}, a result corroborated
theoretically with a simpler jamming model, the Perceptron model from
machine learning, which exhibits a jamming transition as well
\cite{Franz:2015aa,Franz:2015ab}. This result further indicates the
importance of the jamming transition to general CSPs. The full-RSB
values are seemingly in disagreement with the scaling relations from
marginal stability in the presence of localized modes, since they
predict $\theta_{\rm l}=0.17..$. However, based on simulation results
it has been shown that the probability of localized modes decreases
exponentially with dimension and thus they do not contribute to the
full RSB solution for $d\to\infty$ \cite{Charbonneau:2015aa}. Thus, in
3d simulations the so called bucklers (particles with all forces
except one, usually the smallest, approximately aligned in a plane)
are removed from the distribution decreasing the small force counting
and changing the exponent from 0.17 to 0.42 in agreement with the full
RSB replica theory. As a consequence, $\theta=\theta_{\rm e}$ in
agreement with the scaling relations.

High-resolution measurements of the contact network in 3d allow for
the experimental determination of the exponents $\theta$ and $\gamma$,
see Table~\ref{table:kyeyune} and Fig. \ref{Fig:kyeyune}
\cite{Kyeyune-Nyombi:2018aa}. Here, the value of the small force
exponent can be estimated due to the high resolution of contact
detection. Values in the range $\theta\approx 0.11-0.17$ below the
full RSB prediction are found even when bucklers are removed.  Instead
of the equality Eq.~(\ref{Ch2:gammatheta}), the inequality $\gamma\ge
1/(2+\theta)$ is still observed, except for one packing A which is
presumably hyperstatic.  On the other limit of sparse graphs, replica
symmetry calculations gives $\theta=0$ in the thermodynamic limit
using population dynamics implying that RS calculations do not capture
the full physics of the jamming point \cite{Bo:2014aa} (discussed in
Sec.~\ref{Ch5:Sec:cavity}).

\begin{table}[ht]
\centering
\begin{tabular}{| c | c | c | c | c | c | c |}
\hline
Packing  &  $N$ & $z$ & $\phi$ & $\theta$ & $\gamma$ & $1/(2+\theta)$ \\ 
\hline \hline
A & 1393 & 7.57 & 0.66(8) & 0.110(5) & 0.42(2) & 0.474(1)\\
B & 1263 & 6.79 & 0.62(4) & 0.143(4) & 0.62(2) & 0.467(1)\\
C & 1486 & 6.64 & 0.64(7) & 0.170(6) & 0.75(3) & 0.461(1)\\
\hline
\end{tabular}
\caption{\label{table:kyeyune} Structural properties of a 3d
  colloidal packing near marginal stability using high-resolution
  measurements of the contact network
  \cite{Kyeyune-Nyombi:2018aa}. Three slightly different packing
  protocols have been used. Instead of the
  equality~\eqref{Ch2:gammatheta}, the weak force exponent $\theta$
  (Eq.~\eqref{Eq:theta}) and the small gap exponent $\gamma$
  (Eq.~\eqref{Ch2:gammaexp}) are found to satisfy the inequality
  $\gamma\ge 1/(2+\theta)$ \cite{Wyart:2012aa} (except for Packing A
  which might be hyperstatic). The exponent $\theta$ does not change
  appreciable whether bucklers are included or not.}
\end{table}

\subsection{Test of ergodicity and the uniform measure in the Edwards ensemble}
\label{Sec_Edwardstest}

Assuming ergodicity for a jammed system of grains as proposed by
Edwards (see Sec.~\ref{Ch2:Sec:edwardsens}) seems contradictory at
first, but has become meaningful in the first place in light of
certain seminal compaction experiments developed over the years
starting from the work of Nowak {\it et at.} in the 90's
\cite{Knight:1995aa,Nowak:1997aa,Nowak:1998aa,Chakravarty:2003aa,Richard:2005aa,Makse:2005aa,Philippe:2002aa,Brujic:2005aa}.

\begin{center}
\begin{figure}
\includegraphics[width=0.85\columnwidth]{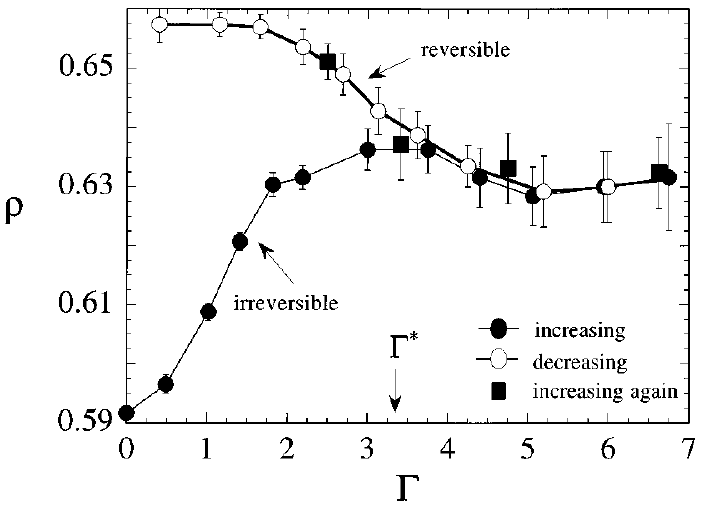} 
\caption{\label{Fig_Nowak1998density} The packing fraction $\rho$
plotted as a function of the shaking intensity $\Gamma$ from
experiments of granular packings undergoing vertical
tapping \cite{Nowak:1998aa}. The intensity is defined as the ratio
of the peak acceleration during a single tap to the gravitational
acceleration. The system is prepared initially at low packing
fraction and subjected to taps of increasing intensity.  The tapping
intensity is then successively reduced, and the system falls on a
reversible branch, where the system retraces the density versus
intensity behavior upon subsequent increases and decreases of the
intensity. From \cite{Nowak:1998aa}.}
\end{figure}
\end{center}

Nowak, {\it et al.} \cite{Nowak:1997aa,Nowak:1998aa} performed a set
of experiments of the compaction of spherical glass beads as a
function of increasing and decreasing vertical tapping intensity.
Figure~\ref{Fig_Nowak1998density} shows their results for the packing
fraction $\rho$ versus the tapping intensity $\Gamma$ (normalized by
the acceleration due to gravity). The key observation is that the
system, after initial transient behavior on the `irreversible branch',
reaches a 'reversible branch' on which it retraces the variation of the
packing fraction upon increasing and decreasing the intensity.  The
initial tapping breaks the frictional contacts that support loose
packed configurations and store information about the system
preparation. On the reversible branch, small tapping intensities
induce denser packings with packing fractions slightly above random
close packing for equal-sized spheres.

In principle, we can interpret the reversible packings as
equilibrium-like states, in which the details of the microscopic
configurations and the compaction protocol are irrelevant, as
demonstrated by the reversible nature of the states evidenced by the
unique branch traveled by the system as the external intensity is
increased and decreased. These are the states for which we expect, in
principle, a statistical mechanical formalism to hold. The existence
of such a reversible branch has been corroborated in a number of
experimental systems with different compaction techniques, e.g., under
mechanical oscillations and vibrations, shearing, or pressure
waves~\cite{Philippe:2002aa,Brujic:2005aa,Chakravarty:2003aa} and
studied with theory and modelling
\cite{Krapivsky:1994aa,Caglioti:1997aa,Nicodemi:1999aa,Prados:2000aa,Nicodemi:1997ac,Nicodemi:1997aa,Nicodemi:1999ab,Nicodemi:1997ab}. However,
this interpretation has been challenged in a number of studies of
ergodicity in jammed matter.

Systems that are subjected to a constant drive such as
infinitesimal tapping or also small shear are able to
explore their phase space dynamically, such that ergodicity can be
tested directly by comparing time averages and averages with respect
to the constant volume ensemble. We stress here, that only
infinitesimal driving forces should be applied to test equiprobable
states (see discussion in Sec. \ref{Sec:outlook}). An agreement of the
two averages has indeed been observed in simple models
\cite{Berg:2002aa,Gradenigo:2015aa}, as well as soft sphere systems
with a small number of particles $N=30$
\cite{Wang:2012aa,Wang:2010aa}.

Some recent systematic results are more controversial though,
motivating a continued investigation of this fascinating concept
\cite{Irastorza:2013aa}. A very detailed and rigorous numerical
analysis confirms that at low tapping intensities, the system can not
be considered to be ergodic: Two different realizations of the same
preparation protocol do not correspond to the same stationary
distribution, indicated by a statistical test of data for both the
packing density \cite{Paillusson:2012aa,Paillusson:2015aa} using
volume histograms sample over time
\cite{McNamara:2009ab,McNamara:2009aa}, and the trace of the
force-moment tensor \cite{Gago:2016aa}. When considering the fraction
of persistent contacts as a function of tapping intensity, one
observes that the non-ergodic regime coincides with a larger
percentage of persistent contacts, while such contacts are almost
absent in the ergodic regime \cite{Gago:2016aa}. The picture that
emerges is that the breakdown of ergodicity is connected to the
presence of contacts that do not break under the effect of the
tapping. In accordance with physical intuition, the system can then
not sample its whole phase space, but is stuck in specific regions
with the consequent breaking of ergodicity. An additional reason to
doubt the validity of ergodicity is the violation of the time reversal
symmetry due to dissipation \cite{Dauchot:2007aa}.

Ergodicity is also intimately related to the existence of
non-equilibrium fluctuation-dissipation relations (FDR) characterized
by an effective temperature \cite{Cugliandolo:2011aa}. For equilibrium systems, the FDR is a
very general result relating time correlations and responses through
the temperature of the thermal environment. Non-equilibrium FDRs have
been shown to hold in a wide range of systems starting with the work
of Ref.~\cite{Cugliandolo:1997aa}, e.g., for glassy systems
\cite{Bellon:2002aa,Crisanti:2003aa,Leuzzi:2009aa} and
models of driven matter \cite{Berthier:2000aa,Loi:2008aa} (see also
the review~\cite{Marconi:2008aa}). It has recently also been
demonstrated in single molecule DNA driven out of equilibrium by an
optical tweezer \cite{Dieterich:2015aa}. Non-equilibrium FDRs and
effective temperatures are often linked to the slow modes of the
relaxation in a glassy phase \cite{Cugliandolo:1997aa}. In granular
compaction, the relaxation to the final density is similarly slow,
following, e.g., an inverse logarithmic law under tapping
\cite{Nowak:1997aa,Nowak:1998aa,Krapivsky:1994aa} and a Kohlrausch-Williams-Watts law under shear \cite{Lu:2008aa}. The fluctuations
induced by the continuous driving allow for the definition of an
effective temperature, which, in an ergodic system, should agree with
the granular temperature associated with the canonical volume
ensemble \cite{Cugliandolo:2011aa}. This allows for an indirect test of ergodicity, which has
been established in a number of systems, both toy models
\cite{Nicodemi:1999aa,Lefevre:2002aa,Lefevre:2002ab,Prados:2002aa,Brey:2000aa,Barrat:2000aa,Dean:2001aa,Nicodemi:2004aa,Coniglio:2004ac,Fierro:2002aa,Fierro:2003aa,Tarjus:2004aa}
and more realistic ones using MD simulation of slowly sheared granular
materials \cite{Makse:2002aa}, as well as experiments measuring
effective temperatures in colloidal jammed systems \cite{Song:2005aa}
and slowly sheared granular materials in a vertical Couette cell
\cite{Wang:2006aa,Wang:2008aa,Potiguar:2006aa} and vibrating cells
\cite{Ribiere:2007aa}. The observation of ratcheting in packings of polygonal particles under cyclic load \cite{Alonso-Marroquin:2004aa} sheds however some doubts about the exploration of configuration space due to systematic irreversible displacements on the grain scale: not only is time reversibility violated, but a steady state does not seem to be reached.

The concept of granular temperature or compactivity
  $X$ raises the question whether it is a well defined
  quantity at all. There are essentially two different methods to calculate
  $X$ from packing data: (i) From the statistics of elementary volume
  cells. Exploiting the analogy with equilibrium statistical
  mechanics, $X$ can be derived by thermodynamic integration over the
  inverse volume fluctuations
  \cite{Nowak:1998aa,Schroter:2005aa,Lechenault:2006aa,Ribiere:2007aa,Briscoe:2008aa,Jin:2010aa}. Alternatively,
  one can use analytical expressions either for the volume
  distribution, such as the $\Gamma$-distribution
  \cite{Aste:2007aa,Aste:2008aa,Aste:2008ab} or for $X$ itself,
  derived e.g. from idealized solutions using quadrons
  \cite{Blumenfeld:2003aa,Blumenfeld:2012aa}. (ii) Using an
  overlapping histograms approach
  \cite{Dean:2003aa,McNamara:2009aa}. The protocol independence of $X$
  obtained from a fit to the quadron solution has been shown in
  \cite{Becker:2015aa}. In \cite{Zhao:2014aa} four different ways of
  measuring $X$ from the same experimental data set of a binary disk
  packing have been systematically compared. Interestingly, only two
  of the methods have been shown to agree quantitatively once the
  density of states is also included as an experimental input. This
  highlights possible inconsistencies between different definitions of
  $X$.

The equilibration of the
temperature-like parameters in Edwards statistical mechanics has been
demonstrated in experiments \cite{Schroter:2005aa,Jorjadze:2011aa,Puckett:2013aa}. However, in \cite{Puckett:2013aa} only the angoricity and not the compactivity has been shown
to equilibrate. An upper bound on the Edwards
entropy in frictional hard-sphere packings has recently been suggested
\cite{Baranau:2016aa}.

Recent criticism in \cite{Blumenfeld:2016aa} has claimed that the
volume function is {\it per se} not suitable as the central concept
for a statistical mechanical approach, since the volume is defined by
the boundary particles and $\mathcal{W}$ is thus independent of the
configurations of bulk particles, i.e., $\partial \mathcal{W}/\partial
\mathbf{q}_i=0$ for these degrees of freedom. As a consequence, the
resulting entropy would be miscalculated due to miscounting of these
configurations. However, in \cite{Becker:2017aa} it
  has been shown that the vanishing derivatives are still consistent
  with statistical mechanics. Even if $\mathcal{W}$ is independent of
  some degrees of freedom, the resulting partition function still
  takes these into account and thus allows the correct calculation of
  macroscopic observables in terms of expectation values.

Related to ergodicity, the second controversial
  concept underlying Edwards statistical mechanics is the assumption
  of equiprobability of jammed microstates, Fig.~\ref{fig:PD2}a. Since
  Edwards' initial conjecture, most studies have focused on testing
  the validity of the consequences of this assumption rather than
  testing it directly. On the other hand, a direct test requires the
  evaluation of all possible jammed configurations and counting the
  occurrence of distinct microstates, which is possible in model
  systems \cite{Bowles:2011aa,Slobinsky:2015aa,Slobinsky:2015ab}. For
  more realistic packings, such a direct test has long been restricted
  to small numbers of particles due to the prohibitively large number
  of resulting jammed states. Some of the first direct tests for up
  $N=14$ particles have shown a highly non-uniform distribution,
  suggesting that the structural and mechanical properties of dense
  granular media are not dominated equally by all possible
  configurations as Edwards assumed, but by the most frequent ones
  \cite{Xu:2005aa,Gao:2006aa,Gao:2009aa}. It has been argued that the
  non-uniformity, which is manifest in a broad distribution of basin
  volumes in the energy landscape that identify jammed states, is due
  to the fast quench into the energy minima
  \cite{Wang:2012aa}. Moreover, it is not clear if the non-uniformity
  survives for larger system sizes.

\begin{figure}[]
\begin{center}
\includegraphics[width=0.85\columnwidth]{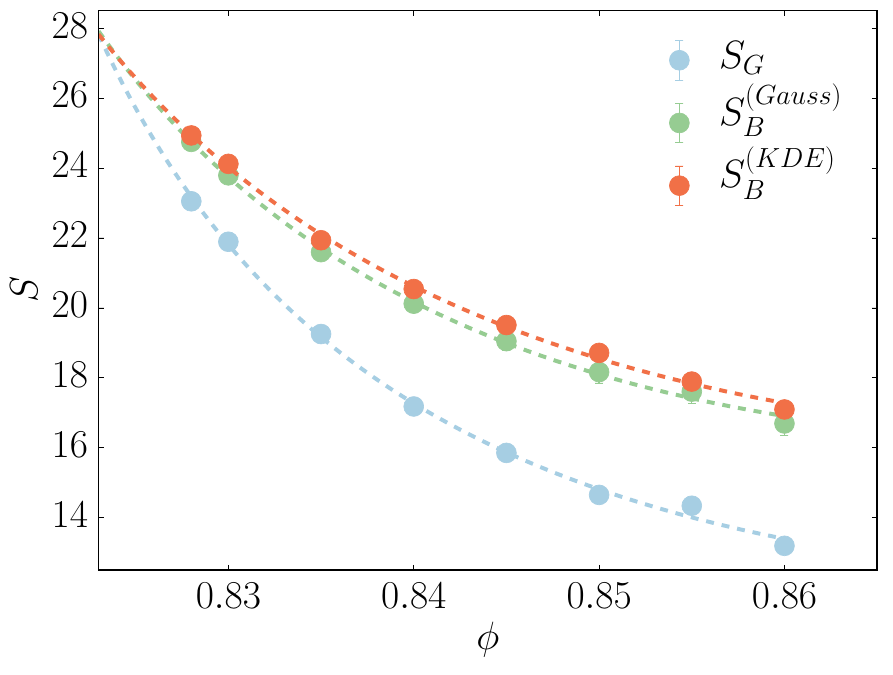}
\caption{(Colors online) Recent numerical results confirm Edwards equiprobability assumption at the jamming transition \cite{Martiniani:2017aa}. Gibbs entropy Eq.~\eqref{Ch3:entropy} and Boltzmann entropy $S^*_B=\log \Omega(V)-\log N!$ demonstrating equiprobability at $\phi^*\approx 0.82$ for $N=64$ particles. $S^*_B$ is computed parametrically (``Gauss") and non-parametrically using a kernel density estimate (``KDE"). From \cite{Martiniani:2017aa}.}
\label{Fig:directtest}
\end{center}
\end{figure}

Remarkable recent progress has been able to
  conclusively validate Edwards' equiprobability assumption for
  realistic system sizes. Advances in numerical methods have enabled a
  direct computation of basin volumes of distinct jammed states of up
  to $N=128$ polydisperse frictionless spheres in both 2d and 3d with a hard core and soft
  shell
  \cite{Xu:2011aa,Asenjo:2014aa,Martiniani:2016aa,Martiniani:2016ab,Martiniani:2017aa,Martiniani:2016ad}. The spheres are jammed by equilibrating the fluid phase, inflating the
  particles and then minimizing the energy to produce mechanically
  stable packings at a given packing density. The minimization
  procedure finds individual packings with a probability $p_i$
  proportional to the volume $v_i$ of their basin of attraction. The
  number of jammed states is $\Omega(\phi)=V_J(\phi)/\left<
  v\right>(\phi)$, where $\left< v\right>(\phi)$ is the average basin
  volume and $V_J(\phi)$ the total phase space volume. The observation
  that different basins have different volumes for a range of $\phi$
  values already implies that they will not be equally populated and
  thus equiprobability breaks down for these densities. However, as
  shown in \cite{Asenjo:2014aa}, the granular entropy still satisfies
  extensivity if one considers the Gibbs entropy \be
\label{Ch3:entropy}
S_G^*=-\sum_ip_i\log p_i-\log N!  \ee The subtracted term $\log N!$
ensures that two systems in identical macrostates are in equilibrium
under an exchange of particles and is required for extensivity
\cite{Swendsen:2006aa,Frenkel:2014aa,Cates:2015aa}. In order to test
equiprobability one can compare $S^*_G$ with the likewise modified
Boltzmann expression $S^*_B=\log \Omega(V)-\log N!$. The Gibbs entropy
satisfies $S^*_G\le S^*_B$ with equality when all $p_i$ are equal,
$p_i=1/\Omega$. Remarkably, $S^*_G$ indeed approaches $S_B^*$ as
$\phi\to\phi^*$ for a specific packing density $\phi^*$ (see
Fig.~\ref{Fig:directtest}) \cite{Martiniani:2017aa}. At $\phi^*$ the
basin volumes decorrelate from structural observables such as
pressure, coordination number, etc. Furthermore using a finite size
scaling analysis one can show that $\phi^*$ coincides with the density
at which pressure fluctuations diverge as $N\to\infty$, which is only
possible at the jamming transition $\phi^J$:
$\phi^*_{N\to\infty}=\phi^J_{N\to\infty}$. The comprehensive study in
\cite{Martiniani:2017aa} thus demonstrates that Edwards assumption of
equiprobability indeed holds at the jamming transition, which
corresponds to the point of maximum entropy. Moreover, it is shown
that equiprobability is still satisfied over the whole range of $\phi$
values if one conditions on a fixed value of the pressure indicating
that the generalized stress-volume Edwards ensemble is also a robust
description.

In general, it is important to keep in mind that
  equiprobability will not hold for all possible packing
  algorithms. For example, the protocol used in \cite{Atkinson:2014aa}
  to generate maximally random jammed monodisperse disk packings based
  on a linear programming algorithm \cite{Torquato:2010ac} samples a
  particular subset of all possible jammed states, which have only a
  very low probability of occurrence in the Edwards ensemble. In
  \cite{Charbonneau:2017aa} it is shown that the configurational
  entropy of jammed packings resulting from adiabatic compression of
  glassy states is systematically smaller than the one obtained from
  Edwards uniform measure. Hence, this protocol generates exponentially
  fewer packings than are possible. A framework to include protocol
  dependence in an Edwards-type ensemble has been suggested
  \cite{Paillusson:2015aa}. Even without such an extension, recent
  theoretical work has shown that the predictions resulting from
  Edwards assumptions are indeed in excellent agreement with empirical
  data, confirming, e.g., the critical properties of hard spheres at
  jamming \cite{Charbonneau:2017aa} (see Sec.~\ref{Ch2:Sec:rcp}), and
  jamming densities in a wide range of different systems as reviewed
  in the next Section~\ref{Sec:volume}. Conceptually, it is possible
  to resolve the problem of protocol dependence if one starts from the
  very beginning by defining the metastable jammed states and not the
  protocols, then one avoids the whole question of the ergodic
  hypothesis or protocol dependence or similar issues, which are not
  really essential for Edwards' statistics. We will discuss in detail
  this line of reasoning in Section~\ref{Sec:constraints} by
  exploiting an analogy between metastable jammed states with
  the metastable states of spin-glass systems.

\section{Edwards volume ensemble}
\label{Sec:volume}

In this chapter we focus on the Voronoi convention to define the
microscopic volume function of an assembly of jammed particles. As we
discuss in detail, Edwards statistical mechanics of a restricted
volume ensemble can then be cast into a predictive framework to
determine packing densities for both spherical and non-spherical
particles. In the next sections we outline the
mean-field statistical mechanical approach based on a coarse-graining
of the Voronoi volume function Eq.~(\ref{Ch2:voronoidef}). In
Secs.~\ref{Ch4:Sec:dimension}--\ref{Ch4:Sec:adhesion}, we discuss
different aspects of packings of spheres, such as the effects of
dimensionality, bidispersity, and adhesion. In
Sec.~\ref{Ch4:Sec:nonsphere} we focus on packings of non-spherical
shapes. A comprehensive phase diagram classifying packings of
frictional, frictionless, adhesive spheres and non-spherical shapes is
presented in Sec.~\ref{Ch4:phasediagram}.

\subsection{Mean-field calculation of the microscopic volume function}
\label{Ch4:Sec:volume3d}

The key question is how analytical progress can be made with the
volume function Eq.~(\ref{Ch2:voronoidef}). The global minimization in
the definition of $l_i(\mathbf{ \hat{c}})$, Eq.~(\ref{Ch2:globmin}),
implies that the volume function is a complicated non-local function. This global
character indicates the existence of strong correlations and greatly
complicates the calculation of, e.g., the partition function in the
Edwards ensemble approach. In order to circumvent these difficulties,
we review here a mean-field geometrical viewpoint developed in a
series of papers
\cite{Song:2008aa,Song:2010aa,Wang:2011aa,Wang:2010aa,Briscoe:2010aa,Meyer:2010aa,Wang:2010ab,Wang:2010ac,Briscoe:2008aa,Baule:2013aa,Liu:2015aa,Baule:2014aa,Portal:2013aa,Bo:2014aa},
where the central quantity is not the exact microscopic volume
function, but rather the average or coarse-grained volume of an
individual cell in the Voronoi tessellation. The packing density
$\phi$ of a system of monodisperse particle of volume $V_0$ is given
by \be
\label{Ch4:phi}
\phi=\frac{NV_0}{\sum_{i=1}^N{\cal
    W}_i}=\frac{V_0}{\frac{1}{N}\sum_{i=1}^N{\cal W}_i}.  \ee In the
limit $N\to\infty$ we replace the denominator by the ensemble averaged
volume of an individual cell $\overline{W}=\left<{\cal W}_i\right>_i$: $\frac{1}{N}\sum_{i=1}^N{\cal W}_i \longrightarrow \overline{W}$ as $N\to\infty$. As a result the volume fraction is simply
\be\phi=V_0/\overline{W}.\ee Considering Eq.~(\ref{Ch2:voronoidef}), we
can perform an ensemble average to obtain: \be
\overline{W}&=&\left<\frac{1}{d}\oint\D \mathbf{\hat{c}}\,
l_i(\mathbf{\hat{c}})^d\right>_i=\frac{1}{d}\oint\D \mathbf{\hat{c}}\,
\left<l_i(\mathbf{\hat{c}})^d\right>_i
\nonumber\\ &=&\frac{1}{d}\oint\D
\mathbf{\hat{c}}\int_{c^*(\mathbf{\hat{c}})}^\infty \D c\,c^d
p(\mathbf{c},z).
\label{Ch4:wav}
\ee In the last step we have introduced the pdf $p(\mathbf{c},z)$
which is the probability density to find the Voronoi boundary VB at a
value $c$ in the direction $\mathbf{\hat{c}}$. This involves a lower
cut-off $c^*$ in the direction $\mathbf{\hat{c}}$ due to the hard-core
boundary of the particles. Crucially, we assume that the pdf is a
function of $\mathbf{c}$ and the coordination number $z$ only rather
than a function of the exact particle configurations in the
packing. This is the key step in the coarse-graining procedure, which
replaces the exact microscopic information contained in
$l_i(\mathbf{\hat{c}})$ by a probabilistic quantity. In the following,
we focus on spheres, where $p(\mathbf{c},z)=p(c,z)$ and
$c^*(\mathbf{\hat{c}})=R$ due to the statistical isotropy of the
packing and the isotropy of the reference particle itself. More
complicated shapes will be treated in subsequent sections.

\begin{figure}
\begin{center}
\includegraphics[width=6cm]{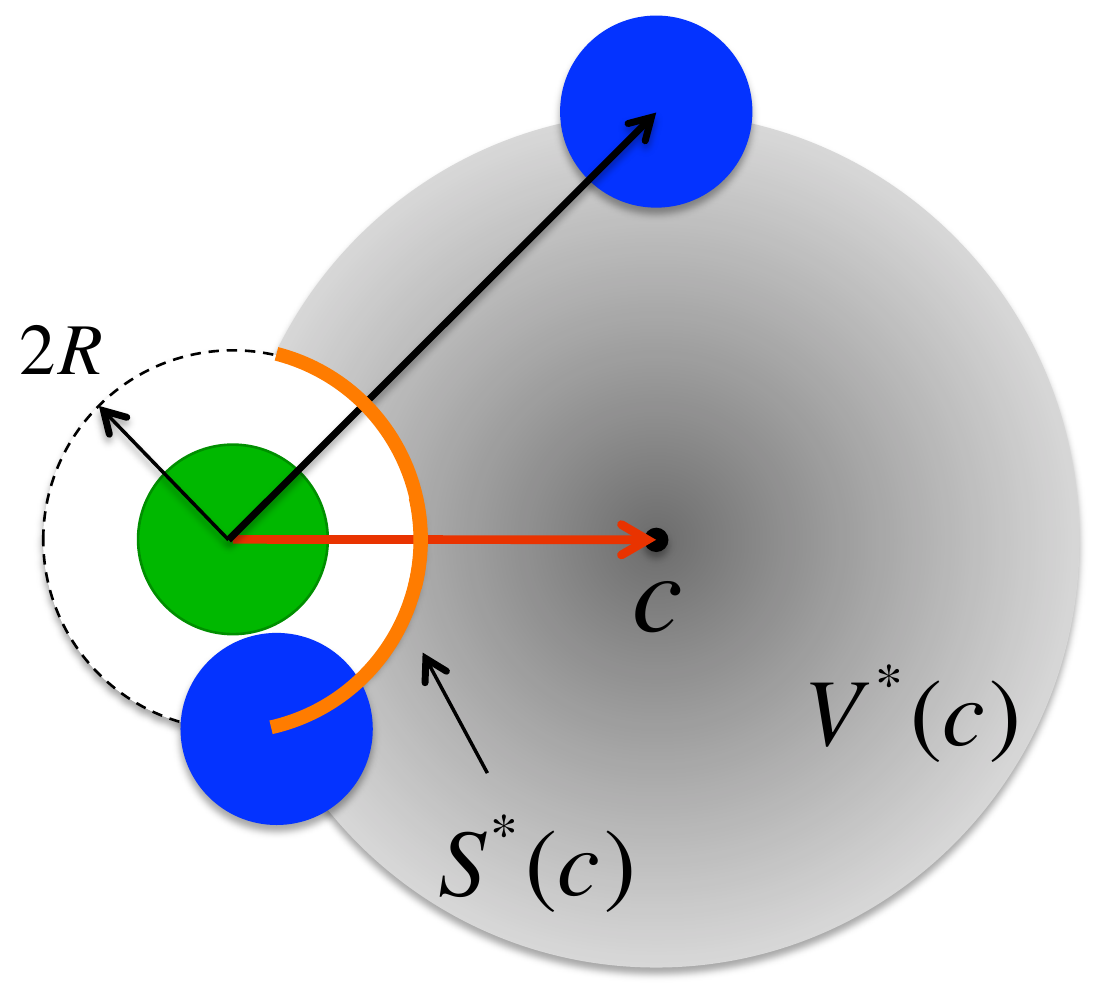}
\caption{\label{Ch4:Fig_voronoi}(Colors online) The condition to have
  the VB in the direction $\hat{s}$ from the reference particle (green
  sphere) at the value $c$ is geometrically related to the exclusion
  volume $\Omega$ for all other particles (blue spheres). Taking into
  account the conventional hard-core excluded volume leads to the
  Voronoi excluded volume Eq.~(\ref{Ch4:vexdef}) (the Moon phase -
  grey volume $V^*$) and Voronoi excluded surface
  Eq.~(\ref{Ch4:vexdef}) (orange line).}
\end{center}
\end{figure}

We now introduce the cumulative distribution function (CDF) $P_>(c,z)$
via the usual definition $p(c,z)=-\frac{\D}{\D
  c}P_>(c,z)$. Eq.~(\ref{Ch4:wav}) becomes then in 3d \be
\label{Ch4:integral}
\overline{W}(z)&=&\frac{4\pi}{3}\int_{R}^\infty \D c\,c^3
p(c,z)\nonumber\\ &=&V_0+4\pi\int_R^\infty \D c\,c^2 \,P_>(c,z), \ee
where $V_0=\frac{4\pi}{3}R^3$. The advantage of using the CDF $P_>$
rather than the pdf, is that the CDF has a simple geometrical
interpretation. We notice first that $P_>$ contains the probability to
find the VB in a given direction $\mathbf{\hat{c}}$ at a value larger
than $c$, given $z$ contacting particles. But this probability equals
the probability that $N-1$ particles are outside a volume $\Omega$
centered at $\mathbf{c}$ relative to the reference particle
(Fig.~\ref{Ch4:Fig_voronoi}). Otherwise, if they were inside that
volume, they would contribute a VB smaller than $c$. The volume
$\Omega$ is thus defined as \be
\label{Ch4:omega}
\Omega(\mathbf{c})=\int \D
\mathbf{r}\,\Theta(c-s(\mathbf{r},\mathbf{\hat{c}}))\Theta(s(\mathbf{r},\mathbf{\hat{c}})),
\ee where $s(\mathbf{r},\mathbf{\hat{c}})$ parametrizes the VB in the
direction $\mathbf{\hat{c}}$ for two spheres of relative position
$\mathbf{r}$. $\Theta(x)$ denotes the usual Heavyside step
function. Due to the isotropy of spheres, the direction
$\mathbf{\hat{c}}$ can be chosen arbitrarily. We refer to $\Omega$ as
the \textit{Voronoi excluded volume}, which extends the standard
concept of the hard-core excluded volume $V_{\rm ex}$ that dominates
the phase behavior of interacting particle systems at thermal
equilibrium \cite{Onsager:1949aa}.

This geometrical interpretation allows us to connect $P_>(c,z)$ with
the $N$-particle pdf
$P_N(\{\mathbf{r}_1,\mathbf{r}_2,...,\mathbf{r}_N\})$ in an exact
way. Without loss of generality we denote the reference particle $i$
as particle $1$. Then, $P_>(c,z)=P_>(\mathbf{r}_1;\Omega)$, i.e., the
probability that the $N-1$ particles apart from particle $1$ are
outside the volume $\Omega$. Since
$P_N(\{\mathbf{r}_1,\mathbf{r}_2,...,\mathbf{r}_N\})$ expresses the
probability to find particle $1$ at $\mathbf{r}_1$, particle $2$ at
$\mathbf{r}_2$, etc., we have \cite{Jin:2010ab} \be
\label{Ch4:cdfex}
P_>(\mathbf{r}_1;\Omega)&=&\mathcal{C}\int\D\mathbf{r}^{N-1}P_N(\{\mathbf{r}_1,\mathbf{r}_2,...,\mathbf{r}_N\})\nonumber\\ &&\times
\prod_{i=2}^N\left[1-m(\mathbf{r}_i-\mathbf{r}_1;\Omega\right], \ee
where $\mathcal{C}$ ensures proper normalization. The indicator
function $m(\mathbf{r};\Omega)$ is given by \be
m(\mathbf{r};\Omega)=\left\{\begin{matrix}1, & \mathbf{r}\in
\Omega\\ & \\ 0, & \mathbf{r}\notin \Omega\end{matrix}\right.  \ee
Equation (\ref{Ch4:cdfex}) is the starting point for the calculation
of $P_>(c,z)$ from a systematic treatment of the particle correlations
as discussed in Sec.~\ref{Ch4:Sec:disks} for 2d packings
\cite{Jin:2014aa} and in Sec.~\ref{Ch4:Sec:dimension} for
high-dimensional packings \cite{Jin:2010ab}. Here, we proceed with a phenomenological approach based on an exact treatment in 1d which is used as an approximation to the 3d case, as originally developed in
\cite{Song:2008aa}.

We can first separate contributions to $P_>$ stemming from bulk and
contacting particles. We introduce two CDFs, the bulk contribution
$P_B$ and the contact contribution $P_C$:
\begin{itemize}
\item $P_{\rm B}$ denotes the probability that spheres in the bulk are
  located outside the Moon-phase grey volume $V^*$ in
  Fig.~\ref{Ch4:Fig_voronoi}. The volume $V^*$ is the volume excluded
  by $\Omega$ for bulk particles and takes into account the overlap
  between $\Omega$ and the hard-core excluded volume $V_{\rm ex}$: \be
\label{Ch4:vexdef}
V^*&=&\Omega-\Omega\cap V_{\rm
  ex}\nonumber\\ &=&\int\D\mathbf{r}\,\Theta(r-2R)\Theta(c-s(\mathbf{r},\mathbf{\hat{c}}))\Theta(s(\mathbf{r},\mathbf{\hat{c}})).
\ee We call $V^*$ the Voronoi excluded volume.
\item $P_{\rm C}$ denotes the probability that contacting spheres are
  located outside the boundary of the grey area indicated in orange in
  Fig.~\ref{Ch4:Fig_voronoi} and denoted $S^*$. The surface $S^*$ is
  the surface excluded by $\Omega$ for contacting particles: \be
\label{Ch4:sexdef}
S^*&=&\partial V_{\rm ex}\cap\Omega\nonumber\\
&=&\left.\oint\D\mathbf{\hat{r}}\,\Theta(c-s(\mathbf{r},\mathbf{\hat{c}}))\Theta(s(\mathbf{r},\mathbf{\hat{c}}))\right|_{r=2R}\, ,
\ee
where $\partial V_{\rm ex}$ denotes the boundary of $V_{\rm ex}$.
\end{itemize}
A key assumption to make analytical progress is to assume $P_{\rm B}$
and $P_{\rm C}$ to be statistically independent, thus $P_> = P_{\rm B}
P_{\rm C}$ . There is no a priori reason why this should be the case,
so the independence should be checked a posteriori from simulation
data. For spheres and non-spherical particles close to the spherical
aspect ratio, it has been verified that independence is a reasonable
assumption \cite{Song:2008aa,Baule:2013aa}. It is then natural to
consider only $P_{\rm C}$ to be a function of $z$. Therefore, \be
\label{Ch4:Pcfactor}
P_>(c,z) =P_{\rm B}(c)  \times P_{\rm C}(c,z).
\ee

We now derive a functional form of the $P_{\rm B}$ term. In 1d, the
distribution of possible arrangements of $N$ hard rods in a volume $V$
can be mapped to the distribution of ideal gas particles by removing
the occupied volume $NV_0$
\cite{Renyi:1958aa,Palasti:1960aa,Krapivsky:1994aa,Tarjus:2004aa}. The
probability to locate one particle at random outside the volume $V^*$
in a system of volume $V-NV_0$ is then $P_>(1)=1-V^*/(V-NV_0)$. For
$N$ ideal particles, we obtain \be
P_>(N)=\left(1-\frac{V^*}{V-NV_0}\right)^N.  \ee The particle density
is $\tilde{\rho}=N/(V-NV_0)$. Therefore \be
\label{Ch4:expo}
\lim_{N\to\infty}P_>(N) = \lim_{N\to\infty}\left(1-\frac{\tilde{\rho} V^*}{N}\right)^N=e^{-\tilde{\rho} V^*}.
\ee
In the thermodynamic limit the probability to observe $N$ particles outside the volume $V^*$ is given by a Boltzmann-like exponential distribution. In this limit, the particle density becomes
\be
\label{Ch4:rho}
\tilde{\rho}=\lim_{N\to\infty}\frac{1}{\frac{1}{N}\sum_{i=1}^N{\cal
    W}_i-V_0}=\frac{1}{\overline{W}-V_0}. \ee
While the above derivation is exact in 1d, the extension to higher dimensions is an
approximation: Even if there is a void with a large enough volume, it
might not be possible to insert a particle due to the constraint
imposed by the geometrical shape of the particles (which does not
exist in 1d). Nevertheless, in what follows, we assume the exponential
distribution of Eq.~(\ref{Ch4:expo}) to be valid in 3d as well and
write \be
\label{Ch4:pb}
P_{\rm B}(c)=e^{-\tilde{\rho} V^*(c)},
\ee
where the Voronoi excluded volume can be calculated explicitly from Eq.~(\ref{Ch4:vexdef}):
\be
\label{Ch4:vexsphere}
V^*(c)&=&V_0\left(\left(\frac{c}{R}\right)^3-4+3\frac{R}{c}\right).
\ee

Furthermore, we also assume $P_{\rm C}$ to have the same exponential
form as Eq.~(\ref{Ch4:pb}), despite not having the large number
approximation leading to it (the maximum coordination is the kissing number 12). Introducing a surface density $\sigma(z)$, we write \be
\label{Ch4:pc}
P_{\rm C}(c)=e^{-\sigma(z) S^*(c)},
\ee
where the Voronoi excluded surface follows from Eq.~(\ref{Ch4:sexdef}):
\be
\label{Ch4:sexsphere}
S^*(c)&=&2S_0\left(1-\frac{R}{c}\right), \ee where $S_0=4\pi R^2$. To
obtain an expression for $\sigma(z)$ we calculate the average
$\left<S^*\right>$ with respect to the pdf $-\frac{\D}{\D c}P_{\rm
  C}(c)$, which yields a simple result
\cite{Song:2008aa,Song:2010aa,Wang:2011aa} \be
\label{Ch4:sigmasrel}
\left< S^*\right> &\approx& 1/\sigma(z).  \ee In turn, $\left<
S^*\right>$ is defined as the average of the solid angles of the gaps
left between $z$ contacting spheres around the reference sphere. An
alternative operational definition assuming an isotropic distribution
of contact particles is:
\begin{enumerate}
\item[(i)] Generate $z$ contacting particles at random.
\item[(ii)] For a given direction $\mathbf{\hat{c}}$, determine the
  minimal value of the VB, denoted by $c_m$.
\item[(iii)] The average $\left< S^*\right>$ follows as a Monte-Carlo
  average in the limit.
\end{enumerate}
\be
\label{Ch4:smc}
\left< S^*\right>=\lim_{n\to\infty}\frac{1}{n}\sum_{i=1}^nS^*(c_{m,i}),
\ee
where $c_{m,i}$ is the $c_m$ value of the $i$th sample. Simulations following this procedure and considering $z=1$ up to the kissing number $z=12$ suggest that
\be
\label{Ch4:sigma}
\sigma(z) \approx \frac{z}{4\pi}\sqrt{3}, \qquad z>1, \ee for a chosen
radius $R=1/2$. The exact constants appearing in this expression are
motivated from an exact treatment of the single particle case plus
corrections due to the occupied surface of contact particles
\cite{Song:2010aa,Wang:2011aa}.

Due to the dependence of $\tilde{\rho}$ on $\overline{W}$, the CDF
$P_>$ is thus \be
\label{Ch4:cdffinal}
P_>(c,z)= \exp \left[-\frac{V^*(c)}{\overline{W}-V_0}-\sigma(z)S^*(c)
  \right], \ee 
where $V^*$, $S^*$, and $\sigma$ are given by
Eqs.~(\ref{Ch4:vexsphere},\ref{Ch4:sexsphere},\ref{Ch4:sigma}).
Overall, Eq.~(\ref{Ch4:cdffinal}) with Eq.~(\ref{Ch4:integral}) leads
to a self-consistent equation to determine $\overline{W}$ as a
function of $z$:
\be
\label{Ch4:self3d}
\overline{W}(z)&=&V_0+4\pi\int_R^\infty \D c\,c^2 \,\exp\left[-\frac{V_0}{\overline{W}(z)-V_0}\right.\times\nonumber\\
&&\times\left.\left(\frac{c^3}{R^3}-4+3\frac{R}{c}\right)-\sigma(z)2S_0\left(1-\frac{R}{c}\right)\right]
\ee for which, remarkably, an analytical solution can be found.  By using
Eqs. (\ref{Ch4:vexsphere},\ref{Ch4:sexsphere}),
Eq.~(\ref{Ch4:self3d}) is satisfied when \cite{Song:2008aa}:
\be \frac{\D}{\D
  c}\left(\frac{1}{w}\left(3\frac{R}{c}\right)+\sigma(z)S^*(c)\right)=0,
\ee where the free volume is $w\equiv(\overline{W}-V_0)/V_0.$ Then, with
Eq.~(\ref{Ch4:sexsphere}) we obtain the solution for $w$ \be
\label{Ch4:wsol}
w(z)=\frac{3}{2S_0\sigma(z)}=\frac{2\sqrt{3}}{z}, \ee using
Eq.~(\ref{Ch4:sigma}) and setting $R=1/2$ for consistency.

As the final result of this section, we arrive at the coarse-grained
mesoscopic volume function \be
\label{Ch4:volfunct}
\overline{W}(z)=V_0+\frac{2\sqrt{3}}{z}V_0, \ee which is a function of
the observable coordination number $z$ rather than the microscopic
configurations of all the particles in the packing. With
Eq.~(\ref{Ch4:phi}), we also obtain the packing density as a function
of $z$ \be
\label{Ch4:phiz}
\phi(z)=\frac{V_0}{\overline{W}}=\frac{z}{z+2\sqrt{3}}.  \ee Equation
~(\ref{Ch4:phiz}) can be interpreted as an equation of state of
disordered sphere packings. In the next section we will show that it
corresponds to the equation of state in $z$--$\phi$ space in the limit
of infinite compactivity.

\subsection{Packing of jammed spheres}


\label{Ch4:Sec:ensemble}

In the hard sphere limit angoricity can be neglected, such that the
statistical mechanics of the packing is described by the volume
function alone. The partition function is then given by Edwards'
canonical one, Eq.~(\ref{Ch2:pfcan}). With the result on the
coarse-grained volume function it is possible to go over from the
fully microscopic partition function Eq.~(\ref{Ch2:pfcan}) to a
mesoscopic one \cite{Song:2008aa,Wang:2011aa}. To this end we change
the integration variables in Eq.~(\ref{Ch2:pfcan}) from the set of
microscopic configurations
$\mathbf{q}=\{\mathbf{q}_1,...,\mathbf{q}_N\}$ (positions and
orientations of the $N$ particles) to the volumes
$\mathcal{W}_i(\mathbf{q})$, Eq.~\ref{Ch2:voronoidef}, of each cell in
the Voronoi tessellation. Since the microscopic volume function is
given as a superposition of the individual cells,
Eq.~(\ref{Ch2:volsum}), the partition function Eq.~(\ref{Ch2:pfcan})
can be expressed as \be
\mathcal{Z}=\prod_{i=1}^N\int\D\mathcal{W}_i\,g(\boldsymbol{\mathcal{W}})e^{-\sum_{i=1}^N\mathcal{W}_i/X}\Theta_{\rm
  jam}.  \ee Here, the function $g(\boldsymbol{\mathcal{W}})$ for
$\boldsymbol{\mathcal{W}}=\{\mathcal{W}_1,...,\mathcal{W}_N\}$ denotes
the density of states. In the coarse-grained picture all the volume
cells are non-interacting and effectively replaced by the volume
function Eq.~(\ref{Ch4:volfunct}). The partition function thus
factorizes $\mathcal{Z}=\mathcal{Z}_i^N$, where \be
\label{Ch4:Q}
\mathcal{Z}_i(X)=\left(\int\D W\,g(W)e^{-W/X}\Theta_{\rm jam}\right)^N
\ee
Averages over the volume ensemble as well as all
thermodynamic information is thus accessible via Eq.~(\ref{Ch4:Q}). The crucial step to go from the full microscopic partition function Eq.~(\ref{Ch2:pfcan}) to Eq.~(\ref{Ch4:Q}) is to introduce
the density of states $g(W)$ for a given volume $W$. Although this
step formally simplifies the integral, the complexity of the problem
is now transferred to determining $g(W)$, which is in principle as difficult to
solve as the model itself. In Eq.~(\ref{Ch4:Q}), $X$ is the compactivity measured in units of the
particle volume $V_0$, and $\Theta_{\rm jam}$ imposes the condition of
jamming. 

In the mean-field view developed in the previous section, $W$ is
directly related to the geometrical coordination number $z$ via
Eq.~(\ref{Ch4:volfunct}). Therefore, we map $g(W)$ to $g(z)$, the
density of states for a given $z$ via a change of variables $g(W)=\int P(W|z)g(z)\D z$, where $P(W|z)$ is the conditional
probability of a volume $W$ for a given $z$, which, with
Eq.~(\ref{Ch4:volfunct}), is given by $P(W|z)=\delta(W-\overline{W}(z))$, where we have neglected
fluctuations in $z$, see \cite{Wang:2010ac}. Substituting these two equations into
Eq.~(\ref{Ch4:Q}) effectively changes the integration variable from
$W$ to $z$ leading to the single particle (isostatic) partition
function \be
\label{Ch4:Qz}
\mathcal{Z}_{\rm iso}(X,Z_{\rm m}) =\int_{Z_{\rm m}}^6 g(z)
\,\exp\left[-\frac{2\sqrt{3}}{zX}\right]\D z.  \ee The jamming
condition is now absorbed into the integration range, which constrains
the coordination number to isostatic packings (therefore the name
isostatic partition function). Notice that in this mesoscopic
mean-field approach the force and torque balance jamming conditions
from $\Theta_{\rm jam}$ Eq.~(\ref{eq:thetajam}) are incorporated when
we set the coordination number to the isostatic value. Thus, in this
way, we circumvent the most difficult problem of implementing the
force jamming condition Eq.~(\ref{eq:thetajam}).

More precisely, the geometric and force/torque constraints from
Eq.~(\ref{eq:thetajam}) imply that there are two types of coordination
numbers: {\it (i)} the geometrical coordination number $z$, parametrizing the
free volume function Eq.~(\ref{Ch4:wsol}) as a function of all
contacting particles, constraining the position of the particle via
the hard-core geometrical interaction Eq.~(\ref{eq:hardcore}). {\it (ii)} The mechanical coordination number $Z_{\rm m}$, counting
only the geometrical contacts $z$ that at the same time carry non-zero
force \cite{Oron:1998aa,Oron:1999aa} and therefore takes into account the force and torque balance
conditions Eqs. (\ref{Ch2:forceb})-(\ref{Ch2:third}) via the
isostatic condition.

From the definition we have $z\ge Z_m$ since there could be a geometric contact
that constraints the motion of the particle but carries no force. This
distinction makes sense when there is friction in the packing.  For
instance, imagine a frictionless particle at the isostatic point
$z=Z_m=6$ (although isostatic is a global property).  Now add friction
to the interactions. The mechanical coordination number can be as low
as $Z_m=4$, but still $z=6$;  the geometrical constraints are the
same, only two forces have been set to zero, allowing for tangential
forces to appear in the remaining 4 contacts.

\begin{figure*}
\begin{center}
\includegraphics[width=16cm]{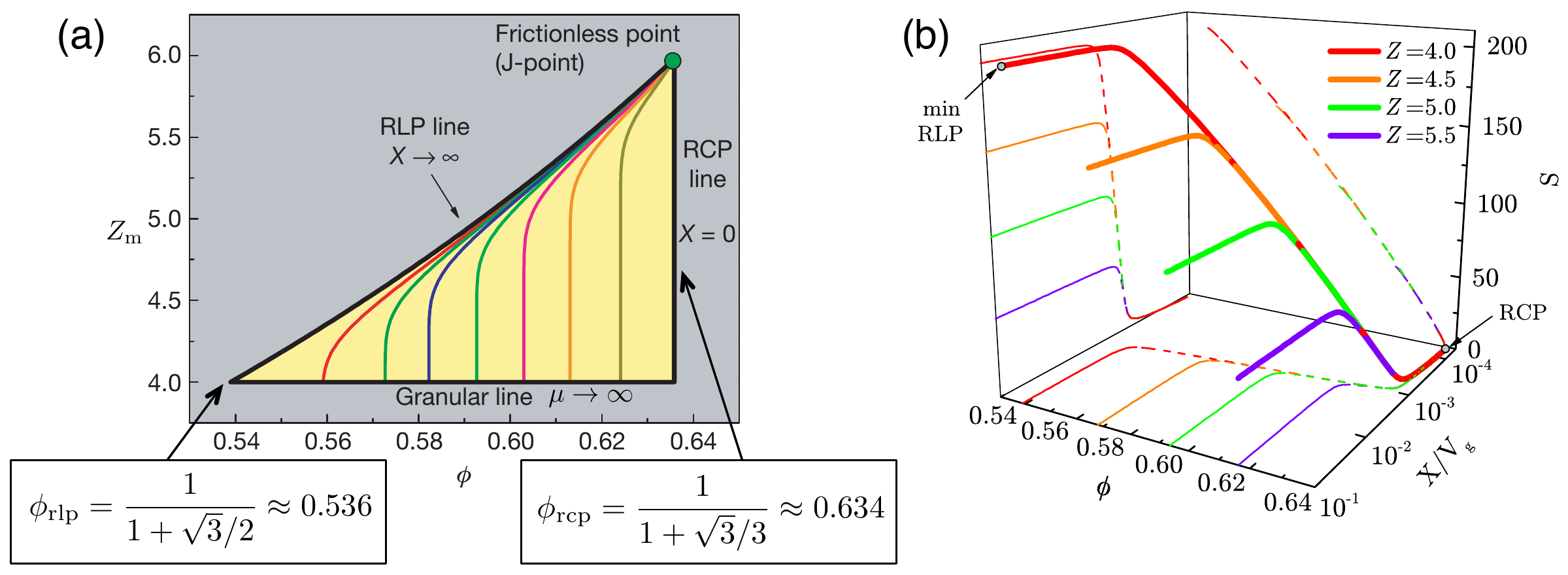}
\caption{\label{Ch4:Fig_xlines}(Colors online) (a) Theoretical
  prediction of the statistical theory Eq.~(\ref{Ch4:phiav}). All
  disordered packings of spheres lie within the yellow triangle
  demarcated by the RCP line at $\phi_{\rm rcp} =  0.634..$, the RLP
  line parametrized by Eq.~(\ref{Ch4:phirlp}) and the lower limit for
  stable packings at $Z=4$ (granular line) for $\mu\to\infty$. Lines
  of constant finite compactivity $X$ are in colour. Packings are
  forbidden in the grey area. (b) Predictions of the equation of state
  of jammed matter in the $(X,\phi,s)$--space determined with
  Eq.~(\ref{Ch4:entropy}). Each line corresponds to a different system
  with $Z_{\rm m}(\mu)$ as indicated. The projections in the
  $(\phi,s)$ and $(X,s)$ planes show that RCP ($X=0$) is less
  disordered than RLP ($X\to\infty$). Adapted from \cite{Song:2008aa}.}
\end{center}
\end{figure*}

For frictionless packings, we have $z=Z_m$. Furthermore, in the limit
of infinite compactivity, where the entropy of the packings is maximum
and therefore, the packings are the most probable to find in
experiments, we will see that again $z=Z_m$ and the distinction
between mechanical and geometrical coordination number disappears.  In
what follows, we will consider the consequences of considering the two
coordination numbers only for the following 3d monodisperse system of
spheres.  The distinction between $z$ and $Z_m$ will allow us to
describe the phase diagram for all compactivities as in
Fig.~\ref{Ch4:Fig_xlines}a, below.  In the remaining sections where we
treat non-spherical particles and others, either we will assume
frictionless particles or packings at infinite compactivity for which
we simply set $z=Z_m$ and get a single equation of state rather than
the yellow area in Fig.~\ref{Ch4:Fig_xlines}a.

The mechanical coordination $Z_{\rm m}$ defines isostatic packings,
which, strictly applies only to the two limits $Z_{\rm m}=2d=6$ for
frictionless particles with friction $\mu\to 0$ and $Z_{\rm m}=d+1=4$
for infinitely rough particles $\mu\to\infty$. An important assumption
is that $Z_{\rm m}$ varies continuously as a function of $\mu$ \be
4\le Z_{\rm m}(\mu) \le z \le 6.  \ee In fact, a universal $Z_{\rm
  m}(\mu)$ curve has been observed for a range of different packing
protocols \cite{Song:2008aa} and calculated analytically in
\cite{Bo:2014aa}. The upper bound of $z$ is the frictionless isostatic
limit. This effectively excludes from the ensemble the partially
crystalline packings, which are characterized by larger $z$. 

The remaining unknown is the density of states $g(z)$, which can be determined using analogies with a quantum mechanical system (see appendix~\ref{App:densitystates}) leading to
\be
\label{Ch4b:gz}
g(z)= (h_z)^{z-\mathcal{D}},  \ee
where $\mathcal{D}$ is the dimension per particle of the
configuration space and $h_z$ a typical distance between jammed configurations in this space. Note that the factor
$(h_z)^{-\mathcal{D}}$ will drop out when performing ensemble
averages. Physically, we expect $h_z\ll 1$. The exact value of $h_z$
can be determined by a fitting of the theoretical values to the
simulation data, but it is not important as long as we take the limit
at the end: $h_z\to 0$.

Having defined the jammed ensemble via the partition function
$\mathcal{Z}_{\rm iso}$, we can calculate the ensemble averaged
packing density $\phi(X,Z_{\rm m})=\left<\phi(z)\right>$ as \be
\label{Ch4:phiav}
\phi(X,Z_{\rm m})=\frac{1}{\mathcal{Z}_{\rm iso}}\int_{Z_{\rm
    m}}^6\frac{z}{z+2\sqrt{3}}e^{-\frac{2\sqrt{3}}{zX}+z\log h_z}\D z.
\ee 

Equation ~(\ref{Ch4:phiav}) gives predictions on the packing densities
as a function of $X$ over the whole range of friction values
$\mu\in[0,\infty)$ since $Z_m(\mu)$ is determined by friction
  \cite{Song:2008aa}.  We can identify three distinct regimes (see
  Fig.~\ref{Ch4:Fig_xlines}):
\begin{enumerate}
\item In the limit of vanishing compactivity ($X\to 0$), only the
  minimum volume at $z=6$ contributes. The density is the RCP limit $\phi_{\rm
    rcp}=\phi(X=0,Z)$: \be
\label{Ch4:phircp}
\phi_{\rm rcp} = \frac{1}{1+1/\sqrt{3}} = 0.634..,\qquad Z_{\rm
  m}(\mu)\in[4,6], \ee and the corresponding RCP free volume is \be\label{Ch4:wrcp}
w_{\rm rcp} = \frac{1}{\sqrt{3}}.  \ee
$\phi_{\rm rcp}$ defines a vertical line in the phase diagram ending
at the $J$-point: (0.634, 6). Here, RCP is identified as the ground
state of the jammed ensemble with maximal density and coordination
number. Notice that this result is also obtained from
Eq.~(\ref{Ch4:phiz}) at $z=6$.

\item In the limit of infinite compactivity ($X \to \infty$), the
  Boltzmann factor $\exp[-2 \sqrt{3}/(z X)] \to 1$, and the average in
  Eq.~(\ref{Ch4:phiav}) is taken over all states with equal
  probability. The $X \to \infty$ limit defines the random loose
  packing equation of state $ \phi_{\rm rlp}(Z) = \phi(X\to
  \infty,Z_{\rm m})$ as a function of $Z_{\rm m}$: \be \phi_{\rm
    rlp}(Z_{\rm m}) &=&\frac{1}{\mathcal{Z}_{\rm iso}(\infty,Z_{\rm
      m})}\int_{Z_{\rm m}}^6 \frac{z}{z+2\sqrt{3}}e^{z \ln h_z}\D
  z\nonumber\\ &\approx&\frac{Z_{\rm m}}{Z_{\rm m}+2\sqrt{3}},\qquad
  Z_{\rm m}(\mu)\in[4,6].
\label{Ch4:phirlp}
\ee The approximation comes from $h_z\to 0$. For small but finite
$h_z\ll 1$, an interesting regime appears of negative compactivity
\cite{Briscoe:2010aa}, yet unstable, leading to the limit of RLP when
$X\to 0^-$ which has been termed as the random very loose packing
\cite{Ciamarra:2008aa}. Thus, $\phi_{\rm rlp}$ spans a whole line in
the phase diagram between the frictionless value $\phi_{\rm rcp}$ upto
the limit $\mu\to\infty$ at:

\be \phi_{\rm rlp}^{\rm min}=\frac{1}{1+\sqrt{3}/2}=
0.536..,\qquad{\rm for} \,\,Z_m=4.\ee The corresponding RLP
free-volume is $w_{\rm rlp}^{\rm min}=\sqrt{3}/{2}$. These values are interpreted as the minimal density of mechanically
stable sphere packings appearing at $Z_m=4$.  We notice that
Eq.~(\ref{Ch4:phirlp}) can be obtained from the single particle
Eq.~(\ref{Ch4:phiz}), by setting $z=Z_m$. Indeed, in the limit of
infinite compactivity the mechanical coordination takes the value of
the geometrical one.

\item Finite compactivity $X$ defines the packings inside the triangle
  bounded by the RCP and RLP lines and the limit for isostaticity
  $Z_{\rm m}=4$ as $\mu\to\infty$ (granular line) are
  characterized. In this case, Eq.~(\ref{Ch4:phiav}) can be solved
  numerically. Figure \ref{Ch4:Fig_xlines}a shows the lines of constant
  compactivity plotted parametrically as a function of $Z_{\rm m}$.

\end{enumerate}

Further thermodynamic characterisation is obtained by considering the
entropy of the jammed configurations, which can be identified by
analogy with the equilibrium framework. In equilibrium statistical
mechanics we have $F=E-TS$, such that $S=E/T+\ln \mathcal{Z}$ using
the free energy expression $F=-T\ln \mathcal{Z}$ (setting $k_{\rm B}$
to unity).  By analogy we obtain the entropy density of the jammed
configuration $s(X,Z_{\rm m})$ (entropy per particle)
\cite{Brujic:2007aa,Briscoe:2008aa,Briscoe:2010aa}: \be
\label{Ch4:entropy}
s(X,Z_{\rm m})&=&\left<W\right>/X+\ln\mathcal{Z}_{\rm iso} \ee
substituting the partition function Eq.~(\ref{Ch4:Qz}) in the last
step. In Fig.~\ref{Ch4:Fig_xlines}b each curve corresponds to a
packing with a different $Z_m$ value determined by
Eq.~(\ref{Ch4:entropy}). The projections $s(\phi)$ and $s(X)$
characterize the nature of randomness in the packings. When comparing
all the packings, the maximum entropy is at $\phi_{\rm rlp}$ for
$X\to\infty$, while the entropy is minimum at $\phi_{\rm rcp}$ for
$X\to 0$. Following the granular line in the phase diagram we obtain
the entropy for infinitely rough spheres showing a larger entropy for
the RLP than the RCP. The same conclusion is obtained for the other
packings at finite friction ($4<Z_{\rm m}<6$). We conclude that the
RLP states are more disordered than the RCP states. 


As stated, in the following results we will focus always on the
$X\to\infty$ regime, where the volume function that is obtained from
the solution of the self-consistent equation is also the equation of
state, since we simply have $z\to Z_m$ for $X\to\infty$ when
calculating the ensemble averaged packing density (compare
Eqs.~(\ref{Ch4:phiz}) and (\ref{Ch4:phirlp})). Therefore, we can drop
the distinction between $Z_m$ and $z$ (for simplicity we consider
$z$), while keeping in mind that there exist further packing states
for finite $X$ that are implied but not explicitly discussed in the
next sections (e.g., in the full phase diagram
Fig.~\ref{Ch4:Fig_phasediagram}).

\subsection{Packing of high-dimensional spheres}

\label{Ch4:Sec:dimension}

According to Eq.~(\ref{Ch4:integral}), the key quantity to calculate
exactly the average volume $\overline{W}$ is the CDF
$P_>(\mathbf{r}_1;\Omega)$ as defined in Eq.~(\ref{Ch4:cdfex}).  This
CDF has been approximated in the work of \cite{Song:2008aa} reviewed
in previous Section \ref{Ch4:Sec:volume3d} by using a simple one
dimensional gas-like model which is analogous in 1d to a parking lot
model
\cite{Renyi:1958aa,Palasti:1960aa,Krapivsky:1994aa,Tarjus:2004aa},
leading to the exponential form (\ref{Ch4:cdffinal}). It turns out
that in the opposite limit of infinite dimensions (mean-field), a
closed form of $P_>$ can be obtained as well, based on general
considerations of correlations in liquid state theory. In this
mean-field high-$d$ limit, the form obtained in \cite{Song:2008aa} can
be determined as a limiting case, with the added possibility to develop
a systematic expansion of $P_>$ in terms of pair distribution functions
allowing to include higher order correlations which were neglected in
\cite{Song:2008aa}. Furthermore, the high-$d$ limit is important to
compare the predictions of the Edwards ensemble to other mean-field
theories such as the RSB solution of hard-sphere packings
\cite{Parisi:2010aa}. The high-dimensional limit is treated next
\cite{Jin:2010ab}.

In large dimensions, the effect of metastability between amorphous and
crystalline phases is strongly reduced, because nucleation is
increasingly suppressed for large $d$
\cite{Skoge:2006aa,Meel:2009aa,Meel:2009ab}. Moreover, mean-field
theory becomes exact for $d\to\infty$, because each degree of freedom
interacts with a large number of neighbours \cite{Parisi:1988aa}
opening up the possibility for exact solutions.

In the following, we discuss the mean-field high-dimensional limit of
the coarse-grained Voronoi volume theory starting from liquid state
theory. We only sketch the main steps in the calculation, for full
details we refer to \cite{Jin:2010ab}. Assuming translational
invariance of the system, Eq.~(\ref{Ch4:cdfex}) can be rewritten as
\be &&P_>(\mathbf{r}_1;\Omega) = 1+\sum_{k=1}^{N-1}(-1)^k
\frac{\rho^k}{k!}\nonumber\\ &&\times\int_{\Omega}
g_{k+1}(\mathbf{r}_{12},\ldots,\mathbf{r}_{1(k+1)}
)d\mathbf{r}_{1i}\cdots d\mathbf{r}_{1(k+1)},
\label{Ch4:Pg}
\ee where $g_n$ denotes the $n$-particle correlation function \be
&&g_n(\mathbf{r}_{12},\mathbf{r}_{13},...,\mathbf{r}_{1n})\nonumber\\ &&=\frac{N!}{\rho^n(N-n)!}\int
P_N(\mathbf{r}^n,\mathbf{r}^{N-n})\D \mathbf{r}^{N-n}, \ee with
$\rho=N/V$ the particle density. The integrals in Eq.~(\ref{Ch4:Pg})
express the probabilities of finding a pair, triplet, etc., of spheres
within the volume $\Omega$. For an exact calculation of $P_>$, we thus
need the exact form of $g_n(\mathbf{r}_{12},\mathbf{r}_{13}\ldots
\mathbf{r}_{1n})$ to all orders, which is not available. However,
assuming the generalized Kirkwood superposition approximation from
liquid theory \cite{Kirkwood:1935aa}, we can approximate $g_n$ in high
dimensions by a simple factorized form \cite{Jin:2010ab}: \be
\label{Ch4:gnsimple}
g_n(\mathbf{r}_{12},\mathbf{r}_{13},\ldots, \mathbf{r}_{1n}) \approx
\prod_{i=2}^{n}g_2(\mathbf{r}_{1i}), \ee where $g_2$ is the pair
correlation function.

Equation ~(\ref{Ch4:gnsimple}) indicates that spheres $2,...,n$ are
correlated with the central sphere $1$ but not with each other, which
is reasonable for large $d$ since the sphere surface is then large
compared with the occupied surface. The term $S_{d-1}$ in
Eq.~(\ref{Ch4:g2simple}) denotes the surface of a $d$-dimensional
sphere with radius $2R$. Substituting Eq.~(\ref{Ch4:gnsimple}) in
Eq.~(\ref{Ch4:Pg}) yields \be
P_>(\mathbf{r}_1;\Omega)&=&\sum_{k=0}^{N-1}(-1)^k\frac{\rho^k}{k!}\left(
\int_{\Omega}g_2(\mathbf{r})d\mathbf{r}
\right)^k\nonumber\\ &=&\exp\left[-\rho \int_{\Omega}
  g_2(\mathbf{r})d\mathbf{r}\right],
\label{Ch4:Pg2}
\ee in the limit $N\rightarrow \infty$ ($\rho\to1/\overline{W}$).

Thus, we see that calculating the CDF $P_>$ reduces to know the form
of the pair correlation function. Indeed, the exponential form
calculated in Section \ref{Ch4:Sec:volume3d} using a 1d model,
Eq.~(\ref{Ch4:cdffinal}), is obtained from Eq.~(\ref{Ch4:Pg2}) by assuming
the following simplified pair correlation function (which has been
considered also in \cite{Torquato:2006aa}): \be
\label{Ch4:g2simple}
g_2(r) = \frac{z}{\rho S_{d-1}}\delta(r-2R)+\Theta(r-2R). \ee

\begin{figure}
\includegraphics[width=10cm]{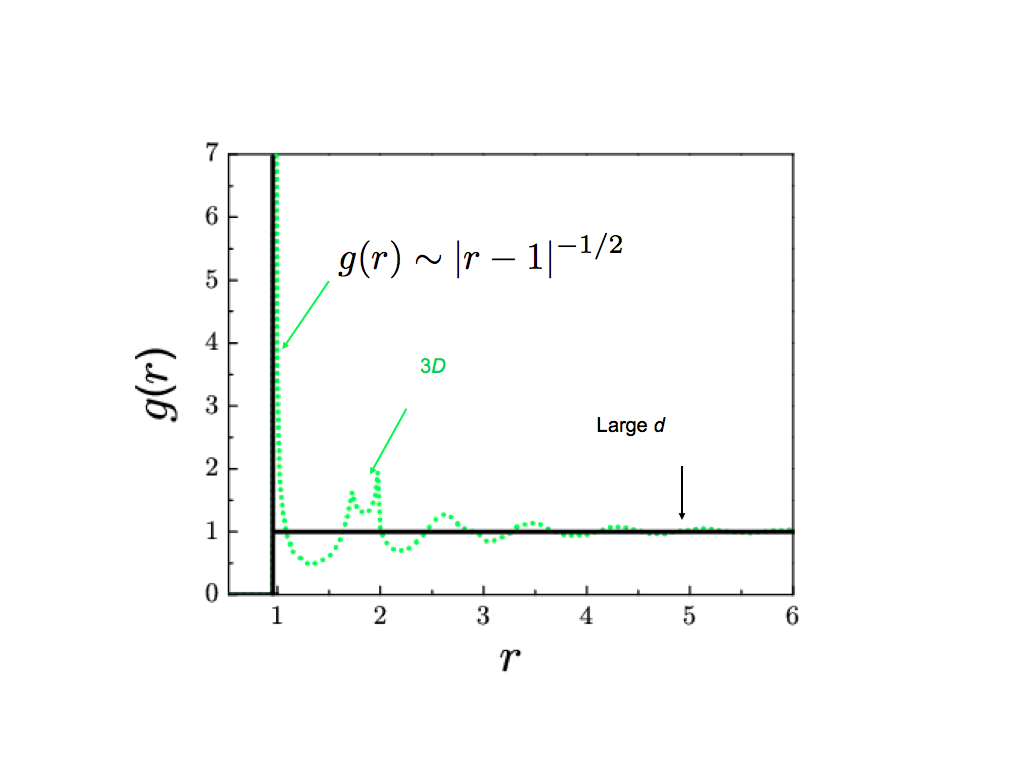}
\caption{\label{Fig:g2}(Colors online) At the core of the mean-field
  approach developed in \cite{Song:2008aa} to calculate the volume
  fraction of 3d packings is the approximation of the real pair
  correlation function (green curve) with its characteristic peaks
  indicating short-range correlations in the packing and the power-law
  decay of the near contacting particles, Eq.~(\ref{Ch2:gammaexp}), by
  a simple delta-function (black curve) at the contacting point plus a
  flat distribution charactering a gas-like bulk of uncorrelated
  particles. Surprisingly, such an approximation, which is expected to
  work better at high dimensions than at low dimensions, gives
  accurate results for the volume fraction in 3d, as shown in Section
  \ref{Ch4:Sec:volume3d}. High-dimensional analyses allow to treat
  higher-order correlations neglected in \cite{Song:2008aa} to improve
  the theoretical predictions in a systematic way as shown in
  Secs. \ref{Ch4:Sec:dimension}, \ref{Ch4:Sec:disks} and
  \ref{Ch4:Sec:adhesion}.}
\end{figure}

This form corresponds to assuming a set of $z$ contacting particles
contributing to the delta-peak at $2R$ plus a set of uncorrelated bulk
particles contributing to a flat (gas-like) distribution characterized
by the $\Theta$-function. This form, depicted in Fig.~\ref{Fig:g2},
further assumes the factorization of the contact and bulk distribution
and represents the simplest form of the pair correlation function,
yet, it gives rise to accurate results for the predicted packing
densities. The important point is that the high-d result
Eq.~(\ref{Ch4:Pg2}) allows to express more accurate pair correlation
functions than Eq.~(\ref{Ch4:g2simple}) into the formalism to
systematically capture higher order features in the correlations, thus
allowing for an improvement of the theoretical results. Such
improvements are treated in Sections \ref{Ch4:Sec:disks} and
\ref{Ch4:Sec:adhesion}.

Using Eq.~(\ref{Ch4:g2simple}) and the definition of $\Omega$,
Eq.~(\ref{Ch4:omega}), we see that the volume integral $\int_{\Omega}
g_2(\mathbf{r})d\mathbf{r}$ becomes \be \int_{\Omega}
g_2(\mathbf{r})d\mathbf{r}=\frac{z S^*(c)}{\rho S_{d-1}}+V^*(c), \ee
where $V^*$ and $S^*$ are the Voronoi excluded volume and surface,
Eqs.~(\ref{Ch4:vexdef},\ref{Ch4:sexdef}), for general $d$. We thus
recover the same factorized form of the CDF as in 3d,
Eq.~(\ref{Ch4:cdffinal}), but now generalized to any dimension $d$,
separating bulk and contact contributions \be
\label{Ch4:CDFd}
P_>(c,z) &=& \exp \left[ - \rho V^*(c) -\frac{z S^*(c)}{S_{d-1}}
  \right], \ee whose validity should increase with increasing
dimension. The Voronoi excluded
volume and surface, $V^*$ and $S^*$, can be calculated with
Eqs.~(\ref{Ch4:vexdef},\ref{Ch4:sexdef}) for general $d$. The term
$z/S_{d-1}$ can be interpreted as the surface density $\sigma(z)$ in
the 3d theory.

The $d$-dimensional generalization of Eq.~(\ref{Ch4:integral}) is
\be
\label{Ch4:integrald}
\overline{W}=V_0^{(d)}+\frac{V_0^{(d)}d}{R^d}\int_R^\infty\D
c\,c^{d-1}P_>(c,z).  \ee For large $d$ an analytical solution of
Eq.~(\ref{Ch4:integrald}) can be obtained. In terms of
$w=(\overline{W}-V^{(d)}_0)/V^{(d)}_0$ one obtains the following
asymptotic predictions of the Edwards ensemble in high-d
\cite{Jin:2010ab} for the free volume: $w_{\rm Edw}= \frac{3}{4d}2^d$,
and the volume fraction in the Edwards ensemble is
\be
\label{highd}
\phi_{\rm Edw}=\frac{4}{3}d\,2^{-d}.
\ee
The scaling $\phi\sim d \, 2^{-d}$ is also found in other approaches for jammed
spheres in high dimensions. In principle, it satisfies the Minkowski
lower bound \cite{Torquato:2010aa}:
\be \phi_{\rm Mink} = \frac{\zeta(d)}{2}2^{-d},\ee
where $\zeta(d)$ is the Riemann zeta function, $\zeta(d) = \sum_{k=1}^\infty
\frac{1}{k^d}$, although this can be regarded as a minimal
requirement. Density functional theory predicts
\cite{Kirkpatrick:1987aa}: \be \phi_{\rm dft} \sim 4.13\, d \, 2^{-d}.\ee
Mode-coupling theory with a Gaussian correction predicts
\cite{Kirkpatrick:1987aa,Ikeda:2010aa}: \be \phi_{\rm mct} \sim
8.26 \, d \, 2^{-d}.\ee 
Replica symmetry breaking theory at the 1 step
predicts \cite{Parisi:2010aa}
\be\phi^{\rm 1RSB}_{\rm th} \sim 6.26 \, d \, 2^{-d},
\label{rsb-d}
\ee
and the full RSB solution predicts \cite{Charbonneau:2014aa}
\be
\phi^{\rm fullRSB}_{\rm th} \sim 6.85 \, d \, 2^{-d}
\ee
as the lower limit of jamming in the J-line ($\phi_j\in[\phi_{\rm th},\phi_{\rm GCP}$).

In general, we see that the Edwards prediction has the same
asymptotic dependence on $d$, Eq.~(\ref{highd}), as the competing
theories. However the prefactors are in disagreement, especially with
the 1RSB calculation. While Edwards ensemble predicts a prefactor
$4/3$, the 1RSB prediction is $6.26$. A comparison of the
large $d$ results for $P_{\rm B}$ and $P_{\rm C}$ with those in 3d indicates that the low $d$
corrections are primarily manifest in the expressions for particle
density $\rho$ and the surface density $\sigma(z)=z/S_{d-1}$ \cite{Jin:2010ab}. In 3d,
the density exhibits van der Waals like corrections due to the
particle volume: $\rho\to
\tilde{\rho}=1/(\overline{W}-V_0)$. Likewise, there are small
corrections to the surface density $z/4\pi\to
\left<S^*\right>^{-1}\approx(z/4\pi)\sqrt{3}$. The origin of the
additional $\sqrt{3}$ factor is not clear. In 2d, further corrections
are needed to obtain agreement of the theory with simulation data, a
case that is treated next.

\subsection{Packing of disks }

\label{Ch4:Sec:disks}

The high-dimensional treatment discussed in the previous section shows
that improvements on the mean field approach of \cite{Song:2008aa} can
be achieved through better approximations to the pair distribution
function by including neglected correlations between neighboring
particles.  These correlations become crucial in low-dimensional
systems, in particular in 2d systems of disk packings.  Interestingly,
below we show that the 2d case allows for a systematic improvement of
the predictions based on a systematic layer expansion of the pair
distribution function through a dimensional reduction of the problem
to a one-dimensional one, as treated next.

In principle, disordered packings of monodisperse disks are difficult
to investigate in 2d, since crystallization typically prevents the
formation of an amorphous jammed state. In \cite{Berryman:1983aa} the
density of jammed disks has been estimated as $\phi_{\rm
  rcp}=0.82\pm0.02$ by extrapolating from the liquid phase. Only
recently, MRJ states of disks have been generated in simulations using
a linear programming algorithm \cite{Torquato:2010ac}. These packings
achieve a packing fraction of $\phi_{\rm mrj}=0.826$ including
rattlers and exhibit an isostatic jammed backbone
\cite{Atkinson:2014aa}. By comparison, the densest crystalline
arrangement of disks is a triangular lattice with
$\phi=\frac{\pi}{\sqrt{12}}\approx 0.9069$, which has already been
proven by Thue \cite{Thue:1892aa}. For disordered packings,
replica theory predicts the J-line in 2d from $\phi_{\rm th}=0.8165$
to the maximum density of glass close packing at $\phi_{\rm
  GCP}=0.8745$ \cite{Parisi:2010aa}, although these values have a large error bar due to the liquid theory approximation used in the calculation. A recent theory based on the
geometric structure approach estimates $\phi_{\rm mrj}=0.834$
\cite{Tian:2015aa}.

In order to elucidate the 2d problem from the viewpoint of the Edwards
ensemble, one can adapt as a first approach the same statistical
theory developed for 3d spheres in Sec.~\ref{Ch4:Sec:volume3d} to the
2d case. This would lead to a self-consistent equation for the average
Voronoi volume as in Eq.~(\ref{Ch4:integral}) \cite{Meyer:2010aa}: \be
\label{Ch4:self2d}
\overline{W}(z)&=&V_0+2\pi\int_R^\infty\D c\,c\,P_>(c,z), \ee where
$P_>(c,z)$ has the form of Eq.~(\ref{Ch4:cdffinal}) with $V_0=\pi R^2$
and the 2d analogues of $V^*$ and $S^*$ are easily calculated. The
surface density $\sigma(z)$ follows from simulations of local
configurations via Eq.~(\ref{Ch4:sigmasrel}). In the relevant $z$
range between the isostatic frictionless value $z=2d=4$ and the lower
limit $z=d+1=3$ for frictional disks, $\sigma(z)$ is found to be
approximately linear: $\sigma(z)=(z-0.5)/\pi$ for $R=1/2$
\cite{Meyer:2010aa}.

Overall, such an implementation would predict a RCP density of 2d
frictionless disks of $\phi_{\rm rcp}\approx 0.89$ greatly exceeding
the empirical values. The reason for the discrepancy are much stronger
correlations between the contact and bulk particles in low dimensions,
such that the assumed independence of the CDFs $P_{\rm B}$ and $P_{\rm
  C}$ in Eq.~(\ref{Ch4:Pcfactor}) is no longer valid. A
phenomenological way to quantify the correlations by coupling bulk and surface terms has been discussed
in \cite{Meyer:2010aa} leading to better agreement with simulation data.

\begin{figure}
\begin{center}
\includegraphics[width=7cm]{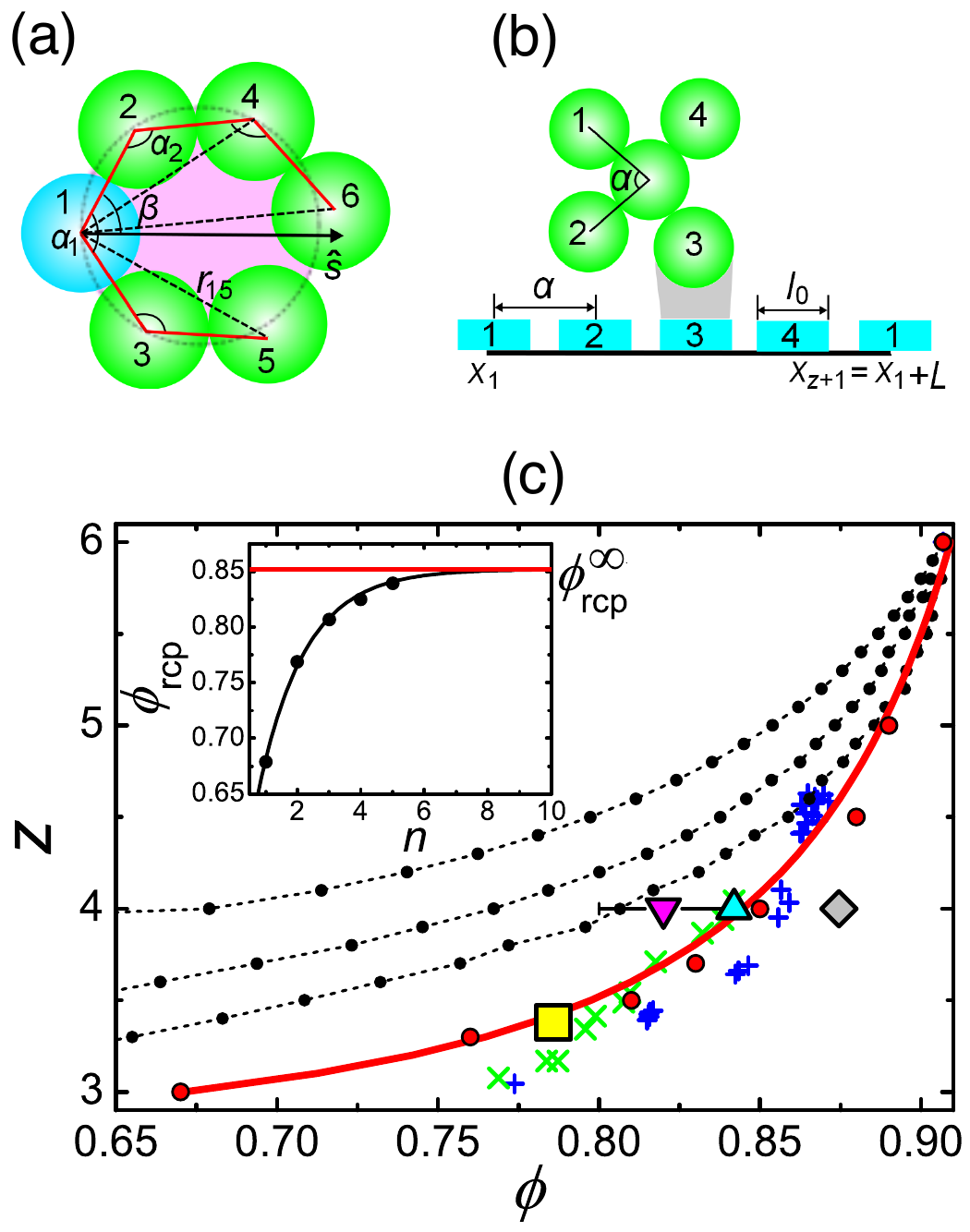}\\
\caption{\label{Ch4:Fig_disks}(Colors online) (a) An illustration of
  the geometrical quantities used in the calculation of $P_>$,
  Eq.~(\ref{Ch4:angles2d}). The $\alpha_j$ are the angles between any
  two Voronoi particles for a given $\hat{s}$. (b) Mapping
  monodisperse contact disks to 1d rods. The 2d exclusive angle
  $\alpha$ corresponds to the 1d gap. (c) Phase diagram of 2d
  packings. Theoretical results for $n = 1,2,3$ (line points, from
  left to right) and $\phi_{\rm rcp}^\infty$ (red) are compared to (i)
  values in the literature: \cite{Berryman:1983aa} (down triangle),
  \cite{Parisi:2010aa} (diamond), and \cite{OHern:2002aa} (up
  triangle), (ii) simulations of $10^4$ monodisperse disks (crosses),
  and polydisperse disks (pluses), and (iii) experimental data of
  frictional disks (square). (Inset) The theoretical RCP volume
  fraction $\phi_{\rm rcp}(n)$ as a function of $n$. The points are
  fitted to a function $\phi(n)= \phi_{\rm rcp}^\infty -k_1e^{-k_2n}$,
  where $k_1=0.34\pm0.02$, $k_2=0.67\pm 0.06$ and $\phi_{\rm
    rcp}^\infty=0.85\pm0.01$. Adapted from \cite{Jin:2014aa}.}
\end{center}
\end{figure}

A systematic way of dealing with the correlations can be
developed by focusing only on particles close to the direction
$\mathbf{\hat{c}}$, i.e., particles that could contribute a VB, and
then constructing a layer expansion into coordination shells
\cite{Jin:2014aa}. We denote these particles as Voronoi particles. In
the exact Eq.~(\ref{Ch4:cdfex}), one can then consider the exclusion
condition
$\prod_{i=2}^{\tilde{n}}\left[1-m(\mathbf{r}_i-\mathbf{r}_1;\Omega\right]$
over $\tilde{n}$ Voronoi particles (including the reference particle)
rather than all $N$ particles in the packing. In 2d, the Voronoi
particles are located on the two closest branches to the direction
$\mathbf{\hat{c}}$ and can be described by a correlation function of
angles $G_{\tilde{n}}(\alpha_1,\alpha_2,...,\alpha_n)$. Using angles
instead of the position coordinates is a suitable parametrization of
the Voronoi particles provided the underlying contact network is
assumed fixed only allowing fluctuations in the angles without
destroying contacts. For such a fixed contact network the degree of
freedom per particle is thus reduced by one and allows to map the
$\tilde{n}-1$ position vectors
$\mathbf{r}_{12},\mathbf{r}_{13},...,\mathbf{r}_{1\tilde{n}}$ onto the
angles $\alpha_1,\alpha_2,...,\alpha_n$ of contacting Voronoi
particles plus the angle $\beta$ describing the direction
$\mathbf{\hat{c}}$ (see Fig.~\ref{Ch4:Fig_disks}). This requires
$\tilde{n}-1=n+1$. Transforming variables from
$(\mathbf{r}_{12},\mathbf{r}_{13},...,\mathbf{r}_{1\tilde{n}})$ to
$(\beta,\alpha_1,\alpha_2,...,\alpha_n)$ in Eq.~(\ref{Ch4:cdfex})
leads to \cite{Jin:2014aa} \be
\label{Ch4:angles2d}
P_>(c)&=&\lim_{n\to\infty}\mathcal{C}'\int\cdots\int\Theta(\alpha_1-\beta)G_n(\alpha_1,...,\alpha_{n})\nonumber\\ &&\times\prod_{j=2}^{n+2}\Theta\left(\frac{r_{1j}}{2\mathbf{\hat{c}}\cdot\mathbf{\hat{r}}_{1j}}-c\right)\D\beta\D\alpha_1\cdots\D\alpha_n,
\ee where the constant $\mathcal{C}'=z/L$ with $L=2\pi$ ensures the
normalization $P_>(R)=1$. Equation ~(\ref{Ch4:angles2d}) becomes exact
as $n\to\infty$ and provides a systematic approximation for finite
$n$. In particular, $n$ can be related to the coordination layers
above and below $\mathbf{\hat{c}}$.

One can then make two key assumptions to make this approach tractable
\cite{Jin:2014aa}. Firstly, one applies the Kirkwood superposition
approximation as in the high-dimensional case for $G_n$: $G_n(\alpha_1,...,\alpha_{n})\approx\prod_{j=1}^nG(\alpha_j)$. Secondly, the system of contacting Voronoi particles is mapped onto a
system of 1d interacting hard rods with an effective potential $V(x)$
(see Fig.~\ref{Ch4:Fig_disks}). Considering the particles in the first
coordination shell (Fig.~\ref{Ch4:Fig_disks}b) leads to a set of $z$
rods at positions $x_i$, $i=1,...,z$, where the rods are of length
$l_0=\pi/3$ and the system size is $L =6l_0$ with periodic boundary
conditions. In addition, the local jamming condition requires that
each particle has at least $d + 1$ contacting neighbours, which can
not all be in the same ``hemisphere". In 2d, this implies that $z\ge
3$ and $\alpha_j\le \pi$. In the rod system, this constraint induces
an upper limit $3 l_0$ on possible rod separations. Thus, the jamming
condition is equivalent to introducing an infinite square-well
potential between two hard rods. Crucially,
the partition function $Q(L,z)$ can then be calculated exactly in 1d \cite{Jin:2014aa}: \be
Q(L,z) &=&\sum_{k=0}^{\lfloor
    \frac{L/l_0-z}{2}\rfloor}(-1)^k
  \binom{z}{k}\frac{[L/l_0-z-2k]^{z-1}}{(z-1)!}\nonumber\\ &&\times
  \Theta(L/l_0-z)\Theta(3z-L/l_0), \ee where $\lfloor x \rfloor$ is
  the integer part of $x$ and the inverse temperature has been set to
  unity since it is irrelevant. This allows to determine the
  distribution of angles (gaps) $G(\alpha)=\langle
  \delta(x_2-x_1-\alpha)\rangle$ \be G(\alpha)  &=&
  \frac{Q(\alpha,1)Q(L-\alpha, z-1)}{Q(L,z)}.
\label{Ch4:singleG}
\ee In the limit $a\to\infty$ the system becomes the classical Tonks
gas of 1d hard rods \cite{Tonks:1936aa}. In the thermodynamic limit
($L\rightarrow \infty$ and $z\rightarrow \infty$), the gap
distribution is $G_{\rm HR} (\alpha) = \rho_f e^{-\rho_f
  (\alpha/l_0-1)}$, where $\rho_f = z/(L/l_0-z)$ is the free density.

The density of 2d disk packings follows by solving
Eq.~(\ref{Ch4:self2d}) with
Eqs.~(\ref{Ch4:angles2d},\ref{Ch4:singleG})
numerically using Monte-Carlo (Fig.~\ref{Ch4:Fig_disks}c).  The
formalism reproduces the highest density of 2d spheres in a triangular
lattice at $\phi \approx 0.91$ for $z=6$. For disordered packings one
obtains the RCP volume fraction: \be \phi^{2d}_{\rm rcp} = 0.85 \pm
0.01, \qquad \mbox{\rm for $z=4$}, \ee and the RLP volume fraction as:
\be \phi^{2d}_{\rm rlp}= 0.67 \pm 0.01, \qquad\mbox{\rm for
  $z=3$}. \ee We see that the prediction of the frictionless RCP point
is close to the numerical results and the result of the 1RSB theory
$\phi_{\rm th}=0.8165$, while a new prediction of RLP at the infinite
friction limit is obtained.

\subsection{Packing of bidisperse spheres}

\label{Ch4:Sec:binary}

Polydispersity with a smooth distribution of sizes typically occurs in
industrial particle synthesis and thus affects packings in many
applications. Qualitatively, one expects an increase in packing
densities due to size variations: The smaller particles can fill those
voids that are not accessible by the larger particles leading to more
efficient packing arrangements, which is indeed observed empirically \cite{Santiso:2002aa,Brouwers:2006aa,Sohn:1968aa,Desmond:2014aa}. Simulations have shown that the
jamming density in polydisperse systems depends also on the compression
rate without crystallization \cite{Hermes:2010aa} and the skewness of the size distribution \cite{Desmond:2014aa}. Since these issues are important in technological applications, as for instance the proportioning of concrete, very efficient phenomenological models have been developed to predict volume fractions of mixtures of various types of grains \cite{Larrard:1999aa}. For size distributions following a power-law, space-filling packings can be constructed \cite{Herrmann:1990aa}.
On the theoretical
side, a 'granocentric' model has been shown to reproduce the packing
characteristics of polydisperse emulsion droplets
\cite{Clusel:2009aa,Corwin:2010aa,Newhall:2011aa,Jorjadze:2011aa,Puckett:2011aa}. Here,
the packing generation is modelled as a random walk in the first
coordination shell with only two parameters, the available solid angle
around each particle and the ratio of contacts to neighbors, which can
both be calibrated to experimental data.

The simpler case of a bidisperse packing with two types of spheres with different
radii has been investigated in \cite{Clarke:1987aa,Santiso:2002aa,Lange-Kristiansen:2005aa,Hopkins:2013aa} using simulations. Here, one can generally observe packing densities that increase from the monodisperse value as both the size ratio and concentration of small spheres is varied.  In \cite{Hopkins:2013aa} mechanically stable packings with a large range of densities $0.634\le \phi\le 0.829$ have been generated using a linear programming algorithm. Interestingly, for a given size ratio, the density is non-monotonic, exhibiting a peak at a specific concentration. A theoretical approach that is able to reproduce the density peak in the bidisperse case has been developed in \cite{Danisch:2010aa} based on the volume
ensemble. The key idea is to treat the spheres of radii $R_1<R_2$ as
different species $1$ and $2$ with independent statistical
properties. If we denote by $x_1$ the fraction of small spheres $1$,
then $x_1=N_1/(N_1+N_2)$, with $N_i$ the number of spheres $i$ in the
packing. Likewise, $x_2=1-x_1$. The overall packing density is \be
\phi=\frac{\overline{V}_g}{\overline{W}},\qquad
\overline{V}_g=\sum_{i=1}^2x_iV_g^{(i)} \ee where
$V_g^{(i)}=\frac{4\pi}{3}R_i^3$ and $\overline{W}$ is the average
volume of a Voronoi cell as before. The average now includes averaging
over the different species, so that \be
\label{Ch4:selfbi}
\overline{W}&=&\sum_{i=1}^2x_i\overline{W}_i,\\ \overline{W}_i&=&V_g^{(i)}+4\pi\int_{R_i}^\infty
\D c\,c^2 \,P^{(i)}_>(c,z),\qquad i=1,2 \ee as a straightforward
extension of Eq.~(\ref{Ch4:integral}). The CDF $P^{(i)}_>(c,z)$
contains the probability that, for a Voronoi cell of species $i$, the
boundary is found at a value larger than $c$. This probability
depends, of course, on both species. Assuming statistical independence
we can introduce a factorization into bulk and contact particles of
both species \cite{Danisch:2010aa} analogously to the monodisperse case
Eq.~(\ref{Ch4:Pcfactor}):

\be P^{(i)}_>(c,z)=P^{(i1)}_{\rm B}(c)P^{(i1)}_{\rm
  C}(c,z)P^{(i2)}_{\rm B}(c)P^{(i2)}_{\rm C}(c,z).  \ee Here,
$P^{(ij)}_{\rm B}$ denotes the CDF due to contributions of bulk
particles of species $j$ to a Voronoi cell of species $i$. Likewise
$P^{(ij)}_{\rm C}$ refers to the contact particles. We express each of
these terms in analogy to the monodisperse case, i.e.,
Eqs.~(\ref{Ch4:pb},\ref{Ch4:pc}), \be P^{(ij)}_{\rm
  B}&=&\exp\left[-\tilde{\rho}_jV_{ij}^*(c)\right],\\ P^{(ij)}_{\rm
  C}&=&\exp\left[-\sigma_{ij}(z)S_{ij}^*(c)\right].  \ee The Voronoi
excluded volume and surface, $V_{ij}^*$ and $S_{ij}^*$, are defined by
Eqs.~(\ref{Ch4:vexdef},\ref{Ch4:sexdef}), where now
$s(\mathbf{r},\mathbf{\hat{c}})$ denotes the VB between spheres of
radii $R_i$ and $R_j$, as parametrized by
Eq.~(\ref{Ch2:VBsphere}). The particle densities $\tilde{\rho}_j$ are
given by \be
\tilde{\rho}_j=\frac{x_j}{\overline{W}-\overline{V}_g},\qquad j=1,2.
\ee

\begin{figure}
\begin{center}
(a)\hspace{4cm}(b)\\
\includegraphics[width=3.5cm]{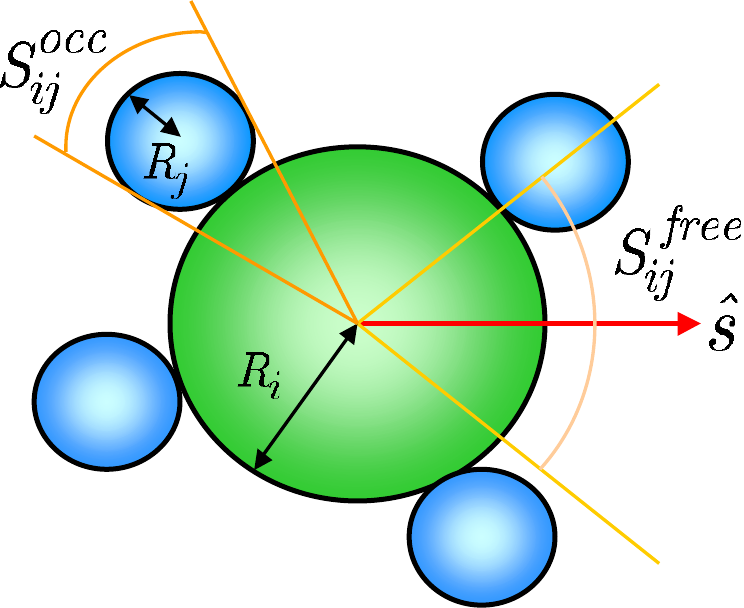}
\includegraphics[width=4.0cm]{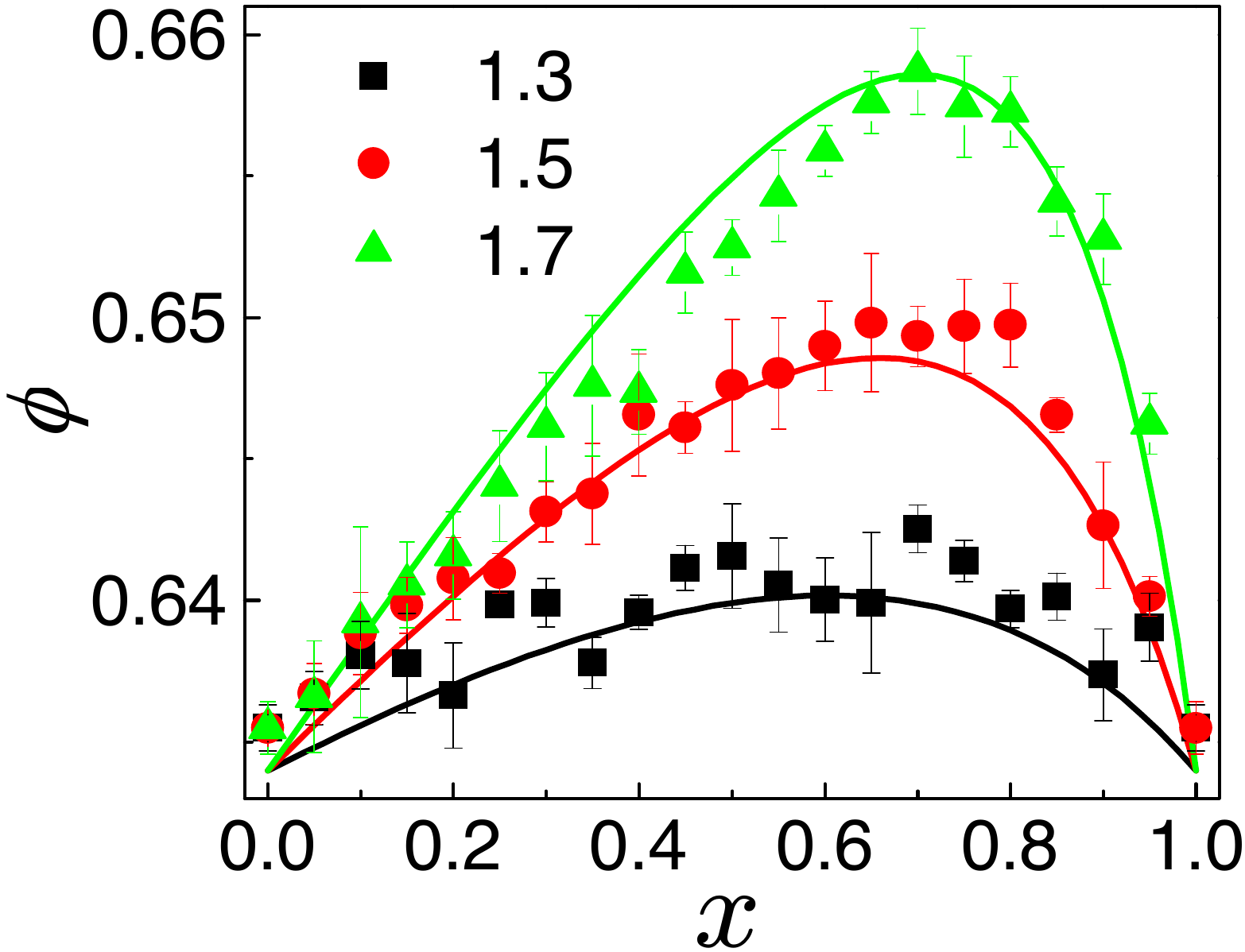}\\
\caption{\label{Ch4:Fig_rcpbi}(Colors online) (a) The occupied surface
  Eq.~(\ref{Ch4:occ}) and the Voronoi excluded surface $S^*_{ij}$. (b)
  Comparison between theory and numerical simulations of Hertzian
  packings at RCP vs the concentration x of small spheres. Different symbols denote
  different ratios $R_1/R_2$. Adapted from \cite{Danisch:2010aa}.}
\end{center}
\end{figure}

The main challenge is to obtain an expression for the surface density
$\sigma_{ij}(z)$. For this, it is first necessary to distinguish
different average contact numbers: $z_{ij}$ is the average number of
spheres $j$ in contact with a sphere $i$. It follows that the average
number of contacts of sphere $i$, denoted by $z_i$, is \be
z_i=z_{i1}+z_{i2},\qquad z=\sum_{i=1}^2x_iz_i.  \ee By relating the
contact numbers $z_i$ to the average occupied surface on sphere $i$,
$\left<S_i^{\rm occ}\right>$, one can obtain the following equations
to relate $z_{ij}$ with $z$ \be z_1=\frac{z}{x_1 + x_2 \frac{\left<
    S^{\rm occ}_1\right>}{\left< S^{occ}_2\right>}}, \quad z_2
=\frac{z}{x_1 \frac{\left< S^{\rm occ}_2\right>}{\left< S^{\rm
      occ}_1\right>} +x_2 }.
\label{Ch4:zis}
\ee
and
\be
\label{Ch4:zij1}
z_{11}=\frac{z_1^2 x_1}{z}, &\qquad& z_{12}=\frac{z_1 z_2 x_2}{z},\\
z_{21}=\frac{z_1 z_2 x_1}{z}, &\qquad& z_{22}=\frac{z_2^2 x_2}{z}.
\label{Ch4:zij2}
\ee where $\left< S_i^{\rm occ} \right>$ is approximated as $\left<
S_i^{\rm occ} \right>=\sum_{j=1}^2x_jS_{ij}^{\rm occ}$ with the exact
expression for the occupied surface (see Fig.~\ref{Ch4:Fig_rcpbi}a)
\be S^{\rm occ}_{ij}
=2\pi\left(1-\sqrt{1-\left(\frac{R_j}{R_i+R_j}\right)^2}\right).
\label{Ch4:occ}
\ee Eqs.~(\ref{Ch4:zis}--\ref{Ch4:zij2}) imply that we
can express $z_{ij}$ as a function of $z$: $z_{ij}=z_{ij}(z)$. As
before, $\sigma_{ij}$ can in principle be obtained from simulations
using Eq.~(\ref{Ch4:sigmasrel}). However, a direct simulation of
$\left<S_{ij}^*\right>$ as a function of $z$ contacting particles
ignores the dependence of the different species that is not resolved
in $z$. Therefore, $\tilde{\sigma}_{ij}$ is introduced via \be
\sigma_{ij}(z)=\tilde{\sigma}_{ij}(z_{ij}(z)).  \ee In turn, we obtain
$\tilde{\sigma}_{ij}=\left<S_{ij}^*\right>^{-1}$ as a function of
$z_{ij}$ by generating configurations around sphere $i$ with the
proportions $z_{i1}/z_i$ of spheres $1$ and $z_{i2}/z_i$ of spheres
$2$. $\left<S_{ij}^*\right>$ follows operationally again as the
Monte-Carlo average Eq.~(\ref{Ch4:smc}).

Overall, the packing density of the bi-disperse packing of spheres can
be calculated by solving the following self-consistent equation for
the free volume $w=\overline{W}-\overline{V}_g$ \be
\label{Ch4:selfbi2}
w&=&4\pi\sum_{i=1}^2x_i\int_{R_i}^\infty \D
c\,c^2\nonumber\\ &&\times\exp\left\{-\sum_{j=1}^2\left[\frac{x_j}{w}V^*_{ij}(c)+\sigma_{ij}(z)S^*_{ij}(c)\right]\right\}.
\ee 

We notice that Eq.~(\ref{Ch4:selfbi2}) is the generalization of
Eq.~(\ref{Ch4:self3d}) from monodisperse to bidisperse packings.
While the monodisperse self-consistent Eq.~(\ref{Ch4:self3d}) admits
a closed analytical solution, the bidisperse Eq.~(\ref{Ch4:selfbi2})
does not. Thus, we resort to a numerical solution of this equation,
and therefore the equation of state $w(z)$ is obtained numerically in
these cases rather than in closed form as obtained for monodisperse
spheres Eq.~(\ref{Ch4:wsol}).

Calculations for all systems (from spheres to non-spheres,
monodisperse or polydisperse and beyond) that use the present
mean-field theory in the Edwards ensemble will end up with a
self-consistent equation for the free volume of the form
Eq.~(\ref{Ch4:self3d}) or Eq.~(\ref{Ch4:selfbi2}). However, so far,
the only self-consistent equation that admits a closed analytical
solution is the 3d monodisperse case leading to
Eq.~(\ref{Ch4:wsol}). The remaining equations of state for all systems
studied so far are too involved and need to be solved numerically.

Results of numerical solutions of Eq.~(\ref{Ch4:selfbi2}) are shown in
Fig.~\ref{Ch4:Fig_rcpbi}b demonstrating good agreement with simulation
data as well as the predictions of the 1RSB hard-sphere glasses
calculations \cite{Biazzo:2009aa}. We observe the pronounced peak as a
function of the species concentration $x=x_1\in [0,1]$. The extension
of the theory to higher-order mixtures is straightforward in
principle. The main challenge is to obtain the generalizations of
Eqs.~(\ref{Ch4:zis},\ref{Ch4:zij1},\ref{Ch4:zij2}). Determining
$\tilde{\sigma}_{ij}(z_{ij})$ from simulations of local packing
configurations becomes also an increasingly complex task.

\subsection{Packing of attractive colloids}

\label{Ch4:Sec:adhesion}

Packings of particles with diameters of around $10 \mu m$ or smaller
enter the domain of colloids and are often dominated by adhesive van
der Waals forces in addition to friction and hard-core
interactions. In fact, packings of adhesive colloidal particles appear
in many areas of engineering as well biological systems
\cite{Marshall:2014aa,Jorjadze:2011aa} and exhibit different
macroscopic structural properties compared with non-adhesive packings
of large grains treated so far, where attractive van der Waals forces
are negligible in comparison with gravity. In \cite{Lois:2008aa} the mechanical response at the jamming
transition has been studied and two second-order
transitions are found in the attractive systems \cite{Lois:2008aa}: a
connectivity percolation transition and a rigidity percolation
transition, where a rigid backbone forms without floppy modes.

Numerical studies of adhesive granular systems have found a range of
packing fractions as a function of particle sizes $\phi\approx
0.1-0.6$
\cite{Yang:2000aa,Valverde:2004aa,Blum:2006aa,Head:2007aa,Martin:2008aa,Kadau:2011aa,Parteli:2014aa}. The
effect of varying the force of adhesion has been systematically
investigated in \cite{Liu:2015aa,Chen:2016aa,Liu:2017aa} using a DEM framework specifically
developed for the ballistic deposition of adhesive Brownian soft
spheres with sliding twisting and rolling friction
\cite{Marshall:2014aa}. A dimensionless adhesion parameter $Ad$,
defined as the ratio between interparticle adhesion work and particle
inertia \cite{Li:2007aa}, can be used to quantify the combined effect
of size and deposition velocity. In the case of $Ad<1$, particle
inertia dominates the adhesion and frictions exhibiting a broad range
of densities and coordination numbers. At $Ad\approx 1$ the isostatic
value $z=4$ for infinitely rough spheres is observed, indicating that
weak adhesion has a similar effect on the packing as strong
friction. However, when $Ad>1$, an adhesion-controlled regime is
observed with a unique curve in the $z$--$\phi$ diagram. The lowest
packing density achieved numerically is $\phi=0.154$ with $z=2.25$ for
$Ad\approx48$. The lowest density agrees well with the data from a
random ballistic deposition experiment \cite{Blum:2006aa} and other
DEM simulations \cite{Yang:2000aa,Parteli:2014aa}.

An analytical representation of the adhesive equation of state can be
derived within the framework of the mean-field Edwards volume function
Eq.~(\ref{Ch4:integral}), where the CDF $P_>$ is defined by
Eq.~(\ref{Ch4:cdfex}). Assuming the same factorization of the
$n$-point correlation function as in high dimensions leads to the
approximation Eq.~(\ref{Ch4:Pg2}), which allows us to relate $P_>$
with the structural properties of the packing expressed in the pair
distribution function $g_2$. We then model $g_2$ by extending the
simple form considered so far for 3d hard-spheres in
Eq.~(\ref{Ch4:g2simple}) in terms of four distinct contributions
following the results of available simulations of hard-sphere packings
and metastable hard-sphere glasses. We consider: 

{\it (i)} A delta-peak due to contacting particles
\cite{Donev:2005aa,Torquato:2006aa,Song:2008aa}; 

{\it (ii)} A power-law peak
as given by Eq.~(\ref{Ch2:gammaexp}) over a range $\epsilon$ due to
near contacting particles \cite{Donev:2005aa,Wyart:2012aa};

{\it (iii)} A step function due to bulk particles
\cite{Torquato:2006aa,Song:2008aa} mimicking a uniform density of bulk
particles;

{\it (iv)} A gap of width $b$ separating bulk and (near) contacting
particles. This gap captures the effect of correlations due to
adhesion and is assumed to depend on $z$: $b=b(z)$. In this way we
model the increased porosity at a given $z$ compared with
adhesion-less packings. Overall, we obtain \be
\label{Ch4:g2ad}
g_2(\mathbf{r},z)&=&\frac{z}{\rho
  \lambda}\delta(r-2R)+\sigma(r-2R)^{-\nu}\Theta(2R+\epsilon-r)\nonumber\\ &&+\Theta(r-(2R+b(z))).
\ee 

For the power law term we assume $\nu=0.38$ from
\cite{Lerner:2013aa} and a width of $\epsilon=0.1R$, which is
approximately the range over which the peak decreases to the bulk
value unity as observed in \cite{Donev:2005aa}. The value $\sigma$ is
then fixed by continuity with the step function term in the absence of
a gap. 

Next, we have to determine the gap of width function $b(z)$ which is
the crucial assumption of the theory. $b(z)$ needs to satisfy a set of
constraints that we impose purely on physical grounds:
\begin{enumerate}[(i)]
\item $b(z)$ is a smooth monotonically decreasing function of
  $z$. Here, the physical picture is that for small $z$ (corresponding
  to looser packings), the gap width is larger due to the increased
  porosity of the packing.
\item At the isostatic limit $z=6$, the gap disappears,
  $b(6)=\epsilon$, and we expect to recover the frictionless RCP
  value, since this value of $z$ represents a maximally dense
  disordered packing of spheres. We obtain from Eq.~(\ref{Ch4:g2ad})
  indeed the prediction for $\phi_{\rm Edw}$, Eq.~(\ref{Ch4:phircp}),
  by choosing an appropriate value of $\lambda$ and accounting for low
  dimensional corrections due to the hard-core excluded volume of the
  reference sphere, such that $\rho\to
  \overline{\rho}=1/(\overline{W}-V_0)$. This constraint thus fixes
  $\rho$ and $\lambda$, as well as one of the parameters in $b(Z)$.
\item In addition, we conjecture the existence of an asymptotic {\it
  adhesive loose packing} (ALP) at $z=2$ and $\phi=1/2^3$ which yields
  $b(2)=1.47$ and fixes a second parameter in $b(z)$. This is
  motivated by the fact that $\phi=1/2^d$ is the lower bound density
  of saturated sphere packings of congruent spheres in $d$ dimensions
  for all $d$ \cite{Torquato:2006aa}.  A saturated packing of
  congruent spheres of unit diameter satisfies that each point in space
  lies within a unit distance from the center of some sphere.
  Moreover, $z=2$ is the lowest possible value for a physical packing:
  If $z<2$ there are more spheres with a single contact (i.e., dimers)
  than with three or more contacts, which identifies that the ALP
  point is only asymptotic.
\end{enumerate}

\begin{figure}
\begin{center}
\includegraphics[width=0.85\columnwidth]{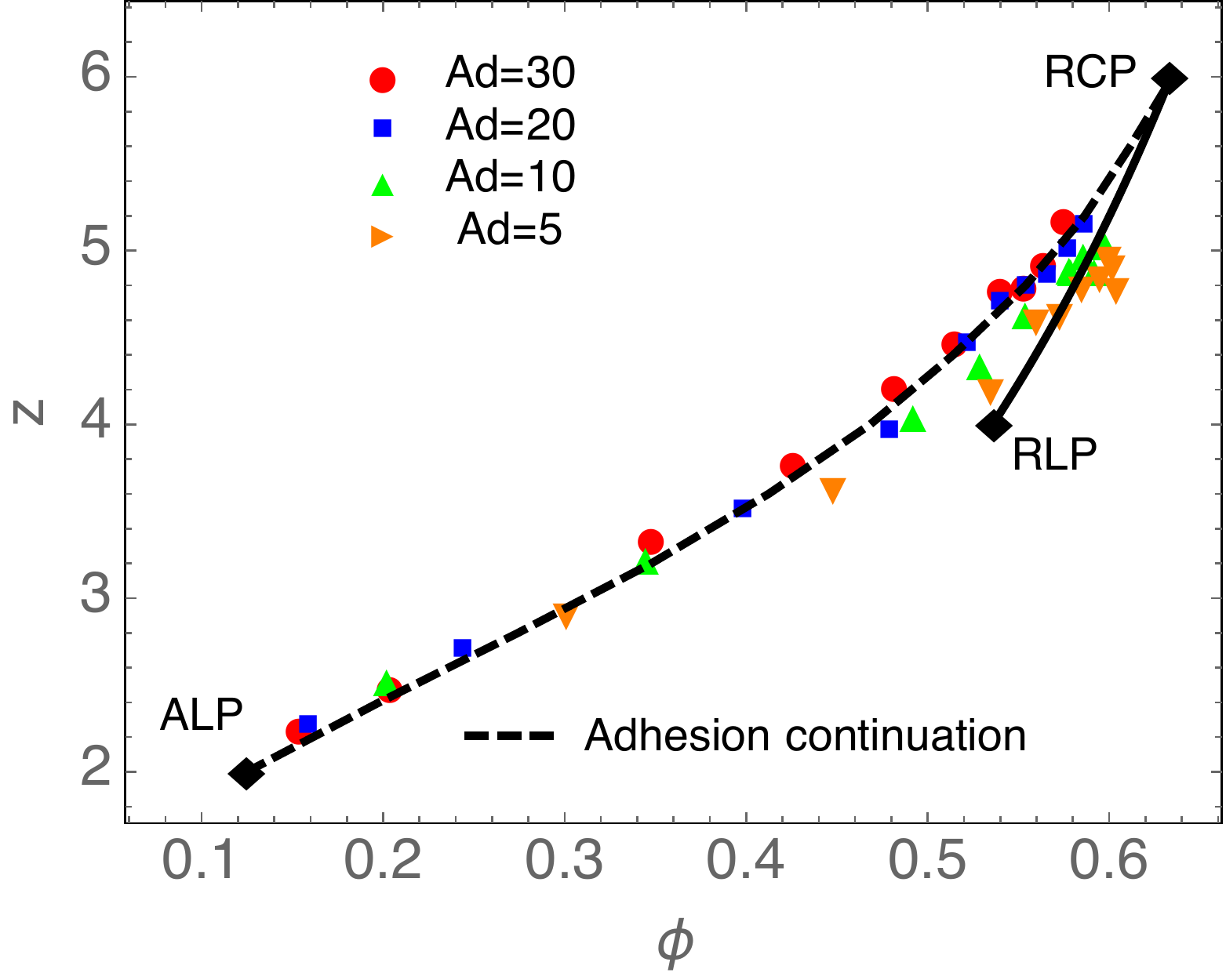}
\caption{\label{Ch4:Fig_adcont}(Colors online) Plot of high $Ad$
  simulation data in the $z$--$\phi$ plane \cite{Liu:2015aa}. The
  adhesive continuation with an exponential $b(z)$ connects the RCP at
  $\phi_{\rm Edw}$ and $z=6$ with the conjectured adhesive loose
  packing point (ALP) at $\phi=2^{-3}$ and $z=2$. The black solid line
  is the RLP line of Fig.~\ref{Ch4:Fig_xlines}(b). Adapted from \cite{Liu:2015aa}.}
\end{center}
\end{figure}

Clearly, $b(z)$ is a smoothly decreasing function, so that we can
assume, e.g., the simple parametric form $b(z)=c_1+c_2e^{-c_3z}$,
such that one fitting parameter is left after the two constraints
$b(6)=\epsilon$ and $b(2)=1.47$ are imposed. Figure
~\ref{Ch4:Fig_adcont} highlights that the exponential decay of $b(z)$
provides an excellent fit to the simulation data providing the
equation of state $\phi(z)$ for adhesive packings. Moreover, the
resulting $P(c,z)$ also agrees well with the empirically measured CDF
over a large range of $Ad$ values \cite{Liu:2015aa}. This means that
including $b(z)$ captures well the essential structural features of
the packing. It is quite intriguing that such a simple modification of
the non-adhesive theory, motivated on physical grounds, leads to such
good agreement not only in the low density regime, but also for mid to
high densities.

These results highlight that attraction in (spherical) particles leads
to a lower density limit for percolation at the ALP with
$\phi_c=1/2^3$. The equivalent $\phi_c$ in attractive colloids is
observed empirically over a range of densities $\phi_c\approx 0.1 -
0.2$ depending on the mechanism for the suppression of
phase-separation \cite{Zaccarelli:2007aa}, e.g., due to an interrupted liquid-gas phase separation \cite{Trappe:2001aa,Lu:2008ab}. The situation is thus
reminiscent of the adhesion-less and frictionless range of densities
$\phi\in[\phi_{\rm th},\phi_{\rm GCP}]$ of the J-line (see
Sec.~\ref{Sec:constraints}).

\subsection{Packing of non-spherical particles}

\label{Ch4:Sec:nonsphere}

The question of optimizing the density of packings made of particles
of a particular shape is an outstanding scientific
problem occupying scientists since the time of Apollonius of Perga \cite{Andrade:2005aa,Herrmann:1990aa,Thomas:1941aa} and Kepler \cite{Kepler:1611aa,Weaire:2008aa}, and still of great practical
importance for all industries involved in granular processing.  In addition, the complex structures that result
from their assembly become increasingly important for the design of
new functional materials
\cite{Glotzer:2007aa,Jaeger:2015aa,Damasceno:2012aa,Baule:2014aa}.

\begin{figure}[!htb]
\begin{center}
\includegraphics[width=8cm]{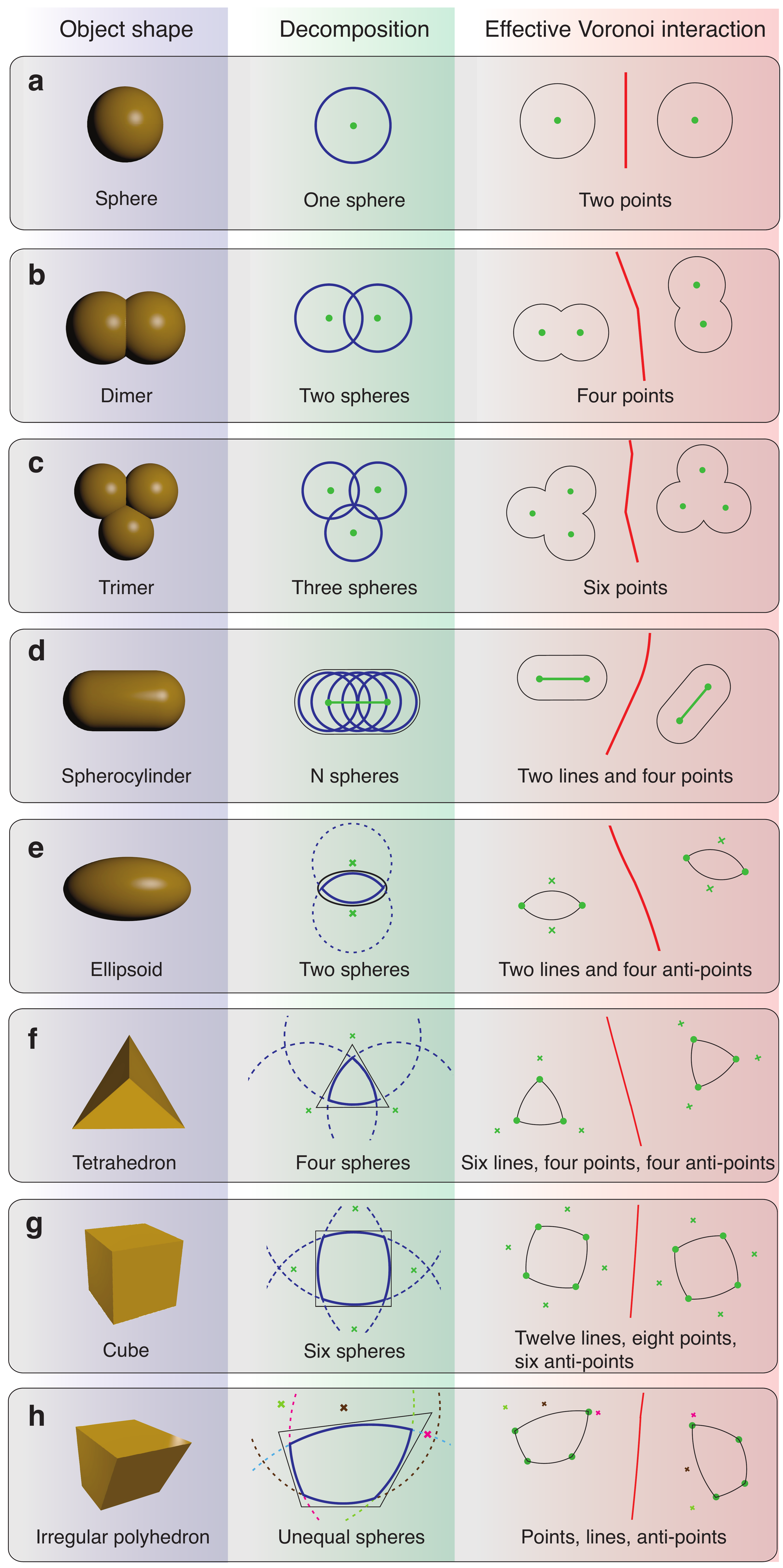}
\caption{\label{Ch4:Fig_shapes}(Colors online) Table of different
  shapes and their VBs. (a--d) For shapes composed
  of spheres, the VB arises due to the effective interaction of the
  points at the centres of the spheres. Since spherocylinders are
  represented by a dense overlap of spheres, the effective interaction
  is that of two lines and four points. (e--h) For more complicated
  shapes that would in principle be modelled by a dense overlap of
  sphere with different radii, we propose approximations in terms of
  intersections of spheres leading to effective interactions between
  `anti-points'. For both classes of shapes, the VB follows an exact
  algorithm leading to analytical expressions (see
  Fig.~\ref{Ch4:Fig_algorithm}). From \cite{Baule:2013aa}. }
\end{center}
\end{figure}

In the absence of theory, searches for the optimal
random packing of non-spherical shapes have focused on empirical
studies on a case-by-case basis.
Table~\ref{Ch4:table1} presents an overview of the maximal packing
densities for a variety of shapes obtained in simulations, experiments
and theory. Recent simulations have found the densest random
packing fraction of, e.g., prolate ellipsoids at $\phi\approx 0.735$ \cite{Donev:2004aa};
spherocylinders at $\phi\approx 0.772$ \cite{Zhao:2012aa} and 2d dimers at $\phi \approx 0.885$
\cite{Schreck:2010aa}. The densest random tetrahedra packing has been
found in simulations with $\phi=0.7858$
\cite{Haji-Akbari:2009aa}. More systematic investigations of the self-assembly
of hard truncated polyhedra families has been done in
\cite{Damasceno:2012aa,Chen:2014aa}. The organizing principles of ordered packings of Platonic and Archimedean solids and other convex and non-convex shapes have been investigated in \cite{Torquato:2009aa,Torquato:2012aa}. Interesting shapes have
been considered also in a systematic way: superballs
\cite{Jiao:2010aa}, puffy tetrahedra \cite{Kallus:2011aa}, polygons \cite{Wang:2015aa} and truncated
vertices \cite{Damasceno:2012aa,Gantapara:2013aa}. A caveat of some
empirical studies is the strong protocol dependence of the final close
packed state even for the same shape: recent studies of spherocylinder
packings, e.g., exhibit a large variance depending on the algorithm
used
\cite{Jiao:2011aa,Williams:2003aa,Zhao:2012aa,Abreu:2003aa,Jia:2007aa,Bargiel:2008aa,Wouterse:2009aa,Lu:2010aa,Kyrylyuk:2011aa}. A
generic theoretical insight is needed if one wants to search over more
extended regions of parameter space of object shapes.

It is empirically clear that non-spherical shapes can
generally achieve denser maximal packing densities than spheres. In
fact, a conjecture attributed to Ulam (recorded in the book
\cite{Gardner:2001aa}) in the context of regular packings, recently
also formulated for random packings \cite{Jiao:2011aa}, states that the sphere is, indeed, the worst packing
object among all convex shapes. In \cite{Kallus:2016aa} it has been shown for random packings that all sufficiently spherical shapes pack more densely than spheres. However, one should notice the
local character of such a conjecture for random packings: Onsager
already proved that elongated spaghetti-like thin rods pack randomly
much worse than spheres \cite{Onsager:1949aa}.

From a numerical point of view, a promising approach to find the best
shape has been put forward by Jaeger and collaborators
\cite{Miskin:2013aa,Miskin:2014aa,Jaeger:2015aa,Roth:2016aa} who used
genetic algorithms (GA) to map the possible space of the constitutive
particle shapes. They consider non-spherical composite particles
formed by gluing spherical particles of different sizes rigidly
connected into a polymer-like non-branched shape. A genetic algorithm
starts with a given shape and perform `mutations' to the constitutive
particles until a desired property, for instance, maximal strength or
maximal packing fraction is achieved. This reverse engineering
approach can generate novel materials with desired properties but of
limited shapes: within this framework, the limits to granular
materials design are the limits to computation \cite{Jaeger:2015aa},
since GA relies heavily on dynamically simulating (e.g., with MD or
MC) the packings to be optimized. Thus, computational limitations are
expected in more complicated shapes such as tetrahedra or irregular
polyhedra, in general.

On the theoretical side, there are successful theories of high density liquids
that have been extended to encompass non-spherical particles, such as
mode-coupling theory \cite{Gotze:2009aa} and density functional theory
\cite{Hansen-Goos:2009aa,Hansen-Goos:2010aa,Marechal:2013aa}. However, they do not apply to the jamming
regime. On the other hand, successful approaches to jamming based on replica
theory so far only apply to spherical particles \cite{Parisi:2010aa,Charbonneau:2017aa}
(see Sec.~\ref{Sec:constraints}). The difficulty to extend replica theory calculations from
  spheres to non-spherical particles stems from the fact that the
  system is not rotationally invariant, which adds more degrees of
  freedom to the description of the cage motion. Replica calculations
  also rely on liquid equations of state, which are typically not
  available in analytical form for non-spherical particles. These
  difficulties can be overcome in principle with numerics, but this is
  most likely cumbersome, and has not been accomplished so far. On the
  other hand, the Edwards approach can be generalized theoretically
  much more easily to non-spherical shapes.

The advantage of the mean-field Edwards approach is that it is based entirely on
the geometry of the particles; its building block is directly the
shape of the constitutive particle. Therefore, Edwards ensemble can be applied in a straightforward way to arbitrary shapes. Such a generalization, providing a comprehensive framework to describe packings of non-spherical particles, has recently been developed \cite{Baule:2013aa}. A drawback of
employing a general theoretical approach rather than direct simulations using, e.g., artificial evolution \cite{Jaeger:2015aa}, is that current theories are at the mean-field level and thus only approximate. However, both approaches can be complementing: A mean-field theory
could identify a reduced region in the space of optimal parameters,
which can then be tackled with more detail using more focused reverse
engineering techniques.

As discussed in the previous sections, the central quantity to
calculate is the average Voronoi volume $\overline{W}$ as a function
of $z$. In the case of frictionless spheres, $z$ is fixed by
isostaticity providing the prediction Eq.~(\ref{Ch4:phircp}) for
RCP. The situation is somewhat more complicated for frictionless
non-spherical particles: Here, both $z$ and $\overline{W}$ depend
independently on the particle shape. For simplicity, we assume
rotationally symmetric particles in the following, where deviations
from the sphere can be parametrized by a single parameter, e.g., the
aspect ratio $\alpha$ measuring length over width. As a consequence,
if we are interested in obtaining the function $\phi(\alpha)$ at RCP,
we need to combine the dependencies $\overline{W}_\alpha(z)$ and
$z(\alpha)$: \be
\phi(\alpha)=\frac{V_0}{\overline{W}_\alpha(z(\alpha))}. \ee We
discuss next how to obtain $\overline{W}_\alpha(z)$ by extending the
framework of the coarse-grained Voronoi volume to non-spherical
particles. A quantitative approach to describe $z(\alpha)$ is
discussed in Sec.~\ref{Ch4:Sec:degenerate}, which requires a
quantitative evaluation of the occurrence of degenerate
configurations.

\begin{table*}[ht]
\centering
\begin{tabular}{|c | c | c | c|}
\hline
Shape  & $\phi_{\rm max}$ simulation & $\phi_{\rm max}$ experiment & $\phi_{\rm max}$ theory \\ 
\hline \hline
disks (2d) & 0.826 \cite{Atkinson:2014aa} & & 0.85 \cite{Jin:2014aa} \\
& & & 0.874 \cite{Parisi:2010aa} \\
& & & 0.834 \cite{Tian:2015aa} \\
Sphere &  0.645 \cite{Skoge:2006aa} & 0.64 \cite{Bernal:1960aa} & 0.634 \cite{Song:2008aa} \\
& & & 0.68 \cite{Parisi:2010aa}  \\

\hline
M\&M candy & & 0.665 \cite{Donev:2004aa} & \\
Dimer & 0.703 \cite{Faure:2009aa} &  &  0.707 \cite{Baule:2013aa} \\
Ellipse (2d) & 0.895 \cite{Delaney:2005aa} & & \\
Oblate ellipsoid & 0.707 \cite{Donev:2004aa} &  &  \\
Prolate ellipsoid & 0.716 \cite{Donev:2004aa}  & &   \\
Spherocylinder & 0.722 \cite{Zhao:2012aa} &  & 0.731 \cite{Baule:2013aa}  \\
Lens-shaped particle & & & 0.736 \cite{Baule:2013aa}   \\
\hline
Tetrahedron & 0.7858 \cite{Haji-Akbari:2009aa} & 0.76 \cite{Jaoshvili:2010aa} & \\
Cube & & 0.67 \cite{Baker:2010aa} & \\
Octahedron & 0.697 \cite{Jiao:2011aa} & 0.64 \cite{Baker:2010aa} & \\
Dodecahedron & 0.716 \cite{Jiao:2011aa} & 0.63 \cite{Baker:2010aa} & \\
Icosahedron & 0.707 \cite{Jiao:2011aa} & 0.59 \cite{Baker:2010aa} & \\

\hline
General ellipsoid & 0.735 \cite{Donev:2004aa} & 0.74 \cite{Man:2005aa} &   \\
Superellipsoid & 0.758 \cite{Delaney:2010aa} & & \\
Superball & 0.674 \cite{Jiao:2010aa} & & \\
Trimer & 0.729 \cite{Roth:2016aa} & & \\
\hline
\end{tabular}
\caption{\label{Ch4:table1}Overview of maximal packing fractions $\phi_{\rm max}$ for a selection of regular shapes in disordered packings obtained with a variety of different packing protocols. Note that the $\phi_{\rm max}$ value is achieved for the aspect ratio, where $\phi$ is maximal, so every value is at a different aspect ratio.}
\end{table*}

\subsubsection{Coarse-grained Voronoi volume of non-spherical shapes}
\label{Ch4:Sec:voronoins}

The key for the mean-field approach to the statistical mechanical
ensemble based on the coarse-grained volume function is
Eq.~(\ref{Ch4:wav}), which replaces the exact global minimization to
obtain the Voronoi boundary $l_i(\mathbf{\hat{c}})$ in the direction
$\mathbf{\hat{c}}$ by the pdf $p(\mathbf{c},z)$. For a general
particle-shape the cut-off $c^*$ describes just the particle surface
parametrized by $\mathbf{\hat{c}}$. Transforming Eq.~(\ref{Ch4:wav})
to the CDF $P_>$ using $p(\mathbf{c},z)=-\frac{\D}{\D
  c}P_>(\mathbf{c},z)$ leads to the volume integral
\cite{Baule:2013aa} \be
\label{Ch4:wav2}
\overline{W}(z)=\int\D\mathbf{c}\,P_>(\mathbf{c},z), \ee where $P_>$
is again interpreted as the probability that $N-1$ particles are
outside a volume $\Omega$ centered at $\mathbf{c}$, since otherwise
they would contribute a shorter VB. $\Omega$ is in principle defined
as in Eq.~(\ref{Ch4:omega}), but is no longer a spherical volume due
to the non-spherical interactions manifest in the parametrization of
the VB. The VB now also depends on the relative orientation
$\mathbf{\hat{t}}$ of the two particles suggesting the definition: \be
\label{Ch4:omegans}
\Omega(\mathbf{c},\mathbf{\hat{t}})=\int \D
\mathbf{r}\,\Theta(c-s(\mathbf{r},\mathbf{\hat{t}},\mathbf\mathbf{\hat{c}}))\Theta(s(\mathbf{r},\mathbf{\hat{t}},\mathbf{\hat{c}})),
\ee for a fixed relative orientation $\mathbf{\hat{t}}$.

So far, the description of $\overline{W}$ is exact within the
statistical mechanical approach. In order to solve the formalism, we
introduce the following mean-field minimal model of the translational
and orientational correlations in the packing \cite{Baule:2013aa}:

\begin{enumerate}

\item Following Onsager \cite{Onsager:1949aa}, we treat particles of
  different orientations as belonging to different species. This is
  the key assumption to treat orientational correlations within a
  mean-field approach. Thus, the problem for non-spherical particles
  can be mapped to that of polydisperse spheres for which $P_>$
  factorizes into the contributions of the different radii (see
  Sec.~\ref{Ch4:Sec:binary}).

\item Translational correlations are treated as in the spherical case
  for high dimensions (see Sec.~\ref{Ch4:Sec:dimension}). Here, the
  Kirkwood superposition approximation leads to a factorization of the
  $n$-point correlation function into a product of pair-correlation
  functions, Eq.~(\ref{Ch4:gnsimple}). Including also the
  factorization of orientations provides the form \be
\label{Ch4:pns}
P_>(\mathbf{c},z)=\exp\left\{-\rho\int\D\mathbf{\hat{t}}\int_{\Omega(\mathbf{c},\mathbf{\hat{t}})}\D\mathbf{r}\,g_2(\mathbf{r},\mathbf{\hat{t}})\right\}.
\ee

\item The pair correlation function is modelled by a delta function
  plus step function as for spheres, Eq.~(\ref{Ch4:g2simple}). This
  form captures the contacting particles and treats the remaining
  particles as an ideal gas-like background: \be
\label{Ch4:g2ns}
g_2(\mathbf{r},\mathbf{\hat{t}})&=&\frac{1}{4\pi}\left[\frac{\sigma(z)}{\rho}\delta\left(r-r^*(\mathbf{\hat{r}},\mathbf{\hat{t}})\right)\right.\nonumber\\ &&\left.+\Theta(r-r^*(\mathbf{\hat{r}},\mathbf{\hat{t}}))\right].
\ee Here, the prefactor $1/4\pi$ describes the density of
orientations, which we assume isotropic. The contact radius $r^*$
denotes the value of $r$ in a direction $\mathbf{\hat{r}}$ for which
two particles are in contact without overlap. In the case of equal
spheres the contact radius is simply
$r^*(\mathbf{\hat{r}},\mathbf{\hat{t}})=2R$. For non-spherical
objects, $r^*$ depends on the object shape and the relative
orientation.

\end{enumerate}

Combining Eq.~(\ref{Ch4:g2ns}) with Eq.~(\ref{Ch4:pns}) recovers the
product form of the CDF $P_>$: \be
\label{Ch4:cdfns}
P_>(\mathbf{c},z)=\exp\left\{-\rho\,\overline{V}^*(\mathbf{c})-\sigma(z)\,\overline{S}^*(\mathbf{c})\right\},
\ee where $\overline{V}^*$ and $\overline{S}^*$ are now
orientationally averaged excluded volume and surface:
$\overline{V}^*=\left<\Omega-\Omega\cap V_{\rm
  ex}\right>_{\mathbf{\hat{t}}}$ and $\overline{S}^*=\left<\partial
V_{\rm ex}\cap\Omega\right>_{\mathbf{\hat{t}}}$ (compare with
Eqs.~(\ref{Ch4:vexdef},\ref{Ch4:sexdef})). The orientational average
is defined as
$\left<...\right>_{\mathbf{\hat{t}}}=\frac{1}{4\pi}\oint...\D\mathbf{\hat{t}}$. Substituting
Eq.~(\ref{Ch4:cdfns}) into Eq.~(\ref{Ch4:wav2}) leads to a
self-consistent equation for $\overline{W}$ due to the dependence of
$\rho$ on $\overline{W}$. In order to be consistent with the spherical
limit, we use $\rho\to\rho_f=1/(\overline{W}-V_0)$ due to the low
dimensional corrections discussed in Sec.~\ref{Ch4:Sec:volume3d}.

In accordance with the treatment of the surface density term
$\sigma(z)$ for 3d spheres, we obtain $\sigma(z)$ by simulating random
local configurations of $z$ contacting particles around a reference
particle and determining the average available free surface. This
surface is given by $\overline{S}^*(\mathbf{c}_{\rm m})$, where
$c_{\rm m}$ is the minimal contributed VB among the $z$ contacts in
the direction $\mathbf{\hat{c}}$. Averaging over many realizations
with a uniform distribution of orientations and averaging also over
all directions $\mathbf{\hat{c}}$ provides the surface density in the
form of a Monte-Carlo average $\sigma(z)=\left<\left<\overline{S}^*(\mathbf{c}_{\rm
    m})\right>\right>_{\mathbf{\hat{c}}}^{-1}$.  
 In this way we can
only calculate $\sigma(z)$ for integer values of $z$. For fractional
$z$ that are predicted from the evaluation of degenerate
configurations in the next section, we use a linear interpolation to
obtain $\overline{W}(z)$.

The theory developed so far captures the effect of particle shape on
the average Voronoi volume as a function of a given $z$. The particle
shape is taken into account in three quantities: {\it (i)}
$c^*(\mathbf{\hat{c}})$, parametrizing the surface of the shape; {\it
  (ii)} $s(\mathbf{r},\mathbf{\hat{t}},\mathbf{\hat{c}})$,
parametrizing the VB between two particles of relative position
$\mathbf{r}$ and orientation $\mathbf{\hat{t}}$; and {\it (iii)} the
contact radius $r^*(\mathbf{\hat{r}},\mathbf{\hat{t}})$. In the
spherical limit, all these quantities simplify considerably and the
spherical theory is recovered, which is analytically solvable as
discussed in Sec.~\ref{Ch4:Sec:volume3d}. For non-spherical shapes,
the VB Point {\it (ii)} above is in general not known in closed
form. In the next section, we discuss a class of shapes for which the
VB can be expressed in exact analytical form. For these shapes, the
theory can be applied in a relatively straightforward way, solving
$\overline{V}^*$ and $\overline{S}^*$ numerically and providing also
$\overline{W}(z)$ in numerical form. In Sec.~\ref{Ch4:Sec:degenerate}
we then discuss the missing part in the theory so far, the dependence
of $z$ itself on the particle shape.

\subsubsection{Parametrization of  non-spherical shapes}
\label{Ch4:Sec:vbnonsp}

In Sec.~\ref{Ch2:Sec:space} the precise definition of the VB between
two particles has been given. We have seen that the VB between two
equal spheres is identical to the VB between two points and is a flat
plane perpendicular to the separation vector. Finding the VB for more
complicated shapes is a challenging problem in computational geometry,
which is typically only solved numerically
\cite{Boissonat:2006aa}. Already for ellipsoids, one of the simplest
non-spherical shape, there is no exact expression for the VB. We
nevertheless approach this problem analytically by considering a
decomposition of the shape into overlapping spheres (see
Fig.~\ref{Ch4:Fig_shapes}a--d). Such a decomposition is trivial for
dimers, trimers, and $n$--mers, where the VB arises effectively due to
the interaction of four, six and $2n$ points. It also applies exactly,
e.g., to spherocylinders, which can be represented as dense overlaps
of spheres. In this case, the VB arises due to the effective
interaction of two lines and four points.

The Voronoi decomposition used for $n$--mers and spherocylinders can
be generalized to arbitrary shapes by using a dense filling of spheres
with unequal radii \cite{Phillips:2012aa}. However, even though this
approach is algorithmically well defined, it may become practically
tedious for dense unions of polydisperse spheres. An alternative
approach that is analytically tractable has been proposed in
\cite{Baule:2013aa}: Convex shapes are approximated by intersections
of a finite number of spheres. An oblate ellipsoid, e.g., is
approximated by a lens-shaped particle, which consists of the
intersection of two spheres \cite{Cinacchi:2015aa}. Likewise, an intersection of four spheres
can be considered an approximation to a tetrahedra, and six spheres
that of a cube (see Fig.~\ref{Ch4:Fig_shapes}e--h). The main insight is that the effective Voronoi interaction of these
shapes is governed by a symmetry: Points map to 'anti-points' (since
the interactions between spheres is inverted). The VB of
ellipsoid-like objects arises from the interaction between four
anti-points and four points in two dimensions or lines in three
dimensions, and thus falls into the same class as spherocylinders. The
VB between two tetrahedra is then due to the interaction between the
vertices (leading to four point interactions), the edges (leading to
six line interactions), and the faces (leading to four anti-point
interactions). For cubes the effective interaction is that of twelve
lines, eight points and six anti-points. This approach can be
generalized to arbitrary polyhedra.

With such a decomposition into overlapping and intersecting spheres,
we can study a large space of particle shapes using Edwards ensemble.
The resulting VBs can be parametrized analytically following an exact
algorithm \cite{Baule:2013aa} (see appendix~\ref{App:VB}).

\subsubsection{Dependence of coordination number on particle shape} 

\label{Ch4:Sec:degenerate}

\begin{figure}
\begin{center}
\includegraphics[width=4cm]{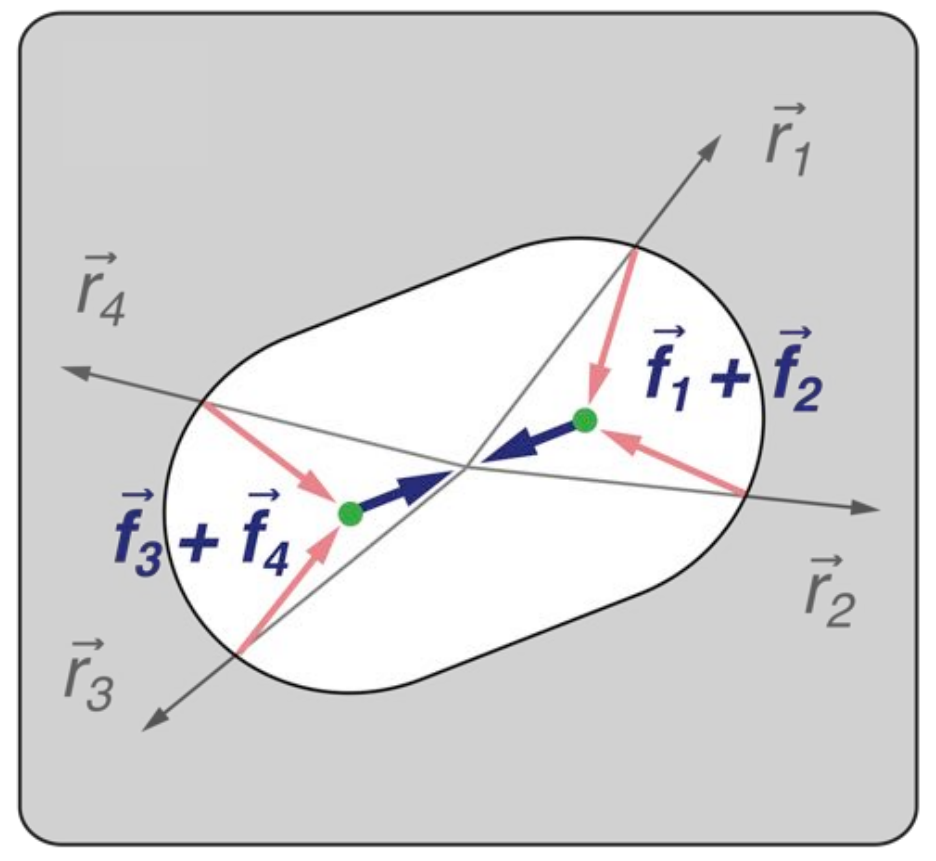}
\caption{\label{Ch4:Fig_degenerate}(Colors online) Illustration of a
  degenerate configuration of a spherocylinder. Vectors
  $\mathbf{r}_1,...,\mathbf{r}_4$ indicate contacts on the spherical
  caps. The normal vector projects the contact forces
  $\mathbf{f}_1,...,\mathbf{f}_4$ onto the centres of the spherical
  caps. Due to the symmetry of the two centres, the respective force
  arms are equal and force balance automatically implies torque
  balance. The force and torque balance
  equations~(\ref{Ch2:forceb},\ref{Ch2:torqueb}) are thus degenerate. From \cite{Baule:2013aa}.}
\end{center}
\end{figure}

As discussed in Sec.~\ref{Ch2:Sec:stability} the physical conditions
of mechanical stability and assuming minimal correlations motivate the
isostatic conjecture Eq.~(\ref{Ch2:iso}) $z=2d_{\rm f}$ in the
frictionless case. While isostaticity is well-satisfied for spheres,
packings of non-spherical objects are in general hypoconstrained with
$z<2 d_{\rm f}$, where $z(\alpha)$ increases smoothly from the
spherical value for $\alpha>1$
\cite{Donev:2004aa,Wouterse:2009aa,Donev:2007aa,Baule:2013aa}. The
fact that these packings are still in a mechanically stable state can
be understood in terms of the occurrence of stable degenerate
configurations, which have so far been shown to occur in packings of
ellipses, ellipsoids, dimers, spherocylinders, and lens-shaped
particles \cite{Chaikin:2006aa,Donev:2007aa,Baule:2013aa}. In the case
of ellipses, one needs in general four contacts to fix (jam) the
ellipse locally such that no displacement is possible
\cite{Alexander:1998aa}. However, it is possible to construct
configurations, where only three contacts are sufficient, namely when
the normal vectors from the points of contact meet at the same point
and the curvature on at least one of the contacts is flat enough to
prevent rotations \cite{Chaikin:2006aa}. Such a configuration is
degenerate since force balance automatically implies torque balance
such that the force and torque balance
equations~(\ref{Ch2:forceb}--\ref{Ch2:torqueb}) are no longer linearly
independent. Despite the fact that these configurations should have
measure zero in the space of all possible configurations, they are
believed to appear more frequently in simulation algorithms such as
the LS algorithm \cite{Donev:2007aa}.

For spherocylinders, the degeneracy appears due to the spherical caps,
which project the normal forces onto the end points of the central
line of the cylindrical part. If all of the contacts are on the
spherical caps, which will frequently occur for small aspect ratios,
force balance will then always imply torque balance, since the force
arms of the two points are identical (see
Fig.~\ref{Ch4:Fig_degenerate}). A similar argument applies to dimers
and lens-shaped particles, and can possibly be extended to other
smooth shapes. In the case of spherocylinders, a degeneracy also
appears for very large aspect ratios, because then all contacts will
predominantly be on the cylindrical part. As a consequence, the normal
vectors are all coplanar and the number of linear independent force
and torque balance equations is reduced by one predicting the contact
number $z \to 8$ as $\alpha \to\infty$, which is indeed observed in
simulations \cite{Zhao:2012aa,Wouterse:2009aa}.

\begin{figure*}
\begin{center}
\includegraphics[width=17cm]{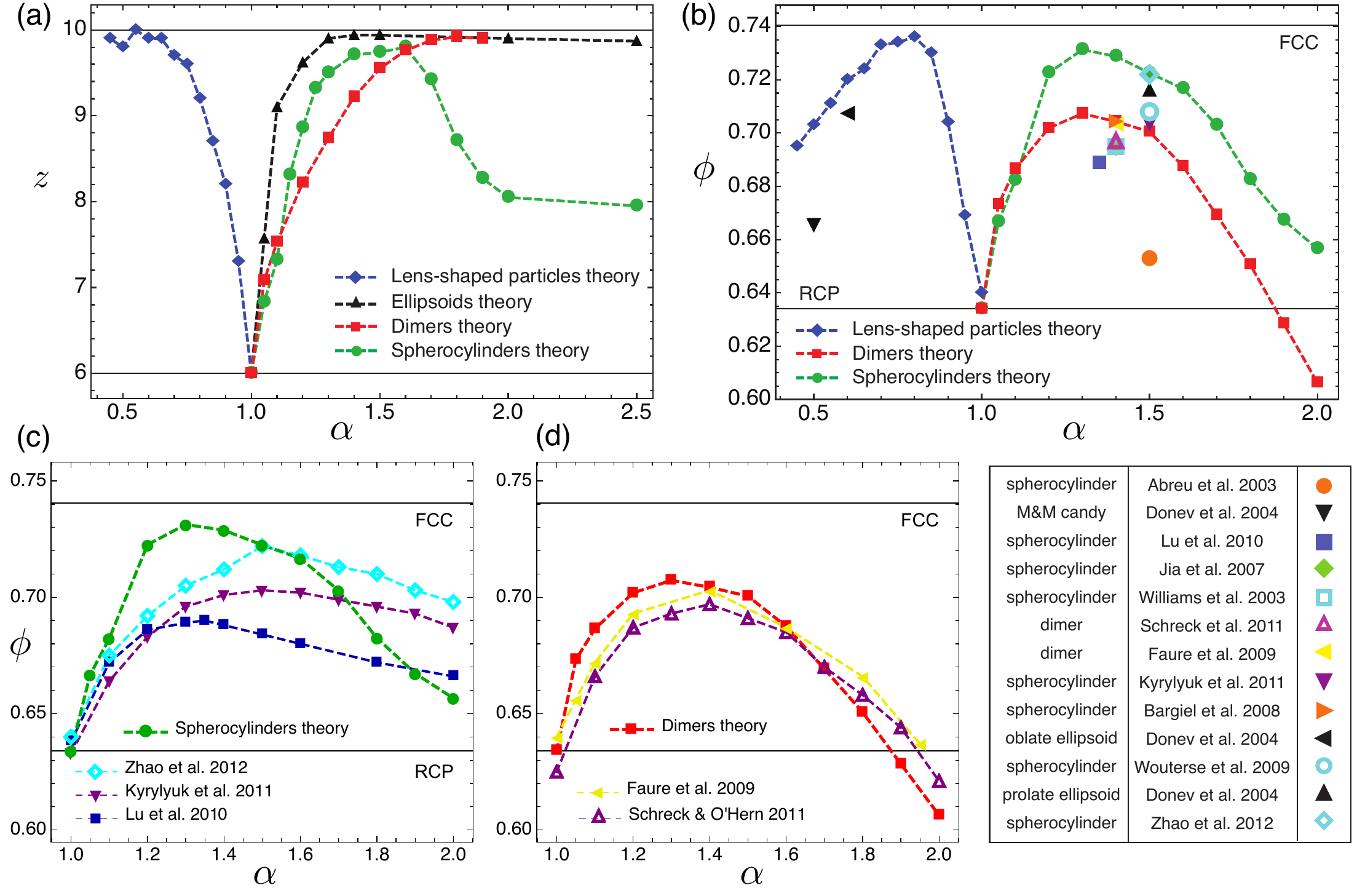}
\caption{\label{Ch4:Fig_results}(Colors online) Theoretical
  predictions for packings of non-spherical particles \cite{Baule:2013aa}. (a) The variation $z(\alpha)$ obtained by
  evaluating the occurrence of degenerate configurations for dimers,
  spherocylinders, ellipsoids of revolution, and lens-shaped
  particles. A smooth increase is obtained in agreement with
  simulation data. For spherocylinders, $z$ decreases to the value $8$
  as $\alpha\to\infty$. (b) Combining $z(\alpha)$ with the results on
  $\overline{W}_\alpha$ from the volume ensemble leads to theoretical
  predictions for $\phi(\alpha)$ exhibiting a density peak for dimers,
  spherocylinders, and lens-shaped particles. Results on $\phi_{\rm
    max}$ for the three shapes from simulations are indicated by
  symbols. The theory captures well both the location of the peak and
  the maximum density. (c) Detailed comparison of theory and
  simulations for spherocylinders \cite{Zhao:2012aa,Kyrylyuk:2011aa,Lu:2010aa}. The theoretical peak is slightly
  shifted to the left and more pronounced than in the empirical
  data. (d) Detailed comparison of theory and simulations for dimers \cite{Faure:2009aa,Schreck:2011aa}
  showing excellent agreement. Figs.~(a,b) from \cite{Baule:2013aa}.}
\end{center}
\end{figure*}

A quantitative method to estimate the probability of these degenerate
configurations is based on the assumption that a particle is always
found in an orientation such that the redundancy in the mechanical
equilibrium conditions is maximal \cite{Baule:2013aa}. This condition
allows us to associate the number of linearly independent equations
involved in mechanical equilibrium with the set of contact
directions. Averaging over the possible sets of contact directions
then yields the average effective number of degrees of freedom
$\tilde{d}_{\rm f}(\alpha)$, from which the coordination number
follows as $z(\alpha)=2\tilde{d}_{\rm f}(\alpha)$
\cite{Baule:2013aa}. This approach recovers the continuous transition
of $z(\alpha)$ from the isostatic spherical value $z=6$ at $\alpha=1$,
to the isostatic value $z=10$, for aspect ratios above $\approx 1.5$
observed in ellipsoids of revolution, spherocylinders, dimers, and
lens-shaped particles, Fig.~\ref{Ch4:Fig_results}a. The trend compares
well to known data for ellipsoids \cite{Donev:2004aa} and
spherocylinders \cite{Zhao:2012aa,Wouterse:2009aa}.

Combining these results on $z(\alpha)$ with the results of
Sec.~\ref{Ch4:Sec:voronoins} on the average Voronoi volume
$\overline{W}_\alpha$ leads to a close theoretical prediction for the
packing density $\phi(\alpha)=V_0/\overline{W}_\alpha(z(\alpha))$
which does not contain any adjustable parameters.  Figure
\ref{Ch4:Fig_results}b presents the results for dimers, spherocylinders
and lenses showing that the theory is an upper bound of the maximal
densities measured in simulations. The theory predicts the maximum
density of spherocylinders at $\alpha = 1.3$ with a density $\phi_{\rm
  max} = 0.731$ and that of dimers at $\alpha= 1.3$ with $\phi_{\rm
  max} =0.707$. For lens-shaped particles a density of $\phi_{\rm
  max}=0.736$ is obtained for $\alpha=0.8$, representing the densest
random packing of an axisymmetric shape known so far. The theoretical
predictions of $\phi(\alpha)$ compare quite well with the available
numerical data for spherocylinders and dimers
(Figs. \ref{Ch4:Fig_results}c, d). The numerical results are obtained with a range of different packing algorithms and show a large variance in terms of the maximal packing densities obtained, for the same shape. The appearance of such a range of densities is understood in detail for the case of spheres, see the discussion in Sec.~\ref{Ch2:Sec:rcp}. As for spheres, the single RCP value calculated within the Edwards ensemble for a given shape is interpreted as a maximum entropy value.

By plotting $z$ against $\phi$
parametrically as a function of $\alpha$, we can also include our
results in the $z$--$\phi$ phase diagram, which is thus extended from
spheres to non-spherical particles and discussed next.  By
plotting $(\phi,z)$ the apparent cusp-like singularity at the
spherical point $\alpha=1$ in $z(\alpha)$ and $\phi(\alpha)$
(Figs.~\ref{Ch4:Fig_results}a, b) disappears and the spherical RCP
point becomes as any other point in the phase diagram.

\subsection{Towards an Edwards phase diagram for all jammed matter}

\label{Ch4:phasediagram}

\begin{figure*}
\begin{center}
\includegraphics[width=16cm]{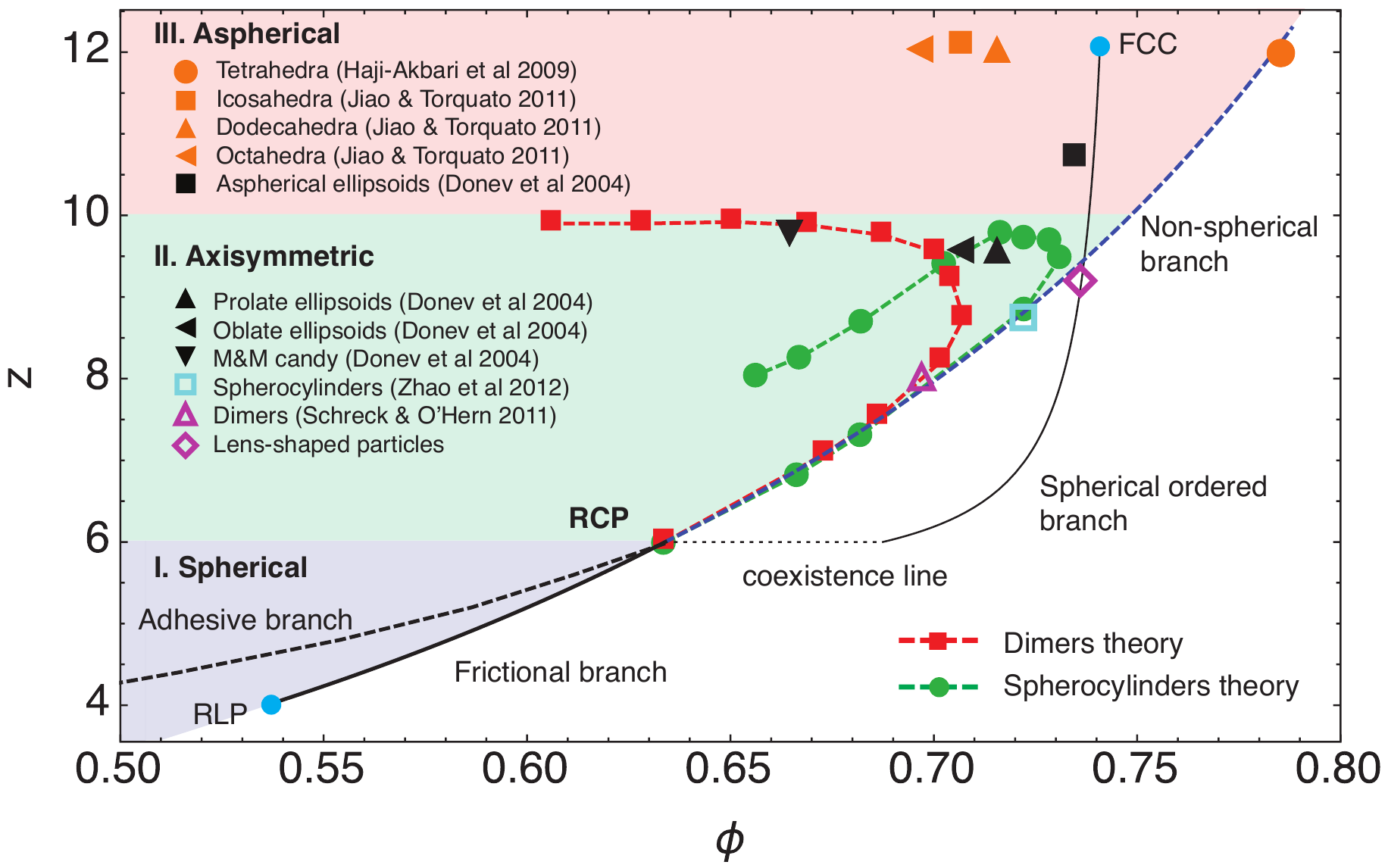}
\caption{\label{Ch4:Fig_phasediagram}(Colors online) Unifying phase diagram in
  the $z$--$\phi$ plane resulting from the Edwards volume
  ensemble theory. Theoretical results on the equations of state for spheres
  with and without adhesion and dimers/spherocylinders are plotted
  together with empirical results on maximal packing densities for
  non-spherical shapes from the literature (where $z$ and $\phi$ have
  been determined in the same simulation). Different phases are
  identified by the symmetry of the constituents. Different equations
  of state due to friction, adhesion, shape, and (partial) order all
  come together at the RCP point. Indicated are the frictional branch \cite{Song:2008aa}, spherical ordered branch \cite{Jin:2010aa}, non-spherical branch \cite{Baule:2013aa}, and adhesive branch \cite{Liu:2015aa}. Adapted from \cite{Baule:2014aa}.}
\end{center}
\end{figure*}

The results from Secs.~\ref{Ch4:Sec:ensemble}, \ref{Ch4:Sec:adhesion},
and \ref{Ch4:Sec:degenerate} are combined in a phase diagram of jammed
matter that can guide our understanding of how random arrangements of
particles fill space as shown in Fig.~\ref{Ch4:Fig_phasediagram}. The
representation in the $z$--$\phi$ plane is in a way the most natural
choice, since both $\phi$ and $z$ are macroscopic observables that characterize the thermodynamic state of the packing. They
can also be measured in simulations in a straightforward way. Although
Fig.~\ref{Ch4:Fig_phasediagram} is far from complete, we observe clear
classifications of packings based on the {\it symmetry} and {\it
  surface properties} of the constituents. Horizontal phase boundaries
are identified by the isostatic condition for frictionless particles,
predicting $z=6$ for isotropic shapes and $z=10$ ($z=12$) for
rotationally symmetric (fully asymmetric) shapes respectively. The
frictionless RCP point at $\phi_{\rm Edw}=0.634...$ and $z=6$ plays a
prominent role in the phase diagram, despite that it contracts the
J-line. It splits up (although in a continuous manner, except for
ordering) the equation of state into four different branches governed
by friction, shape, adhesion, and order, as follows:

{\it Frictional branch}. The infinite compactivity RLP branch connects
the RCP point $(0.634,6)$ with the minimal RLP point at
$(0.536,4)$. This branch is the upper limit of the triangle of
mechanically stable disordered sphere packings depicted in the phase
diagram for 3d monodisperse spheres in Fig.~\ref{Ch4:Fig_xlines}. The
RLP branch is parametrized by varying the friction $\mu$ and thus $z$
in the equation of state~(\ref{Ch4:phirlp}).

{\it Non-spherical branch}. Surprisingly, we find that both dimer and
spherocylinder packings appear as smooth continuations of spherical
packings. The analytic form of this continuation from the spherical
random branch can be derived (blue dashed line in
Fig.~\ref{Ch4:Fig_phasediagram}) by solving the self-consistent
equation~(\ref{Ch4:wav2}) perturbatively for small aspect ratios
\cite{Baule:2013aa}.

 A comparison of our theoretical results with
empirical data for a large variety of shapes indicates that the
analytic continuation provides an upper bound of density on the
$z$--$\phi$ phase diagram for a fixed $z$. Maximally dense disordered
packings appear to the left of this boundary, while the packings to
the right of it are partially ordered. We observe that the maximally
dense packings of dimers, spherocylinders, lens-shaped particles and
tetrahedra all lie surprisingly close to the analytic continuation of
RCP. Whether there is any deeper geometrical meaning to this remains
an open question. Recent exact local expansions from the spherical RCP
point to arbitrary shapes agree very well with our results and may
shed further light on this question \cite{Kallus:2016aa}.  We also
notice that the frictional and non-spherical branches are continuous
at the spherical RCP point suggesting that a variation in friction
might be analogous to varying shape in the phase diagram.

{\it Adhesive branch}. The non-spherical branch can also be continued
into the adhesive branch of spheres, which splits off at RCP. The
adhesive branch describes the universal high adhesion regime for
$Ad>1$ reaching the adhesive loose packing (ALP) point at $\phi=1/2^3$
and $z=2$ (see Sec.~\ref{Ch4:Sec:adhesion}).

{\it Spherical ordered branch}. As discussed in
Sec.~\ref{Ch2:Sec:rcp}, the RCP point has been associated with the
freezing point of a first order phase transition between a fully
disordered packing of spheres and the crystalline FCC phase
\cite{Radin:2008aa,Jin:2010aa}. The signature of this disorder-order
transition is a discontinuity in the entropy density of jammed
configurations as a function of the compactivity. Experiments on hard sphere packings indeed confirm the first order transition scenario, observing the onset of crystallization at $\phi_{\rm f}\approx 0.64$ at the end of the frictional branch, as well as the coexistence line \cite{Francois:2013aa,Hanifpour:2014aa,Hanifpour:2015aa}. The spherical ordered branch provides
another boundary, which separates tetrahedra from all other shapes:
Tetrahedra are the only shape that pack in a disordered way denser
than spheres in a FCC crystal.

The picture that emerges from this phase diagram is that spherical
packings can be generated on the frictional branch between the RLP and
RCP limits by variation of the inter-particle friction and along the
adhesion branch by varying interparticle attraction. Beyond RCP, these
two lines can be continued smoothly by deforming the sphere into
elongated shapes. The ordered branch does not connect smoothly to any
of these branches, instead appears through a first order phase
transition with a coexistence regime. It suggests that introducing
order is a more drastic modification than modifying the particle
interactions due to geometry or surface frictional properties. This
distinction is similar to the one between discontinuous first and
continuous higher-order phase transitions. 

Overall, it seems that the central importance historically given to
the spherical RCP point may not be justified. In the whole share of
things, the spherical point appears as any other inconsequential point
in a continuous variation of jammed states driven by friction,
attraction and shape. It is as though each jammed state (ranging from
spherical to dimers, trimers, polymers, spherocylinders, ellipsoids,
tetrahedra and cubes, from frictionless to frictional and adhesive
grains) carries the features of one great single organizing principle
in which all the jammed states organize, too; so that everything links
to everything else, moved by one organizing idea which is the
universal physical principle in nature \cite{Schopenhauer:1974aa}.

Such an organizing principle is captured by the phase diagram in
Fig.~\ref{Ch4:Fig_phasediagram} where the volume fraction as a
function of $\alpha$ for non-spherical particles appears as an
analytical continuation of the equation of state for the spherical
particles. It is as though the sphere system with friction can be made
analogous to a non-spherical system without friction by following the
continuation branch. Likewise, the RCP point bifurcates into other
equations of state following the appearance of adhesion between
particles as seen in Fig.~\ref{Ch4:Fig_phasediagram}. We may
conjecture that all these packings with different interactions (from
hard-spheres to attraction and friction) and different shapes (from
spheres to ellipsoids, etc.)  can be made part of an organizing
principle embodied in the statistical mechanical laws.

\section{Jamming Satisfaction Problem, JSP}
\label{Sec:constraints}

We close our review by providing a novel understanding of the jamming
criticality under the Edwards ensemble as the phase transition between
the satisfiable and the unsatisfiable phases of the Jamming
Satisfaction Problem. At the very end we suggest a unifying view of
the Edwards ensemble of grains with the statistical mechanics of
spin-glasses.

As we explained in Sec. \ref{Ch2:Sec:stability}, a packing can be
described as an ensemble of particles with given positions and
orientations, satisfying a set of geometrical and mechanical
constraints. As such, it is an instance of a constraint satisfaction
problem: the Jamming Satisfaction Problem (JSP). Solving the JSP, in
general, is a very complicated task, and one needs to resort to some
approximations. The first main approximation that we applied across
this review consisted in decoupling the geometrical problem of
determining the contact network of the packing from the mechanical
problem of finding the force distribution. Thus, in
Sec.~\ref{Sec:volume} we developed the Edwards volume ensemble that
considers in detail the volume ensemble, but does not directly
consider the full force ensemble, which is only taken into account by
the global isostatic constraint on the average coordination number
establishing force balance.

Below, we consider another reduced JSP where one now fixes the
geometry of the packing considering it as a random graph (thus, fixing
the volume ensemble), and then considering the full force ensemble on
these random graphs to find the force distribution
\cite{Bo:2014aa}. An ensemble average over all possible random graphs
consistent with prescribed (local) conditions of jamming and excluded
volume on the positions of neighbouring particles is performed to
obtain the force distribution. Such a reduced JSP is therefore
amenable to be solved for sparse networks by the cavity method from
spin-glass theory \cite{Mezard:2001aa,Mezard:2009aa}, where one
considers the geometric configuration of the particles in the packing
as fixed, and then finds the force distribution \cite{Bo:2014aa}.

This force distribution is nothing but the uniform Edwards' measure
${\Theta_{\rm jam}}$ over all possible solutions of the JSP Eq.~(\ref{eq:thetajam}) where the hard-core constraint is relaxed, being
automatically satisfied because we are considering the contact network
fixed. To emphasize the dependence of ${\Theta_{\rm jam}}$ solely on
the force configuration $\{\mathbf{f}\}$ for a given realization of
the contact network $\{\mathbf{d}\}$, we use the notation
${\Theta_{\rm jam}}(\{\mathbf{f}\}|\{\mathbf{d}\})=P(\{\mathbf{f}\})$,
with the normalization or partition function $\mathcal{Z}$ is  the
number of solutions of this JSP.  The important point is that if
$\mathcal{Z}\geq1$ then there exists a solution to the JSP, i.e., it
is satisfiable (SAT). Conversely, if $\mathcal{Z}<1$ there are no
solutions to the JSP, i.e., it is unsatisfiable (UNSAT)
\cite{Kirkpatrick:1994aa}.

The SAT/UNSAT threshold of the JSP is marked by the coordination
number $z_c^{\rm min}(\mu)$ that separates the region where solutions
do exist (i.e. where $\mathcal{Z}>1$) from the region without
solutions (where $\mathcal{Z}<1$), corresponding to an
underdetermined/overdetermined set of equations, respectively
\cite{Bo:2014aa}.  In the limiting case of frictionless particles,
$z_c^{\rm min}(\mu)$ should be compared with the naive Maxwell
counting isostatic condition: $z_c^{\rm min}(\mu=0)=2d_f$, although
the JSP takes into account the full set of constraints,
Eqs. (\ref{eq:thetajam}), rather than only force balance as in Maxwell
counting. The JSP thus extends this naive counting to the full set of
constraints including friction $\mu$. A jammed isostatic assembly of
particles lies exactly on the edge between these two phases, i.e.,
where a solution to the JSP first appears as one increases the average
coordination number $z(\mu)$.  Figure \ref{Ch5:Fig_zmin_linbo} shows
the average coordination number $z_c^{\rm min}(\mu)$ at the jamming
transition as a function of the friction coefficient $\mu$ in a 2d
sphere packing, obtained by solving the JSP through the cavity method
as explained next~\cite{Bo:2014aa}. Results are consistent with
existing numerical simulations
\cite{Makse:2000aa,Silbert:2002ab,Kasahara:2004aa,Shundyak:2007aa,Silbert:2010aa,Papanikolaou:2013aa,Shen:2014aa,Song:2008aa}.

\subsection{Cavity approach to JSP}
\label{Ch5:Sec:cavity}

Solving the JSP amounts to compute the single force distributions
$P(\mathbf{f}^i_a)$ at the contacts $a$'s of the particle
$i$'s. However, calculating these single force distributions
$P(\mathbf{f}^i_a)$ from the joint distribution $P(\{\mathbf{f}\})$
Eq.~(\ref{eq:thetajam}) is still a very demanding computational task,
which requires some additional mean-field approximations to be solved.

\begin{figure}
\includegraphics[width=\columnwidth]{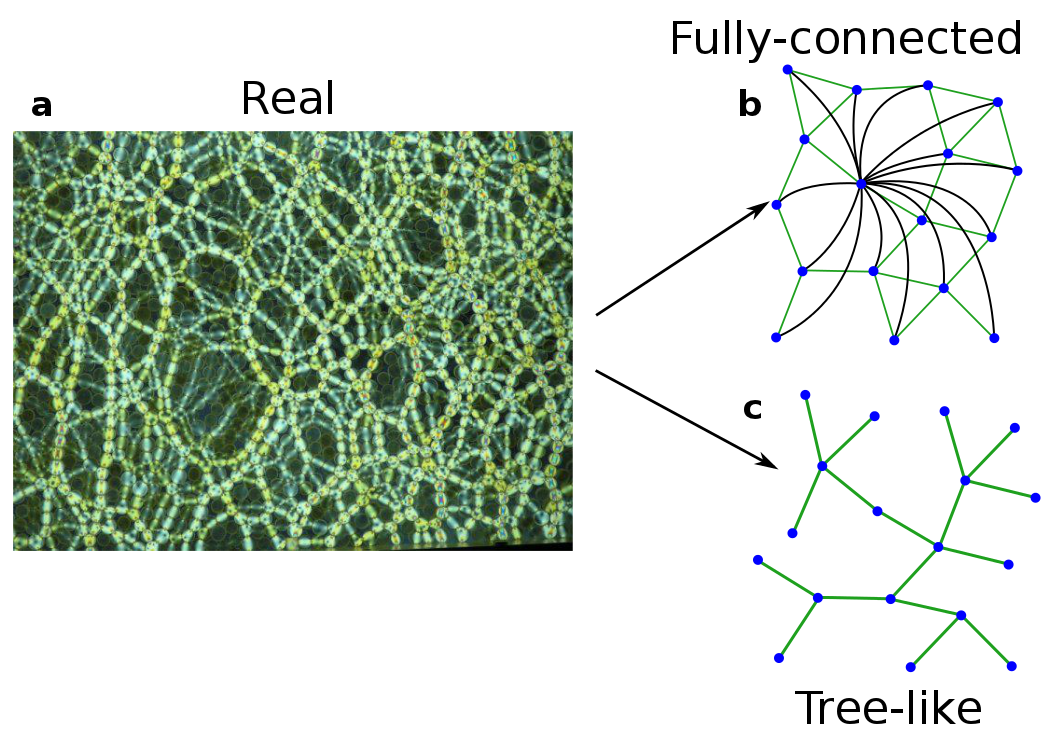}
\caption{\label{Ch5:Fig_bethe_bragg} (a) A real finite-dimensional
  packing is composed of strongly correlated force chains and
  geometrical loops at short scale (image reprinted with permission
  from the Behringer Group, Duke University). However,
  state-of-the-art theoretical approaches to describe this correlated
  structure rely upon mean-field infinite-dimensional approximate
  treatments of such a packing as a: (b) Fully-connected packing where
  every single particle interacts with any other particle in the
  packing; the real interaction network is approximated by a complete
  graph, i.e., each node is connected with all other nodes as shown
  for one of them.  (c) Locally-tree like packing where the real
  network is approximated by a sparse random graph that locally looks
  like a tree structure with no loops, i.e., loops in the network are
  neglected, except at relatively large scales that diverge with
  system size, although very slowly as $\ell \sim \ln N$.}
\end{figure}

There are two preferred mean-field theories (both of infinite
dimensional nature): the first one is the infinite range model, which
assumes that each particle is in contact with every other particle in
the packing. The archetypical model is the Sherrington-Kirkpatrick
(SK) model of fully connected spin-glasses \cite{Sherrington:1975aa}
which has been adapted to the hard-sphere case in \cite{Parisi:2010aa}
(see Secs.~\ref{Ch2:Sec:rcp}).  As a result of this approximation
scheme, the real finite dimensional contact network
Fig.~\ref{Ch5:Fig_bethe_bragg}a is substituted by a fully-connected
network of possible interactions, i.e., a complete graph as shown in
Fig.~\ref{Ch5:Fig_bethe_bragg}b. The solution of such a model is
possible since, in a complete graph, each interaction becomes very
weak, rendering a fully connected model into a weakly connected system
that can be solved exactly under the hierarchy of replica symmetry
breaking schemes \cite{Mezard:2009aa,Parisi:2010aa}. A simpler version
than the SK model, yet showing all the phenomenology of jamming, is a
model adapted from machine learning; the perceptron recently studied
in \cite{Franz:2015aa,Franz:2015ab}.

A second mean-field theory of choice consists in approximating the
contact network by a sparse random graph \cite{Mezard:2001aa}, which
allows one to preserve an essential property of real finite
dimensional packings: the finite coordination number $z$.  The sparse
random graph scheme assumes that the local contact network around each
particle can be approximated by a tree-like structure, i.e. it
neglects the strong local correlations of loops and force chains of a
real packing Fig.~\ref{Ch5:Fig_bethe_bragg}a by a locally tree-like
structure, Fig.~\ref{Ch5:Fig_bethe_bragg}c.  Under this approximation
the JSP can be solved by a method known as cavity method
\cite{Mezard:2001aa,Mezard:2009aa}, which we explain next.

It should be noticed that, although the cavity approach is a mean
field theory valid for infinite dimensions, a dimensional dependence
appears in the non-overlap condition in the definition of the network
ensemble, see \cite{Bo:2014aa} for details. The crucial quantity to
consider in the cavity method is not the single force distribution
itself $P(\mathbf{f}^i_a)$, but a modified one, called the cavity
force distribution and denoted by $P_{i\to a}(\mathbf{f}^i_a)$.
Physically, $P_{i\to a}(\mathbf{f}^i_a)$ is the probability
distribution of the force $\mathbf{f}^i_a$ at the contact $a$ in a
modified packing where the particle $j$ touching the particle $i$ at
the contact $a$ has been removed (from where the name cavity derives).
The rationale to consider $P_{i\to a}(\mathbf{f}^i_a)$ instead of the
``true" force distribution $P(\mathbf{f}^i_a)$ is that for the cavity
distributions it is possible to derive a set of self-consistent
equations if one neglects the correlation between $P_{i\to
  a}(\mathbf{f}^i_a)$ and $P_{j\to a}(\mathbf{f}^j_a)$ (hence the need
of a tree-like network) \cite{Bo:2014aa}.

\begin{figure}
\includegraphics[width=\columnwidth]{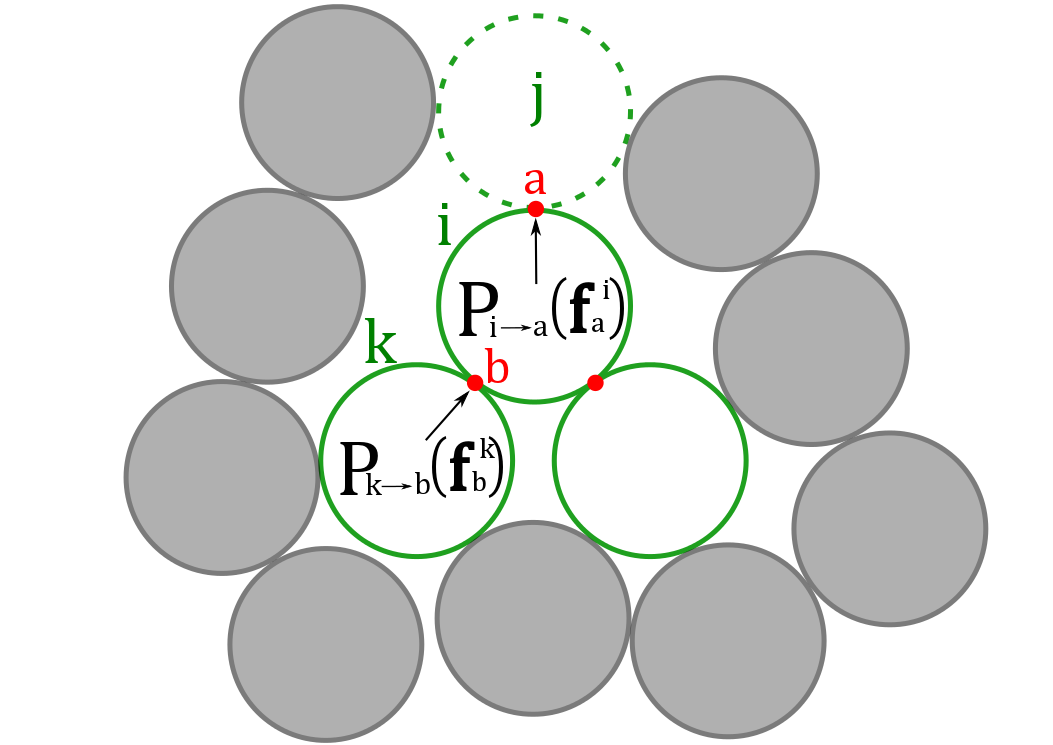}
\caption{Calculation of the cavity force distribution $P_{i\to a}$. 
First particle $j$ (dashed contour) is virtually removed from the packing. 
Then $P_{i\to a}$ for particle $i$ is computed by convoluting the distributions 
$P_{k\to b}$ of the neighboring particles $k$ with the local 
mechanical constraint $\chi_i$ enforcing force and torque balances. 
\label{Ch5:Fig_messpass} }
\end{figure}

For example, the cavity equation for $P_{i\to a}(\mathbf{f}^i_a)$ can
be obtained by simply convoluting the cavity force distributions
$P_{k\to b}(\mathbf{f}^k_b)$ of the particles $k\neq j$ neighbors of
particle $i$ with the local mechanical constraint $\chi_i$, as
depicted in Fig.~\ref{Ch5:Fig_messpass}, and mathematically expressed
as follows: \be P_{i\to a}(\mathbf{f}^i_a)\propto \int
\prod_{b\in\partial i\setminus a} d\mathbf{f}^k_b\ \chi_i
\prod_{k\in\partial b\setminus i}P_{k\to b}(\mathbf{f}^k_b),
\label{eq:cavity_eqs}
\ee
where the symbol $\propto$ implies a normalization factor, and the 
mechanical constraint $\chi_i$ on particle $i$ is given by:
\begin{align}
\chi_i\Big(\{\mathbf{f}_a^i\}_{a\in\partial i}\Big) &=
\delta\left(\sum_{a\in\partial i}\mathbf{f}_a^i\right)
\delta\left(\sum_{a\in\partial
  i}\mathbf{d}_a^i\times\mathbf{f}_a^i\right)\nonumber\\ &\times
\prod_{a\in\partial i}\theta\left(\mu f^i_{a,n} -
|\mathbf{f}^i_{a,\tau}|\right) \theta\left(
-\mathbf{d}_a^i\cdot\mathbf{f}_a^i \right).
\end{align}
Notice that the contact directions $\{\mathbf{d}_a^i\}$ are kept
fixed: they represent the "quenched" disorder introduced by the
underlying contact network, which is kept fixed.

Once the set of cavity equations \eqref{eq:cavity_eqs} has been
solved--- e.g. by iteration under the Replica Symmetric (RS)
assumption \cite{Bo:2014aa}--- one can reconstruct back the original
force distribution at contact $a$ by simply multiplying the cavity
force distributions $P_{i\to a}(\mathbf{f}^i_a)$ and $P_{j\to
  a}(\mathbf{f}^j_a)$ coming from the two particles $i$ and $j$ in
contact at $a$: \be P(\mathbf{f}^i_a)\propto\ P_{i\to
  a}(\mathbf{f}^i_a)P_{j\to a}(\mathbf{f}^j_a).  \ee
 The result shows an exponential decay at
large forces and a non-zero value for $P(f)$ at $f=0$, i.e., it gives
an exponent at the RS level
\be
\label{Ch5:thetars}
\theta_{\rm RS}=0 \ee for the small force scaling $P(f)\sim
f^{\theta}$, Eq.~(\ref{Eq:theta}). This last prediction is
inconsistent with simulation results, which find a nonzero value of
the exponent $\theta$ in the interval $0.2\leq \theta\leq 0.5$. It
should be noted that Eq.~(\ref{Ch5:thetars}) is obtained exactly at
the thermodynamic limit, so no finite size effects are expected.

The discrepancy could be in principle due to the abundance of short
loops in the real finite-dimensional contact network that are
neglected by the locally tree-like contact network structure
considered by the cavity method. However, it is known that the
fraction of short force loops decreases with dimension at jamming--- a
results valid for any random network in infinite dimensions--- yet,
the non-zero weak force power-law exponent is obtained in the high
dimensional calculations in the fully connected case
\cite{Charbonneau:2012aa}. In this case, the complexity lost by the
consideration of a uniform fully connected network is somehow overcome
by the fractal complexity provided by the fullRSB solution, which in
this case, gives rise to the concomitant non-zero small-force
exponent.  Whether a zero exponent result is the byproduct of the
cavity calculation being done at the RS level or of the absence of
loops in the structure is to be determined.

A similar situation appears in the replica approach to the problem:
The original 1RSB calculation under the replica approach of the force
distribution for hard sphere glasses done in
\cite{Parisi:2010aa} led to a trivial scaling \be \theta_{\rm 1RSB} =
0,
\label{1rsb-force}
\ee while the non-zero exponent was only obtained when the full RSB
calculation was performed \cite{Charbonneau:2014aa} \be \theta_{\rm
  fullRSB}=0.42...
\label{fullrsb-force}  \ee It should be noticed, though, that 1RSB level
calculations and above are substantially more difficult to perform
with the cavity method than with replicas (e.g., no
  calculation exists above 1RSB with the cavity method for any model,
  although it has recently been conjectured how the cavity method
  could be used to describe the full RSB scenario
  \cite{Parisi:2017aa}).

\begin{figure}
\includegraphics[width=0.85\columnwidth]{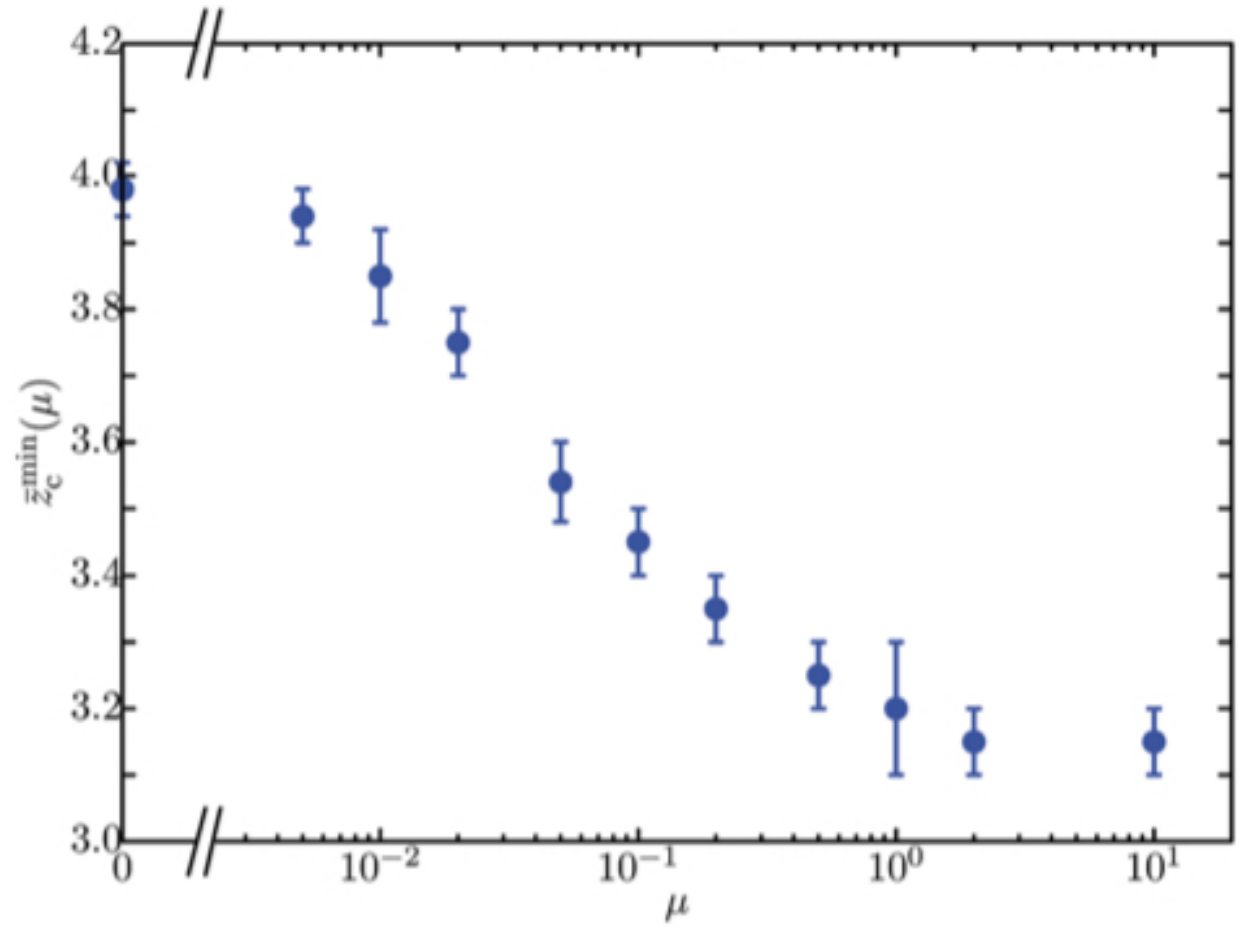}
\caption{Linear-log plot of average coordination number $z_c^{\rm
    min}(\mu)$ at the jamming transition as a function of the friction
  coefficient $\mu$ in 2-D sphere packing calculated with the cavity
  method. The curve $z_c^{\rm min}(\mu)$ separates the SAT/UNSAT
  phases of jamming. For $z>z_c^{\rm min}(\mu)$, the force balance
  equations are satisfied while they are not when $z<z_c^{\rm
    min}(\mu)$. At the transition $z_c^{\rm min}(\mu)$ for a given
  $\mu$ a jammed critical state exists separating the SAT from the
  UNSAT phases.  $z_c^{\rm min}(\mu)$ shows a monotonic decrease with
  increasing $\mu$ from the isostatic Maxwell estimation $z_c^{\rm
    min}(\mu=0)=2D=4$ to $z_c^{\rm min}(\mu=\infty)\geq D+1=3$.  Error
  bar indicates the range from the largest $z_c^{\rm min}(\mu)$ having
  no solution to the smallest $z_c^{\rm min}(\mu)$ having
  solution. Data points represents the mean of the range. From \cite{Bo:2014aa}.
\label{Ch5:Fig_zmin_linbo} }
\end{figure}

Despite these discrepancies, the main result of the cavity approach is
the detection of the SAT/UNSAT transition of the JSP for sphere
packings with arbitrary friction coefficient, and a lower bound
estimate of the critical coordination number $z_c^{\rm min}(\mu)$ at
the jamming transition as a function of the friction coefficient
$\mu$, as shown in Fig.~\ref{Ch5:Fig_zmin_linbo}. Moreover, the cavity
method seems a promising way to study JSPs for packings with particles
of arbitrary shapes, which are difficult to perform with replicas.

\subsection{Edwards uniform measure hypothesis in the Edwards-Anderson spin-glass model}

The main goal of this section is to investigate Edwards' conjecture of
equiprobable jammed states in the spin-glass model first introduced by
Edwards together with Anderson \cite{Edwards:1975aa}, thus, bringing
together two of the most significance contributions of Edwards:
spin-glasses \cite{Edwards:1975aa} and granular matter
\cite{Edwards:1989aa}.  We leverage some rigorous results
~\cite{Newman:1999aa} to understand what is effectively right and what
may go wrong with that hypothesis by precisely stating it in terms of
metastable states in spin-glasses and jamming.  We will see how this
definition of metastable jammed states leads to the most precise test
so far of the Edwards uniform measure hypothesis in the exactly
solvable SK model~\cite{Sherrington:1975aa}, which we propose to
perform in Sec.~\ref{pandora}.

The Ising spin-glass on the $d$-dimensional cubic lattice $Z^d$, also
known as the Edwards-Anderson model, is described by the following
Hamiltonian~\cite{Edwards:1975aa}:
\begin{equation}
\mathcal{H}(\vec{\sigma}) = - \sum_{\langle ij\rangle}J_{ij}\s_i\s_j\ ,
\label{eq:SGHamilt}
\end{equation} 
where $i$ are the sites of $Z^d$, the spins $\s_i=\pm1$, and the
sum is over nearest neighbor spins. The couplings $J_{ij}$ are independent
identically distributed random variables, and we assume their common
distribution to be continuous and to have a finite mean.

\begin{figure}[h!]
\begin{center}
\includegraphics[width=.7\columnwidth]{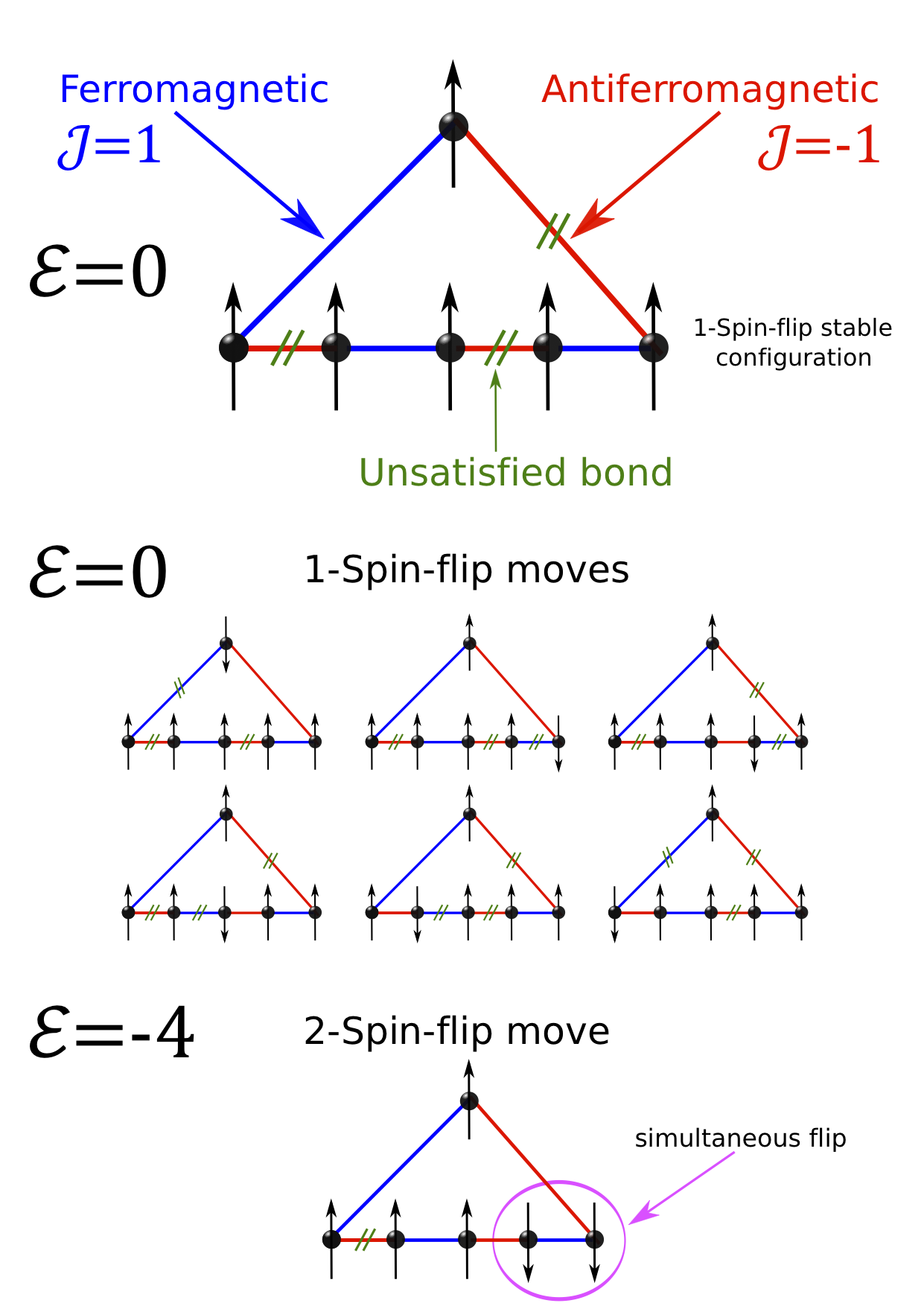}
\caption{Example of a $1$-Spin flip stable configuration.}
\label{fig:1-SF}
\end{center}
\end{figure}

A distinguishing property of spin glasses, which pertains to many
complex systems including granular media, is that they feature a
``rugged energy (or free energy) landscape". 
To be more clear, let us consider a
zero-temperature dynamics, where at each time step a spin is randomly
chosen and flips if it lowers the energy, otherwise it does not move,
until no more spins will flip.
At variance with a pure ferromagnet, in the spin glass this dynamics
arrests very quickly, and also at a quite high-energy state, the
reason being due to, precisely, the abundance of metastable states.
The type of metastable states concerned in this specific case are
$1$-SF metastable states, discussed in Section~\ref{Sec:meta} and
Fig.~\ref{fig:PD2}a, since they are reached following a dynamics that
flips one spin at a time: when the system arrives in one of these
configurations, no single spin can lower the energy by flipping, but
if two neighboring spins are allowed to flip simultaneously, then
lower energy states are available. In other words, $1$-SF states are
stable against a single spin-flip, but not necessarily against two 
(or more) simultaneous spins-flip. An example of one-spin-flip 
metastable state is shown in Fig.~\ref{fig:1-SF} along with a possible 
two-spin-flip move (shown in the lowest panel) needed to escape the 
$1$-SF metastable trap. 
As discussed in Table~\ref{Ch2:table1} these 1-SF 
metastable states are analogous to the locally jammed states introduced 
by~\cite{Torquato:2001aa} and called 1-PD in the table.

The concept of $1$-SF metastable states can be easily extended to
$k$-spin-flip ($k$-SF) metastable states, even without resorting to a
specific dynamics, but using solely the Hamiltonian of the system
Eq.~\eqref{eq:SGHamilt} \cite{Biroli:2000aa}. We define a $k$-spin-flip metastable state as
a (infinite volume) configuration whose energy cannot be lowered by
flipping any connected subset of $1,2,\dots, k$ spins.  In particular,
the ground states of the system correspond to configurations whose
energy cannot be lowered by flipping any finite number of spins,
i.e., they are found in the limit $k\to\infty$, hence the ground
state of the spin-glass is the $\infty$-SF state, Fig.~\ref{fig:PD2}a.

The $k$-SF metastable
states are analogous to the $k$-PD metastable collective jamming states
defined in Table \ref{Ch2:table1} that generalize the concept of
collective jamming in \cite{Torquato:2001aa}. The corresponding ground
state of jamming is then the $\infty$-PD state.  We, thus, end up with
a nice analogy between spin-glasses and jamming which we can leverage
to harness the nature of metastable jammed states in terms of exact
results for spin-glass metastable states obtained by
~\cite{Newman:1999aa}.  

It is important to see that the $k$-PD or $k$-SF states are hierarchically
organized one inside another as seen in Fig. \ref{fig:PD2}a.  For
instance, 2-PD (2-SF) metastable states form a subset of the 1-PD (1-SF)
metastable states, since states which are 2-SF-stable are automatically
1-SF-stable, but the converse is not necessarily true. 
Also, the energies of 2-SF metastable states may cross, in principle, the
energies of 1-SF metastable states, Fig.~\ref{fig:PD2}a. This hierarchy
defines the k-SF-core metastable states and the k-SF-shell: the 1-SF-shell 
consists of 1-SF metastable states which are not in 2-SF-core. In
general, the $k$-SF-shell consists of $k$-SF metastable states which are not 
in the $k+1$-SF-core. The $\infty$-SF core is then the ground state.

Now we may ask: how do we visit the $k$-SF metastable states for $k>1$?
To answer this question we need to introduce more precisely the
concept of {\it dynamics}.

A $k$-spin-flips dynamics is defined in such a way that rigid flips of
all lattice animals (finite connected subset of $Z^d$) up to $k$ spins
can occur. For example, in the case $k=2$ both single-spin flips and
rigid flips of all nearest neighbor pairs of spins are allowed (see 
the bottom panel in Fig.~\ref{fig:1-SF} as an example of a $2$-SF move).
At each step of the dynamics a lattice animal of size $\ell\leq k$ is
chosen at random with probability $p_\ell$ and it flips if the
resulting configuration has lower energy, otherwise it does not flip.
We denote by $\omega_k$ a given realization of this $k$-SF dynamics
\cite{Newman:1999aa} and the ending metastable configuration of such a
path as $\vec{\s}^\infty_k$.

Having defined the $k$-SF dynamics, we can now state an important
rigorous result obtained by Newman and Stein \cite{Newman:1999aa}:
{\it every end state} $\vec{\s}^\infty_k$ {\it of a dynamics} $\omega_k$
{\it has the same energy density} $e_k$ (energy per site), which thus
depends only on the choice of the $k$-SF dynamics.
Therefore, once a given $k$-SF dynamics is chosen, almost all
realizations $\omega_k$ of this dynamics will end in configurations
$\vec{\s}^\infty_k$ having the same energy density.  Furthermore, if
we focus only on the states of energy $e_k$ reachable by the dynamics
we chose (which may not be all the available states with that energy),
can we say something about the way they are sampled by the dynamics?
The answer is yes, in that all these final states not only have the
same energy, but they are {\it equiprobable}, i.e., they are reachable
with the same probability as rigorously proved by
\cite{Newman:1999aa}.  Due to the fact that the states reachable by
the dynamics may not represent all the available states with that
energy, then, this rigorous proof represent a weak proof of the
Edwards uniform measure.  The strong proof would imply that
all states available at energy $e_k$ are indeed accessed by the dynamics. 
We can explain graphically this point with the aid of 
Fig.~\ref{fig:PD2}a. Consider a given energy $\epsilon_k$ and the 
corresponding set of $k$-SF/PD metastable states with energy $\epsilon_k$, 
i.e. the ones with complexity $\Sigma_{k\rm SF}(\epsilon_k)$. 
The whole set of {\it available} $k$-SF/PD states with energy $\epsilon_k$ 
forms the $k$-SF-core. Thus, the strong proof of the Edwards uniform 
measure would imply all the states in the $k$-SF-core to be {\it accessible} 
by the $k$-SF-dynamics.

We thus arrive to the following important conclusions:
\begin{enumerate}
\item For a given choice of the dynamics, we can never visit all the available 
$k$-SF/PD metastable states, because they span a continuous range 
of energies (or volume fractions) and, evidently, it does not make much sense 
to ask if we visit those states with equal probability, without further specifying 
their energy (or volume fraction).
\item If a given $k$-SF/PD dynamics visits all the metastable states in the 
$k$-SF/PD-core, then these states are also visited with the same probability. 

\end{enumerate}

In light of the conclusion 1. we may reformulate Edwards€™ hypothesis for a particular $k$-PD state rather than for all the states (all $k$-PD states) together, saying that ``when $N$ grains occupy a volume $V$, 
they do so in such a way that all the $k$-PD metastable states 
corresponding to that volume $V$ are equally weighted".

From conclusion 2. we arrive to the real meaningful question 
and related Edwards' conjecture, which is: {\it does a given dynamics, which 
terminates always in configurations having the same energy (or volume fraction), 
sample uniformly ALL the available metastable states at that given energy, i.e., 
the whole $k$SF/PD-core?}

As discussed in Sec.~\ref{Sec_Edwardstest} there exist certain protocols that do not
sample packing states with a uniform probability, therefore, Edwards
hypothesis may not be provable correct for all possible protocols. Likewise, simulations of
jammed states, for instance using LS algorithms
\cite{Lubachevsky:1990aa}, may not be able to provide an answer to
this question for systems large enough to be of definitive value.
Thus, in the next section we propose an exact calculation to test
Edwards ergodic assumption in the exactly solvable
Sherrington-Kirkpatrick model~\cite{Sherrington:1975aa}, which is a
mean-field model of a spin glass where the metastable states can be
mathematically and precisely defined and allows for a rigorous test of
Edwards hypothesis.

The Edwards hypothesis in a more general sense applies to granular
matter and spin glasses and hard sphere glasses as well. Thus we
explore this analogy in the next section to test Edwards ergodic
hypothesis in more detail.

\subsection{Opening Pandora's box: Test of Edwards uniform measure in 
the Sherrington-Kirkpatrick spin-glass model}
\label{pandora}

As explained across this review, four recent (and not so recent)
remarkable results have been achieved that support the validity of the
uniform measure hypothesis for jammed states as proposed by Edwards:

\begin{enumerate}
\item The state-of-the-art simulations done in
  \cite{Martiniani:2017aa} allowing a direct computation of basin
  volumes of distinct jammed states, which confirm the validity of
  Edwards ergodicity at the jamming transition (Section
  \ref{Sec_Edwardstest} and Fig. \ref{Fig:directtest}).

\item The exact solution of the jammed ground state in
  infinite-dimensional fully-connected hard-sphere model done in
  \cite{Charbonneau:2014aa} using full replica symmetry breaking. The
  $\infty$-PD ground states stable under $k$-PD displacements with
  $k\to \infty$ and $N\to\infty$ and finite $\alpha=k/N$ define the
  J-line ranging from $\alpha=0$ to $\alpha=1$ (see
  Fig.~\ref{fig:PD2}a), and are obtained using the Edwards uniform
  measure.

\item The analytical study in \cite{Sharma:2016aa} of zero-temperature
  metastable minima in classical Heisenberg spin glass in a random
  magnetic field. Such a study confirms that the energy reached
  dynamically is in agreement with a computation of metastable states
  using Edwards equiprobability, see Eq.~(12) in \cite{Sharma:2016aa}.

\item The rigorous results of Newman and Stein \cite{Newman:1999aa}
  probing a weaker formulation of Edwards uniform measure: the final
  states that a zero-temperature dynamics in spin-glass model arrive
  at a given energy are solely determined by the dynamical protocol
  and are accessed with equal probability for a given energy. The
  important fact is that for every protocol there are certain states
  with a given energy that are achievable and those states are equally
  probable.  Although, the final states visited by the protocol may
  not be all the available states with that energy, hence the weak
  Edwards formulation.

\end{enumerate}

Armed with these four results, we now propose to perform a fifth exact
calculation to integrate them and provide another (most probably
penultimate, perhaps final) test to the long-standing saga on the
validity of the Edwards uniform measure (Fig.~\ref{fig:PD2}b).  The
test consists to validate the Edwards measure in the metastable states
as done in \cite{Sharma:2016aa}, following the use of the Edwards
assumption to calculate the ground state of the hard sphere model in
\cite{Charbonneau:2014aa} and using the exact results of
\cite{Newman:1999aa}. This test can be done for the 1-SF metastable
state in the exactly solvable Sherrington-Kirkpatrick (SK) spin-glass
model~\cite{Sherrington:1975aa}, which is the canonical mean field
model of spin glasses. The interest in considering this particular
model stems from the fact that it allows one to calculate analytically
the metastable states using Edwards uniform measure. The results of
this calculation then can be compared with the corresponding
quantities measured in dynamical simulations of the SK
model. Comparing exact measurements in the Edwards ensemble with
dynamics provides the ideal testing ground to examine the
applicability of Edwards predictions.

\subsubsection{Penultimate test of Edwards in the SK model}

The SK model is the infinite dimensional limit of the Edwards-Anderson
model, whose Hamiltonian is akin to the one given in
Eq.~(\ref{eq:SGHamilt}), but the sum runs over all $N(N-1)/2$ pairs of
distinct spins, becoming a solvable mean-field model:

\begin{equation}
\mathcal{H}_{\rm SK}(\vec{\sigma}) = -
\frac{1}{\sqrt{N}}\sum_{i,j=1}^{N}J_{ij}\s_i\s_j\ .
\label{sk}
\end{equation}

A key quantity which can be calculated exactly in the SK model is the
`complexity' $\Sigma(\epsilon)$ as a function of the energy density,
$\epsilon$, as schematically shown in Fig.~\ref{fig:PD2}a (we only
consider the system at zero
temperature)~\cite{Bray:1980aa}. Physically, the complexity
$\Sigma(\epsilon)$ is defined as the logarithmic scaled number of
metastable states $\mathcal{N}_N(\epsilon)$ of a given energy density
$e$:
\begin{equation}
\Sigma(\epsilon) = \lim_{N\to\infty}\frac{\log\mathcal{N}_N(\epsilon)}{N}\ ,
\label{sigma}
\end{equation}
where $N$ is the size of the system (i.e. the number of spins). The
word `scaled' indicates that $\Sigma(\epsilon)$ is the logarithm of
$\mathcal{N}_N(\epsilon)$ scaled by $N$.

We propose to solve the SK model for the 1-SF metastable states to
obtain analytically their number $\mathcal{N}_N(\epsilon)$.
From the `dynamic' point of view, we consider a 1-SF dynamics at zero 
temperature, starting from a random
initial configuration, sampled, for example, from a symmetric
Bernoulli distribution.  We can then apply the general results
discussed above.  Specifically, the 1-SF dynamics will
arrest always in states (i.e. configurations) having the same energy
\cite{Newman:1999aa}, say $\epsilon$, and the number of such states,
which we denote by $\Gamma_N(\epsilon)$, is exponentially large in the
system size $N$. On the other side, from the `static' point of view,
we can calculate analytically the total number of available 1-SF
metastable states of energy $\epsilon$ under the Edwards uniform
measure from Eq.~(\ref{sigma}), which is given precisely by
$\mathcal{N}_N(\epsilon)\sim
e^{N\Sigma(\epsilon)}$~\cite{Bray:1980aa}.

The Edwards ergodic hypothesis is: does the
dynamically generated $\Gamma_N(\epsilon)$ equal the static uniform
averaged $\mathcal{N}_N(\epsilon)$:
\begin{equation}
\Gamma_N(\epsilon) \overset{{\rm Edw}}{=} \mathcal{N}_N(\epsilon)\ ?
\end{equation}
And, if so, does the dynamics pick up all the
$\mathcal{N}_N(\epsilon)$ states with the same probability?

If Edwards hypothesis is correct, then the answer to both these
questions is affirmative. Actually, the first condition, 
i.e.  $\Gamma_N(\epsilon) = \mathcal{N}_N(\epsilon)$, is also sufficient for 
the second to be true according to the exact results of Newman and Stein,
point 4 above~\cite{Newman:1999aa}.
However, measuring $\Gamma_N(\epsilon)$ from the dynamics is not an
easy task, and hence we have to resort to another convenient quantity.
A suitable, and easily measurable, observable to test Edwards
hypothesis is the distribution of local fields $P(h)$. The local field
$h_i$ acting on spin $i$ is defined as $h_i=\frac{1}{\sqrt{N}}\sum_{j\neq
  i}J_{ij}\sigma_j$, and, in a 1-SF stable configuration, all these
local fields satisfy the condition $h_i\sigma_i>0$ for any $i$ [see
\cite{Roberts:1981aa,Bray:1980aa} and Eq.~(12) in
\cite{Sharma:2016aa}].
 
Thus, we arrive at a mathematically tractable definition of metastable
1-SF state in the SK model, which can be incorporated into the
partition function of the SK model. This has been done in
\cite{Roberts:1981aa} by considering the 1-SF condition
$h_i\sigma_i>0$ by adding the constraint $\Theta(\sum_{j\neq i}
\sigma_i J_{ij} \sigma_j)$ in the partition function. Thus, the exact
mean-field solution for $P(h)$ for this 1-SF metastable state under
the Edwards uniform measure can be obtained. We notice {\it en
  passant} that the work \cite{Roberts:1981aa} predates by a decade
the Edwards formulation. Indeed, the
validity of Edwards uniform measure has been debated in the spin glass
community \cite{Mezard:2003aa} earlier than in the granular community.

The number of 1-SF metastable states is then obtained from:
\begin{equation}
\mathcal{N}_N(\epsilon) = \sum_{\sigma}\delta\hspace{-.1cm}
 \left(\epsilon
+\frac{1}{\sqrt{N}}\sum_{i,j=1}^{N}J_{ij}\s_i\s_j\right) 
\prod_{i=1}^{N}\hspace{-.1cm}
\Theta\hspace{-.1cm}
\left[\sigma_i\hspace{-.1cm}\sum_{j\neq i} J_{ij} 
\sigma_j\right]\hspace{-.1cm}.
\label{1sf}
\end{equation}

Such a prediction can be then compared with the states dynamically
obtained under a 1-SF dynamics from the SK model by using, for
instance, a single-spin-flip Glauber dynamics as done in
\cite{Eastham:2006aa}. Thus, a precise analytical test of Edwards
ergodicity can be achieved in the SK model for metastable states.  To
perform similar test in a realistic model of granular matter would
require a mathematical definition of 1-PD locally metastable states
for jammed hard spheres analogous to 1-SF in the SK model, which
eventually might be incorporated into the Edwards partition function
of hard-spheres to test Edwards hypothesis in such a jammed
model. Such an approach has already proven to be fruitful. In
\cite{Muller:2015aa}, corresponding properties of the SK model and
jammed hard spheres based on marginal stability have been derived by
exploiting the analogy between a spin flip and the opening or closing
of a particle contact.

Specifically, the test consists to compare the form of $P(h)$ measured
at the ending configurations of the 1-SF dynamics with the one
predicted by Edwards uniform measure, in particular for small values
of the local fields $h\sim0$, which assumes the scaling form in
analogy with the force distribution, Eq. (\ref{Eq:theta}):
 \begin{equation}P(h) \sim h^{\alpha}, \,\,\,\,\, \mbox{for $h\to 0$}, 
\label{alpha}
\end{equation}

We note that a lower bound on the exponent $\alpha$ can be already
derived by imposing the stability of 1-SF metastable states with
respect to single spin-flips. The argument goes as follows: consider
two spins $\sigma_i$ and $\sigma_j$, along with their local fields
$h_i$ and $h_j$ and their coupling $J_{ij}$. The energy cost to flip
one spin, say $\sigma_i$, is given by $\Delta E=2|h_i| -
2J_{ij}\sigma_i\sigma_j$.  The non trivial case is realized when the
bond $J_{ij}$ is satisfied, i.e. when $J_{ij}\sigma_i\sigma_j > 0$, so
that we have $\Delta E=2|h_i| - 2|J_{ij}|$.  Since this condition must
be satisfied even by the smallest possible field $h_i\sim
N^{-1/(1+\alpha)}$, and since $|J_{ij}|\sim N^{-1/2}$, then the
stability condition $\Delta E>0$ of the 1-SF metastable state gives
$\alpha\geq1$.  Therefore, the distribution $P(h)$ must vanish at
small fields like $h^{\alpha}$ with an exponent $\alpha$ not smaller
than one. A direct dynamical measurement of $P(h)$ in the final
configurations of a 1-SF dynamics shows that $P(h)$ indeed vanishes
linearly for $h\to0$ \cite{Eastham:2006aa}:
\begin{equation}
P(h)\sim h,  \,\,\,\,\,\,\,  \mbox{dynamics,}
\label{linear}
\end{equation}
i.e. the lower bound $\alpha\geq1$ is actually saturated.

On the other side, what is the form of $P(h)$ calculated by using
Edwards hypothesis on the equiprobability of all the available 1-SF
metastable states of energy $\epsilon$ from Eq.~(\ref{1sf})?

The exact calculation of $P(h)$ for the 1-SF metastable states using
Edwards ensemble can be carried out. In fact, at the present, $P(h)$
has been already obtained using the Edwards partition function
Eq.~(\ref{1sf}) but only at the replica symmetry (RS) level in
~\cite{Roberts:1981aa,Eastham:2006aa}. This calculation gives for
$h\to 0$, $P(0) \propto \mbox{const} > 0$ in contradiction with the
dynamical result Eq.~(\ref{linear}). This result has led the authors
of \cite{Eastham:2006aa} to claim the failure of the Edwards
hypothesis in the Sherrington-Kirkpatrick spin glass.

However, there is an inconsistency in the RS calculation of $P(h)$
performed in~\cite{Roberts:1981aa,Eastham:2006aa} in the fact that the
RS calculation is exact only above a certain energy density
$\epsilon_c \sim -0.672...$~\cite{Bray:1980aa} (to the left of the
full RSB transition at $\alpha=0$ in Fig. \ref{fig:PD2}a), and ceases
to be valid below that energy. But the energy $\epsilon$ of the states
selected by the 1-SF dynamics leading to Eq.~(\ref{linear}) (and any
protocol we are aware of) lies below the critical energy $\epsilon_c$
($\epsilon<\epsilon_c$), where the RS calculation of $P(h)$ is not
correct.  As a consequence, also the RS value of the intercept $P(0)$
obtained in \cite{Eastham:2006aa} is wrong.  Therefore, the correct
calculation to predict $P(h)$ for energies $\epsilon<\epsilon_c$ to
obtain the exponent $\alpha$ in Eq.~(\ref{alpha}) to be compared to
the dynamical result $\alpha=1$ needs to be done by taking into
account the effect of full RSB, as in the low temperature phase of the
SK model to the right of the full RSB transition in
Fig. \ref{fig:PD2}a.  This calculation has not been carried out yet
(mainly because of its algebraic complexity) and could represent a
strong theoretical test of the Edwards uniform measure at the
mean-field level for 1-SF metastable states.

It should be noted that the analog of $P(h)$ is the distribution of
inter-particle forces $P(f)$ in the hard-sphere model,
Eq.~(\ref{Eq:theta}).
Now, in the hard-spheres model, a RS calculation of $P(f)$ gives
$P(0)>0$ \cite{Bo:2014aa}, i.e., a finite intercept at zero force, and
even the 1-RSB solution (i.e. the solution accounting for just the
first level in the hierarchical breaking of replica symmetry) gives
$P(0)>0$ as well \cite{Parisi:2010aa}, as discussed in
Eq.~(\ref{1rsb-force}). Only at the full-RSB level one finds the
correct behavior \cite{Charbonneau:2014aa}: $P(f) \sim f^{\theta}$
with $\theta=0.42$, Eq.~(\ref{fullrsb-force}), and $P(0)=0$.

In light of these results, we expect that the full RSB calculation of
$P(h)$ for 1-SF in the SK model will be needed as well to obtain the
correct scaling. This calculation is based on similar calculations
done by Bray and Moore in \cite{Eastham:2006aa} that goes back to old
controversies regarding equiprobability of metastable states in the
spin-glass field that started with \cite{Roberts:1981aa}, see Fig.~7 in 
\cite{Mezard:2003aa}.  We recognize that the behavior of $P(f)$
is not a direct measure of the equiprobability. However, $P(f)$ is
the most accessible calculation that can be done to test the
predictions of Edwards theory.

\section{Conclusions and outlook}
\label{Sec:outlook}

More than 25 years after Edwards original hypothesis on the entropy of
granular matter, it becomes increasingly evident that the consequences
of Edwards simple statement are far reaching. For one, it allows us to
understand the properties of jammed granular matter --- one of the
paradigms of athermal matter states --- by analogy with thermal
equilibrium systems. The first-order transition of jammed spheres
identified within Edwards' thermodynamics \cite{Jin:2010aa} is
reminiscent of the entropy induced phase transition of equilibrium
hard spheres, which is found at $\phi=0.494$ and $\phi=0.545$,
respectively. Despite this analogy, the physical origins of these two
transitions are fundamentally different: the equilibrium phase
transition is a consequence of the maximization of the conventional
entropy, while the transition at RCP of jammed spheres is driven by
the competition between volume minimization and maximization of the
entropy of jammed configurations, Eq.~(\ref{Ch2:grentropy}).

Such an analogy can probably be extended to other disorder-order phase
transition observed in equilibrium systems. Anisotropic elongated
particles, e.g., exhibit transitions between isotropic and nematic
phases: For large $\alpha$, Onsager's theory of equilibrium hard rods
predicts a first order isotropic-nematic transition with freezing
point at the rescaled density $\phi \alpha=3.29$ and melting point at
$\phi \alpha=4.19$ \cite{Onsager:1949aa}. By analogy with the case of
jammed spheres, one might wonder whether packings of non-spherical
particles exhibit similar transitions that could be characterized in
the $z$--$\phi$ phase diagram. Packings of hard thin rods indeed
satisfy a scaling law, where the RCP has been experimentally
identified at $\phi\alpha \approx 5.4$
\cite{Philipse:1996aa}. Dynamically, transitions to orientationally
ordered states can be induced in rod systems by shaking
\cite{Yadav:2013aa}, but the entropic characterization of such
transitions remains an open problem.

For colloidal suspensions of more complex shapes like polyhedra, both
liquid crystalline as well as plastic crystalline and even
quasicrystalline phases have been found
\cite{Haji-Akbari:2009aa,Agarwal:2011aa,Damasceno:2012aa,Marechal:2013aa}. Entropic
concepts based on shape are only starting to be explored even for
equilibrium systems
\cite{Anders:2014aa,Escobedo:2014aa,Cohen:2016aa}. In the jammed
regime, the behavior of packing density as a function of shape has
been shown to be exceedingly complex \cite{Chen:2014aa}. Edwards
granular entropy might be the key to understand such empirical data on
a more fundamental level.

Our approach based on the self-consistent equation~(\ref{Ch4:wav2})
can be applied to a large variety of both convex and non-convex
shapes. The key is to parametrize the Voronoi boundary between two
such shapes, which allows for the calculation of the Voronoi excluded
volume and surface. In fact, analytical expressions for the Voronoi
boundary can be derived following an exact algorithm for arbitrary
shapes by decomposing the shape into overlapping and intersecting
spheres (see
Figs.~\ref{Ch4:Fig_shapes} and \ref{Ch4:Fig_algorithm}). Therefore, a
systematic search for maximally dense packings in the space of given
object shapes can be performed using our framework. Extensions to
mixtures and polydisperse packings can also be formulated. This might
elucidate in particular the validity of Ulam's conjecture that the
sphere is the worst packing object in 3d \cite{Gardner:2001aa}, which
has also been formulated in a random version \cite{Jiao:2011aa}
locally around the sphere shape \cite{Kallus:2016aa}.

Thus, the Edwards' approach could help generally to elucidate how
macroscopic properties of granular matter arise from the anisotropy of
the constituents -- one of the central questions in present day
materials science \cite{Glotzer:2007aa}. A better understanding of
this problem will facilitate, e.g., the engineering of new functional
materials with particular mechanical responses by tuning the shape of
the building blocks \cite{Athanassiadis:2014aa,Jaeger:2015aa} or to
new ways to construct space filling tilings
\cite{Herrmann:1990aa,Andrade:2005aa}. Edwards statistical mechanics
might be the key to tackle these problems guided by theory rather than
direct simulations.

We postulate that a unifying theoretical framework can predict not
only the structural properties (volume fraction and coordination
number), but also mechanical properties (vibrational density of states
and yield stress) and dissipative properties (damping) as a function
of the shape and interaction properties (e.g., friction) of the
constitutive particles. If such an approach is possible, then one
could envision to span the large parameter space of the problem from a
theoretical point of view to obtain predictions of optimal packings
with desired properties. The penalty for approaching the problem
theoretically rather than by a direct numerical generation of the
packings as with reverse-engineering evolutionary algorithms
\cite{Miskin:2013aa} is that results are obtain theoretically at the
mean-field level. Thus, predictions of the resulting optimal shapes
can only be approximate.

On the other hand, it might be possible to develop a theory versatile
enough to encompass a large portion of the parameter space which
cannot be easily accessed by the direct simulation of packing
protocols in reverse engineering. Such a theory might explore
particles made by rigidly gluing spheres in arbitrary shapes, and also
other generic shapes such as (a) union of spheres of arbitrary radius,
(b) intersection of spheres of arbitrary radius leading to
tetrahedral-like particles and in general (c) any irregular polyhedra,
Fig.~\ref{Ch4:Fig_shapes}.  Another advantage is the ability to
possibly span over more than one relevant property of granular
materials, not only density but also yield stress and
dissipation. Furthermore, such an approach would include interparticle
friction, a property that was not considered before, yet, it is of
crucial importance in granular packings.

Additional insight can be provided by analytically solvable models
that take into account realistic excluded volume effects due to
non-spherical shapes. The recent solution of the `Paris car parking
problem', e.g., reveals the existence of two shape universality
classes that are manifest in different exponents in the
asymptotic approach to jamming \cite{Baule:2017aa}.

On the more fundamental side of things, the controversy on the
validity of Edwards statistical mechanics has been caused by different
interpretations of Edwards' laconic statement \cite{Edwards:1994aa}:
{\it ``We assume that when $N$ grains occupy a volume $V$ they do so
  in such a way that all configurations are equally weighted.  We
  assume this; it is the analog of the ergodic hypothesis of
  conventional thermal physics.''}

As regards the veracity of this statement, it is not rigorously
established not disproved yet. We have reviewed the recent encouraging
results of
\cite{Martiniani:2017aa,Charbonneau:2014aa,Sharma:2016aa,Newman:1999aa}
and have proposed a calculation for the 1-SF states in the SK model.
Besides, one must not be fooled by believing that a statistical
mechanics description of granular media is a least well-founded branch
of theoretical physics, if only one remembers that almost every branch
of theoretical physics is lacking `rigorous proofs', although this is
not considered as an inappropriate foundation for such branches. The
main issue with Edwards' statement, and the reason why it will be
likely hard to reach an end to the diatribe, is that the statement, as
it stands, is incomplete.

From a broad standpoint, the problem is whether it is possible to
describe the properties of the asymptotic states of the dynamics by
using only static features of the system. In Edwards' statement there
is no reference at all to which are those asymptotic dynamic states.
To solve this issue, we have proposed a rigorous definition of jammed
states as those configurations satisfying the geometrical hard-core
and mechanical force and torque balances constraints. Then we have
further classified those jammed states on the basis of their stability
properties under $k$-Particle-Displacements, inspired by an analogous
characterization of (energetically) metastable states in spin glasses
through the concept of $k$-Spin-Flips.  With this definition of the
asymptotic dynamic states, we redefined (in italics) Edwards' ensemble
by the following proposition:

\begin{itemize}
\item[]
  ``We assume that when $N$ grains occupy a volume $V$ they do so in
such a way that all stable jammed configurations {\it in a given
  $k$-PD jamming category (i.e. at given volume fraction)} are equally
weighted. We assume this; it is the analogue of the ergodic hypothesis
of conventional thermal physics ({\it and also out-of-equilibrium spin
  glasses and hard-sphere glasses})."
\end{itemize}

This statement also clarifies the role of the protocol, i.e. of the
dynamics, in the Edwards' ensemble.  A ``legal" protocol is the one
for which the asymptotic dynamic states are in a given $k$-PD-core. 
This is, again, motivated by a spin-glass
analogy. In this case an example of correct protocol is, for instance,
a single-spin-flip Glauber dynamics, for which the asymptotic dynamic
states are in the $1$-SF-core and all have the same energy.
In the granular framework this is equivalent to say that the
asymptotic jammed states of a legal protocol are only the $k$-PD
metastable states (with a fixed $k$, for instance the $1$-PD), and they
(presumably) have the same volume.
Then the question of whether these states are statistically equivalent
(i.e. equiprobable) remains still open, and we have suggested a model
(SK) where an end-to-end comparison between the results of dynamics
and a static computation can be performed, in principle, in an exact
analytical way.

An ``illegal" protocol is one that mixes different $k$-PD metastable
states, i.e., whose asymptotic dynamic states have different values of
$k$, and hence different stability properties. Nothing can be claimed
for such illegal protocols.  In the case of legal protocols, it has
been rigorously proved in spin glasses that statistical equivalence of
the asymptotic dynamic states of the given protocol holds true, i.e.,
the $k$-SF visited by a given dynamics are indeed equiprobable
\cite{Newman:1999aa}.  Whether this statement is also rigorous for
jammed states is an open question, but the correctness in spin glasses
points towards an affirmative answer. The stronger claim that the
asymptotic dynamic states are also the totality of $k$-PD ($k$-SF)
metastable states with given volume fraction (energy density) is not
analytically proved or disproved for any model we are aware of.

Conversely, in the strong tapping regime, the statistical equivalence
of the asymptotic dynamic states cannot be claimed. Notwithstanding,
this does not preclude the use of Edwards' ensemble as a very
principled approximation supposedly more justified than other
mean-field approaches. {\it A fortiori}, the great advantage of
Edwards' approach is that it leads to concrete quantitative
predictions for realistic packing scenarios. As we discuss in detail
in Sec.~\ref{Sec:volume}, the volume ensemble in the Voronoi
convention allows us to treat packings of frictional and frictionless
particles, adhesive and non-adhesive, granular and colloidal sizes,
mono-disperse and poly-disperse, in 2d, 3d and beyond, as well as
spherical and non-spherical shapes within a unified framework.  Such a
comprehensive treatment is currently out of reach for any other
approach that can treat glassy and/or jammed systems analytically,
such as mode-coupling theory \cite{Gotze:2009aa} or replica theory
\cite{Parisi:2010aa,Charbonneau:2014aa}.  Moreover, the analytical
efforts needed to extend these theories to incorporate, for instance,
friction or anisotropies may be unsurmountable.  The verdict on
Edwards' Alexandrian solution to this Gordian Knot, as on every
physical theory, should be returned, ultimately, on the goodness of
its predictions when compared with experimental data and practical
applications.

\begin{acknowledgements}

AB acknowledges funding under EPSRC grant EP/L020955/1. FM and HAM
acknowledge funding from NSF (Grant No. DMR-1308235) and DOE
Geosciences Division (Grant No. DE-FG02-03ER15458). We are grateful to
the following scientists whom, over the years, have shaped our vision
of the granular problem: J. S. Andrade Jr., L. Bo, T. Boutreux,
J. Bruji\'c, S. F. Edwards, P.-G. de Gennes, N. Gland, S. Havlin,
J. T. Jenkins, Y. Jin, D. L. Johnson, J. Kurchan, S. Li, G. Parisi,
R. Mari, L. La Ragione, M. Shattuck, C. Song, H. E. Stanley,
M. S. Tomassone, J. J. Valenza, K. Wang, and P. Wang. We are grateful
for comments on the review by: R. Blumenfeld, J.-P. Bouchaud,
B. Chakraborty, P. Charbonneau, S. Franz, G. Gradenigo, S. Martiniani,
M. Moore, C. O'Hern, G. Parisi, M. Saadatfar, M. Shattuck, M. Sperl,
M. Wyart, A. Zaccone, and F. Zamponi. We also thank B. Behringer, S. Martiniani and S. Nagel for the permission to use
their images.

\end{acknowledgements}

\begin{appendix}

\section{Bounds on the average coordination number}

\label{App:isostatic}

A packing is geometrically rigid if it can not be deformed under any
translation or rotation of the particles without deforming the
particles or breaking any of the contacts \cite{Alexander:1998aa}. In
$d$ dimensions, there are $d$ force balance equations
Eq.~(\ref{Ch2:forceb}) and $d(d-1)/2$ torque balance equations
Eq.~(\ref{Ch2:torqueb}). The number of equations can in general be
associated with the configurational degrees of freedom (dofs), so that
per particle we have in total $d_{\rm f}=d(d+1)/2$ configurational
dofs.

Geometrical rigidity requires that all $Nd_{\rm f}$
degrees of freedom in the packing are constrained by contacts (assuming periodic boundary conditions). For
frictional particles there are $d$ force components at contact and
since all contacts are shared by two particles we thus require
$Nd\,z/2\ge Nd_{\rm f}$ or \be
\label{Ch2:fb1}
z\ge 2d_{\rm f}/d=d+1.  \ee For frictionless particles there is only a
single force component at each contact due to Eq.~(\ref{Ch2:normalf}):
The normal unit vector is fixed by $\mathbf{d}_a^i$. The equivalent
rigidity condition is thus $Nz/2\ge Nd_{\rm f}$ or \be
\label{Ch2:fb2}
z\ge 2d_{\rm f}.  \ee For frictionless spheres the normal unit vector is
parallel to $\mathbf{d}_a^i$ so that Eqs.~(\ref{Ch2:torqueb}) are
always trivially satisfied. In this case $d_{\rm f}=d$, which corresponds to
the translational dofs since rotations are irrelevant.

If Eqs.~(\ref{Ch2:fb1},\ref{Ch2:fb2}) are not satisfied there exist
zero energy modes (so called floppy modes) that can deform the packing
without any energy cost. If the equalities hold, i.e., $z=d+1$ for
frictional particles and $z=2d_{\rm f}$ for frictionless particles,

On the other hand, we can obtain an upper bound on $z$ by imposing
that a generic disordered packing will have the minimal number of
contacts. If any two particles precisely touch at a single point
without deformation, we find that a single contact fixes one component
of the vector connecting the two center of masses. Overall, there are
then $Nz/2$ constraints on the configurational dofs from touching
contacts. From the constraint $Nz/2\le Nd_{\rm f}$ we obtain \be
\label{Ch2:fb3}
z\le 2d_{\rm f} \ee for both frictional and frictionless particles. Note
that for particles interacting with a soft potential the touching
condition can only be satisfied at zero pressure. Likewise,
realistic hard particles usually suffer slight deformations when
jammed, complicating the analysis \cite{Roux:2000aa,Donev:2007aa}

\section{Density of states $g(z)$}
\label{App:densitystates}

The density of states $g(z)$ can be calculated using analogies with a quantum mechanical system in three steps:

{\it (i)} First, we consider that the packing of hard spheres is
jammed in a $\infty-$PD configuration where there can be no collective
motion of any contacting subset of particles leading to unjamming when
including the normal and tangential forces between the particles. As
discussed in the introduction, this jammed state is the ground state
and corresponds to the collectively jammed category proposed in \cite{Torquato:2001aa}. While the degrees of freedom are
continuous, the fact that the packing is collectively jammed implies
that the jammed configurations in the volume space are not
continuous. Otherwise there would be a continuous transformation in
the position space that would unjam the system contradicting the fact
that the packing is collectively jammed. Thus, we consider that the
configuration space of jammed matter is discrete, since we cannot
change one configuration to another in a continuous way. A similar
consideration of discreteness has been studied in
\cite{Torquato:2001aa}. 

{\it (ii)} Second, we refer to the dimension per particle of the
configuration space as $\mathcal{D}$ and consider that the distance
between two jammed configurations is not broadly distributed (meaning
that the average distance is well-defined). We call the typical
(average) distance between configurations in the configuration space as
$h_z$, and therefore the number of configurations per particle is
proportional to $(h_z)^{-\mathcal{D}}$. The constant $h_z$ plays the
role of Planck's constant in quantum mechanics which sets the
discreteness of the phase space via the uncertainty principle.

{\it (iii)} Third, we add $z$ constraints per particle due to the fact
that the particle is jammed by $z$ contacts. Thus, there are $Nz$
position constraints ($|r_{ij}|=2R$) for a jammed state of hard
spheres as compared to the unjammed ``gas" state. Therefore, the
number of degrees of freedom is reduced to $\mathcal{D}-z$, and the
number of configurations is then $1/(h_z)^{\mathcal{D}-z}$ leading to
\be
\label{Ch4:gz}
g(z)= (h_z)^{z-\mathcal{D}}.  \ee Note that the factor
$(h_z)^{-\mathcal{D}}$ will drop out when performing ensemble
averages. Physically, we expect $h_z\ll 1$. The exact value of $h_z$
can be determined by a fitting of the theoretical values to the
simulation data, but it is not important as long as we take the limit
at the end: $h_z\to 0$.

\section{Algorithm to calculate Voronoi boundaries analytically}
\label{App:VB}

\begin{figure}
\begin{center}
\includegraphics[width=8cm]{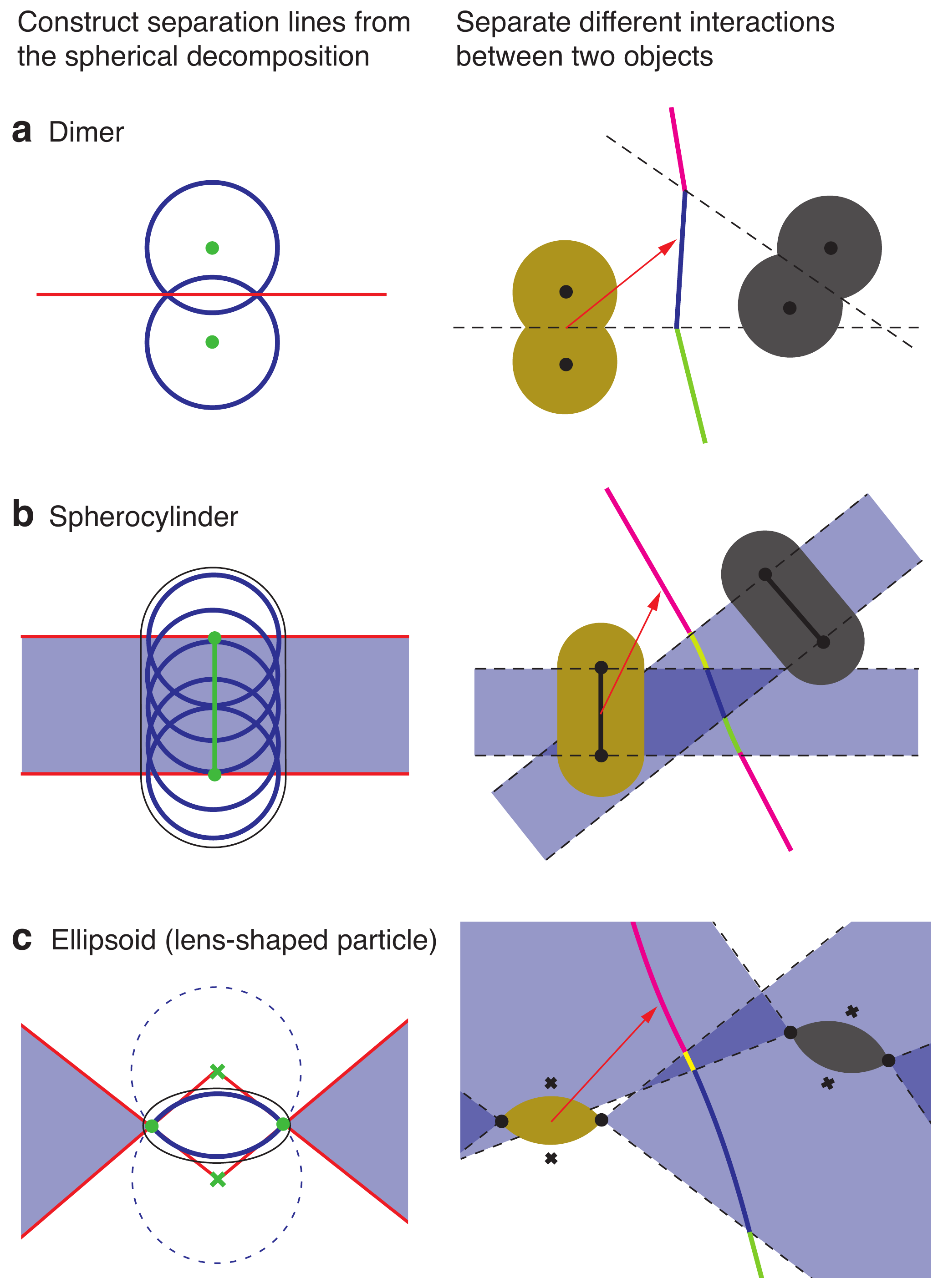}
\caption{\label{Ch4:Fig_algorithm}(Colors online) Exact algorithm to
  obtain analytical expressions for the VB from the construction of
  separation lines \cite{Baule:2013aa}. (a) For dimers, the two
  separation lines identify the correct surface out of four possible
  ones. The pink part of the VB, e.g., is the VB between the two upper
  spheres. (b) For spherocylinders, the line-line, line-point,
  point-line, and point-point interactions lead to nine different
  surfaces that are separated by four lines. The yellow part of the
  VB, e.g., is due to the upper point on spherocylinder 1 and the line
  of 2. Regions of line interactions are indicated by blue shades. (c)
  For lens-shaped particles the separation lines are given by radial
  vectors through the intersection line of the sphere segments (shown
  as points in 2d). The different point and line interactions are
  separated analogously to spherocylinders, as shown. From \cite{Baule:2013aa}. }
\end{center}
\end{figure}

Every segment of the VB arises due to
the Voronoi interaction between a particular sphere on each of the two
particles, reducing the problem to identifying the correct spheres that
interact (see Fig.~\ref{Ch4:Fig_algorithm}). The spheres that interact
are determined by separation lines given as the VBs between the
spheres in the filling. For dimers, there is one separation line for
each object, tesselating space into four areas, in which only one
interaction is correct (Fig.~\ref{Ch4:Fig_algorithm}a). The dense
overlap of spheres in spherocylinders leads to a line as effective
Voronoi interaction at the centre of the cylindrical part. This line
interaction has to be separated from the point interactions due to the
centres of the spherical caps as indicated. Overall, the two
separation lines for each object lead to a tessellation of space into
nine different areas, where only one of the possible line-line,
line-point, point-line, and point-point interactions is possible
(Fig.~\ref{Ch4:Fig_algorithm}b). 

The spherical decomposition of ellipsoid-like lens-shaped particles is
analogous to dimers, only that now the opposite sphere centres
interact (``anti-points"). In addition, the positive curvature at the
intersection point leads to an additional line interaction, which is a
circle in 3d (a point in 2d) and indicated here by two points. The
separation lines are then given by radial vectors through the
intersection point/line. The Voronoi interaction between two
ellipsoids is thus given by two pairs of two anti-points and a line,
which is the same class of interactions as spherocylinders. The
different point and line interactions are separated analogously to
spherocylinders, as shown in Fig.~\ref{Ch4:Fig_algorithm}c.

\end{appendix}


\begin{thebibliography}{451}%
\makeatletter
\providecommand \@ifxundefined [1]{%
 \@ifx{#1\undefined}
}%
\providecommand \@ifnum [1]{%
 \ifnum #1\expandafter \@firstoftwo
 \else \expandafter \@secondoftwo
 \fi
}%
\providecommand \@ifx [1]{%
 \ifx #1\expandafter \@firstoftwo
 \else \expandafter \@secondoftwo
 \fi
}%
\providecommand \natexlab [1]{#1}%
\providecommand \enquote  [1]{``#1''}%
\providecommand \bibnamefont  [1]{#1}%
\providecommand \bibfnamefont [1]{#1}%
\providecommand \citenamefont [1]{#1}%
\providecommand \href@noop [0]{\@secondoftwo}%
\providecommand \href [0]{\begingroup \@sanitize@url \@href}%
\providecommand \@href[1]{\@@startlink{#1}\@@href}%
\providecommand \@@href[1]{\endgroup#1\@@endlink}%
\providecommand \@sanitize@url [0]{\catcode `\\12\catcode `\$12\catcode
  `\&12\catcode `\#12\catcode `\^12\catcode `\_12\catcode `\%12\relax}%
\providecommand \@@startlink[1]{}%
\providecommand \@@endlink[0]{}%
\providecommand \url  [0]{\begingroup\@sanitize@url \@url }%
\providecommand \@url [1]{\endgroup\@href {#1}{\urlprefix }}%
\providecommand \urlprefix  [0]{URL }%
\providecommand \Eprint [0]{\href }%
\providecommand \doibase [0]{http://dx.doi.org/}%
\providecommand \selectlanguage [0]{\@gobble}%
\providecommand \bibinfo  [0]{\@secondoftwo}%
\providecommand \bibfield  [0]{\@secondoftwo}%
\providecommand \translation [1]{[#1]}%
\providecommand \BibitemOpen [0]{}%
\providecommand \bibitemStop [0]{}%
\providecommand \bibitemNoStop [0]{.\EOS\space}%
\providecommand \EOS [0]{\spacefactor3000\relax}%
\providecommand \BibitemShut  [1]{\csname bibitem#1\endcsname}%
\let\auto@bib@innerbib\@empty
\bibitem [{\citenamefont {Abreu}\ \emph {et~al.}(2003)\citenamefont {Abreu},
  \citenamefont {Tavares},\ and\ \citenamefont {Castier}}]{Abreu:2003aa}%
  \BibitemOpen
  \bibfield  {author} {\bibinfo {author} {\bibnamefont {Abreu}, \bibfnamefont
  {C.}}, \bibinfo {author} {\bibfnamefont {F.}~\bibnamefont {Tavares}}, \ and\
  \bibinfo {author} {\bibfnamefont {M.}~\bibnamefont {Castier}}} (\bibinfo
  {year} {2003}),\ \href@noop {} {\bibfield  {journal} {\bibinfo  {journal}
  {Powder Technol.}\ }\textbf {\bibinfo {volume} {134}},\ \bibinfo {pages}
  {167}}\BibitemShut {NoStop}%
\bibitem [{\citenamefont {Agarwal}\ and\ \citenamefont
  {Escobedo}(2011)}]{Agarwal:2011aa}%
  \BibitemOpen
  \bibfield  {author} {\bibinfo {author} {\bibnamefont {Agarwal}, \bibfnamefont
  {U.}}, \ and\ \bibinfo {author} {\bibfnamefont {F.~A.}\ \bibnamefont
  {Escobedo}}} (\bibinfo {year} {2011}),\ \href
  {http://dx.doi.org/10.1038/nmat2959} {\bibfield  {journal} {\bibinfo
  {journal} {Nature Mater.}\ }\textbf {\bibinfo {volume} {10}},\ \bibinfo
  {pages} {230}}\BibitemShut {NoStop}%
\bibitem [{\citenamefont {Alexander}(1998)}]{Alexander:1998aa}%
  \BibitemOpen
  \bibfield  {author} {\bibinfo {author} {\bibnamefont {Alexander},
  \bibfnamefont {S.}}} (\bibinfo {year} {1998}),\ \href {\doibase
  http://dx.doi.org/10.1016/S0370-1573(97)00069-0} {\bibfield  {journal}
  {\bibinfo  {journal} {Phys. Rep.}\ }\textbf {\bibinfo {volume} {296}},\
  \bibinfo {pages} {65 }}\BibitemShut {NoStop}%
\bibitem [{\citenamefont {Alonso-Marroqu\'{\i}n}\ and\ \citenamefont
  {Herrmann}(2004)}]{Alonso-Marroquin:2004aa}%
  \BibitemOpen
  \bibfield  {author} {\bibinfo {author} {\bibnamefont {Alonso-Marroqu\'{\i}n},
  \bibfnamefont {F.}}, \ and\ \bibinfo {author} {\bibfnamefont {H.~J.}\
  \bibnamefont {Herrmann}}} (\bibinfo {year} {2004}),\ \href {\doibase
  10.1103/PhysRevLett.92.054301} {\bibfield  {journal} {\bibinfo  {journal}
  {Phys. Rev. Lett.}\ }\textbf {\bibinfo {volume} {92}},\ \bibinfo {pages}
  {054301}}\BibitemShut {NoStop}%
\bibitem [{\citenamefont {van Anders}\ \emph {et~al.}(2014)\citenamefont {van
  Anders}, \citenamefont {Klotsa}, \citenamefont {Ahmed}, \citenamefont
  {Engel},\ and\ \citenamefont {Glotzer}}]{Anders:2014aa}%
  \BibitemOpen
  \bibfield  {author} {\bibinfo {author} {\bibnamefont {van Anders},
  \bibfnamefont {G.}}, \bibinfo {author} {\bibfnamefont {D.}~\bibnamefont
  {Klotsa}}, \bibinfo {author} {\bibfnamefont {N.~K.}\ \bibnamefont {Ahmed}},
  \bibinfo {author} {\bibfnamefont {M.}~\bibnamefont {Engel}}, \ and\ \bibinfo
  {author} {\bibfnamefont {S.~C.}\ \bibnamefont {Glotzer}}} (\bibinfo {year}
  {2014}),\ \href {\doibase 10.1073/pnas.1418159111} {\bibfield  {journal}
  {\bibinfo  {journal} {Proc. Nat. Acad. Sci.}\ }\textbf {\bibinfo {volume}
  {111}},\ \bibinfo {pages} {E4812}}\BibitemShut {NoStop}%
\bibitem [{\citenamefont {Andrade}\ \emph {et~al.}(2005)\citenamefont
  {Andrade}, \citenamefont {Herrmann}, \citenamefont {Andrade},\ and\
  \citenamefont {da~Silva}}]{Andrade:2005aa}%
  \BibitemOpen
  \bibfield  {author} {\bibinfo {author} {\bibnamefont {Andrade}, \bibfnamefont
  {J.~S.}}, \bibinfo {author} {\bibfnamefont {H.~J.}\ \bibnamefont {Herrmann}},
  \bibinfo {author} {\bibfnamefont {R.~F.~S.}\ \bibnamefont {Andrade}}, \ and\
  \bibinfo {author} {\bibfnamefont {L.~R.}\ \bibnamefont {da~Silva}}} (\bibinfo
  {year} {2005}),\ \href {\doibase 10.1103/PhysRevLett.94.018702} {\bibfield
  {journal} {\bibinfo  {journal} {Phys. Rev. Lett.}\ }\textbf {\bibinfo
  {volume} {94}},\ \bibinfo {pages} {018702}}\BibitemShut {NoStop}%
\bibitem [{\citenamefont {Anikeenko}\ and\ \citenamefont
  {Medvedev}(2007)}]{Anikeenko:2007aa}%
  \BibitemOpen
  \bibfield  {author} {\bibinfo {author} {\bibnamefont {Anikeenko},
  \bibfnamefont {A.~V.}}, \ and\ \bibinfo {author} {\bibfnamefont {N.~N.}\
  \bibnamefont {Medvedev}}} (\bibinfo {year} {2007}),\ \href {\doibase
  10.1103/PhysRevLett.98.235504} {\bibfield  {journal} {\bibinfo  {journal}
  {Phys. Rev. Lett.}\ }\textbf {\bibinfo {volume} {98}},\ \bibinfo {pages}
  {235504}}\BibitemShut {NoStop}%
\bibitem [{\citenamefont {Anikeenko}\ \emph {et~al.}(2008)\citenamefont
  {Anikeenko}, \citenamefont {Medvedev},\ and\ \citenamefont
  {Aste}}]{Anikeenko:2008aa}%
  \BibitemOpen
  \bibfield  {author} {\bibinfo {author} {\bibnamefont {Anikeenko},
  \bibfnamefont {A.~V.}}, \bibinfo {author} {\bibfnamefont {N.~N.}\
  \bibnamefont {Medvedev}}, \ and\ \bibinfo {author} {\bibfnamefont
  {T.}~\bibnamefont {Aste}}} (\bibinfo {year} {2008}),\ \href {\doibase
  10.1103/PhysRevE.77.031101} {\bibfield  {journal} {\bibinfo  {journal} {Phys.
  Rev. E}\ }\textbf {\bibinfo {volume} {77}},\ \bibinfo {pages}
  {031101}}\BibitemShut {NoStop}%
\bibitem [{\citenamefont {Aristoff}\ and\ \citenamefont
  {Radin}(2009)}]{Aristoff:2009aa}%
  \BibitemOpen
  \bibfield  {author} {\bibinfo {author} {\bibnamefont {Aristoff},
  \bibfnamefont {D.}}, \ and\ \bibinfo {author} {\bibfnamefont
  {C.}~\bibnamefont {Radin}}} (\bibinfo {year} {2009}),\ \href {\doibase
  10.1007/s10955-009-9722-4} {\bibfield  {journal} {\bibinfo  {journal} {J.
  Stat. Phys.}\ }\textbf {\bibinfo {volume} {135}},\ \bibinfo {pages}
  {1}}\BibitemShut {NoStop}%
\bibitem [{\citenamefont {Asenjo}\ \emph {et~al.}(2014)\citenamefont {Asenjo},
  \citenamefont {Paillusson},\ and\ \citenamefont {Frenkel}}]{Asenjo:2014aa}%
  \BibitemOpen
  \bibfield  {author} {\bibinfo {author} {\bibnamefont {Asenjo}, \bibfnamefont
  {D.}}, \bibinfo {author} {\bibfnamefont {F.}~\bibnamefont {Paillusson}}, \
  and\ \bibinfo {author} {\bibfnamefont {D.}~\bibnamefont {Frenkel}}} (\bibinfo
  {year} {2014}),\ \href {\doibase 10.1103/PhysRevLett.112.098002} {\bibfield
  {journal} {\bibinfo  {journal} {Phys. Rev. Lett.}\ }\textbf {\bibinfo
  {volume} {112}},\ \bibinfo {pages} {098002}}\BibitemShut {NoStop}%
\bibitem [{\citenamefont {Aste}(2005)}]{Aste:2005ab}%
  \BibitemOpen
  \bibfield  {author} {\bibinfo {author} {\bibnamefont {Aste}, \bibfnamefont
  {T.}}} (\bibinfo {year} {2005}),\ \href {\doibase
  10.1088/0953-8984/17/24/001} {\bibfield  {journal} {\bibinfo  {journal} {J.
  Phys. Cond. Mat.}\ }\textbf {\bibinfo {volume} {17}},\ \bibinfo {pages}
  {S2361}}\BibitemShut {NoStop}%
\bibitem [{\citenamefont {Aste}(2006)}]{Aste:2006aa}%
  \BibitemOpen
  \bibfield  {author} {\bibinfo {author} {\bibnamefont {Aste}, \bibfnamefont
  {T.}}} (\bibinfo {year} {2006}),\ \href {\doibase
  10.1103/PhysRevLett.96.018002} {\bibfield  {journal} {\bibinfo  {journal}
  {Phys. Rev. Lett.}\ }\textbf {\bibinfo {volume} {96}},\ \bibinfo {pages}
  {018002}}\BibitemShut {NoStop}%
\bibitem [{\citenamefont {Aste}\ and\ \citenamefont
  {Coniglio}(2004)}]{Aste:2004aa}%
  \BibitemOpen
  \bibfield  {author} {\bibinfo {author} {\bibnamefont {Aste}, \bibfnamefont
  {T.}}, \ and\ \bibinfo {author} {\bibfnamefont {A.}~\bibnamefont {Coniglio}}}
  (\bibinfo {year} {2004}),\ \href
  {http://stacks.iop.org/0295-5075/67/i=2/a=165} {\bibfield  {journal}
  {\bibinfo  {journal} {Europhys. Lett.}\ }\textbf {\bibinfo {volume} {67}},\
  \bibinfo {pages} {165}}\BibitemShut {NoStop}%
\bibitem [{\citenamefont {Aste}\ and\ \citenamefont
  {Di~Matteo}(2008{\natexlab{a}})}]{Aste:2008aa}%
  \BibitemOpen
  \bibfield  {author} {\bibinfo {author} {\bibnamefont {Aste}, \bibfnamefont
  {T.}}, \ and\ \bibinfo {author} {\bibfnamefont {T.}~\bibnamefont
  {Di~Matteo}}} (\bibinfo {year} {2008}{\natexlab{a}}),\ \href {\doibase
  10.1103/PhysRevE.77.021309} {\bibfield  {journal} {\bibinfo  {journal} {Phys.
  Rev. E}\ }\textbf {\bibinfo {volume} {77}},\ \bibinfo {pages}
  {021309}}\BibitemShut {NoStop}%
\bibitem [{\citenamefont {Aste}\ and\ \citenamefont
  {Di~Matteo}(2008{\natexlab{b}})}]{Aste:2008ab}%
  \BibitemOpen
  \bibfield  {author} {\bibinfo {author} {\bibnamefont {Aste}, \bibfnamefont
  {T.}}, \ and\ \bibinfo {author} {\bibfnamefont {T.}~\bibnamefont
  {Di~Matteo}}} (\bibinfo {year} {2008}{\natexlab{b}}),\ \href {\doibase
  10.1140/epjb/e2008-00224-8} {\bibfield  {journal} {\bibinfo  {journal} {Eur.
  Phys. J. B}\ }\textbf {\bibinfo {volume} {64}},\ \bibinfo {pages}
  {511}}\BibitemShut {NoStop}%
\bibitem [{\citenamefont {Aste}\ \emph {et~al.}(2007)\citenamefont {Aste},
  \citenamefont {Di~Matteo}, \citenamefont {Saadatfar}, \citenamefont {Senden},
  \citenamefont {Schr\"oter},\ and\ \citenamefont {Swinney}}]{Aste:2007aa}%
  \BibitemOpen
  \bibfield  {author} {\bibinfo {author} {\bibnamefont {Aste}, \bibfnamefont
  {T.}}, \bibinfo {author} {\bibfnamefont {T.}~\bibnamefont {Di~Matteo}},
  \bibinfo {author} {\bibfnamefont {M.}~\bibnamefont {Saadatfar}}, \bibinfo
  {author} {\bibfnamefont {T.~J.}\ \bibnamefont {Senden}}, \bibinfo {author}
  {\bibfnamefont {M.}~\bibnamefont {Schr\"oter}}, \ and\ \bibinfo {author}
  {\bibfnamefont {H.~L.}\ \bibnamefont {Swinney}}} (\bibinfo {year} {2007}),\
  \href {\doibase 10.1209/0295-5075/79/24003} {\bibfield  {journal} {\bibinfo
  {journal} {Europhys. Lett.}\ }\textbf {\bibinfo {volume} {79}},\ \bibinfo
  {pages} {24003}}\BibitemShut {NoStop}%
\bibitem [{\citenamefont {Aste}\ \emph {et~al.}(2004)\citenamefont {Aste},
  \citenamefont {Saadatfar}, \citenamefont {Sakellariou},\ and\ \citenamefont
  {Senden}}]{Aste:2004ab}%
  \BibitemOpen
  \bibfield  {author} {\bibinfo {author} {\bibnamefont {Aste}, \bibfnamefont
  {T.}}, \bibinfo {author} {\bibfnamefont {M.}~\bibnamefont {Saadatfar}},
  \bibinfo {author} {\bibfnamefont {A.}~\bibnamefont {Sakellariou}}, \ and\
  \bibinfo {author} {\bibfnamefont {T.}~\bibnamefont {Senden}}} (\bibinfo
  {year} {2004}),\ \enquote {\bibinfo {title} {Investigating the geometrical
  structure of disordered sphere packings},}\ in\ \href {\doibase
  http://dx.doi.org/10.1016/j.physa.2004.03.034} {\emph {\bibinfo {booktitle}
  {Proceedings of the International Conference New Materials and
  Complexity}}},\ Vol.\ \bibinfo {volume} {339},\ pp.\ \bibinfo {pages} {16 --
  23}\BibitemShut {NoStop}%
\bibitem [{\citenamefont {Aste}\ \emph {et~al.}(2005)\citenamefont {Aste},
  \citenamefont {Saadatfar},\ and\ \citenamefont {Senden}}]{Aste:2005aa}%
  \BibitemOpen
  \bibfield  {author} {\bibinfo {author} {\bibnamefont {Aste}, \bibfnamefont
  {T.}}, \bibinfo {author} {\bibfnamefont {M.}~\bibnamefont {Saadatfar}}, \
  and\ \bibinfo {author} {\bibfnamefont {T.}~\bibnamefont {Senden}}} (\bibinfo
  {year} {2005}),\ \href {\doibase 10.1103/PhysRevE.71.061302} {\bibfield
  {journal} {\bibinfo  {journal} {Phys. Rev. E}\ }\textbf {\bibinfo {volume}
  {71}},\ \bibinfo {pages} {061302}}\BibitemShut {NoStop}%
\bibitem [{\citenamefont {Aste}\ \emph {et~al.}(2006)\citenamefont {Aste},
  \citenamefont {Saadatfar},\ and\ \citenamefont {Senden}}]{Aste:2006ab}%
  \BibitemOpen
  \bibfield  {author} {\bibinfo {author} {\bibnamefont {Aste}, \bibfnamefont
  {T.}}, \bibinfo {author} {\bibfnamefont {M.}~\bibnamefont {Saadatfar}}, \
  and\ \bibinfo {author} {\bibfnamefont {T.~J.}\ \bibnamefont {Senden}}}
  (\bibinfo {year} {2006}),\ \href {\doibase 10.1088/1742-5468/2006/07/P07010}
  {\bibinfo  {journal} {J. Stat. Mech.}\ ,\ \bibinfo {pages}
  {P07010}}\BibitemShut {NoStop}%
\bibitem [{\citenamefont {Athanassiadis}\ \emph {et~al.}(2014)\citenamefont
  {Athanassiadis}, \citenamefont {Miskin}, \citenamefont {Kaplan},
  \citenamefont {Rodenberg}, \citenamefont {Lee}, \citenamefont {Merritt},
  \citenamefont {Brown}, \citenamefont {Amend}, \citenamefont {Lipson},\ and\
  \citenamefont {Jaeger}}]{Athanassiadis:2014aa}%
  \BibitemOpen
\bibfield  {journal} {  }\bibfield  {author} {\bibinfo {author} {\bibnamefont
  {Athanassiadis}, \bibfnamefont {A.~G.}}, \bibinfo {author} {\bibfnamefont
  {M.~Z.}\ \bibnamefont {Miskin}}, \bibinfo {author} {\bibfnamefont
  {P.}~\bibnamefont {Kaplan}}, \bibinfo {author} {\bibfnamefont
  {N.}~\bibnamefont {Rodenberg}}, \bibinfo {author} {\bibfnamefont {S.~H.}\
  \bibnamefont {Lee}}, \bibinfo {author} {\bibfnamefont {J.}~\bibnamefont
  {Merritt}}, \bibinfo {author} {\bibfnamefont {E.}~\bibnamefont {Brown}},
  \bibinfo {author} {\bibfnamefont {J.}~\bibnamefont {Amend}}, \bibinfo
  {author} {\bibfnamefont {H.}~\bibnamefont {Lipson}}, \ and\ \bibinfo {author}
  {\bibfnamefont {H.~M.}\ \bibnamefont {Jaeger}}} (\bibinfo {year} {2014}),\
  \href {\doibase 10.1039/C3SM52047A} {\bibfield  {journal} {\bibinfo
  {journal} {Soft Matter}\ }\textbf {\bibinfo {volume} {10}},\ \bibinfo {pages}
  {48}}\BibitemShut {NoStop}%
\bibitem [{\citenamefont {Atkinson}\ \emph {et~al.}(2014)\citenamefont
  {Atkinson}, \citenamefont {Stillinger},\ and\ \citenamefont
  {Torquato}}]{Atkinson:2014aa}%
  \BibitemOpen
  \bibfield  {author} {\bibinfo {author} {\bibnamefont {Atkinson},
  \bibfnamefont {S.}}, \bibinfo {author} {\bibfnamefont {F.~H.}\ \bibnamefont
  {Stillinger}}, \ and\ \bibinfo {author} {\bibfnamefont {S.}~\bibnamefont
  {Torquato}}} (\bibinfo {year} {2014}),\ \href {\doibase
  10.1073/pnas.1408371112} {\bibfield  {journal} {\bibinfo  {journal} {Proc.
  Nat. Acad. Sci.}\ }\textbf {\bibinfo {volume} {111}},\ \bibinfo {pages}
  {18436}}\BibitemShut {NoStop}%
\bibitem [{\citenamefont {Aurenhammer}(1991)}]{Aurenhammer:1991aa}%
  \BibitemOpen
  \bibfield  {author} {\bibinfo {author} {\bibnamefont {Aurenhammer},
  \bibfnamefont {F.}}} (\bibinfo {year} {1991}),\ \href@noop {} {\bibfield
  {journal} {\bibinfo  {journal} {ACM Computing Surveys}\ }\textbf {\bibinfo
  {volume} {23}},\ \bibinfo {pages} {345}}\BibitemShut {NoStop}%
\bibitem [{\citenamefont {Bagi}(1997)}]{Bagi:1997aa}%
  \BibitemOpen
  \bibfield  {author} {\bibinfo {author} {\bibnamefont {Bagi}, \bibfnamefont
  {K.}}} (\bibinfo {year} {1997}),\ \enquote {\bibinfo {title} {Analysis of
  micro-variables through entropy principles},}\ in\ \href@noop {} {\emph
  {\bibinfo {booktitle} {Powders and Grains 97}}},\ \bibinfo {editor} {edited
  by\ \bibinfo {editor} {\bibfnamefont {R.~P.}\ \bibnamefont {Behringer}}\ and\
  \bibinfo {editor} {\bibfnamefont {J.~T.}\ \bibnamefont {Jenkins}}},\ pp.\
  \bibinfo {pages} {251--254}\BibitemShut {NoStop}%
\bibitem [{\citenamefont {Bagi}(2003)}]{Bagi:2003aa}%
  \BibitemOpen
  \bibfield  {author} {\bibinfo {author} {\bibnamefont {Bagi}, \bibfnamefont
  {K.}}} (\bibinfo {year} {2003}),\ \href {\doibase 10.1007/s10035-002-0123-5}
  {\bibfield  {journal} {\bibinfo  {journal} {Granular Matter}\ }\textbf
  {\bibinfo {volume} {5}},\ \bibinfo {pages} {45}}\BibitemShut {NoStop}%
\bibitem [{\citenamefont {Baker}\ and\ \citenamefont
  {Kudrolli}(2010)}]{Baker:2010aa}%
  \BibitemOpen
  \bibfield  {author} {\bibinfo {author} {\bibnamefont {Baker}, \bibfnamefont
  {J.}}, \ and\ \bibinfo {author} {\bibfnamefont {A.}~\bibnamefont {Kudrolli}}}
  (\bibinfo {year} {2010}),\ \href {\doibase 10.1103/PhysRevE.82.061304}
  {\bibfield  {journal} {\bibinfo  {journal} {Phys. Rev. E}\ }\textbf {\bibinfo
  {volume} {82}},\ \bibinfo {pages} {061304}}\BibitemShut {NoStop}%
\bibitem [{\citenamefont {Ball}\ and\ \citenamefont
  {Blumenfeld}(2002)}]{Ball:2002aa}%
  \BibitemOpen
  \bibfield  {author} {\bibinfo {author} {\bibnamefont {Ball}, \bibfnamefont
  {R.}}, \ and\ \bibinfo {author} {\bibfnamefont {R.}~\bibnamefont
  {Blumenfeld}}} (\bibinfo {year} {2002}),\ \href {\doibase
  10.1103/PhysRevLett.88.115505} {\bibfield  {journal} {\bibinfo  {journal}
  {Phys. Rev. Lett.}\ }\textbf {\bibinfo {volume} {88}},\ \bibinfo {pages}
  {115505}}\BibitemShut {NoStop}%
\bibitem [{\citenamefont {Baranau}\ \emph {et~al.}(2016)\citenamefont
  {Baranau}, \citenamefont {Zhao}, \citenamefont {Scheel}, \citenamefont
  {Tallarek},\ and\ \citenamefont {Schr\"oter}}]{Baranau:2016aa}%
  \BibitemOpen
  \bibfield  {author} {\bibinfo {author} {\bibnamefont {Baranau}, \bibfnamefont
  {V.}}, \bibinfo {author} {\bibfnamefont {S.-C.}\ \bibnamefont {Zhao}},
  \bibinfo {author} {\bibfnamefont {M.}~\bibnamefont {Scheel}}, \bibinfo
  {author} {\bibfnamefont {U.}~\bibnamefont {Tallarek}}, \ and\ \bibinfo
  {author} {\bibfnamefont {M.}~\bibnamefont {Schr\"oter}}} (\bibinfo {year}
  {2016}),\ \href {\doibase 10.1039/C6SM00567E} {\bibfield  {journal} {\bibinfo
   {journal} {Soft Matter}\ }\textbf {\bibinfo {volume} {12}},\ \bibinfo
  {pages} {3991}}\BibitemShut {NoStop}%
\bibitem [{\citenamefont {Bargiel}(2008)}]{Bargiel:2008aa}%
  \BibitemOpen
  \bibfield  {author} {\bibinfo {author} {\bibnamefont {Bargiel}, \bibfnamefont
  {M.}}} (\bibinfo {year} {2008}),\ \href@noop {} {\bibfield  {journal}
  {\bibinfo  {journal} {Computational Science--ICCS2008}\ }\textbf {\bibinfo
  {volume} {5102}},\ \bibinfo {pages} {126}}\BibitemShut {NoStop}%
\bibitem [{\citenamefont {Barrat}\ \emph {et~al.}(2000)\citenamefont {Barrat},
  \citenamefont {Kurchan}, \citenamefont {Loreto},\ and\ \citenamefont
  {Sellitto}}]{Barrat:2000aa}%
  \BibitemOpen
  \bibfield  {author} {\bibinfo {author} {\bibnamefont {Barrat}, \bibfnamefont
  {A.}}, \bibinfo {author} {\bibfnamefont {J.}~\bibnamefont {Kurchan}},
  \bibinfo {author} {\bibfnamefont {V.}~\bibnamefont {Loreto}}, \ and\ \bibinfo
  {author} {\bibfnamefont {M.}~\bibnamefont {Sellitto}}} (\bibinfo {year}
  {2000}),\ \href {\doibase 10.1103/PhysRevLett.85.5034} {\bibfield  {journal}
  {\bibinfo  {journal} {Phys. Rev. Lett.}\ }\textbf {\bibinfo {volume} {85}},\
  \bibinfo {pages} {5034}}\BibitemShut {NoStop}%
\bibitem [{\citenamefont {Baule}(2017)}]{Baule:2017aa}%
  \BibitemOpen
  \bibfield  {author} {\bibinfo {author} {\bibnamefont {Baule}, \bibfnamefont
  {A.}}} (\bibinfo {year} {2017}),\ \href {\doibase
  10.1103/PhysRevLett.119.028003} {\bibfield  {journal} {\bibinfo  {journal}
  {Phys. Rev. Lett.}\ }\textbf {\bibinfo {volume} {119}},\ \bibinfo {pages}
  {028003}}\BibitemShut {NoStop}%
\bibitem [{\citenamefont {Baule}\ and\ \citenamefont
  {Makse}(2014)}]{Baule:2014aa}%
  \BibitemOpen
  \bibfield  {author} {\bibinfo {author} {\bibnamefont {Baule}, \bibfnamefont
  {A.}}, \ and\ \bibinfo {author} {\bibfnamefont {H.~A.}\ \bibnamefont
  {Makse}}} (\bibinfo {year} {2014}),\ \href {\doibase 10.1039/c3sm52783b}
  {\bibfield  {journal} {\bibinfo  {journal} {Soft Matter}\ }\textbf {\bibinfo
  {volume} {10}},\ \bibinfo {pages} {4423}}\BibitemShut {NoStop}%
\bibitem [{\citenamefont {Baule}\ \emph {et~al.}(2013)\citenamefont {Baule},
  \citenamefont {Mari}, \citenamefont {Bo}, \citenamefont {Portal},\ and\
  \citenamefont {Makse}}]{Baule:2013aa}%
  \BibitemOpen
  \bibfield  {author} {\bibinfo {author} {\bibnamefont {Baule}, \bibfnamefont
  {A.}}, \bibinfo {author} {\bibfnamefont {R.}~\bibnamefont {Mari}}, \bibinfo
  {author} {\bibfnamefont {L.}~\bibnamefont {Bo}}, \bibinfo {author}
  {\bibfnamefont {L.}~\bibnamefont {Portal}}, \ and\ \bibinfo {author}
  {\bibfnamefont {H.~A.}\ \bibnamefont {Makse}}} (\bibinfo {year} {2013}),\
  \href {\doibase 10.1038/ncomms3194} {\bibfield  {journal} {\bibinfo
  {journal} {Nature Commun.}\ }\textbf {\bibinfo {volume} {4}},\ \bibinfo
  {pages} {2194}}\BibitemShut {NoStop}%
\bibitem [{\citenamefont {Becker}\ and\ \citenamefont
  {Kassner}(2015)}]{Becker:2015aa}%
  \BibitemOpen
  \bibfield  {author} {\bibinfo {author} {\bibnamefont {Becker}, \bibfnamefont
  {V.}}, \ and\ \bibinfo {author} {\bibfnamefont {K.}~\bibnamefont {Kassner}}}
  (\bibinfo {year} {2015}),\ \href {\doibase 10.1103/PhysRevE.92.052201}
  {\bibfield  {journal} {\bibinfo  {journal} {Phys. Rev. E}\ }\textbf {\bibinfo
  {volume} {92}},\ \bibinfo {pages} {052201}}\BibitemShut {NoStop}%
\bibitem [{\citenamefont {Becker}\ and\ \citenamefont
  {Kassner}(2017)}]{Becker:2017aa}%
  \BibitemOpen
  \bibfield  {author} {\bibinfo {author} {\bibnamefont {Becker}, \bibfnamefont
  {V.}}, \ and\ \bibinfo {author} {\bibfnamefont {K.}~\bibnamefont {Kassner}}}
  (\bibinfo {year} {2017}),\ \href {\doibase 10.1103/PhysRevLett.119.039801}
  {\bibfield  {journal} {\bibinfo  {journal} {Phys. Rev. Lett.}\ }\textbf
  {\bibinfo {volume} {119}},\ \bibinfo {pages} {039801}}\BibitemShut {NoStop}%
\bibitem [{\citenamefont {Bellon}\ and\ \citenamefont
  {Ciliberto}(2002)}]{Bellon:2002aa}%
  \BibitemOpen
  \bibfield  {author} {\bibinfo {author} {\bibnamefont {Bellon}, \bibfnamefont
  {L.}}, \ and\ \bibinfo {author} {\bibfnamefont {S.}~\bibnamefont
  {Ciliberto}}} (\bibinfo {year} {2002}),\ \href {\doibase
  http://dx.doi.org/10.1016/S0167-2789(02)00520-1} {\bibfield  {journal}
  {\bibinfo  {journal} {Physica D}\ }\textbf {\bibinfo {volume} {168--169}},\
  \bibinfo {pages} {325 }}\BibitemShut {NoStop}%
\bibitem [{\citenamefont {Berg}\ \emph {et~al.}(2002)\citenamefont {Berg},
  \citenamefont {Franz},\ and\ \citenamefont {Sellitto}}]{Berg:2002aa}%
  \BibitemOpen
  \bibfield  {author} {\bibinfo {author} {\bibnamefont {Berg}, \bibfnamefont
  {J.}}, \bibinfo {author} {\bibfnamefont {S.}~\bibnamefont {Franz}}, \ and\
  \bibinfo {author} {\bibfnamefont {M.}~\bibnamefont {Sellitto}}} (\bibinfo
  {year} {2002}),\ \href {\doibase 10.1140/epjb.e20020099} {\bibfield
  {journal} {\bibinfo  {journal} {Eur. Phys. J. B}\ }\textbf {\bibinfo {volume}
  {26}},\ \bibinfo {pages} {349}}\BibitemShut {NoStop}%
\bibitem [{\citenamefont {Bernal}(1960)}]{Bernal:1960ab}%
  \BibitemOpen
  \bibfield  {author} {\bibinfo {author} {\bibnamefont {Bernal}, \bibfnamefont
  {J.~D.}}} (\bibinfo {year} {1960}),\ \href
  {http://dx.doi.org/10.1038/185068a0} {\bibfield  {journal} {\bibinfo
  {journal} {Nature}\ }\textbf {\bibinfo {volume} {185}},\ \bibinfo {pages}
  {68}}\BibitemShut {NoStop}%
\bibitem [{\citenamefont {Bernal}(1964)}]{Bernal:1964aa}%
  \BibitemOpen
  \bibfield  {author} {\bibinfo {author} {\bibnamefont {Bernal}, \bibfnamefont
  {J.~D.}}} (\bibinfo {year} {1964}),\ \href@noop {} {\bibfield  {journal}
  {\bibinfo  {journal} {Proc. Roy. Soc. London Ser. A}\ }\textbf {\bibinfo
  {volume} {280}},\ \bibinfo {pages} {299}}\BibitemShut {NoStop}%
\bibitem [{\citenamefont {Bernal}\ and\ \citenamefont
  {Mason}(1960)}]{Bernal:1960aa}%
  \BibitemOpen
  \bibfield  {author} {\bibinfo {author} {\bibnamefont {Bernal}, \bibfnamefont
  {J.~D.}}, \ and\ \bibinfo {author} {\bibfnamefont {J.}~\bibnamefont {Mason}}}
  (\bibinfo {year} {1960}),\ \href {http://dx.doi.org/10.1038/188910a0}
  {\bibfield  {journal} {\bibinfo  {journal} {Nature}\ }\textbf {\bibinfo
  {volume} {188}},\ \bibinfo {pages} {910}}\BibitemShut {NoStop}%
\bibitem [{\citenamefont {Berryman}(1983)}]{Berryman:1983aa}%
  \BibitemOpen
  \bibfield  {author} {\bibinfo {author} {\bibnamefont {Berryman},
  \bibfnamefont {J.~G.}}} (\bibinfo {year} {1983}),\ \href {\doibase
  10.1103/PhysRevA.27.1053} {\bibfield  {journal} {\bibinfo  {journal} {Phys.
  Rev. A}\ }\textbf {\bibinfo {volume} {27}},\ \bibinfo {pages}
  {1053}}\BibitemShut {NoStop}%
\bibitem [{\citenamefont {Berthier}\ \emph {et~al.}(2000)\citenamefont
  {Berthier}, \citenamefont {Barrat},\ and\ \citenamefont
  {Kurchan}}]{Berthier:2000aa}%
  \BibitemOpen
  \bibfield  {author} {\bibinfo {author} {\bibnamefont {Berthier},
  \bibfnamefont {L.}}, \bibinfo {author} {\bibfnamefont {J.-L.}\ \bibnamefont
  {Barrat}}, \ and\ \bibinfo {author} {\bibfnamefont {J.}~\bibnamefont
  {Kurchan}}} (\bibinfo {year} {2000}),\ \href {\doibase
  10.1103/PhysRevE.61.5464} {\bibfield  {journal} {\bibinfo  {journal} {Phys.
  Rev. E}\ }\textbf {\bibinfo {volume} {61}},\ \bibinfo {pages}
  {5464}}\BibitemShut {NoStop}%
\bibitem [{\citenamefont {Bi}\ \emph {et~al.}(2015)\citenamefont {Bi},
  \citenamefont {Henkes}, \citenamefont {Daniels},\ and\ \citenamefont
  {Chakraborty}}]{Bi:2015aa}%
  \BibitemOpen
  \bibfield  {author} {\bibinfo {author} {\bibnamefont {Bi}, \bibfnamefont
  {D.}}, \bibinfo {author} {\bibfnamefont {S.}~\bibnamefont {Henkes}}, \bibinfo
  {author} {\bibfnamefont {K.~E.}\ \bibnamefont {Daniels}}, \ and\ \bibinfo
  {author} {\bibfnamefont {B.}~\bibnamefont {Chakraborty}}} (\bibinfo {year}
  {2015}),\ \href {\doibase 10.1146/annurev-conmatphys-031214-014336}
  {\bibfield  {journal} {\bibinfo  {journal} {Annual Review of Condensed Matter
  Physics}\ }\textbf {\bibinfo {volume} {6}},\ \bibinfo {pages}
  {63}}\BibitemShut {NoStop}%
\bibitem [{\citenamefont {Biazzo}\ \emph {et~al.}(2009)\citenamefont {Biazzo},
  \citenamefont {Caltagirone}, \citenamefont {Parisi},\ and\ \citenamefont
  {Zamponi}}]{Biazzo:2009aa}%
  \BibitemOpen
  \bibfield  {author} {\bibinfo {author} {\bibnamefont {Biazzo}, \bibfnamefont
  {I.}}, \bibinfo {author} {\bibfnamefont {F.}~\bibnamefont {Caltagirone}},
  \bibinfo {author} {\bibfnamefont {G.}~\bibnamefont {Parisi}}, \ and\ \bibinfo
  {author} {\bibfnamefont {F.}~\bibnamefont {Zamponi}}} (\bibinfo {year}
  {2009}),\ \href {\doibase 10.1103/PhysRevLett.102.195701} {\bibfield
  {journal} {\bibinfo  {journal} {Phys. Rev. Lett.}\ }\textbf {\bibinfo
  {volume} {102}},\ \bibinfo {pages} {195701}}\BibitemShut {NoStop}%
\bibitem [{\citenamefont {Biroli}\ and\ \citenamefont
  {Monasson}(2000)}]{Biroli:2000aa}%
  \BibitemOpen
  \bibfield  {author} {\bibinfo {author} {\bibnamefont {Biroli}, \bibfnamefont
  {G.}}, \ and\ \bibinfo {author} {\bibfnamefont {R.}~\bibnamefont {Monasson}}}
  (\bibinfo {year} {2000}),\ \href
  {http://stacks.iop.org/0295-5075/50/i=2/a=155} {\bibfield  {journal}
  {\bibinfo  {journal} {EPL (Europhysics Letters)}\ }\textbf {\bibinfo {volume}
  {50}}~(\bibinfo {number} {2}),\ \bibinfo {pages} {155}}\BibitemShut {NoStop}%
\bibitem [{\citenamefont {Blum}\ \emph {et~al.}(2006)\citenamefont {Blum},
  \citenamefont {Schr{\"a}pler}, \citenamefont {Davidsson},\ and\ \citenamefont
  {Trigo-Rodr{\'\i}guez}}]{Blum:2006aa}%
  \BibitemOpen
  \bibfield  {author} {\bibinfo {author} {\bibnamefont {Blum}, \bibfnamefont
  {J.}}, \bibinfo {author} {\bibfnamefont {R.}~\bibnamefont {Schr{\"a}pler}},
  \bibinfo {author} {\bibfnamefont {B.~J.~R.}\ \bibnamefont {Davidsson}}, \
  and\ \bibinfo {author} {\bibfnamefont {J.~M.}\ \bibnamefont
  {Trigo-Rodr{\'\i}guez}}} (\bibinfo {year} {2006}),\ \href
  {http://stacks.iop.org/0004-637X/652/i=2/a=1768} {\bibfield  {journal}
  {\bibinfo  {journal} {The Astrophysical Journal}\ }\textbf {\bibinfo {volume}
  {652}},\ \bibinfo {pages} {1768}}\BibitemShut {NoStop}%
\bibitem [{\citenamefont {Blumenfeld}\ \emph {et~al.}(2016)\citenamefont
  {Blumenfeld}, \citenamefont {Amitai}, \citenamefont {Jordan},\ and\
  \citenamefont {Hihinashvili}}]{Blumenfeld:2016aa}%
  \BibitemOpen
  \bibfield  {author} {\bibinfo {author} {\bibnamefont {Blumenfeld},
  \bibfnamefont {R.}}, \bibinfo {author} {\bibfnamefont {S.}~\bibnamefont
  {Amitai}}, \bibinfo {author} {\bibfnamefont {J.~F.}\ \bibnamefont {Jordan}},
  \ and\ \bibinfo {author} {\bibfnamefont {R.}~\bibnamefont {Hihinashvili}}}
  (\bibinfo {year} {2016}),\ \href {\doibase 10.1103/PhysRevLett.116.148001}
  {\bibfield  {journal} {\bibinfo  {journal} {Phys. Rev. Lett.}\ }\textbf
  {\bibinfo {volume} {116}},\ \bibinfo {pages} {148001}}\BibitemShut {NoStop}%
\bibitem [{\citenamefont {Blumenfeld}\ and\ \citenamefont
  {Edwards}(2003)}]{Blumenfeld:2003aa}%
  \BibitemOpen
  \bibfield  {author} {\bibinfo {author} {\bibnamefont {Blumenfeld},
  \bibfnamefont {R.}}, \ and\ \bibinfo {author} {\bibfnamefont
  {S.}~\bibnamefont {Edwards}}} (\bibinfo {year} {2003}),\ \href {\doibase
  10.1103/PhysRevLett.90.114303} {\bibfield  {journal} {\bibinfo  {journal}
  {Phys. Rev. Lett.}\ }\textbf {\bibinfo {volume} {90}},\
  10.1103/PhysRevLett.90.114303}\BibitemShut {NoStop}%
\bibitem [{\citenamefont {Blumenfeld}\ and\ \citenamefont
  {Edwards}(2006)}]{Blumenfeld:2006aa}%
  \BibitemOpen
  \bibfield  {author} {\bibinfo {author} {\bibnamefont {Blumenfeld},
  \bibfnamefont {R.}}, \ and\ \bibinfo {author} {\bibfnamefont
  {S.}~\bibnamefont {Edwards}}} (\bibinfo {year} {2006}),\ \href {\doibase
  10.1140/epje/e2006-00014-7} {\bibfield  {journal} {\bibinfo  {journal} {Eur.
  Phys. J. E}\ }\textbf {\bibinfo {volume} {19}},\ \bibinfo {pages}
  {23}}\BibitemShut {NoStop}%
\bibitem [{\citenamefont {Blumenfeld}\ and\ \citenamefont
  {Edwards}(2009)}]{Blumenfeld:2009aa}%
  \BibitemOpen
  \bibfield  {author} {\bibinfo {author} {\bibnamefont {Blumenfeld},
  \bibfnamefont {R.}}, \ and\ \bibinfo {author} {\bibfnamefont {S.~F.}\
  \bibnamefont {Edwards}}} (\bibinfo {year} {2009}),\ \href {\doibase
  10.1021/jp809768y} {\bibfield  {journal} {\bibinfo  {journal} {J. Phys. Chem.
  B}\ }\textbf {\bibinfo {volume} {113}},\ \bibinfo {pages} {3981}}\BibitemShut
  {NoStop}%
\bibitem [{\citenamefont {Blumenfeld}\ \emph {et~al.}(2012)\citenamefont
  {Blumenfeld}, \citenamefont {Jordan},\ and\ \citenamefont
  {Edwards}}]{Blumenfeld:2012aa}%
  \BibitemOpen
  \bibfield  {author} {\bibinfo {author} {\bibnamefont {Blumenfeld},
  \bibfnamefont {R.}}, \bibinfo {author} {\bibfnamefont {J.~F.}\ \bibnamefont
  {Jordan}}, \ and\ \bibinfo {author} {\bibfnamefont {S.~F.}\ \bibnamefont
  {Edwards}}} (\bibinfo {year} {2012}),\ \href {\doibase
  10.1103/PhysRevLett.109.238001} {\bibfield  {journal} {\bibinfo  {journal}
  {Phys. Rev. Lett.}\ }\textbf {\bibinfo {volume} {109}},\ \bibinfo {pages}
  {238001}}\BibitemShut {NoStop}%
\bibitem [{\citenamefont {Bo}\ \emph {et~al.}(2014)\citenamefont {Bo},
  \citenamefont {Mari}, \citenamefont {Song},\ and\ \citenamefont
  {Makse}}]{Bo:2014aa}%
  \BibitemOpen
  \bibfield  {author} {\bibinfo {author} {\bibnamefont {Bo}, \bibfnamefont
  {L.}}, \bibinfo {author} {\bibfnamefont {R.}~\bibnamefont {Mari}}, \bibinfo
  {author} {\bibfnamefont {C.}~\bibnamefont {Song}}, \ and\ \bibinfo {author}
  {\bibfnamefont {H.~A.}\ \bibnamefont {Makse}}} (\bibinfo {year} {2014}),\
  \href {\doibase 10.1039/C4SM00667D} {\bibfield  {journal} {\bibinfo
  {journal} {Soft Matter}\ }\textbf {\bibinfo {volume} {10}},\ \bibinfo {pages}
  {7379}}\BibitemShut {NoStop}%
\bibitem [{\citenamefont {Bohannon}(2017)}]{Bohannon:2017aa}%
  \BibitemOpen
  \bibfield  {author} {\bibinfo {author} {\bibnamefont {Bohannon},
  \bibfnamefont {J.}}} (\bibinfo {year} {2017}),\ \href {\doibase
  10.1126/science.355.6324.470} {\bibfield  {journal} {\bibinfo  {journal}
  {Science}\ }\textbf {\bibinfo {volume} {355}}~(\bibinfo {number} {6324}),\
  \bibinfo {pages} {470}}\BibitemShut {NoStop}%
\bibitem [{\citenamefont {Boissonat}\ \emph {et~al.}(2006)\citenamefont
  {Boissonat}, \citenamefont {Wormser},\ and\ \citenamefont
  {Yvinec}}]{Boissonat:2006aa}%
  \BibitemOpen
  \bibfield  {author} {\bibinfo {author} {\bibnamefont {Boissonat},
  \bibfnamefont {J.~D.}}, \bibinfo {author} {\bibfnamefont {C.}~\bibnamefont
  {Wormser}}, \ and\ \bibinfo {author} {\bibfnamefont {M.}~\bibnamefont
  {Yvinec}}} (\bibinfo {year} {2006}),\ in\ \href@noop {} {\emph {\bibinfo
  {booktitle} {Effective Computational Geometry for Curves and Surfaces}}},\
  \bibinfo {series and number} {Mathematics and Visualization},\ \bibinfo
  {editor} {edited by\ \bibinfo {editor} {\bibfnamefont {J.~D.}\ \bibnamefont
  {Boissonnat}}\ and\ \bibinfo {editor} {\bibfnamefont {M.}~\bibnamefont
  {Teillaud}}}\ (\bibinfo  {publisher} {Springer})\ p.~\bibinfo {pages}
  {67}\BibitemShut {NoStop}%
\bibitem [{\citenamefont {Borzsonyi}\ and\ \citenamefont
  {Stannarius}(2013)}]{Borzsonyi:2013aa}%
  \BibitemOpen
  \bibfield  {author} {\bibinfo {author} {\bibnamefont {Borzsonyi},
  \bibfnamefont {T.}}, \ and\ \bibinfo {author} {\bibfnamefont
  {R.}~\bibnamefont {Stannarius}}} (\bibinfo {year} {2013}),\ \href@noop {}
  {\bibfield  {journal} {\bibinfo  {journal} {Soft Matter}\ }\textbf {\bibinfo
  {volume} {9}},\ \bibinfo {pages} {7401}}\BibitemShut {NoStop}%
\bibitem [{\citenamefont {Bouchaud}(2002)}]{Bouchaud:2002aa}%
  \BibitemOpen
  \bibfield  {author} {\bibinfo {author} {\bibnamefont {Bouchaud},
  \bibfnamefont {J.-P.}}} (\bibinfo {year} {2002}),\ \enquote {\bibinfo {title}
  {Granular media: Some ideas from statistical physics},}\ in\ \href@noop {}
  {\emph {\bibinfo {booktitle} {Slow Relaxations and Nonequilibrium Dynamics in
  Condensed Matter}}},\ \bibinfo {series} {Les Houches Lecture Notes},
  Vol.~\bibinfo {volume} {77},\ \bibinfo {editor} {edited by\ \bibinfo {editor}
  {\bibfnamefont {J.-L.}\ \bibnamefont {Barrat}}, \bibinfo {editor}
  {\bibfnamefont {M.}~\bibnamefont {Feigelman}}, \bibinfo {editor}
  {\bibfnamefont {J.}~\bibnamefont {Kurchan}}, \ and\ \bibinfo {editor}
  {\bibfnamefont {J.}~\bibnamefont {Dalibard}}},\ Chap.~\bibinfo {chapter} {4}\
  (\bibinfo  {publisher} {Springer-Verlag})\ pp.\ \bibinfo {pages}
  {131--197}\BibitemShut {NoStop}%
\bibitem [{\citenamefont {{Bovet}}\ \emph {et~al.}(2016)\citenamefont
  {{Bovet}}, \citenamefont {{Morone}},\ and\ \citenamefont
  {{Makse}}}]{Bovet:2016aa}%
  \BibitemOpen
  \bibfield  {author} {\bibinfo {author} {\bibnamefont {{Bovet}}, \bibfnamefont
  {A.}}, \bibinfo {author} {\bibfnamefont {F.}~\bibnamefont {{Morone}}}, \ and\
  \bibinfo {author} {\bibfnamefont {H.~A.}\ \bibnamefont {{Makse}}}} (\bibinfo
  {year} {2016}),\ \href@noop {} {\bibfield  {journal} {\bibinfo  {journal}
  {ArXiv e-prints}\ }}\Eprint {http://arxiv.org/abs/1610.01587}
  {arXiv:1610.01587} \BibitemShut {NoStop}%
\bibitem [{\citenamefont {Bowles}\ and\ \citenamefont
  {Ashwin}(2011)}]{Bowles:2011aa}%
  \BibitemOpen
  \bibfield  {author} {\bibinfo {author} {\bibnamefont {Bowles}, \bibfnamefont
  {R.~K.}}, \ and\ \bibinfo {author} {\bibfnamefont {S.~S.}\ \bibnamefont
  {Ashwin}}} (\bibinfo {year} {2011}),\ \href {\doibase
  10.1103/PhysRevE.83.031302} {\bibfield  {journal} {\bibinfo  {journal} {Phys.
  Rev. E}\ }\textbf {\bibinfo {volume} {83}},\ \bibinfo {pages}
  {031302}}\BibitemShut {NoStop}%
\bibitem [{\citenamefont {Bray}\ and\ \citenamefont
  {Moore}(1980)}]{Bray:1980aa}%
  \BibitemOpen
  \bibfield  {author} {\bibinfo {author} {\bibnamefont {Bray}, \bibfnamefont
  {A.~J.}}, \ and\ \bibinfo {author} {\bibfnamefont {M.~A.}\ \bibnamefont
  {Moore}}} (\bibinfo {year} {1980}),\ \href
  {http://stacks.iop.org/0022-3719/13/i=19/a=002} {\bibfield  {journal}
  {\bibinfo  {journal} {Journal of Physics C: Solid State Physics}\ }\textbf
  {\bibinfo {volume} {13}}~(\bibinfo {number} {19}),\ \bibinfo {pages}
  {L469}}\BibitemShut {NoStop}%
\bibitem [{\citenamefont {Brey}\ \emph {et~al.}(2000)\citenamefont {Brey},
  \citenamefont {Prados},\ and\ \citenamefont {Sanchez-Rey}}]{Brey:2000aa}%
  \BibitemOpen
  \bibfield  {author} {\bibinfo {author} {\bibnamefont {Brey}, \bibfnamefont
  {J.}}, \bibinfo {author} {\bibfnamefont {A.}~\bibnamefont {Prados}}, \ and\
  \bibinfo {author} {\bibfnamefont {B.}~\bibnamefont {Sanchez-Rey}}} (\bibinfo
  {year} {2000}),\ \href {\doibase
  http://dx.doi.org/10.1016/S0378-4371(99)00375-1} {\bibfield  {journal}
  {\bibinfo  {journal} {Physica A}\ }\textbf {\bibinfo {volume} {275}},\
  \bibinfo {pages} {310 }}\BibitemShut {NoStop}%
\bibitem [{\citenamefont {Briscoe}\ \emph {et~al.}(2008)\citenamefont
  {Briscoe}, \citenamefont {Song}, \citenamefont {Wang},\ and\ \citenamefont
  {Makse}}]{Briscoe:2008aa}%
  \BibitemOpen
  \bibfield  {author} {\bibinfo {author} {\bibnamefont {Briscoe}, \bibfnamefont
  {C.}}, \bibinfo {author} {\bibfnamefont {C.}~\bibnamefont {Song}}, \bibinfo
  {author} {\bibfnamefont {P.}~\bibnamefont {Wang}}, \ and\ \bibinfo {author}
  {\bibfnamefont {H.~A.}\ \bibnamefont {Makse}}} (\bibinfo {year} {2008}),\
  \href {\doibase 10.1103/PhysRevLett.101.188001} {\bibfield  {journal}
  {\bibinfo  {journal} {Phys. Rev. Lett.}\ }\textbf {\bibinfo {volume} {101}},\
  \bibinfo {pages} {188001}}\BibitemShut {NoStop}%
\bibitem [{\citenamefont {Briscoe}\ \emph {et~al.}(2010)\citenamefont
  {Briscoe}, \citenamefont {Song}, \citenamefont {Wang},\ and\ \citenamefont
  {Makse}}]{Briscoe:2010aa}%
  \BibitemOpen
  \bibfield  {author} {\bibinfo {author} {\bibnamefont {Briscoe}, \bibfnamefont
  {C.}}, \bibinfo {author} {\bibfnamefont {C.}~\bibnamefont {Song}}, \bibinfo
  {author} {\bibfnamefont {P.}~\bibnamefont {Wang}}, \ and\ \bibinfo {author}
  {\bibfnamefont {H.~A.}\ \bibnamefont {Makse}}} (\bibinfo {year} {2010}),\
  \href {\doibase 10.1016/j.physa.2010.05.054} {\bibfield  {journal} {\bibinfo
  {journal} {Physica A}\ }\textbf {\bibinfo {volume} {389}},\ \bibinfo {pages}
  {3978}}\BibitemShut {NoStop}%
\bibitem [{\citenamefont {Brouwers}(2006)}]{Brouwers:2006aa}%
  \BibitemOpen
  \bibfield  {author} {\bibinfo {author} {\bibnamefont {Brouwers},
  \bibfnamefont {H.~J.~H.}}} (\bibinfo {year} {2006}),\ \href {\doibase
  10.1103/PhysRevE.74.031309} {\bibfield  {journal} {\bibinfo  {journal} {Phys.
  Rev. E}\ }\textbf {\bibinfo {volume} {74}},\ \bibinfo {pages}
  {031309}}\BibitemShut {NoStop}%
\bibitem [{\citenamefont {Bruji\'c}\ \emph
  {et~al.}(2003{\natexlab{a}})\citenamefont {Bruji\'c}, \citenamefont
  {Edwards}, \citenamefont {Grinev}, \citenamefont {Hopkinson}, \citenamefont
  {Bruji\'c},\ and\ \citenamefont {Makse}}]{Brujic:2003ab}%
  \BibitemOpen
  \bibfield  {author} {\bibinfo {author} {\bibnamefont {Bruji\'c},
  \bibfnamefont {J.}}, \bibinfo {author} {\bibfnamefont {S.}~\bibnamefont
  {Edwards}}, \bibinfo {author} {\bibfnamefont {D.}~\bibnamefont {Grinev}},
  \bibinfo {author} {\bibfnamefont {I.}~\bibnamefont {Hopkinson}}, \bibinfo
  {author} {\bibfnamefont {D.}~\bibnamefont {Bruji\'c}}, \ and\ \bibinfo
  {author} {\bibfnamefont {H.}~\bibnamefont {Makse}}} (\bibinfo {year}
  {2003}{\natexlab{a}}),\ \href {\doibase 10.1039/b204414e} {\bibfield
  {journal} {\bibinfo  {journal} {Farad. Disc.}\ }\textbf {\bibinfo {volume}
  {123}},\ \bibinfo {pages} {207}}\BibitemShut {NoStop}%
\bibitem [{\citenamefont {Bruji\'c}\ \emph
  {et~al.}(2003{\natexlab{b}})\citenamefont {Bruji\'c}, \citenamefont
  {Edwards}, \citenamefont {Hopkinson},\ and\ \citenamefont
  {Makse}}]{Brujic:2003aa}%
  \BibitemOpen
  \bibfield  {author} {\bibinfo {author} {\bibnamefont {Bruji\'c},
  \bibfnamefont {J.}}, \bibinfo {author} {\bibfnamefont {S.}~\bibnamefont
  {Edwards}}, \bibinfo {author} {\bibfnamefont {I.}~\bibnamefont {Hopkinson}},
  \ and\ \bibinfo {author} {\bibfnamefont {H.}~\bibnamefont {Makse}}} (\bibinfo
  {year} {2003}{\natexlab{b}}),\ \href {\doibase 10.1016/S0378-4371(03)00477-1}
  {\bibfield  {journal} {\bibinfo  {journal} {Physica A}\ }\textbf {\bibinfo
  {volume} {327}},\ \bibinfo {pages} {201}}\BibitemShut {NoStop}%
\bibitem [{\citenamefont {Bruji\'c}\ \emph {et~al.}(2007)\citenamefont
  {Bruji\'c}, \citenamefont {Song}, \citenamefont {Wang}, \citenamefont
  {Briscoe}, \citenamefont {Marty},\ and\ \citenamefont
  {Makse}}]{Brujic:2007aa}%
  \BibitemOpen
  \bibfield  {author} {\bibinfo {author} {\bibnamefont {Bruji\'c},
  \bibfnamefont {J.}}, \bibinfo {author} {\bibfnamefont {C.}~\bibnamefont
  {Song}}, \bibinfo {author} {\bibfnamefont {P.}~\bibnamefont {Wang}}, \bibinfo
  {author} {\bibfnamefont {C.}~\bibnamefont {Briscoe}}, \bibinfo {author}
  {\bibfnamefont {G.}~\bibnamefont {Marty}}, \ and\ \bibinfo {author}
  {\bibfnamefont {H.~A.}\ \bibnamefont {Makse}}} (\bibinfo {year} {2007}),\
  \href {\doibase 10.1103/PhysRevLett.98.248001} {\bibfield  {journal}
  {\bibinfo  {journal} {Phys. Rev. Lett.}\ }\textbf {\bibinfo {volume} {98}},\
  \bibinfo {pages} {248001}}\BibitemShut {NoStop}%
\bibitem [{\citenamefont {Bruji\'c}\ \emph {et~al.}(2005)\citenamefont
  {Bruji\'c}, \citenamefont {Wang}, \citenamefont {Song}, \citenamefont
  {Johnson}, \citenamefont {Sindt},\ and\ \citenamefont
  {Makse}}]{Brujic:2005aa}%
  \BibitemOpen
  \bibfield  {author} {\bibinfo {author} {\bibnamefont {Bruji\'c},
  \bibfnamefont {J.}}, \bibinfo {author} {\bibfnamefont {P.}~\bibnamefont
  {Wang}}, \bibinfo {author} {\bibfnamefont {C.}~\bibnamefont {Song}}, \bibinfo
  {author} {\bibfnamefont {D.~L.}\ \bibnamefont {Johnson}}, \bibinfo {author}
  {\bibfnamefont {O.}~\bibnamefont {Sindt}}, \ and\ \bibinfo {author}
  {\bibfnamefont {H.~A.}\ \bibnamefont {Makse}}} (\bibinfo {year} {2005}),\
  \href {\doibase 10.1103/PhysRevLett.95.128001} {\bibfield  {journal}
  {\bibinfo  {journal} {Phys. Rev. Lett.}\ }\textbf {\bibinfo {volume} {95}},\
  \bibinfo {pages} {128001}}\BibitemShut {NoStop}%
\bibitem [{\citenamefont {Caglioti}\ \emph {et~al.}(1997)\citenamefont
  {Caglioti}, \citenamefont {Loreto}, \citenamefont {Herrmann},\ and\
  \citenamefont {Nicodemi}}]{Caglioti:1997aa}%
  \BibitemOpen
  \bibfield  {author} {\bibinfo {author} {\bibnamefont {Caglioti},
  \bibfnamefont {E.}}, \bibinfo {author} {\bibfnamefont {V.}~\bibnamefont
  {Loreto}}, \bibinfo {author} {\bibfnamefont {H.}~\bibnamefont {Herrmann}}, \
  and\ \bibinfo {author} {\bibfnamefont {M.}~\bibnamefont {Nicodemi}}}
  (\bibinfo {year} {1997}),\ \href {\doibase 10.1103/PhysRevLett.79.1575}
  {\bibfield  {journal} {\bibinfo  {journal} {Phys. Rev. Lett.}\ }\textbf
  {\bibinfo {volume} {79}},\ \bibinfo {pages} {1575}}\BibitemShut {NoStop}%
\bibitem [{\citenamefont {Cates}\ \emph {et~al.}(1998)\citenamefont {Cates},
  \citenamefont {Wittmer}, \citenamefont {Bouchaud},\ and\ \citenamefont
  {Claudin}}]{Cates:1998aa}%
  \BibitemOpen
  \bibfield  {author} {\bibinfo {author} {\bibnamefont {Cates}, \bibfnamefont
  {M.}}, \bibinfo {author} {\bibfnamefont {J.}~\bibnamefont {Wittmer}},
  \bibinfo {author} {\bibfnamefont {J.}~\bibnamefont {Bouchaud}}, \ and\
  \bibinfo {author} {\bibfnamefont {P.}~\bibnamefont {Claudin}}} (\bibinfo
  {year} {1998}),\ \href {\doibase 10.1103/PhysRevLett.81.1841} {\bibfield
  {journal} {\bibinfo  {journal} {Phys. Rev. Lett.}\ }\textbf {\bibinfo
  {volume} {81}},\ \bibinfo {pages} {1841}}\BibitemShut {NoStop}%
\bibitem [{\citenamefont {Cates}\ and\ \citenamefont
  {Manoharan}(2015)}]{Cates:2015aa}%
  \BibitemOpen
  \bibfield  {author} {\bibinfo {author} {\bibnamefont {Cates}, \bibfnamefont
  {M.~E.}}, \ and\ \bibinfo {author} {\bibfnamefont {V.~N.}\ \bibnamefont
  {Manoharan}}} (\bibinfo {year} {2015}),\ \href {\doibase 10.1039/C5SM01014D}
  {\bibfield  {journal} {\bibinfo  {journal} {Soft Matter}\ }\textbf {\bibinfo
  {volume} {11}},\ \bibinfo {pages} {6538}}\BibitemShut {NoStop}%
\bibitem [{\citenamefont {Chaikin}\ \emph {et~al.}(2006)\citenamefont
  {Chaikin}, \citenamefont {Donev}, \citenamefont {Man}, \citenamefont
  {Stillinger},\ and\ \citenamefont {Torquato}}]{Chaikin:2006aa}%
  \BibitemOpen
  \bibfield  {author} {\bibinfo {author} {\bibnamefont {Chaikin}, \bibfnamefont
  {P.~M.}}, \bibinfo {author} {\bibfnamefont {A.}~\bibnamefont {Donev}},
  \bibinfo {author} {\bibfnamefont {W.}~\bibnamefont {Man}}, \bibinfo {author}
  {\bibfnamefont {F.~H.}\ \bibnamefont {Stillinger}}, \ and\ \bibinfo {author}
  {\bibfnamefont {S.}~\bibnamefont {Torquato}}} (\bibinfo {year} {2006}),\
  \href@noop {} {\bibfield  {journal} {\bibinfo  {journal} {Ind. Eng. Chem.
  Res.}\ }\textbf {\bibinfo {volume} {45}},\ \bibinfo {pages}
  {6960}}\BibitemShut {NoStop}%
\bibitem [{\citenamefont {Chakraborty}(2010)}]{Chakraborty:2010aa}%
  \BibitemOpen
  \bibfield  {author} {\bibinfo {author} {\bibnamefont {Chakraborty},
  \bibfnamefont {B.}}} (\bibinfo {year} {2010}),\ \href {\doibase
  10.1039/b927435a} {\bibfield  {journal} {\bibinfo  {journal} {Soft Matter}\
  }\textbf {\bibinfo {volume} {6}},\ \bibinfo {pages} {2884}}\BibitemShut
  {NoStop}%
\bibitem [{\citenamefont {Chakravarty}\ \emph {et~al.}(2003)\citenamefont
  {Chakravarty}, \citenamefont {Edwards}, \citenamefont {Grinev}, \citenamefont
  {Mann}, \citenamefont {Phillipson},\ and\ \citenamefont
  {Walton}}]{Chakravarty:2003aa}%
  \BibitemOpen
  \bibfield  {author} {\bibinfo {author} {\bibnamefont {Chakravarty},
  \bibfnamefont {A.}}, \bibinfo {author} {\bibfnamefont {S.}~\bibnamefont
  {Edwards}}, \bibinfo {author} {\bibfnamefont {D.}~\bibnamefont {Grinev}},
  \bibinfo {author} {\bibfnamefont {M.}~\bibnamefont {Mann}}, \bibinfo {author}
  {\bibfnamefont {T.}~\bibnamefont {Phillipson}}, \ and\ \bibinfo {author}
  {\bibfnamefont {A.}~\bibnamefont {Walton}}} (\bibinfo {year} {2003}),\ in\
  \href@noop {} {\emph {\bibinfo {booktitle} {Proceedings of the Workshop on
  the Quasi-static Deformations of Particulate Materials}}}\BibitemShut
  {NoStop}%
\bibitem [{\citenamefont {Charbonneau}\ \emph
  {et~al.}(2015{\natexlab{a}})\citenamefont {Charbonneau}, \citenamefont
  {Corwin}, \citenamefont {Parisi}, \citenamefont {Poncet},\ and\ \citenamefont
  {Zamponi}}]{Charbonneau:2015ab}%
  \BibitemOpen
  \bibfield  {author} {\bibinfo {author} {\bibnamefont {Charbonneau},
  \bibfnamefont {P.}}, \bibinfo {author} {\bibfnamefont {E.~I.}\ \bibnamefont
  {Corwin}}, \bibinfo {author} {\bibfnamefont {G.}~\bibnamefont {Parisi}},
  \bibinfo {author} {\bibfnamefont {A.}~\bibnamefont {Poncet}}, \ and\ \bibinfo
  {author} {\bibfnamefont {F.}~\bibnamefont {Zamponi}}} (\bibinfo {year}
  {2015}{\natexlab{a}}),\ \href@noop {} {\bibfield  {journal} {\bibinfo
  {journal} {ArXiv e-prints}\ }}\Eprint {http://arxiv.org/abs/1512.09100}
  {arXiv:1512.09100 [cond-mat.dis-nn]} \BibitemShut {NoStop}%
\bibitem [{\citenamefont {Charbonneau}\ \emph {et~al.}(2012)\citenamefont
  {Charbonneau}, \citenamefont {Corwin}, \citenamefont {Parisi},\ and\
  \citenamefont {Zamponi}}]{Charbonneau:2012aa}%
  \BibitemOpen
  \bibfield  {author} {\bibinfo {author} {\bibnamefont {Charbonneau},
  \bibfnamefont {P.}}, \bibinfo {author} {\bibfnamefont {E.~I.}\ \bibnamefont
  {Corwin}}, \bibinfo {author} {\bibfnamefont {G.}~\bibnamefont {Parisi}}, \
  and\ \bibinfo {author} {\bibfnamefont {F.}~\bibnamefont {Zamponi}}} (\bibinfo
  {year} {2012}),\ \href {\doibase 10.1103/PhysRevLett.109.205501} {\bibfield
  {journal} {\bibinfo  {journal} {Phys. Rev. Lett.}\ }\textbf {\bibinfo
  {volume} {109}},\ \bibinfo {pages} {205501}}\BibitemShut {NoStop}%
\bibitem [{\citenamefont {Charbonneau}\ \emph
  {et~al.}(2015{\natexlab{b}})\citenamefont {Charbonneau}, \citenamefont
  {Corwin}, \citenamefont {Parisi},\ and\ \citenamefont
  {Zamponi}}]{Charbonneau:2015aa}%
  \BibitemOpen
  \bibfield  {author} {\bibinfo {author} {\bibnamefont {Charbonneau},
  \bibfnamefont {P.}}, \bibinfo {author} {\bibfnamefont {E.~I.}\ \bibnamefont
  {Corwin}}, \bibinfo {author} {\bibfnamefont {G.}~\bibnamefont {Parisi}}, \
  and\ \bibinfo {author} {\bibfnamefont {F.}~\bibnamefont {Zamponi}}} (\bibinfo
  {year} {2015}{\natexlab{b}}),\ \href {\doibase
  10.1103/PhysRevLett.114.125504} {\bibfield  {journal} {\bibinfo  {journal}
  {Phys. Rev. Lett.}\ }\textbf {\bibinfo {volume} {114}},\ \bibinfo {pages}
  {125504}}\BibitemShut {NoStop}%
\bibitem [{\citenamefont {Charbonneau}\ \emph
  {et~al.}(2014{\natexlab{a}})\citenamefont {Charbonneau}, \citenamefont
  {Kurchan}, \citenamefont {Parisi}, \citenamefont {Urbani},\ and\
  \citenamefont {Zamponi}}]{Charbonneau:2014ab}%
  \BibitemOpen
  \bibfield  {author} {\bibinfo {author} {\bibnamefont {Charbonneau},
  \bibfnamefont {P.}}, \bibinfo {author} {\bibfnamefont {J.}~\bibnamefont
  {Kurchan}}, \bibinfo {author} {\bibfnamefont {G.}~\bibnamefont {Parisi}},
  \bibinfo {author} {\bibfnamefont {P.}~\bibnamefont {Urbani}}, \ and\ \bibinfo
  {author} {\bibfnamefont {F.}~\bibnamefont {Zamponi}}} (\bibinfo {year}
  {2014}{\natexlab{a}}),\ \href
  {http://stacks.iop.org/1742-5468/2014/i=10/a=P10009} {\bibfield  {journal}
  {\bibinfo  {journal} {J. Stat. Mech.}\ }\textbf {\bibinfo {volume} {2014}},\
  \bibinfo {pages} {P10009}}\BibitemShut {NoStop}%
\bibitem [{\citenamefont {Charbonneau}\ \emph
  {et~al.}(2014{\natexlab{b}})\citenamefont {Charbonneau}, \citenamefont
  {Kurchan}, \citenamefont {Parisi}, \citenamefont {Urbani},\ and\
  \citenamefont {Zamponi}}]{Charbonneau:2014aa}%
  \BibitemOpen
  \bibfield  {author} {\bibinfo {author} {\bibnamefont {Charbonneau},
  \bibfnamefont {P.}}, \bibinfo {author} {\bibfnamefont {J.}~\bibnamefont
  {Kurchan}}, \bibinfo {author} {\bibfnamefont {G.}~\bibnamefont {Parisi}},
  \bibinfo {author} {\bibfnamefont {P.}~\bibnamefont {Urbani}}, \ and\ \bibinfo
  {author} {\bibfnamefont {F.}~\bibnamefont {Zamponi}}} (\bibinfo {year}
  {2014}{\natexlab{b}}),\ \href {http://dx.doi.org/10.1038/ncomms4725}
  {\bibfield  {journal} {\bibinfo  {journal} {Nature Commun.}\ }\textbf
  {\bibinfo {volume} {5}},\ \bibinfo {pages} {3725}}\BibitemShut {NoStop}%
\bibitem [{\citenamefont {Charbonneau}\ \emph {et~al.}(2017)\citenamefont
  {Charbonneau}, \citenamefont {Kurchan}, \citenamefont {Parisi}, \citenamefont
  {Urbani},\ and\ \citenamefont {Zamponi}}]{Charbonneau:2017aa}%
  \BibitemOpen
  \bibfield  {author} {\bibinfo {author} {\bibnamefont {Charbonneau},
  \bibfnamefont {P.}}, \bibinfo {author} {\bibfnamefont {J.}~\bibnamefont
  {Kurchan}}, \bibinfo {author} {\bibfnamefont {G.}~\bibnamefont {Parisi}},
  \bibinfo {author} {\bibfnamefont {P.}~\bibnamefont {Urbani}}, \ and\ \bibinfo
  {author} {\bibfnamefont {F.}~\bibnamefont {Zamponi}}} (\bibinfo {year}
  {2017}),\ \href {\doibase 10.1146/annurev-conmatphys-031016-025334}
  {\bibfield  {journal} {\bibinfo  {journal} {Annual Review of Condensed Matter
  Physics}\ }\textbf {\bibinfo {volume} {8}}~(\bibinfo {number} {1}),\ \bibinfo
  {pages} {265}}\BibitemShut {NoStop}%
\bibitem [{\citenamefont {Chaudhuri}\ \emph {et~al.}(2010)\citenamefont
  {Chaudhuri}, \citenamefont {Berthier},\ and\ \citenamefont
  {Sastry}}]{Chaudhuri:2010aa}%
  \BibitemOpen
  \bibfield  {author} {\bibinfo {author} {\bibnamefont {Chaudhuri},
  \bibfnamefont {P.}}, \bibinfo {author} {\bibfnamefont {L.}~\bibnamefont
  {Berthier}}, \ and\ \bibinfo {author} {\bibfnamefont {S.}~\bibnamefont
  {Sastry}}} (\bibinfo {year} {2010}),\ \href {\doibase
  10.1103/PhysRevLett.104.165701} {\bibfield  {journal} {\bibinfo  {journal}
  {Phys. Rev. Lett.}\ }\textbf {\bibinfo {volume} {104}},\ \bibinfo {pages}
  {165701}}\BibitemShut {NoStop}%
\bibitem [{\citenamefont {Chen}\ \emph {et~al.}(2014)\citenamefont {Chen},
  \citenamefont {Klotsa}, \citenamefont {Engel}, \citenamefont {Damasceno},\
  and\ \citenamefont {Glotzer}}]{Chen:2014aa}%
  \BibitemOpen
  \bibfield  {author} {\bibinfo {author} {\bibnamefont {Chen}, \bibfnamefont
  {E.~R.}}, \bibinfo {author} {\bibfnamefont {D.}~\bibnamefont {Klotsa}},
  \bibinfo {author} {\bibfnamefont {M.}~\bibnamefont {Engel}}, \bibinfo
  {author} {\bibfnamefont {P.~F.}\ \bibnamefont {Damasceno}}, \ and\ \bibinfo
  {author} {\bibfnamefont {S.~C.}\ \bibnamefont {Glotzer}}} (\bibinfo {year}
  {2014}),\ \href {\doibase 10.1103/PhysRevX.4.011024} {\bibfield  {journal}
  {\bibinfo  {journal} {Phys. Rev. X}\ }\textbf {\bibinfo {volume} {4}},\
  \bibinfo {pages} {011024}}\BibitemShut {NoStop}%
\bibitem [{\citenamefont {Chen}\ \emph {et~al.}(2016)\citenamefont {Chen},
  \citenamefont {Li}, \citenamefont {Liu},\ and\ \citenamefont
  {Makse}}]{Chen:2016aa}%
  \BibitemOpen
  \bibfield  {author} {\bibinfo {author} {\bibnamefont {Chen}, \bibfnamefont
  {S.}}, \bibinfo {author} {\bibfnamefont {S.}~\bibnamefont {Li}}, \bibinfo
  {author} {\bibfnamefont {W.}~\bibnamefont {Liu}}, \ and\ \bibinfo {author}
  {\bibfnamefont {H.~A.}\ \bibnamefont {Makse}}} (\bibinfo {year} {2016}),\
  \href {\doibase 10.1039/C5SM02403J} {\bibfield  {journal} {\bibinfo
  {journal} {Soft Matter}\ }\textbf {\bibinfo {volume} {12}},\ \bibinfo {pages}
  {1836}}\BibitemShut {NoStop}%
\bibitem [{\citenamefont {Ciamarra}(2007)}]{Ciamarra:2007aa}%
  \BibitemOpen
  \bibfield  {author} {\bibinfo {author} {\bibnamefont {Ciamarra},
  \bibfnamefont {M.~P.}}} (\bibinfo {year} {2007}),\ \href {\doibase
  10.1103/PhysRevLett.99.089401} {\bibfield  {journal} {\bibinfo  {journal}
  {Phys. Rev. Lett.}\ }\textbf {\bibinfo {volume} {99}},\ \bibinfo {pages}
  {089401}}\BibitemShut {NoStop}%
\bibitem [{\citenamefont {Ciamarra}\ and\ \citenamefont
  {Coniglio}(2008)}]{Ciamarra:2008aa}%
  \BibitemOpen
  \bibfield  {author} {\bibinfo {author} {\bibnamefont {Ciamarra},
  \bibfnamefont {M.~P.}}, \ and\ \bibinfo {author} {\bibfnamefont
  {A.}~\bibnamefont {Coniglio}}} (\bibinfo {year} {2008}),\ \href {\doibase
  10.1103/PhysRevLett.101.128001} {\bibfield  {journal} {\bibinfo  {journal}
  {Phys. Rev. Lett.}\ }\textbf {\bibinfo {volume} {101}},\
  10.1103/PhysRevLett.101.128001}\BibitemShut {NoStop}%
\bibitem [{\citenamefont {Ciamarra}\ \emph {et~al.}(2010)\citenamefont
  {Ciamarra}, \citenamefont {Coniglio},\ and\ \citenamefont
  {de~Candia}}]{Ciamarra:2010aa}%
  \BibitemOpen
  \bibfield  {author} {\bibinfo {author} {\bibnamefont {Ciamarra},
  \bibfnamefont {M.~P.}}, \bibinfo {author} {\bibfnamefont {A.}~\bibnamefont
  {Coniglio}}, \ and\ \bibinfo {author} {\bibfnamefont {A.}~\bibnamefont
  {de~Candia}}} (\bibinfo {year} {2010}),\ \href {\doibase 10.1039/C001904F}
  {\bibfield  {journal} {\bibinfo  {journal} {Soft Matter}\ }\textbf {\bibinfo
  {volume} {6}},\ \bibinfo {pages} {2975}}\BibitemShut {NoStop}%
\bibitem [{\citenamefont {Ciamarra}\ \emph {et~al.}(2006)\citenamefont
  {Ciamarra}, \citenamefont {Coniglio},\ and\ \citenamefont
  {Nicodemi}}]{Ciamarra:2006aa}%
  \BibitemOpen
  \bibfield  {author} {\bibinfo {author} {\bibnamefont {Ciamarra},
  \bibfnamefont {M.~P.}}, \bibinfo {author} {\bibfnamefont {A.}~\bibnamefont
  {Coniglio}}, \ and\ \bibinfo {author} {\bibfnamefont {M.}~\bibnamefont
  {Nicodemi}}} (\bibinfo {year} {2006}),\ \href {\doibase
  10.1103/PhysRevLett.97.158001} {\bibfield  {journal} {\bibinfo  {journal}
  {Phys. Rev. Lett.}\ }\textbf {\bibinfo {volume} {97}},\ \bibinfo {pages}
  {158001}}\BibitemShut {NoStop}%
\bibitem [{\citenamefont {Cinacchi}\ and\ \citenamefont
  {Torquato}(2015)}]{Cinacchi:2015aa}%
  \BibitemOpen
  \bibfield  {author} {\bibinfo {author} {\bibnamefont {Cinacchi},
  \bibfnamefont {G.}}, \ and\ \bibinfo {author} {\bibfnamefont
  {S.}~\bibnamefont {Torquato}}} (\bibinfo {year} {2015}),\ \href {\doibase
  http://dx.doi.org/10.1063/1.4936938} {\bibfield  {journal} {\bibinfo
  {journal} {The Journal of Chemical Physics}\ }\textbf {\bibinfo {volume}
  {143}},\ \bibinfo {pages} {224506}}\BibitemShut {NoStop}%
\bibitem [{\citenamefont {Clarke}\ and\ \citenamefont
  {J\'onsson}(1993)}]{Clarke:1993aa}%
  \BibitemOpen
  \bibfield  {author} {\bibinfo {author} {\bibnamefont {Clarke}, \bibfnamefont
  {A.~S.}}, \ and\ \bibinfo {author} {\bibfnamefont {H.}~\bibnamefont
  {J\'onsson}}} (\bibinfo {year} {1993}),\ \href {\doibase
  10.1103/PhysRevE.47.3975} {\bibfield  {journal} {\bibinfo  {journal} {Phys.
  Rev. E}\ }\textbf {\bibinfo {volume} {47}},\ \bibinfo {pages}
  {3975}}\BibitemShut {NoStop}%
\bibitem [{\citenamefont {Clarke}\ and\ \citenamefont
  {Wiley}(1987)}]{Clarke:1987aa}%
  \BibitemOpen
  \bibfield  {author} {\bibinfo {author} {\bibnamefont {Clarke}, \bibfnamefont
  {A.~S.}}, \ and\ \bibinfo {author} {\bibfnamefont {J.~D.}\ \bibnamefont
  {Wiley}}} (\bibinfo {year} {1987}),\ \href {\doibase
  10.1103/PhysRevB.35.7350} {\bibfield  {journal} {\bibinfo  {journal} {Phys.
  Rev. B}\ }\textbf {\bibinfo {volume} {35}},\ \bibinfo {pages}
  {7350}}\BibitemShut {NoStop}%
\bibitem [{\citenamefont {Clusel}\ \emph {et~al.}(2009)\citenamefont {Clusel},
  \citenamefont {Corwin}, \citenamefont {Siemens},\ and\ \citenamefont
  {Bruji\'c}}]{Clusel:2009aa}%
  \BibitemOpen
  \bibfield  {author} {\bibinfo {author} {\bibnamefont {Clusel}, \bibfnamefont
  {M.}}, \bibinfo {author} {\bibfnamefont {E.~I.}\ \bibnamefont {Corwin}},
  \bibinfo {author} {\bibfnamefont {A.~O.~N.}\ \bibnamefont {Siemens}}, \ and\
  \bibinfo {author} {\bibfnamefont {J.}~\bibnamefont {Bruji\'c}}} (\bibinfo
  {year} {2009}),\ \href {\doibase 10.1038/nature08158} {\bibfield  {journal}
  {\bibinfo  {journal} {Nature}\ }\textbf {\bibinfo {volume} {460}},\ \bibinfo
  {pages} {611}}\BibitemShut {NoStop}%
\bibitem [{\citenamefont {Cohen}\ \emph {et~al.}(2016)\citenamefont {Cohen},
  \citenamefont {Dorosz}, \citenamefont {Schofield}, \citenamefont
  {Schilling},\ and\ \citenamefont {Sloutskin}}]{Cohen:2016aa}%
  \BibitemOpen
  \bibfield  {author} {\bibinfo {author} {\bibnamefont {Cohen}, \bibfnamefont
  {A.~P.}}, \bibinfo {author} {\bibfnamefont {S.}~\bibnamefont {Dorosz}},
  \bibinfo {author} {\bibfnamefont {A.~B.}\ \bibnamefont {Schofield}}, \bibinfo
  {author} {\bibfnamefont {T.}~\bibnamefont {Schilling}}, \ and\ \bibinfo
  {author} {\bibfnamefont {E.}~\bibnamefont {Sloutskin}}} (\bibinfo {year}
  {2016}),\ \href {\doibase 10.1103/PhysRevLett.116.098001} {\bibfield
  {journal} {\bibinfo  {journal} {Phys. Rev. Lett.}\ }\textbf {\bibinfo
  {volume} {116}},\ \bibinfo {pages} {098001}}\BibitemShut {NoStop}%
\bibitem [{\citenamefont {Coniglio}\ \emph {et~al.}(2004)\citenamefont
  {Coniglio}, \citenamefont {de~Candia}, \citenamefont {Fierro}, \citenamefont
  {Nicodemi},\ and\ \citenamefont {Tarzia}}]{Coniglio:2004ac}%
  \BibitemOpen
  \bibfield  {author} {\bibinfo {author} {\bibnamefont {Coniglio},
  \bibfnamefont {A.}}, \bibinfo {author} {\bibfnamefont {A.}~\bibnamefont
  {de~Candia}}, \bibinfo {author} {\bibfnamefont {A.}~\bibnamefont {Fierro}},
  \bibinfo {author} {\bibfnamefont {M.}~\bibnamefont {Nicodemi}}, \ and\
  \bibinfo {author} {\bibfnamefont {M.}~\bibnamefont {Tarzia}}} (\bibinfo
  {year} {2004}),\ \href {\doibase 10.1016/j.physa.2004.06.011} {\bibfield
  {journal} {\bibinfo  {journal} {Physica A}\ }\textbf {\bibinfo {volume}
  {344}},\ \bibinfo {pages} {431}}\BibitemShut {NoStop}%
\bibitem [{\citenamefont {Coniglio}\ \emph {et~al.}(2002)\citenamefont
  {Coniglio}, \citenamefont {Fierro},\ and\ \citenamefont
  {Nicodemi}}]{Coniglio:2002ab}%
  \BibitemOpen
  \bibfield  {author} {\bibinfo {author} {\bibnamefont {Coniglio},
  \bibfnamefont {A.}}, \bibinfo {author} {\bibfnamefont {A.}~\bibnamefont
  {Fierro}}, \ and\ \bibinfo {author} {\bibfnamefont {M.}~\bibnamefont
  {Nicodemi}}} (\bibinfo {year} {2002}),\ \href {\doibase
  10.1140/epje/i2002-10079-y} {\bibfield  {journal} {\bibinfo  {journal} {Eur.
  Phys. J. E}\ }\textbf {\bibinfo {volume} {9}},\ \bibinfo {pages}
  {219}}\BibitemShut {NoStop}%
\bibitem [{\citenamefont {Coniglio}\ and\ \citenamefont
  {Herrmann}(1996)}]{Coniglio:1996aa}%
  \BibitemOpen
  \bibfield  {author} {\bibinfo {author} {\bibnamefont {Coniglio},
  \bibfnamefont {A.}}, \ and\ \bibinfo {author} {\bibfnamefont
  {H.}~\bibnamefont {Herrmann}}} (\bibinfo {year} {1996}),\ \href {\doibase
  http://dx.doi.org/10.1016/0378-4371(95)00433-5} {\bibfield  {journal}
  {\bibinfo  {journal} {Physica A}\ }\textbf {\bibinfo {volume} {225}},\
  \bibinfo {pages} {1 }}\BibitemShut {NoStop}%
\bibitem [{\citenamefont {Coniglio}\ and\ \citenamefont
  {Nicodemi}(2000)}]{Coniglio:2000aa}%
  \BibitemOpen
  \bibfield  {author} {\bibinfo {author} {\bibnamefont {Coniglio},
  \bibfnamefont {A.}}, \ and\ \bibinfo {author} {\bibfnamefont
  {M.}~\bibnamefont {Nicodemi}}} (\bibinfo {year} {2000}),\ \href {\doibase
  10.1088/0953-8984/12/29/331} {\bibfield  {journal} {\bibinfo  {journal} {J.
  Phys. Cond. Mat.}\ }\textbf {\bibinfo {volume} {12}},\ \bibinfo {pages}
  {6601}}\BibitemShut {NoStop}%
\bibitem [{\citenamefont {Coniglio}\ and\ \citenamefont
  {Nicodemi}(2001)}]{Coniglio:2001aa}%
  \BibitemOpen
  \bibfield  {author} {\bibinfo {author} {\bibnamefont {Coniglio},
  \bibfnamefont {A.}}, \ and\ \bibinfo {author} {\bibfnamefont
  {M.}~\bibnamefont {Nicodemi}}} (\bibinfo {year} {2001}),\ \href {\doibase
  10.1016/S0378-4371(01)00190-X} {\bibfield  {journal} {\bibinfo  {journal}
  {Physica A}\ }\textbf {\bibinfo {volume} {296}},\ \bibinfo {pages}
  {451}}\BibitemShut {NoStop}%
\bibitem [{\citenamefont {Conway}\ and\ \citenamefont
  {Sloane}(1999)}]{Conway:1999aa}%
  \BibitemOpen
  \bibfield  {author} {\bibinfo {author} {\bibnamefont {Conway}, \bibfnamefont
  {J.}}, \ and\ \bibinfo {author} {\bibfnamefont {N.~J.~A.}\ \bibnamefont
  {Sloane}}} (\bibinfo {year} {1999}),\ \href@noop {} {\emph {\bibinfo {title}
  {Sphere packings, Lattices and Groups}}},\ \bibinfo {edition} {3rd}\ ed.,\
  \bibinfo {series} {A series of comprehensive mathematics}, Vol.\ \bibinfo
  {volume} {290}\ (\bibinfo  {publisher} {Springer},\ \bibinfo {address} {New
  York})\BibitemShut {NoStop}%
\bibitem [{\citenamefont {Corwin}\ \emph {et~al.}(2005)\citenamefont {Corwin},
  \citenamefont {Jaeger},\ and\ \citenamefont {Nagel}}]{Corwin:2005aa}%
  \BibitemOpen
  \bibfield  {author} {\bibinfo {author} {\bibnamefont {Corwin}, \bibfnamefont
  {E.}}, \bibinfo {author} {\bibfnamefont {H.}~\bibnamefont {Jaeger}}, \ and\
  \bibinfo {author} {\bibfnamefont {S.}~\bibnamefont {Nagel}}} (\bibinfo {year}
  {2005}),\ \href {\doibase 10.1038/nature03698} {\bibfield  {journal}
  {\bibinfo  {journal} {Nature}\ }\textbf {\bibinfo {volume} {435}},\ \bibinfo
  {pages} {1075}}\BibitemShut {NoStop}%
\bibitem [{\citenamefont {Corwin}\ \emph {et~al.}(2010)\citenamefont {Corwin},
  \citenamefont {Clusel}, \citenamefont {Siemens},\ and\ \citenamefont
  {Bruji\'c}}]{Corwin:2010aa}%
  \BibitemOpen
  \bibfield  {author} {\bibinfo {author} {\bibnamefont {Corwin}, \bibfnamefont
  {E.~I.}}, \bibinfo {author} {\bibfnamefont {M.}~\bibnamefont {Clusel}},
  \bibinfo {author} {\bibfnamefont {A.~O.~N.}\ \bibnamefont {Siemens}}, \ and\
  \bibinfo {author} {\bibfnamefont {J.}~\bibnamefont {Bruji\'c}}} (\bibinfo
  {year} {2010}),\ \href {\doibase 10.1039/c000984a} {\bibfield  {journal}
  {\bibinfo  {journal} {Soft Matter}\ }\textbf {\bibinfo {volume} {6}},\
  \bibinfo {pages} {2949}}\BibitemShut {NoStop}%
\bibitem [{\citenamefont {Cremona}(1890)}]{Cremona:1890aa}%
  \BibitemOpen
  \bibfield  {author} {\bibinfo {author} {\bibnamefont {Cremona}, \bibfnamefont
  {L.}}} (\bibinfo {year} {1890}),\ \href@noop {} {\emph {\bibinfo {title}
  {Graphical statics: two treatises on the graphical calculus and reciprocal
  figures in graphical statics}}}\ (\bibinfo  {publisher} {Clarendon
  Press})\BibitemShut {NoStop}%
\bibitem [{\citenamefont {Crisanti}\ and\ \citenamefont
  {Leuzzi}(2006)}]{Crisanti:2006aa}%
  \BibitemOpen
  \bibfield  {author} {\bibinfo {author} {\bibnamefont {Crisanti},
  \bibfnamefont {A.}}, \ and\ \bibinfo {author} {\bibfnamefont
  {L.}~\bibnamefont {Leuzzi}}} (\bibinfo {year} {2006}),\ \href {\doibase
  10.1103/PhysRevB.73.014412} {\bibfield  {journal} {\bibinfo  {journal} {Phys.
  Rev. B}\ }\textbf {\bibinfo {volume} {73}},\ \bibinfo {pages}
  {014412}}\BibitemShut {NoStop}%
\bibitem [{\citenamefont {Crisanti}\ and\ \citenamefont
  {Ritort}(2003)}]{Crisanti:2003aa}%
  \BibitemOpen
  \bibfield  {author} {\bibinfo {author} {\bibnamefont {Crisanti},
  \bibfnamefont {A.}}, \ and\ \bibinfo {author} {\bibfnamefont
  {F.}~\bibnamefont {Ritort}}} (\bibinfo {year} {2003}),\ \href
  {http://stacks.iop.org/0305-4470/36/i=21/a=201} {\bibfield  {journal}
  {\bibinfo  {journal} {J. Phys. A}\ }\textbf {\bibinfo {volume} {36}},\
  \bibinfo {pages} {R181}}\BibitemShut {NoStop}%
\bibitem [{\citenamefont {Cugliandolo}(2011)}]{Cugliandolo:2011aa}%
  \BibitemOpen
  \bibfield  {author} {\bibinfo {author} {\bibnamefont {Cugliandolo},
  \bibfnamefont {L.~F.}}} (\bibinfo {year} {2011}),\ \href {\doibase
  10.1088/1751-8113/44/48/483001} {\bibfield  {journal} {\bibinfo  {journal}
  {J. Phys. A}\ }\textbf {\bibinfo {volume} {44}},\ \bibinfo {pages}
  {483001}}\BibitemShut {NoStop}%
\bibitem [{\citenamefont {Cugliandolo}\ \emph {et~al.}(1997)\citenamefont
  {Cugliandolo}, \citenamefont {Kurchan},\ and\ \citenamefont
  {Peliti}}]{Cugliandolo:1997aa}%
  \BibitemOpen
  \bibfield  {author} {\bibinfo {author} {\bibnamefont {Cugliandolo},
  \bibfnamefont {L.~F.}}, \bibinfo {author} {\bibfnamefont {J.}~\bibnamefont
  {Kurchan}}, \ and\ \bibinfo {author} {\bibfnamefont {L.}~\bibnamefont
  {Peliti}}} (\bibinfo {year} {1997}),\ \href {\doibase
  10.1103/PhysRevE.55.3898} {\bibfield  {journal} {\bibinfo  {journal} {Phys.
  Rev. E}\ }\textbf {\bibinfo {volume} {55}},\ \bibinfo {pages}
  {3898}}\BibitemShut {NoStop}%
\bibitem [{\citenamefont {Damasceno}\ \emph {et~al.}(2012)\citenamefont
  {Damasceno}, \citenamefont {Engel},\ and\ \citenamefont
  {Glotzer}}]{Damasceno:2012aa}%
  \BibitemOpen
  \bibfield  {author} {\bibinfo {author} {\bibnamefont {Damasceno},
  \bibfnamefont {P.~F.}}, \bibinfo {author} {\bibfnamefont {M.}~\bibnamefont
  {Engel}}, \ and\ \bibinfo {author} {\bibfnamefont {S.~C.}\ \bibnamefont
  {Glotzer}}} (\bibinfo {year} {2012}),\ \href@noop {} {\bibfield  {journal}
  {\bibinfo  {journal} {Science}\ }\textbf {\bibinfo {volume} {337}},\ \bibinfo
  {pages} {453}}\BibitemShut {NoStop}%
\bibitem [{\citenamefont {Danisch}\ \emph {et~al.}(2010)\citenamefont
  {Danisch}, \citenamefont {Jin},\ and\ \citenamefont
  {Makse}}]{Danisch:2010aa}%
  \BibitemOpen
  \bibfield  {author} {\bibinfo {author} {\bibnamefont {Danisch}, \bibfnamefont
  {M.}}, \bibinfo {author} {\bibfnamefont {Y.}~\bibnamefont {Jin}}, \ and\
  \bibinfo {author} {\bibfnamefont {H.~A.}\ \bibnamefont {Makse}}} (\bibinfo
  {year} {2010}),\ \href {\doibase 10.1103/PhysRevE.81.051303} {\bibfield
  {journal} {\bibinfo  {journal} {Phys. Rev. E}\ }\textbf {\bibinfo {volume}
  {81}},\ \bibinfo {pages} {051303}}\BibitemShut {NoStop}%
\bibitem [{\citenamefont {Dauchot}(2007)}]{Dauchot:2007aa}%
  \BibitemOpen
  \bibfield  {author} {\bibinfo {author} {\bibnamefont {Dauchot}, \bibfnamefont
  {O.}}} (\bibinfo {year} {2007}),\ \enquote {\bibinfo {title} {Glassy
  behaviours in a-thermal systems, the case of granular media: A tentative
  review},}\ in\ \href {\doibase 10.1007/3-540-69684-9\_4} {\emph {\bibinfo
  {booktitle} {Ageing and the Glass Transition}}},\ \bibinfo {series} {Lecture
  Notes in Physics}, Vol.\ \bibinfo {volume} {716},\ \bibinfo {editor} {edited
  by\ \bibinfo {editor} {\bibfnamefont {M.}~\bibnamefont {Henkel}}, \bibinfo
  {editor} {\bibfnamefont {M.}~\bibnamefont {Pleimling}}, \ and\ \bibinfo
  {editor} {\bibfnamefont {R.}~\bibnamefont {Sanctuary}}}\ (\bibinfo
  {publisher} {Springer})\ pp.\ \bibinfo {pages} {161--206}\BibitemShut
  {NoStop}%
\bibitem [{\citenamefont {Dean}\ and\ \citenamefont
  {Lefevre}(2003)}]{Dean:2003aa}%
  \BibitemOpen
  \bibfield  {author} {\bibinfo {author} {\bibnamefont {Dean}, \bibfnamefont
  {D.}}, \ and\ \bibinfo {author} {\bibfnamefont {A.}~\bibnamefont {Lefevre}}}
  (\bibinfo {year} {2003}),\ \href {\doibase 10.1103/PhysRevLett.90.198301}
  {\bibfield  {journal} {\bibinfo  {journal} {Phys. Rev. Lett.}\ }\textbf
  {\bibinfo {volume} {90}},\ 10.1103/PhysRevLett.90.198301}\BibitemShut
  {NoStop}%
\bibitem [{\citenamefont {Dean}\ and\ \citenamefont
  {Lef\`evre}(2001)}]{Dean:2001aa}%
  \BibitemOpen
  \bibfield  {author} {\bibinfo {author} {\bibnamefont {Dean}, \bibfnamefont
  {D.~S.}}, \ and\ \bibinfo {author} {\bibfnamefont {A.}~\bibnamefont
  {Lef\`evre}}} (\bibinfo {year} {2001}),\ \href {\doibase
  10.1103/PhysRevLett.86.5639} {\bibfield  {journal} {\bibinfo  {journal}
  {Phys. Rev. Lett.}\ }\textbf {\bibinfo {volume} {86}},\ \bibinfo {pages}
  {5639}}\BibitemShut {NoStop}%
\bibitem [{\citenamefont {DeGiuli}\ \emph
  {et~al.}(2014{\natexlab{a}})\citenamefont {DeGiuli}, \citenamefont
  {Laversanne-Finot}, \citenamefont {During}, \citenamefont {Lerner},\ and\
  \citenamefont {Wyart}}]{DeGiuli:2014ab}%
  \BibitemOpen
  \bibfield  {author} {\bibinfo {author} {\bibnamefont {DeGiuli}, \bibfnamefont
  {E.}}, \bibinfo {author} {\bibfnamefont {A.}~\bibnamefont
  {Laversanne-Finot}}, \bibinfo {author} {\bibfnamefont {G.}~\bibnamefont
  {During}}, \bibinfo {author} {\bibfnamefont {E.}~\bibnamefont {Lerner}}, \
  and\ \bibinfo {author} {\bibfnamefont {M.}~\bibnamefont {Wyart}}} (\bibinfo
  {year} {2014}{\natexlab{a}}),\ \href {\doibase 10.1039/C4SM00561A} {\bibfield
   {journal} {\bibinfo  {journal} {Soft Matter}\ }\textbf {\bibinfo {volume}
  {10}},\ \bibinfo {pages} {5628}}\BibitemShut {NoStop}%
\bibitem [{\citenamefont {DeGiuli}\ \emph
  {et~al.}(2014{\natexlab{b}})\citenamefont {DeGiuli}, \citenamefont {Lerner},
  \citenamefont {Brito},\ and\ \citenamefont {Wyart}}]{DeGiuli:2014aa}%
  \BibitemOpen
  \bibfield  {author} {\bibinfo {author} {\bibnamefont {DeGiuli}, \bibfnamefont
  {E.}}, \bibinfo {author} {\bibfnamefont {E.}~\bibnamefont {Lerner}}, \bibinfo
  {author} {\bibfnamefont {C.}~\bibnamefont {Brito}}, \ and\ \bibinfo {author}
  {\bibfnamefont {M.}~\bibnamefont {Wyart}}} (\bibinfo {year}
  {2014}{\natexlab{b}}),\ \href {\doibase 10.1073/pnas.1415298111} {\bibfield
  {journal} {\bibinfo  {journal} {Proc. Nat. Acad. Sci.}\ }\textbf {\bibinfo
  {volume} {111}},\ \bibinfo {pages} {17054}}\BibitemShut {NoStop}%
\bibitem [{\citenamefont {DeGiuli}\ \emph {et~al.}(2015)\citenamefont
  {DeGiuli}, \citenamefont {Lerner},\ and\ \citenamefont
  {Wyart}}]{DeGiuli:2015aa}%
  \BibitemOpen
  \bibfield  {author} {\bibinfo {author} {\bibnamefont {DeGiuli}, \bibfnamefont
  {E.}}, \bibinfo {author} {\bibfnamefont {E.}~\bibnamefont {Lerner}}, \ and\
  \bibinfo {author} {\bibfnamefont {M.}~\bibnamefont {Wyart}}} (\bibinfo {year}
  {2015}),\ \href {\doibase 10.1063/1.4918737} {\bibfield  {journal} {\bibinfo
  {journal} {The Journal of Chemical Physics}\ }\textbf {\bibinfo {volume}
  {142}}~(\bibinfo {number} {16}),\ \bibinfo {pages} {164503}}\BibitemShut
  {NoStop}%
\bibitem [{\citenamefont {Delaney}\ \emph {et~al.}(2010)\citenamefont
  {Delaney}, \citenamefont {Di~Matteo},\ and\ \citenamefont
  {Aste}}]{Delaney:2010aa}%
  \BibitemOpen
  \bibfield  {author} {\bibinfo {author} {\bibnamefont {Delaney}, \bibfnamefont
  {G.~W.}}, \bibinfo {author} {\bibfnamefont {T.}~\bibnamefont {Di~Matteo}}, \
  and\ \bibinfo {author} {\bibfnamefont {T.}~\bibnamefont {Aste}}} (\bibinfo
  {year} {2010}),\ \href {\doibase 10.1039/b927490a} {\bibfield  {journal}
  {\bibinfo  {journal} {Soft Matter}\ }\textbf {\bibinfo {volume} {6}},\
  \bibinfo {pages} {2992}}\BibitemShut {NoStop}%
\bibitem [{\citenamefont {Delaney}\ \emph {et~al.}(2005)\citenamefont
  {Delaney}, \citenamefont {Weaire}, \citenamefont {Hutzler},\ and\
  \citenamefont {Murphy}}]{Delaney:2005aa}%
  \BibitemOpen
  \bibfield  {author} {\bibinfo {author} {\bibnamefont {Delaney}, \bibfnamefont
  {G.~W.}}, \bibinfo {author} {\bibfnamefont {D.}~\bibnamefont {Weaire}},
  \bibinfo {author} {\bibfnamefont {S.}~\bibnamefont {Hutzler}}, \ and\
  \bibinfo {author} {\bibfnamefont {S.}~\bibnamefont {Murphy}}} (\bibinfo
  {year} {2005}),\ \href@noop {} {\bibfield  {journal} {\bibinfo  {journal}
  {Philos. Mag. Lett.}\ }\textbf {\bibinfo {volume} {85}},\ \bibinfo {pages}
  {89}}\BibitemShut {NoStop}%
\bibitem [{\citenamefont {Desmond}\ and\ \citenamefont
  {Weeks}(2014)}]{Desmond:2014aa}%
  \BibitemOpen
  \bibfield  {author} {\bibinfo {author} {\bibnamefont {Desmond}, \bibfnamefont
  {K.~W.}}, \ and\ \bibinfo {author} {\bibfnamefont {E.~R.}\ \bibnamefont
  {Weeks}}} (\bibinfo {year} {2014}),\ \href {\doibase
  10.1103/PhysRevE.90.022204} {\bibfield  {journal} {\bibinfo  {journal} {Phys.
  Rev. E}\ }\textbf {\bibinfo {volume} {90}},\ \bibinfo {pages}
  {022204}}\BibitemShut {NoStop}%
\bibitem [{\citenamefont {Dieterich}\ \emph {et~al.}(2015)\citenamefont
  {Dieterich}, \citenamefont {Camunas-Soler}, \citenamefont
  {Ribezzi-Crivellari}, \citenamefont {Seifert},\ and\ \citenamefont
  {Ritort}}]{Dieterich:2015aa}%
  \BibitemOpen
  \bibfield  {author} {\bibinfo {author} {\bibnamefont {Dieterich},
  \bibfnamefont {E.}}, \bibinfo {author} {\bibfnamefont {J.}~\bibnamefont
  {Camunas-Soler}}, \bibinfo {author} {\bibfnamefont {M.}~\bibnamefont
  {Ribezzi-Crivellari}}, \bibinfo {author} {\bibfnamefont {U.}~\bibnamefont
  {Seifert}}, \ and\ \bibinfo {author} {\bibfnamefont {F.}~\bibnamefont
  {Ritort}}} (\bibinfo {year} {2015}),\ \href
  {http://dx.doi.org/10.1038/nphys3435} {\bibfield  {journal} {\bibinfo
  {journal} {Nature Phys.}\ }\textbf {\bibinfo {volume} {11}},\ \bibinfo
  {pages} {971}}\BibitemShut {NoStop}%
\bibitem [{\citenamefont {Digby}(1981)}]{Digby:1981aa}%
  \BibitemOpen
  \bibfield  {author} {\bibinfo {author} {\bibnamefont {Digby}, \bibfnamefont
  {P.~J.}}} (\bibinfo {year} {1981}),\ \href@noop {} {\bibfield  {journal}
  {\bibinfo  {journal} {Journal of Applied Mechanics}\ }\textbf {\bibinfo
  {volume} {48}},\ \bibinfo {pages} {803}}\BibitemShut {NoStop}%
\bibitem [{\citenamefont {Donev}\ \emph {et~al.}(2004)\citenamefont {Donev},
  \citenamefont {Cisse}, \citenamefont {Sachs}, \citenamefont {Variano},
  \citenamefont {Stillinger}, \citenamefont {Connelly}, \citenamefont
  {Torquato},\ and\ \citenamefont {Chaikin}}]{Donev:2004aa}%
  \BibitemOpen
  \bibfield  {author} {\bibinfo {author} {\bibnamefont {Donev}, \bibfnamefont
  {A.}}, \bibinfo {author} {\bibfnamefont {I.}~\bibnamefont {Cisse}}, \bibinfo
  {author} {\bibfnamefont {D.}~\bibnamefont {Sachs}}, \bibinfo {author}
  {\bibfnamefont {E.}~\bibnamefont {Variano}}, \bibinfo {author} {\bibfnamefont
  {F.}~\bibnamefont {Stillinger}}, \bibinfo {author} {\bibfnamefont
  {R.}~\bibnamefont {Connelly}}, \bibinfo {author} {\bibfnamefont
  {S.}~\bibnamefont {Torquato}}, \ and\ \bibinfo {author} {\bibfnamefont
  {P.}~\bibnamefont {Chaikin}}} (\bibinfo {year} {2004}),\ \href@noop {}
  {\bibfield  {journal} {\bibinfo  {journal} {Science}\ }\textbf {\bibinfo
  {volume} {303}},\ \bibinfo {pages} {990}}\BibitemShut {NoStop}%
\bibitem [{\citenamefont {Donev}\ \emph {et~al.}(2007)\citenamefont {Donev},
  \citenamefont {Connelly}, \citenamefont {Stillinger},\ and\ \citenamefont
  {Torquato}}]{Donev:2007aa}%
  \BibitemOpen
  \bibfield  {author} {\bibinfo {author} {\bibnamefont {Donev}, \bibfnamefont
  {A.}}, \bibinfo {author} {\bibfnamefont {R.}~\bibnamefont {Connelly}},
  \bibinfo {author} {\bibfnamefont {F.~H.}\ \bibnamefont {Stillinger}}, \ and\
  \bibinfo {author} {\bibfnamefont {S.}~\bibnamefont {Torquato}}} (\bibinfo
  {year} {2007}),\ \href {\doibase 10.1103/PhysRevE.75.051304} {\bibfield
  {journal} {\bibinfo  {journal} {Phys. Rev. E}\ }\textbf {\bibinfo {volume}
  {75}},\ \bibinfo {pages} {051304}}\BibitemShut {NoStop}%
\bibitem [{\citenamefont {Donev}\ \emph
  {et~al.}(2005{\natexlab{a}})\citenamefont {Donev}, \citenamefont
  {Stillinger},\ and\ \citenamefont {Torquato}}]{Donev:2005ab}%
  \BibitemOpen
  \bibfield  {author} {\bibinfo {author} {\bibnamefont {Donev}, \bibfnamefont
  {A.}}, \bibinfo {author} {\bibfnamefont {F.~H.}\ \bibnamefont {Stillinger}},
  \ and\ \bibinfo {author} {\bibfnamefont {S.}~\bibnamefont {Torquato}}}
  (\bibinfo {year} {2005}{\natexlab{a}}),\ \href {\doibase
  10.1103/PhysRevLett.95.090604} {\bibfield  {journal} {\bibinfo  {journal}
  {Phys. Rev. Lett.}\ }\textbf {\bibinfo {volume} {95}},\ \bibinfo {pages}
  {090604}}\BibitemShut {NoStop}%
\bibitem [{\citenamefont {Donev}\ \emph
  {et~al.}(2005{\natexlab{b}})\citenamefont {Donev}, \citenamefont {Torquato},\
  and\ \citenamefont {Stillinger}}]{Donev:2005aa}%
  \BibitemOpen
  \bibfield  {author} {\bibinfo {author} {\bibnamefont {Donev}, \bibfnamefont
  {A.}}, \bibinfo {author} {\bibfnamefont {S.}~\bibnamefont {Torquato}}, \ and\
  \bibinfo {author} {\bibfnamefont {F.}~\bibnamefont {Stillinger}}} (\bibinfo
  {year} {2005}{\natexlab{b}}),\ \href {\doibase 10.1103/PhysRevE.71.011105}
  {\bibfield  {journal} {\bibinfo  {journal} {Phys. Rev. E}\ }\textbf {\bibinfo
  {volume} {71}},\ \bibinfo {pages} {011105}}\BibitemShut {NoStop}%
\bibitem [{\citenamefont {Drocco}\ \emph {et~al.}(2005)\citenamefont {Drocco},
  \citenamefont {Hastings}, \citenamefont {Reichhardt},\ and\ \citenamefont
  {Reichhardt}}]{Drocco:2005aa}%
  \BibitemOpen
  \bibfield  {author} {\bibinfo {author} {\bibnamefont {Drocco}, \bibfnamefont
  {J.~A.}}, \bibinfo {author} {\bibfnamefont {M.~B.}\ \bibnamefont {Hastings}},
  \bibinfo {author} {\bibfnamefont {C.~J.~O.}\ \bibnamefont {Reichhardt}}, \
  and\ \bibinfo {author} {\bibfnamefont {C.}~\bibnamefont {Reichhardt}}}
  (\bibinfo {year} {2005}),\ \href {\doibase 10.1103/PhysRevLett.95.088001}
  {\bibfield  {journal} {\bibinfo  {journal} {Phys. Rev. Lett.}\ }\textbf
  {\bibinfo {volume} {95}},\ \bibinfo {pages} {088001}}\BibitemShut {NoStop}%
\bibitem [{\citenamefont {During}\ \emph {et~al.}(2013)\citenamefont {During},
  \citenamefont {Lerner},\ and\ \citenamefont {Wyart}}]{During:2013aa}%
  \BibitemOpen
  \bibfield  {author} {\bibinfo {author} {\bibnamefont {During}, \bibfnamefont
  {G.}}, \bibinfo {author} {\bibfnamefont {E.}~\bibnamefont {Lerner}}, \ and\
  \bibinfo {author} {\bibfnamefont {M.}~\bibnamefont {Wyart}}} (\bibinfo {year}
  {2013}),\ \href {\doibase 10.1039/C2SM25878A} {\bibfield  {journal} {\bibinfo
   {journal} {Soft Matter}\ }\textbf {\bibinfo {volume} {9}},\ \bibinfo {pages}
  {146}}\BibitemShut {NoStop}%
\bibitem [{\citenamefont {Eastham}\ \emph {et~al.}(2006)\citenamefont
  {Eastham}, \citenamefont {Blythe}, \citenamefont {Bray},\ and\ \citenamefont
  {Moore}}]{Eastham:2006aa}%
  \BibitemOpen
  \bibfield  {author} {\bibinfo {author} {\bibnamefont {Eastham}, \bibfnamefont
  {P.~R.}}, \bibinfo {author} {\bibfnamefont {R.~A.}\ \bibnamefont {Blythe}},
  \bibinfo {author} {\bibfnamefont {A.~J.}\ \bibnamefont {Bray}}, \ and\
  \bibinfo {author} {\bibfnamefont {M.~A.}\ \bibnamefont {Moore}}} (\bibinfo
  {year} {2006}),\ \href {\doibase 10.1103/PhysRevB.74.020406} {\bibfield
  {journal} {\bibinfo  {journal} {Phys. Rev. B}\ }\textbf {\bibinfo {volume}
  {74}},\ \bibinfo {pages} {020406}}\BibitemShut {NoStop}%
\bibitem [{\citenamefont {Edwards}\ and\ \citenamefont
  {Grinev}(1999{\natexlab{a}})}]{Edwards:1999aa}%
  \BibitemOpen
  \bibfield  {author} {\bibinfo {author} {\bibnamefont {Edwards}, \bibfnamefont
  {S.}}, \ and\ \bibinfo {author} {\bibfnamefont {D.}~\bibnamefont {Grinev}}}
  (\bibinfo {year} {1999}{\natexlab{a}}),\ \href {\doibase
  10.1103/PhysRevLett.82.5397} {\bibfield  {journal} {\bibinfo  {journal}
  {Phys. Rev. Lett.}\ }\textbf {\bibinfo {volume} {82}},\ \bibinfo {pages}
  {5397}}\BibitemShut {NoStop}%
\bibitem [{\citenamefont {Edwards}\ and\ \citenamefont
  {Grinev}(2001)}]{Edwards:2001aa}%
  \BibitemOpen
  \bibfield  {author} {\bibinfo {author} {\bibnamefont {Edwards}, \bibfnamefont
  {S.}}, \ and\ \bibinfo {author} {\bibfnamefont {D.}~\bibnamefont {Grinev}}}
  (\bibinfo {year} {2001}),\ \href {\doibase
  http://dx.doi.org/10.1016/S0009-2509(01)00157-9} {\bibfield  {journal}
  {\bibinfo  {journal} {Chem. Eng. Sci.}\ }\textbf {\bibinfo {volume} {56}},\
  \bibinfo {pages} {5451 }}\BibitemShut {NoStop}%
\bibitem [{\citenamefont {Edwards}\ and\ \citenamefont
  {Oakeshott}(1989)}]{Edwards:1989aa}%
  \BibitemOpen
  \bibfield  {author} {\bibinfo {author} {\bibnamefont {Edwards}, \bibfnamefont
  {S.}}, \ and\ \bibinfo {author} {\bibfnamefont {R.}~\bibnamefont
  {Oakeshott}}} (\bibinfo {year} {1989}),\ \href {\doibase
  10.1016/0378-4371(89)90034-4} {\bibfield  {journal} {\bibinfo  {journal}
  {Physica A}\ }\textbf {\bibinfo {volume} {157}},\ \bibinfo {pages}
  {1080}}\BibitemShut {NoStop}%
\bibitem [{\citenamefont {Edwards}(1991)}]{Edwards:1991aa}%
  \BibitemOpen
  \bibfield  {author} {\bibinfo {author} {\bibnamefont {Edwards}, \bibfnamefont
  {S.~F.}}} (\bibinfo {year} {1991}),\ \enquote {\bibinfo {title} {The aging of
  glass forming liquids},}\ in\ \href@noop {} {\emph {\bibinfo {booktitle}
  {Disorder in Condensed Matter Physics}}},\ \bibinfo {editor} {edited by\
  \bibinfo {editor} {\bibfnamefont {J.}~\bibnamefont {Blackman}}\ and\ \bibinfo
  {editor} {\bibfnamefont {J.}~\bibnamefont {Taguena}}}\ (\bibinfo  {publisher}
  {Oxford University Press},\ \bibinfo {address} {Oxford})\ pp.\ \bibinfo
  {pages} {147--154}\BibitemShut {NoStop}%
\bibitem [{\citenamefont {Edwards}(1994)}]{Edwards:1994aa}%
  \BibitemOpen
  \bibfield  {author} {\bibinfo {author} {\bibnamefont {Edwards}, \bibfnamefont
  {S.~F.}}} (\bibinfo {year} {1994}),\ \enquote {\bibinfo {title} {The role of
  entropy in the specification of a powder},}\ in\ \href@noop {} {\emph
  {\bibinfo {booktitle} {Granular matter: an interdisciplinary approach}}},\
  \bibinfo {editor} {edited by\ \bibinfo {editor} {\bibfnamefont
  {A.}~\bibnamefont {Mehta}}}\ (\bibinfo  {publisher} {Springer, New York})\
  pp.\ \bibinfo {pages} {121--140}\BibitemShut {NoStop}%
\bibitem [{\citenamefont {Edwards}(2008)}]{Edwards:2008aa}%
  \BibitemOpen
  \bibfield  {author} {\bibinfo {author} {\bibnamefont {Edwards}, \bibfnamefont
  {S.~F.}}} (\bibinfo {year} {2008}),\ \href {\doibase
  10.1088/1751-8113/41/32/324019} {\bibfield  {journal} {\bibinfo  {journal}
  {J. Phys. A}\ }\textbf {\bibinfo {volume} {41}},\ \bibinfo {pages}
  {324019}}\BibitemShut {NoStop}%
\bibitem [{\citenamefont {Edwards}\ and\ \citenamefont
  {Anderson}(1975)}]{Edwards:1975aa}%
  \BibitemOpen
  \bibfield  {author} {\bibinfo {author} {\bibnamefont {Edwards}, \bibfnamefont
  {S.~F.}}, \ and\ \bibinfo {author} {\bibfnamefont {P.~W.}\ \bibnamefont
  {Anderson}}} (\bibinfo {year} {1975}),\ \href
  {http://stacks.iop.org/0305-4608/5/i=5/a=017} {\bibfield  {journal} {\bibinfo
   {journal} {Journal of Physics F: Metal Physics}\ }\textbf {\bibinfo {volume}
  {5}}~(\bibinfo {number} {5}),\ \bibinfo {pages} {965}}\BibitemShut {NoStop}%
\bibitem [{\citenamefont {Edwards}\ \emph {et~al.}(2004)\citenamefont
  {Edwards}, \citenamefont {Bruji\'c},\ and\ \citenamefont
  {Makse}}]{Edwards:2004aa}%
  \BibitemOpen
  \bibfield  {author} {\bibinfo {author} {\bibnamefont {Edwards}, \bibfnamefont
  {S.~F.}}, \bibinfo {author} {\bibfnamefont {J.}~\bibnamefont {Bruji\'c}}, \
  and\ \bibinfo {author} {\bibfnamefont {H.~A.}\ \bibnamefont {Makse}}}
  (\bibinfo {year} {2004}),\ \enquote {\bibinfo {title} {A basis for the
  statistical mechanics of granular systems},}\ in\ \href {\doibase
  10.1016/B978-044451607-7/50002-9} {\emph {\bibinfo {booktitle} {Unifying
  Concepts in Granular Media and Glasses}}},\ \bibinfo {editor} {edited by\
  \bibinfo {editor} {\bibfnamefont {A.}~\bibnamefont {Coniglio}}, \bibinfo
  {editor} {\bibfnamefont {A.}~\bibnamefont {Fierro}}, \bibinfo {editor}
  {\bibfnamefont {H.}~\bibnamefont {Herrmann}}, \ and\ \bibinfo {editor}
  {\bibfnamefont {M.}~\bibnamefont {Nicodemi}}}\ (\bibinfo  {publisher}
  {Elsevier Science BV})\ pp.\ \bibinfo {pages} {9--23}\BibitemShut {NoStop}%
\bibitem [{\citenamefont {Edwards}\ and\ \citenamefont
  {Grinev}(1999{\natexlab{b}})}]{Edwards:1999ab}%
  \BibitemOpen
  \bibfield  {author} {\bibinfo {author} {\bibnamefont {Edwards}, \bibfnamefont
  {S.~F.}}, \ and\ \bibinfo {author} {\bibfnamefont {D.~V.}\ \bibnamefont
  {Grinev}}} (\bibinfo {year} {1999}{\natexlab{b}}),\ \href {\doibase
  http://dx.doi.org/10.1063/1.166429} {\bibfield  {journal} {\bibinfo
  {journal} {Chaos}\ }\textbf {\bibinfo {volume} {9}},\ \bibinfo {pages}
  {551}}\BibitemShut {NoStop}%
\bibitem [{\citenamefont {van Eerd}\ \emph {et~al.}(2007)\citenamefont {van
  Eerd}, \citenamefont {Ellenbroek}, \citenamefont {van Hecke}, \citenamefont
  {Snoeijer},\ and\ \citenamefont {Vlugt}}]{Eerd:2007aa}%
  \BibitemOpen
  \bibfield  {author} {\bibinfo {author} {\bibnamefont {van Eerd},
  \bibfnamefont {A.~R.~T.}}, \bibinfo {author} {\bibfnamefont {W.~G.}\
  \bibnamefont {Ellenbroek}}, \bibinfo {author} {\bibfnamefont
  {M.}~\bibnamefont {van Hecke}}, \bibinfo {author} {\bibfnamefont {J.~H.}\
  \bibnamefont {Snoeijer}}, \ and\ \bibinfo {author} {\bibfnamefont {T.~J.~H.}\
  \bibnamefont {Vlugt}}} (\bibinfo {year} {2007}),\ \href {\doibase
  10.1103/PhysRevE.75.060302} {\bibfield  {journal} {\bibinfo  {journal} {Phys.
  Rev. E}\ }\textbf {\bibinfo {volume} {75}},\ \bibinfo {pages}
  {060302}}\BibitemShut {NoStop}%
\bibitem [{\citenamefont {Ellenbroek}\ \emph {et~al.}(2009)\citenamefont
  {Ellenbroek}, \citenamefont {van Hecke},\ and\ \citenamefont {van
  Saarloos}}]{Ellenbroek:2009ab}%
  \BibitemOpen
  \bibfield  {author} {\bibinfo {author} {\bibnamefont {Ellenbroek},
  \bibfnamefont {W.~G.}}, \bibinfo {author} {\bibfnamefont {M.}~\bibnamefont
  {van Hecke}}, \ and\ \bibinfo {author} {\bibfnamefont {W.}~\bibnamefont {van
  Saarloos}}} (\bibinfo {year} {2009}),\ \href {\doibase
  10.1103/PhysRevE.80.061307} {\bibfield  {journal} {\bibinfo  {journal} {Phys.
  Rev. E}\ }\textbf {\bibinfo {volume} {80}},\ \bibinfo {pages}
  {061307}}\BibitemShut {NoStop}%
\bibitem [{\citenamefont {Ellenbroek}\ \emph {et~al.}(2006)\citenamefont
  {Ellenbroek}, \citenamefont {Somfai}, \citenamefont {van Hecke},\ and\
  \citenamefont {van Saarloos}}]{Ellenbroek:2006aa}%
  \BibitemOpen
  \bibfield  {author} {\bibinfo {author} {\bibnamefont {Ellenbroek},
  \bibfnamefont {W.~G.}}, \bibinfo {author} {\bibfnamefont {E.}~\bibnamefont
  {Somfai}}, \bibinfo {author} {\bibfnamefont {M.}~\bibnamefont {van Hecke}}, \
  and\ \bibinfo {author} {\bibfnamefont {W.}~\bibnamefont {van Saarloos}}}
  (\bibinfo {year} {2006}),\ \href {\doibase 10.1103/PhysRevLett.97.258001}
  {\bibfield  {journal} {\bibinfo  {journal} {Phys. Rev. Lett.}\ }\textbf
  {\bibinfo {volume} {97}},\ \bibinfo {pages} {258001}}\BibitemShut {NoStop}%
\bibitem [{\citenamefont {Erikson}\ \emph {et~al.}(2002)\citenamefont
  {Erikson}, \citenamefont {Mueggenburg}, \citenamefont {Jaeger},\ and\
  \citenamefont {Nagel}}]{Erikson:2002aa}%
  \BibitemOpen
  \bibfield  {author} {\bibinfo {author} {\bibnamefont {Erikson}, \bibfnamefont
  {J.~M.}}, \bibinfo {author} {\bibfnamefont {N.~W.}\ \bibnamefont
  {Mueggenburg}}, \bibinfo {author} {\bibfnamefont {H.~M.}\ \bibnamefont
  {Jaeger}}, \ and\ \bibinfo {author} {\bibfnamefont {S.~R.}\ \bibnamefont
  {Nagel}}} (\bibinfo {year} {2002}),\ \href {\doibase
  10.1103/PhysRevE.66.040301} {\bibfield  {journal} {\bibinfo  {journal} {Phys.
  Rev. E}\ }\textbf {\bibinfo {volume} {66}},\ \bibinfo {pages}
  {040301}}\BibitemShut {NoStop}%
\bibitem [{\citenamefont {Escobedo}(2014)}]{Escobedo:2014aa}%
  \BibitemOpen
  \bibfield  {author} {\bibinfo {author} {\bibnamefont {Escobedo},
  \bibfnamefont {F.~A.}}} (\bibinfo {year} {2014}),\ \href {\doibase
  10.1039/C4SM01646G} {\bibfield  {journal} {\bibinfo  {journal} {Soft Matter}\
  }\textbf {\bibinfo {volume} {10}},\ \bibinfo {pages} {8388}}\BibitemShut
  {NoStop}%
\bibitem [{\citenamefont {Farrell}\ \emph {et~al.}(2010)\citenamefont
  {Farrell}, \citenamefont {Martini},\ and\ \citenamefont
  {Menon}}]{Farrell:2010aa}%
  \BibitemOpen
  \bibfield  {author} {\bibinfo {author} {\bibnamefont {Farrell}, \bibfnamefont
  {G.~R.}}, \bibinfo {author} {\bibfnamefont {K.~M.}\ \bibnamefont {Martini}},
  \ and\ \bibinfo {author} {\bibfnamefont {N.}~\bibnamefont {Menon}}} (\bibinfo
  {year} {2010}),\ \href {\doibase 10.1039/C0SM00038H} {\bibfield  {journal}
  {\bibinfo  {journal} {Soft Matter}\ }\textbf {\bibinfo {volume} {6}},\
  \bibinfo {pages} {2925}}\BibitemShut {NoStop}%
\bibitem [{\citenamefont {Faure}\ \emph {et~al.}(2009)\citenamefont {Faure},
  \citenamefont {Lefebvre-Lepot},\ and\ \citenamefont {Semin}}]{Faure:2009aa}%
  \BibitemOpen
  \bibfield  {author} {\bibinfo {author} {\bibnamefont {Faure}, \bibfnamefont
  {S.}}, \bibinfo {author} {\bibfnamefont {A.}~\bibnamefont {Lefebvre-Lepot}},
  \ and\ \bibinfo {author} {\bibfnamefont {B.}~\bibnamefont {Semin}}} (\bibinfo
  {year} {2009}),\ \enquote {\bibinfo {title} {Dynamic numerical investigation
  of random packing for spherical and nonconvex particles},}\ in\ \href@noop {}
  {\emph {\bibinfo {booktitle} {ESAIM: Proceedings}}},\ Vol.~\bibinfo {volume}
  {28},\ \bibinfo {editor} {edited by\ \bibinfo {editor} {\bibfnamefont
  {M.}~\bibnamefont {Ismail}}, \bibinfo {editor} {\bibfnamefont
  {B.}~\bibnamefont {Maury}}, \ and\ \bibinfo {editor} {\bibfnamefont {J.-F.}\
  \bibnamefont {Gerbeau}}},\ pp.\ \bibinfo {pages} {13--32}\BibitemShut
  {NoStop}%
\bibitem [{\citenamefont {Ferenc}\ and\ \citenamefont
  {N{\'e}da}(2007)}]{Ferenc:2007aa}%
  \BibitemOpen
  \bibfield  {author} {\bibinfo {author} {\bibnamefont {Ferenc}, \bibfnamefont
  {J.-S.}}, \ and\ \bibinfo {author} {\bibfnamefont {Z.}~\bibnamefont
  {N{\'e}da}}} (\bibinfo {year} {2007}),\ \href {\doibase
  http://dx.doi.org/10.1016/j.physa.2007.07.063} {\bibfield  {journal}
  {\bibinfo  {journal} {Physica A: Statistical Mechanics and its Applications}\
  }\textbf {\bibinfo {volume} {385}}~(\bibinfo {number} {2}),\ \bibinfo {pages}
  {518 }}\BibitemShut {NoStop}%
\bibitem [{\citenamefont {Fierro}\ \emph
  {et~al.}(2002{\natexlab{a}})\citenamefont {Fierro}, \citenamefont
  {Nicodemi},\ and\ \citenamefont {Coniglio}}]{Fierro:2002aa}%
  \BibitemOpen
  \bibfield  {author} {\bibinfo {author} {\bibnamefont {Fierro}, \bibfnamefont
  {A.}}, \bibinfo {author} {\bibfnamefont {M.}~\bibnamefont {Nicodemi}}, \ and\
  \bibinfo {author} {\bibfnamefont {A.}~\bibnamefont {Coniglio}}} (\bibinfo
  {year} {2002}{\natexlab{a}}),\ \href {\doibase 10.1209/epl/i2002-00173-x}
  {\bibfield  {journal} {\bibinfo  {journal} {Europhys. Lett.}\ }\textbf
  {\bibinfo {volume} {59}},\ \bibinfo {pages} {642}}\BibitemShut {NoStop}%
\bibitem [{\citenamefont {Fierro}\ \emph
  {et~al.}(2002{\natexlab{b}})\citenamefont {Fierro}, \citenamefont
  {Nicodemi},\ and\ \citenamefont {Coniglio}}]{Fierro:2002ab}%
  \BibitemOpen
  \bibfield  {author} {\bibinfo {author} {\bibnamefont {Fierro}, \bibfnamefont
  {A.}}, \bibinfo {author} {\bibfnamefont {M.}~\bibnamefont {Nicodemi}}, \ and\
  \bibinfo {author} {\bibfnamefont {A.}~\bibnamefont {Coniglio}}} (\bibinfo
  {year} {2002}{\natexlab{b}}),\ \href {\doibase 10.1103/PhysRevE.66.061301}
  {\bibfield  {journal} {\bibinfo  {journal} {Phys. Rev. E}\ }\textbf {\bibinfo
  {volume} {66}},\ \bibinfo {pages} {061301}}\BibitemShut {NoStop}%
\bibitem [{\citenamefont {Fierro}\ \emph {et~al.}(2003)\citenamefont {Fierro},
  \citenamefont {Nicodemi},\ and\ \citenamefont {Coniglio}}]{Fierro:2003aa}%
  \BibitemOpen
  \bibfield  {author} {\bibinfo {author} {\bibnamefont {Fierro}, \bibfnamefont
  {A.}}, \bibinfo {author} {\bibfnamefont {M.}~\bibnamefont {Nicodemi}}, \ and\
  \bibinfo {author} {\bibfnamefont {A.}~\bibnamefont {Coniglio}}} (\bibinfo
  {year} {2003}),\ \href {\doibase 10.1088/0953-8984/15/11/331} {\bibfield
  {journal} {\bibinfo  {journal} {J. Phys. Cond. Mat.}\ }\textbf {\bibinfo
  {volume} {15}},\ \bibinfo {pages} {S1095}}\BibitemShut {NoStop}%
\bibitem [{\citenamefont {Finney}(1970)}]{Finney:1970aa}%
  \BibitemOpen
  \bibfield  {author} {\bibinfo {author} {\bibnamefont {Finney}, \bibfnamefont
  {J.~L.}}} (\bibinfo {year} {1970}),\ \href {\doibase 10.1098/rspa.1970.0189}
  {\bibfield  {journal} {\bibinfo  {journal} {Proc. Roy. Soc. London A}\
  }\textbf {\bibinfo {volume} {319}},\ \bibinfo {pages} {479}}\BibitemShut
  {NoStop}%
\bibitem [{\citenamefont {Francois}\ \emph {et~al.}(2013)\citenamefont
  {Francois}, \citenamefont {Saadatfar}, \citenamefont {Cruikshank},\ and\
  \citenamefont {Sheppard}}]{Francois:2013aa}%
  \BibitemOpen
  \bibfield  {author} {\bibinfo {author} {\bibnamefont {Francois},
  \bibfnamefont {N.}}, \bibinfo {author} {\bibfnamefont {M.}~\bibnamefont
  {Saadatfar}}, \bibinfo {author} {\bibfnamefont {R.}~\bibnamefont
  {Cruikshank}}, \ and\ \bibinfo {author} {\bibfnamefont {A.}~\bibnamefont
  {Sheppard}}} (\bibinfo {year} {2013}),\ \href {\doibase
  10.1103/PhysRevLett.111.148001} {\bibfield  {journal} {\bibinfo  {journal}
  {Phys. Rev. Lett.}\ }\textbf {\bibinfo {volume} {111}},\ \bibinfo {pages}
  {148001}}\BibitemShut {NoStop}%
\bibitem [{\citenamefont {Franz}\ and\ \citenamefont
  {Parisi}(2016)}]{Franz:2015ab}%
  \BibitemOpen
  \bibfield  {author} {\bibinfo {author} {\bibnamefont {Franz}, \bibfnamefont
  {S.}}, \ and\ \bibinfo {author} {\bibfnamefont {G.}~\bibnamefont {Parisi}}}
  (\bibinfo {year} {2016}),\ \href
  {http://stacks.iop.org/1751-8121/49/i=14/a=145001} {\bibfield  {journal}
  {\bibinfo  {journal} {J. Phys. A of Physics A: Mathematical and Theoretical}\
  }\textbf {\bibinfo {volume} {49}},\ \bibinfo {pages} {145001}}\BibitemShut
  {NoStop}%
\bibitem [{\citenamefont {Franz}\ \emph {et~al.}(2015)\citenamefont {Franz},
  \citenamefont {Parisi}, \citenamefont {Urbani},\ and\ \citenamefont
  {Zamponi}}]{Franz:2015aa}%
  \BibitemOpen
  \bibfield  {author} {\bibinfo {author} {\bibnamefont {Franz}, \bibfnamefont
  {S.}}, \bibinfo {author} {\bibfnamefont {G.}~\bibnamefont {Parisi}}, \bibinfo
  {author} {\bibfnamefont {P.}~\bibnamefont {Urbani}}, \ and\ \bibinfo {author}
  {\bibfnamefont {F.}~\bibnamefont {Zamponi}}} (\bibinfo {year} {2015}),\ \href
  {\doibase 10.1073/pnas.1511134112} {\bibfield  {journal} {\bibinfo  {journal}
  {Proc. Nat. Acad. Sci.}\ }\textbf {\bibinfo {volume} {112}},\ \bibinfo
  {pages} {14539}}\BibitemShut {NoStop}%
\bibitem [{\citenamefont {Frenkel}(2014)}]{Frenkel:2014aa}%
  \BibitemOpen
  \bibfield  {author} {\bibinfo {author} {\bibnamefont {Frenkel}, \bibfnamefont
  {D.}}} (\bibinfo {year} {2014}),\ \href {\doibase
  10.1080/00268976.2014.904051} {\bibfield  {journal} {\bibinfo  {journal}
  {Molecular Physics}\ }\textbf {\bibinfo {volume} {112}}~(\bibinfo {number}
  {17}),\ \bibinfo {pages} {2325}},\ \Eprint
  {http://arxiv.org/abs/http://dx.doi.org/10.1080/00268976.2014.904051}
  {http://dx.doi.org/10.1080/00268976.2014.904051} \BibitemShut {NoStop}%
\bibitem [{\citenamefont {Frenkel}\ \emph {et~al.}(2008)\citenamefont
  {Frenkel}, \citenamefont {Blumenfeld}, \citenamefont {Grof},\ and\
  \citenamefont {King}}]{Frenkel:2008aa}%
  \BibitemOpen
  \bibfield  {author} {\bibinfo {author} {\bibnamefont {Frenkel}, \bibfnamefont
  {G.}}, \bibinfo {author} {\bibfnamefont {R.}~\bibnamefont {Blumenfeld}},
  \bibinfo {author} {\bibfnamefont {Z.}~\bibnamefont {Grof}}, \ and\ \bibinfo
  {author} {\bibfnamefont {P.~R.}\ \bibnamefont {King}}} (\bibinfo {year}
  {2008}),\ \href {\doibase 10.1103/PhysRevE.77.041304} {\bibfield  {journal}
  {\bibinfo  {journal} {Phys. Rev. E}\ }\textbf {\bibinfo {volume} {77}},\
  10.1103/PhysRevE.77.041304}\BibitemShut {NoStop}%
\bibitem [{\citenamefont {Gago}\ \emph {et~al.}(2016)\citenamefont {Gago},
  \citenamefont {Maza},\ and\ \citenamefont {Pugnaloni}}]{Gago:2016aa}%
  \BibitemOpen
  \bibfield  {author} {\bibinfo {author} {\bibnamefont {Gago}, \bibfnamefont
  {P.~A.}}, \bibinfo {author} {\bibfnamefont {D.}~\bibnamefont {Maza}}, \ and\
  \bibinfo {author} {\bibfnamefont {L.~A.}\ \bibnamefont {Pugnaloni}}}
  (\bibinfo {year} {2016}),\ \href@noop {} {\bibfield  {journal} {\bibinfo
  {journal} {Papers in Physics}\ }\textbf {\bibinfo {volume} {8}},\ \bibinfo
  {pages} {080001}}\BibitemShut {NoStop}%
\bibitem [{\citenamefont {Gantapara}\ \emph {et~al.}(2013)\citenamefont
  {Gantapara}, \citenamefont {de~Graaf}, \citenamefont {van Roij},\ and\
  \citenamefont {Dijkstra}}]{Gantapara:2013aa}%
  \BibitemOpen
  \bibfield  {author} {\bibinfo {author} {\bibnamefont {Gantapara},
  \bibfnamefont {A.~P.}}, \bibinfo {author} {\bibfnamefont {J.}~\bibnamefont
  {de~Graaf}}, \bibinfo {author} {\bibfnamefont {R.}~\bibnamefont {van Roij}},
  \ and\ \bibinfo {author} {\bibfnamefont {M.}~\bibnamefont {Dijkstra}}}
  (\bibinfo {year} {2013}),\ \href {\doibase 10.1103/PhysRevLett.111.015501}
  {\bibfield  {journal} {\bibinfo  {journal} {Phys. Rev. Lett.}\ }\textbf
  {\bibinfo {volume} {111}},\ \bibinfo {pages} {015501}}\BibitemShut {NoStop}%
\bibitem [{\citenamefont {Gao}\ \emph {et~al.}(2006)\citenamefont {Gao},
  \citenamefont {Blawzdziewicz},\ and\ \citenamefont {O'Hern}}]{Gao:2006aa}%
  \BibitemOpen
  \bibfield  {author} {\bibinfo {author} {\bibnamefont {Gao}, \bibfnamefont
  {G.-J.}}, \bibinfo {author} {\bibfnamefont {J.}~\bibnamefont
  {Blawzdziewicz}}, \ and\ \bibinfo {author} {\bibfnamefont {C.~S.}\
  \bibnamefont {O'Hern}}} (\bibinfo {year} {2006}),\ \href {\doibase
  10.1103/PhysRevE.74.061304} {\bibfield  {journal} {\bibinfo  {journal} {Phys.
  Rev. E}\ }\textbf {\bibinfo {volume} {74}},\ \bibinfo {pages}
  {061304}}\BibitemShut {NoStop}%
\bibitem [{\citenamefont {Gao}\ \emph {et~al.}(2009)\citenamefont {Gao},
  \citenamefont {Blawzdziewicz}, \citenamefont {O'Hern},\ and\ \citenamefont
  {Shattuck}}]{Gao:2009aa}%
  \BibitemOpen
  \bibfield  {author} {\bibinfo {author} {\bibnamefont {Gao}, \bibfnamefont
  {G.-J.}}, \bibinfo {author} {\bibfnamefont {J.}~\bibnamefont
  {Blawzdziewicz}}, \bibinfo {author} {\bibfnamefont {C.~S.}\ \bibnamefont
  {O'Hern}}, \ and\ \bibinfo {author} {\bibfnamefont {M.}~\bibnamefont
  {Shattuck}}} (\bibinfo {year} {2009}),\ \href {\doibase
  10.1103/PhysRevE.80.061304} {\bibfield  {journal} {\bibinfo  {journal} {Phys.
  Rev. E}\ }\textbf {\bibinfo {volume} {80}},\ \bibinfo {pages}
  {061304}}\BibitemShut {NoStop}%
\bibitem [{\citenamefont {Gardner}(2001)}]{Gardner:2001aa}%
  \BibitemOpen
  \bibfield  {author} {\bibinfo {author} {\bibnamefont {Gardner}, \bibfnamefont
  {M.}}} (\bibinfo {year} {2001}),\ \href@noop {} {\emph {\bibinfo {title} {The
  Colossal Book of Mathematics: Classic Puzzles, Paradoxes, and Problems}}}\
  (\bibinfo  {publisher} {Norton})\BibitemShut {NoStop}%
\bibitem [{\citenamefont {Gendelman}\ \emph {et~al.}(2016)\citenamefont
  {Gendelman}, \citenamefont {Pollack}, \citenamefont {Procaccia},
  \citenamefont {Sengupta},\ and\ \citenamefont {Zylberg}}]{Gendelman:2016aa}%
  \BibitemOpen
  \bibfield  {author} {\bibinfo {author} {\bibnamefont {Gendelman},
  \bibfnamefont {O.}}, \bibinfo {author} {\bibfnamefont {Y.~G.}\ \bibnamefont
  {Pollack}}, \bibinfo {author} {\bibfnamefont {I.}~\bibnamefont {Procaccia}},
  \bibinfo {author} {\bibfnamefont {S.}~\bibnamefont {Sengupta}}, \ and\
  \bibinfo {author} {\bibfnamefont {J.}~\bibnamefont {Zylberg}}} (\bibinfo
  {year} {2016}),\ \href {\doibase 10.1103/PhysRevLett.116.078001} {\bibfield
  {journal} {\bibinfo  {journal} {Phys. Rev. Lett.}\ }\textbf {\bibinfo
  {volume} {116}},\ \bibinfo {pages} {078001}}\BibitemShut {NoStop}%
\bibitem [{\citenamefont {Glotzer}\ and\ \citenamefont
  {Solomon}(2007)}]{Glotzer:2007aa}%
  \BibitemOpen
  \bibfield  {author} {\bibinfo {author} {\bibnamefont {Glotzer}, \bibfnamefont
  {S.~C.}}, \ and\ \bibinfo {author} {\bibfnamefont {M.~J.}\ \bibnamefont
  {Solomon}}} (\bibinfo {year} {2007}),\ \href
  {http://dx.doi.org/10.1038/nmat1949} {\bibfield  {journal} {\bibinfo
  {journal} {Nature Mater.}\ }\textbf {\bibinfo {volume} {6}},\ \bibinfo
  {pages} {557}}\BibitemShut {NoStop}%
\bibitem [{\citenamefont {Goddard}(2004)}]{Goddard:2004aa}%
  \BibitemOpen
  \bibfield  {author} {\bibinfo {author} {\bibnamefont {Goddard}, \bibfnamefont
  {J.}}} (\bibinfo {year} {2004}),\ \href {\doibase
  http://dx.doi.org/10.1016/j.ijsolstr.2004.05.049} {\bibfield  {journal}
  {\bibinfo  {journal} {Int. J. Solids Struct.}\ }\textbf {\bibinfo {volume}
  {41}},\ \bibinfo {pages} {5851 }}\BibitemShut {NoStop}%
\bibitem [{\citenamefont {Goldbart}\ \emph {et~al.}(2005)\citenamefont
  {Goldbart}, \citenamefont {Goldenfeld},\ and\ \citenamefont
  {Sherrington}}]{Goldbart:2005aa}%
  \BibitemOpen
  \bibfield  {author} {\bibinfo {author} {\bibnamefont {Goldbart},
  \bibfnamefont {P.~M.}}, \bibinfo {author} {\bibfnamefont {N.}~\bibnamefont
  {Goldenfeld}}, \ and\ \bibinfo {author} {\bibfnamefont {D.}~\bibnamefont
  {Sherrington}}} (\bibinfo {year} {2005}),\ \href@noop {} {\emph {\bibinfo
  {title} {Stealing the gold: A celebration of the pioneering physics of Sam
  Edwards}}}\BibitemShut {NoStop}%
\bibitem [{\citenamefont {G\"otze}(2009)}]{Gotze:2009aa}%
  \BibitemOpen
  \bibfield  {author} {\bibinfo {author} {\bibnamefont {G\"otze}, \bibfnamefont
  {W.}}} (\bibinfo {year} {2009}),\ \href@noop {} {\emph {\bibinfo {title}
  {Complex Dynamics of Glass-Forming Liquids: A Mode-Coupling Theory}}}\
  (\bibinfo  {publisher} {Oxford University Press})\BibitemShut {NoStop}%
\bibitem [{\citenamefont {Gradenigo}\ \emph {et~al.}(2015)\citenamefont
  {Gradenigo}, \citenamefont {Ferrero}, \citenamefont {Bertin},\ and\
  \citenamefont {Barrat}}]{Gradenigo:2015aa}%
  \BibitemOpen
  \bibfield  {author} {\bibinfo {author} {\bibnamefont {Gradenigo},
  \bibfnamefont {G.}}, \bibinfo {author} {\bibfnamefont {E.~E.}\ \bibnamefont
  {Ferrero}}, \bibinfo {author} {\bibfnamefont {E.}~\bibnamefont {Bertin}}, \
  and\ \bibinfo {author} {\bibfnamefont {J.-L.}\ \bibnamefont {Barrat}}}
  (\bibinfo {year} {2015}),\ \href {\doibase 10.1103/PhysRevLett.115.140601}
  {\bibfield  {journal} {\bibinfo  {journal} {Phys. Rev. Lett.}\ }\textbf
  {\bibinfo {volume} {115}},\ \bibinfo {pages} {140601}}\BibitemShut {NoStop}%
\bibitem [{\citenamefont {Haji-Akbari}\ \emph {et~al.}(2009)\citenamefont
  {Haji-Akbari}, \citenamefont {Engel}, \citenamefont {Keys}, \citenamefont
  {Zheng}, \citenamefont {Petschek}, \citenamefont {Palffy-Muhoray},\ and\
  \citenamefont {Glotzer}}]{Haji-Akbari:2009aa}%
  \BibitemOpen
  \bibfield  {author} {\bibinfo {author} {\bibnamefont {Haji-Akbari},
  \bibfnamefont {A.}}, \bibinfo {author} {\bibfnamefont {M.}~\bibnamefont
  {Engel}}, \bibinfo {author} {\bibfnamefont {A.~S.}\ \bibnamefont {Keys}},
  \bibinfo {author} {\bibfnamefont {X.}~\bibnamefont {Zheng}}, \bibinfo
  {author} {\bibfnamefont {R.~G.}\ \bibnamefont {Petschek}}, \bibinfo {author}
  {\bibfnamefont {P.}~\bibnamefont {Palffy-Muhoray}}, \ and\ \bibinfo {author}
  {\bibfnamefont {S.~C.}\ \bibnamefont {Glotzer}}} (\bibinfo {year} {2009}),\
  \href@noop {} {\bibfield  {journal} {\bibinfo  {journal} {Nature}\ }\textbf
  {\bibinfo {volume} {462}},\ \bibinfo {pages} {773}}\BibitemShut {NoStop}%
\bibitem [{\citenamefont {Hales}(2005)}]{Hales:2005aa}%
  \BibitemOpen
  \bibfield  {author} {\bibinfo {author} {\bibnamefont {Hales}, \bibfnamefont
  {T.~C.}}} (\bibinfo {year} {2005}),\ \href@noop {} {\bibfield  {journal}
  {\bibinfo  {journal} {Ann. Math.}\ }\textbf {\bibinfo {volume} {162}},\
  \bibinfo {pages} {1065 }}\BibitemShut {NoStop}%
\bibitem [{\citenamefont {Hanifpour}\ \emph {et~al.}(2015)\citenamefont
  {Hanifpour}, \citenamefont {Francois}, \citenamefont {Robins}, \citenamefont
  {Kingston}, \citenamefont {Vaez~Allaei},\ and\ \citenamefont
  {Saadatfar}}]{Hanifpour:2015aa}%
  \BibitemOpen
  \bibfield  {author} {\bibinfo {author} {\bibnamefont {Hanifpour},
  \bibfnamefont {M.}}, \bibinfo {author} {\bibfnamefont {N.}~\bibnamefont
  {Francois}}, \bibinfo {author} {\bibfnamefont {V.}~\bibnamefont {Robins}},
  \bibinfo {author} {\bibfnamefont {A.}~\bibnamefont {Kingston}}, \bibinfo
  {author} {\bibfnamefont {S.~M.}\ \bibnamefont {Vaez~Allaei}}, \ and\ \bibinfo
  {author} {\bibfnamefont {M.}~\bibnamefont {Saadatfar}}} (\bibinfo {year}
  {2015}),\ \href {\doibase 10.1103/PhysRevE.91.062202} {\bibfield  {journal}
  {\bibinfo  {journal} {Phys. Rev. E}\ }\textbf {\bibinfo {volume} {91}},\
  \bibinfo {pages} {062202}}\BibitemShut {NoStop}%
\bibitem [{\citenamefont {Hanifpour}\ \emph {et~al.}(2014)\citenamefont
  {Hanifpour}, \citenamefont {Francois}, \citenamefont {Vaez~Allaei},
  \citenamefont {Senden},\ and\ \citenamefont {Saadatfar}}]{Hanifpour:2014aa}%
  \BibitemOpen
  \bibfield  {author} {\bibinfo {author} {\bibnamefont {Hanifpour},
  \bibfnamefont {M.}}, \bibinfo {author} {\bibfnamefont {N.}~\bibnamefont
  {Francois}}, \bibinfo {author} {\bibfnamefont {S.~M.}\ \bibnamefont
  {Vaez~Allaei}}, \bibinfo {author} {\bibfnamefont {T.}~\bibnamefont {Senden}},
  \ and\ \bibinfo {author} {\bibfnamefont {M.}~\bibnamefont {Saadatfar}}}
  (\bibinfo {year} {2014}),\ \href {\doibase 10.1103/PhysRevLett.113.148001}
  {\bibfield  {journal} {\bibinfo  {journal} {Phys. Rev. Lett.}\ }\textbf
  {\bibinfo {volume} {113}},\ \bibinfo {pages} {148001}}\BibitemShut {NoStop}%
\bibitem [{\citenamefont {Hansen-Goos}\ and\ \citenamefont
  {Mecke}(2009)}]{Hansen-Goos:2009aa}%
  \BibitemOpen
  \bibfield  {author} {\bibinfo {author} {\bibnamefont {Hansen-Goos},
  \bibfnamefont {H.}}, \ and\ \bibinfo {author} {\bibfnamefont
  {K.}~\bibnamefont {Mecke}}} (\bibinfo {year} {2009}),\ \href {\doibase
  10.1103/PhysRevLett.102.018302} {\bibfield  {journal} {\bibinfo  {journal}
  {Phys. Rev. Lett.}\ }\textbf {\bibinfo {volume} {102}},\ \bibinfo {pages}
  {018302}}\BibitemShut {NoStop}%
\bibitem [{\citenamefont {Hansen-Goos}\ and\ \citenamefont
  {Mecke}(2010)}]{Hansen-Goos:2010aa}%
  \BibitemOpen
  \bibfield  {author} {\bibinfo {author} {\bibnamefont {Hansen-Goos},
  \bibfnamefont {H.}}, \ and\ \bibinfo {author} {\bibfnamefont
  {K.}~\bibnamefont {Mecke}}} (\bibinfo {year} {2010}),\ \href
  {http://stacks.iop.org/0953-8984/22/i=36/a=364107} {\bibfield  {journal}
  {\bibinfo  {journal} {Journal of Physics: Condensed Matter}\ }\textbf
  {\bibinfo {volume} {22}}~(\bibinfo {number} {36}),\ \bibinfo {pages}
  {364107}}\BibitemShut {NoStop}%
\bibitem [{\citenamefont {Head}(2007)}]{Head:2007aa}%
  \BibitemOpen
  \bibfield  {author} {\bibinfo {author} {\bibnamefont {Head}, \bibfnamefont
  {D.~A.}}} (\bibinfo {year} {2007}),\ \href {\doibase
  10.1140/epje/e2007-00022-1} {\bibfield  {journal} {\bibinfo  {journal} {Eur.
  Phys. J. E}\ }\textbf {\bibinfo {volume} {22}}~(\bibinfo {number} {2}),\
  \bibinfo {pages} {151}}\BibitemShut {NoStop}%
\bibitem [{\citenamefont {van Hecke}(2010)}]{Hecke:2010aa}%
  \BibitemOpen
  \bibfield  {author} {\bibinfo {author} {\bibnamefont {van Hecke},
  \bibfnamefont {M.}}} (\bibinfo {year} {2010}),\ \href {\doibase
  10.1088/0953-8984/22/3/033101} {\bibfield  {journal} {\bibinfo  {journal} {J.
  Phys. Cond. Mat.}\ }\textbf {\bibinfo {volume} {22}},\ \bibinfo {pages}
  {033101}}\BibitemShut {NoStop}%
\bibitem [{\citenamefont {Henkes}\ and\ \citenamefont
  {Chakraborty}(2005)}]{Henkes:2005aa}%
  \BibitemOpen
  \bibfield  {author} {\bibinfo {author} {\bibnamefont {Henkes}, \bibfnamefont
  {S.}}, \ and\ \bibinfo {author} {\bibfnamefont {B.}~\bibnamefont
  {Chakraborty}}} (\bibinfo {year} {2005}),\ \href {\doibase
  10.1103/PhysRevLett.95.198002} {\bibfield  {journal} {\bibinfo  {journal}
  {Phys. Rev. Lett.}\ }\textbf {\bibinfo {volume} {95}},\ \bibinfo {pages}
  {198002}}\BibitemShut {NoStop}%
\bibitem [{\citenamefont {Henkes}\ and\ \citenamefont
  {Chakraborty}(2009)}]{Henkes:2009aa}%
  \BibitemOpen
  \bibfield  {author} {\bibinfo {author} {\bibnamefont {Henkes}, \bibfnamefont
  {S.}}, \ and\ \bibinfo {author} {\bibfnamefont {B.}~\bibnamefont
  {Chakraborty}}} (\bibinfo {year} {2009}),\ \href {\doibase
  10.1103/PhysRevE.79.061301} {\bibfield  {journal} {\bibinfo  {journal} {Phys.
  Rev. E}\ }\textbf {\bibinfo {volume} {79}},\ \bibinfo {pages}
  {061301}}\BibitemShut {NoStop}%
\bibitem [{\citenamefont {Henkes}\ \emph {et~al.}(2010)\citenamefont {Henkes},
  \citenamefont {van Hecke},\ and\ \citenamefont {van
  Saarloos}}]{Henkes:2010aa}%
  \BibitemOpen
  \bibfield  {author} {\bibinfo {author} {\bibnamefont {Henkes}, \bibfnamefont
  {S.}}, \bibinfo {author} {\bibfnamefont {M.}~\bibnamefont {van Hecke}}, \
  and\ \bibinfo {author} {\bibfnamefont {W.}~\bibnamefont {van Saarloos}}}
  (\bibinfo {year} {2010}),\ \href {\doibase 10.1209/0295-5075/90/14003}
  {\bibfield  {journal} {\bibinfo  {journal} {Europhys. Lett.}\ }\textbf
  {\bibinfo {volume} {90}},\ \bibinfo {pages} {14003}}\BibitemShut {NoStop}%
\bibitem [{\citenamefont {Henkes}\ \emph {et~al.}(2007)\citenamefont {Henkes},
  \citenamefont {O'Hern},\ and\ \citenamefont {Chakraborty}}]{Henkes:2007aa}%
  \BibitemOpen
  \bibfield  {author} {\bibinfo {author} {\bibnamefont {Henkes}, \bibfnamefont
  {S.}}, \bibinfo {author} {\bibfnamefont {C.~S.}\ \bibnamefont {O'Hern}}, \
  and\ \bibinfo {author} {\bibfnamefont {B.}~\bibnamefont {Chakraborty}}}
  (\bibinfo {year} {2007}),\ \href {\doibase 10.1103/PhysRevLett.99.038002}
  {\bibfield  {journal} {\bibinfo  {journal} {Phys. Rev. Lett.}\ }\textbf
  {\bibinfo {volume} {99}},\ \bibinfo {pages} {038002}}\BibitemShut {NoStop}%
\bibitem [{\citenamefont {Hermes}\ and\ \citenamefont
  {Dijkstra}(2010)}]{Hermes:2010aa}%
  \BibitemOpen
  \bibfield  {author} {\bibinfo {author} {\bibnamefont {Hermes}, \bibfnamefont
  {M.}}, \ and\ \bibinfo {author} {\bibfnamefont {M.}~\bibnamefont {Dijkstra}}}
  (\bibinfo {year} {2010}),\ \href@noop {} {\bibfield  {journal} {\bibinfo
  {journal} {EPL}\ }\textbf {\bibinfo {volume} {89}},\ \bibinfo {pages}
  {38005}}\BibitemShut {NoStop}%
\bibitem [{\citenamefont {Herrmann}(1993)}]{Herrmann:1993ab}%
  \BibitemOpen
  \bibfield  {author} {\bibinfo {author} {\bibnamefont {Herrmann},
  \bibfnamefont {H.~J.}}} (\bibinfo {year} {1993}),\ \href@noop {} {\bibfield
  {journal} {\bibinfo  {journal} {Physica A}\ }\textbf {\bibinfo {volume}
  {191}},\ \bibinfo {pages} {263}}\BibitemShut {NoStop}%
\bibitem [{\citenamefont {Herrmann}\ \emph {et~al.}(1990)\citenamefont
  {Herrmann}, \citenamefont {Mantica},\ and\ \citenamefont
  {Bessis}}]{Herrmann:1990aa}%
  \BibitemOpen
  \bibfield  {author} {\bibinfo {author} {\bibnamefont {Herrmann},
  \bibfnamefont {H.~J.}}, \bibinfo {author} {\bibfnamefont {G.}~\bibnamefont
  {Mantica}}, \ and\ \bibinfo {author} {\bibfnamefont {D.}~\bibnamefont
  {Bessis}}} (\bibinfo {year} {1990}),\ \href {\doibase
  10.1103/PhysRevLett.65.3223} {\bibfield  {journal} {\bibinfo  {journal}
  {Phys. Rev. Lett.}\ }\textbf {\bibinfo {volume} {65}},\ \bibinfo {pages}
  {3223}}\BibitemShut {NoStop}%
\bibitem [{\citenamefont {Hiwatari}\ \emph {et~al.}(1984)\citenamefont
  {Hiwatari}, \citenamefont {Saito},\ and\ \citenamefont
  {Ueda}}]{Hiwatari:1984aa}%
  \BibitemOpen
  \bibfield  {author} {\bibinfo {author} {\bibnamefont {Hiwatari},
  \bibfnamefont {Y.}}, \bibinfo {author} {\bibfnamefont {T.}~\bibnamefont
  {Saito}}, \ and\ \bibinfo {author} {\bibfnamefont {A.}~\bibnamefont {Ueda}}}
  (\bibinfo {year} {1984}),\ \href@noop {} {\bibfield  {journal} {\bibinfo
  {journal} {J. Chem. Phys.}\ }\textbf {\bibinfo {volume} {81}},\ \bibinfo
  {pages} {6044}}\BibitemShut {NoStop}%
\bibitem [{\citenamefont {Hopkins}\ \emph {et~al.}(2013)\citenamefont
  {Hopkins}, \citenamefont {Stillinger},\ and\ \citenamefont
  {Torquato}}]{Hopkins:2013aa}%
  \BibitemOpen
  \bibfield  {author} {\bibinfo {author} {\bibnamefont {Hopkins}, \bibfnamefont
  {A.~B.}}, \bibinfo {author} {\bibfnamefont {F.~H.}\ \bibnamefont
  {Stillinger}}, \ and\ \bibinfo {author} {\bibfnamefont {S.}~\bibnamefont
  {Torquato}}} (\bibinfo {year} {2013}),\ \href {\doibase
  10.1103/PhysRevE.88.022205} {\bibfield  {journal} {\bibinfo  {journal} {Phys.
  Rev. E}\ }\textbf {\bibinfo {volume} {88}},\ \bibinfo {pages}
  {022205}}\BibitemShut {NoStop}%
\bibitem [{\citenamefont {Hsu}\ \emph {et~al.}(2009)\citenamefont {Hsu},
  \citenamefont {Johnson}, \citenamefont {Ingale}, \citenamefont {Valenza},
  \citenamefont {Gland},\ and\ \citenamefont {Makse}}]{Hsu:2009aa}%
  \BibitemOpen
  \bibfield  {author} {\bibinfo {author} {\bibnamefont {Hsu}, \bibfnamefont
  {C.-J.}}, \bibinfo {author} {\bibfnamefont {D.~L.}\ \bibnamefont {Johnson}},
  \bibinfo {author} {\bibfnamefont {R.~A.}\ \bibnamefont {Ingale}}, \bibinfo
  {author} {\bibfnamefont {J.~J.}\ \bibnamefont {Valenza}}, \bibinfo {author}
  {\bibfnamefont {N.}~\bibnamefont {Gland}}, \ and\ \bibinfo {author}
  {\bibfnamefont {H.~A.}\ \bibnamefont {Makse}}} (\bibinfo {year} {2009}),\
  \href {\doibase 10.1103/PhysRevLett.102.058001} {\bibfield  {journal}
  {\bibinfo  {journal} {Phys. Rev. Lett.}\ }\textbf {\bibinfo {volume} {102}},\
  \bibinfo {pages} {058001}}\BibitemShut {NoStop}%
\bibitem [{\citenamefont {Hu}\ \emph {et~al.}(2014{\natexlab{a}})\citenamefont
  {Hu}, \citenamefont {Johnson}, \citenamefont {Valenza}, \citenamefont
  {Santibanez},\ and\ \citenamefont {Makse}}]{Hu:2014ab}%
  \BibitemOpen
  \bibfield  {author} {\bibinfo {author} {\bibnamefont {Hu}, \bibfnamefont
  {Y.}}, \bibinfo {author} {\bibfnamefont {D.~L.}\ \bibnamefont {Johnson}},
  \bibinfo {author} {\bibfnamefont {J.~J.}\ \bibnamefont {Valenza}}, \bibinfo
  {author} {\bibfnamefont {F.}~\bibnamefont {Santibanez}}, \ and\ \bibinfo
  {author} {\bibfnamefont {H.~A.}\ \bibnamefont {Makse}}} (\bibinfo {year}
  {2014}{\natexlab{a}}),\ \href@noop {} {\bibfield  {journal} {\bibinfo
  {journal} {Phys. Rev. E}\ }\textbf {\bibinfo {volume} {89}}}\BibitemShut
  {NoStop}%
\bibitem [{\citenamefont {Hu}\ \emph {et~al.}(2014{\natexlab{b}})\citenamefont
  {Hu}, \citenamefont {Makse}, \citenamefont {Valenza},\ and\ \citenamefont
  {Johnson}}]{Hu:2014aa}%
  \BibitemOpen
  \bibfield  {author} {\bibinfo {author} {\bibnamefont {Hu}, \bibfnamefont
  {Y.}}, \bibinfo {author} {\bibfnamefont {H.~A.}\ \bibnamefont {Makse}},
  \bibinfo {author} {\bibfnamefont {J.~J.}\ \bibnamefont {Valenza}}, \ and\
  \bibinfo {author} {\bibfnamefont {D.~L.}\ \bibnamefont {Johnson}}} (\bibinfo
  {year} {2014}{\natexlab{b}}),\ \href {\doibase 10.1190/geo2013-0459.1}
  {\bibfield  {journal} {\bibinfo  {journal} {Geophysics}\ }\textbf {\bibinfo
  {volume} {79}},\ \bibinfo {pages} {L41}}\BibitemShut {NoStop}%
\bibitem [{\citenamefont {Huang}(1987)}]{Huang:1987aa}%
  \BibitemOpen
  \bibfield  {author} {\bibinfo {author} {\bibnamefont {Huang}, \bibfnamefont
  {K.}}} (\bibinfo {year} {1987}),\ \href@noop {} {\emph {\bibinfo {title}
  {Statistical Mechanics}}}\ (\bibinfo  {publisher} {Wiley})\BibitemShut
  {NoStop}%
\bibitem [{\citenamefont {Ikeda}\ and\ \citenamefont
  {Berthier}(2015)}]{Ikeda:2015aa}%
  \BibitemOpen
  \bibfield  {author} {\bibinfo {author} {\bibnamefont {Ikeda}, \bibfnamefont
  {A.}}, \ and\ \bibinfo {author} {\bibfnamefont {L.}~\bibnamefont {Berthier}}}
  (\bibinfo {year} {2015}),\ \href {\doibase 10.1103/PhysRevE.92.012309}
  {\bibfield  {journal} {\bibinfo  {journal} {Phys. Rev. E}\ }\textbf {\bibinfo
  {volume} {92}},\ \bibinfo {pages} {012309}}\BibitemShut {NoStop}%
\bibitem [{\citenamefont {Ikeda}\ \emph {et~al.}(2017)\citenamefont {Ikeda},
  \citenamefont {Berthier},\ and\ \citenamefont {Parisi}}]{Ikeda:2017aa}%
  \BibitemOpen
  \bibfield  {author} {\bibinfo {author} {\bibnamefont {Ikeda}, \bibfnamefont
  {A.}}, \bibinfo {author} {\bibfnamefont {L.}~\bibnamefont {Berthier}}, \ and\
  \bibinfo {author} {\bibfnamefont {G.}~\bibnamefont {Parisi}}} (\bibinfo
  {year} {2017}),\ \href {\doibase 10.1103/PhysRevE.95.052125} {\bibfield
  {journal} {\bibinfo  {journal} {Phys. Rev. E}\ }\textbf {\bibinfo {volume}
  {95}},\ \bibinfo {pages} {052125}}\BibitemShut {NoStop}%
\bibitem [{\citenamefont {Ikeda}\ \emph {et~al.}(2012)\citenamefont {Ikeda},
  \citenamefont {Berthier},\ and\ \citenamefont {Sollich}}]{Ikeda:2012aa}%
  \BibitemOpen
  \bibfield  {author} {\bibinfo {author} {\bibnamefont {Ikeda}, \bibfnamefont
  {A.}}, \bibinfo {author} {\bibfnamefont {L.}~\bibnamefont {Berthier}}, \ and\
  \bibinfo {author} {\bibfnamefont {P.}~\bibnamefont {Sollich}}} (\bibinfo
  {year} {2012}),\ \href {\doibase 10.1103/PhysRevLett.109.018301} {\bibfield
  {journal} {\bibinfo  {journal} {Phys. Rev. Lett.}\ }\textbf {\bibinfo
  {volume} {109}},\ \bibinfo {pages} {018301}}\BibitemShut {NoStop}%
\bibitem [{\citenamefont {Ikeda}\ and\ \citenamefont
  {Miyazaki}(2010)}]{Ikeda:2010aa}%
  \BibitemOpen
  \bibfield  {author} {\bibinfo {author} {\bibnamefont {Ikeda}, \bibfnamefont
  {A.}}, \ and\ \bibinfo {author} {\bibfnamefont {K.}~\bibnamefont {Miyazaki}}}
  (\bibinfo {year} {2010}),\ \href {\doibase 10.1103/PhysRevLett.104.255704}
  {\bibfield  {journal} {\bibinfo  {journal} {Phys. Rev. Lett.}\ }\textbf
  {\bibinfo {volume} {104}},\ \bibinfo {pages} {255704}}\BibitemShut {NoStop}%
\bibitem [{\citenamefont {Irastorza}\ \emph {et~al.}(2013)\citenamefont
  {Irastorza}, \citenamefont {Carlevaro},\ and\ \citenamefont
  {Pugnaloni}}]{Irastorza:2013aa}%
  \BibitemOpen
  \bibfield  {author} {\bibinfo {author} {\bibnamefont {Irastorza},
  \bibfnamefont {R.~M.}}, \bibinfo {author} {\bibfnamefont {C.~M.}\
  \bibnamefont {Carlevaro}}, \ and\ \bibinfo {author} {\bibfnamefont {L.~A.}\
  \bibnamefont {Pugnaloni}}} (\bibinfo {year} {2013}),\ \href
  {http://stacks.iop.org/1742-5468/2013/i=12/a=P12012} {\bibfield  {journal}
  {\bibinfo  {journal} {Journal of Statistical Mechanics: Theory and
  Experiment}\ }\textbf {\bibinfo {volume} {2013}}~(\bibinfo {number} {12}),\
  \bibinfo {pages} {P12012}}\BibitemShut {NoStop}%
\bibitem [{\citenamefont {Jaeger}(2015)}]{Jaeger:2015aa}%
  \BibitemOpen
  \bibfield  {author} {\bibinfo {author} {\bibnamefont {Jaeger}, \bibfnamefont
  {H.~M.}}} (\bibinfo {year} {2015}),\ \href {\doibase 10.1039/C4SM01923G}
  {\bibfield  {journal} {\bibinfo  {journal} {Soft Matter}\ }\textbf {\bibinfo
  {volume} {11}},\ \bibinfo {pages} {12}}\BibitemShut {NoStop}%
\bibitem [{\citenamefont {Jaeger}\ \emph {et~al.}(1996)\citenamefont {Jaeger},
  \citenamefont {Nagel},\ and\ \citenamefont {Behringer}}]{Jaeger:1996aa}%
  \BibitemOpen
  \bibfield  {author} {\bibinfo {author} {\bibnamefont {Jaeger}, \bibfnamefont
  {H.~M.}}, \bibinfo {author} {\bibfnamefont {S.~R.}\ \bibnamefont {Nagel}}, \
  and\ \bibinfo {author} {\bibfnamefont {R.~P.}\ \bibnamefont {Behringer}}}
  (\bibinfo {year} {1996}),\ \href {\doibase 10.1103/RevModPhys.68.1259}
  {\bibfield  {journal} {\bibinfo  {journal} {Rev. Mod. Phys.}\ }\textbf
  {\bibinfo {volume} {68}},\ \bibinfo {pages} {1259}}\BibitemShut {NoStop}%
\bibitem [{\citenamefont {Jaoshvili}\ \emph {et~al.}(2010)\citenamefont
  {Jaoshvili}, \citenamefont {Esakia}, \citenamefont {Porrati},\ and\
  \citenamefont {Chaikin}}]{Jaoshvili:2010aa}%
  \BibitemOpen
  \bibfield  {author} {\bibinfo {author} {\bibnamefont {Jaoshvili},
  \bibfnamefont {A.}}, \bibinfo {author} {\bibfnamefont {A.}~\bibnamefont
  {Esakia}}, \bibinfo {author} {\bibfnamefont {M.}~\bibnamefont {Porrati}}, \
  and\ \bibinfo {author} {\bibfnamefont {P.~M.}\ \bibnamefont {Chaikin}}}
  (\bibinfo {year} {2010}),\ \href {\doibase 10.1103/PhysRevLett.104.185501}
  {\bibfield  {journal} {\bibinfo  {journal} {Phys. Rev. Lett.}\ }\textbf
  {\bibinfo {volume} {104}},\ \bibinfo {pages} {185501}}\BibitemShut {NoStop}%
\bibitem [{\citenamefont {Jaynes}(1957{\natexlab{a}})}]{Jaynes:1957aa}%
  \BibitemOpen
  \bibfield  {author} {\bibinfo {author} {\bibnamefont {Jaynes}, \bibfnamefont
  {E.~T.}}} (\bibinfo {year} {1957}{\natexlab{a}}),\ \href {\doibase
  10.1103/PhysRev.106.620} {\bibfield  {journal} {\bibinfo  {journal} {Phys.
  Rev.}\ }\textbf {\bibinfo {volume} {106}},\ \bibinfo {pages}
  {620}}\BibitemShut {NoStop}%
\bibitem [{\citenamefont {Jaynes}(1957{\natexlab{b}})}]{Jaynes:1957ab}%
  \BibitemOpen
  \bibfield  {author} {\bibinfo {author} {\bibnamefont {Jaynes}, \bibfnamefont
  {E.~T.}}} (\bibinfo {year} {1957}{\natexlab{b}}),\ \href {\doibase
  10.1103/PhysRev.108.171} {\bibfield  {journal} {\bibinfo  {journal} {Phys.
  Rev.}\ }\textbf {\bibinfo {volume} {108}},\ \bibinfo {pages}
  {171}}\BibitemShut {NoStop}%
\bibitem [{\citenamefont {Jenkins}\ \emph {et~al.}(2005)\citenamefont
  {Jenkins}, \citenamefont {Johnson}, \citenamefont {Ragione},\ and\
  \citenamefont {Makse}}]{Jenkins:2005aa}%
  \BibitemOpen
  \bibfield  {author} {\bibinfo {author} {\bibnamefont {Jenkins}, \bibfnamefont
  {J.}}, \bibinfo {author} {\bibfnamefont {D.}~\bibnamefont {Johnson}},
  \bibinfo {author} {\bibfnamefont {L.~L.}\ \bibnamefont {Ragione}}, \ and\
  \bibinfo {author} {\bibfnamefont {H.}~\bibnamefont {Makse}}} (\bibinfo {year}
  {2005}),\ \href {\doibase http://dx.doi.org/10.1016/j.jmps.2004.06.002}
  {\bibfield  {journal} {\bibinfo  {journal} {J. Mech. Phys. Solids}\ }\textbf
  {\bibinfo {volume} {53}},\ \bibinfo {pages} {197 }}\BibitemShut {NoStop}%
\bibitem [{\citenamefont {Jia}\ \emph {et~al.}(2007)\citenamefont {Jia},
  \citenamefont {M.}, \citenamefont {Williams},\ and\ \citenamefont
  {Rhodes}}]{Jia:2007aa}%
  \BibitemOpen
  \bibfield  {author} {\bibinfo {author} {\bibnamefont {Jia}, \bibfnamefont
  {X.}}, \bibinfo {author} {\bibfnamefont {G.}~\bibnamefont {M.}}, \bibinfo
  {author} {\bibfnamefont {R.~A.}\ \bibnamefont {Williams}}, \ and\ \bibinfo
  {author} {\bibfnamefont {D.}~\bibnamefont {Rhodes}}} (\bibinfo {year}
  {2007}),\ \href@noop {} {\bibfield  {journal} {\bibinfo  {journal} {Powder
  Technol.}\ }\textbf {\bibinfo {volume} {174}},\ \bibinfo {pages}
  {10}}\BibitemShut {NoStop}%
\bibitem [{\citenamefont {Jiao}\ \emph {et~al.}(2010)\citenamefont {Jiao},
  \citenamefont {Stillinger},\ and\ \citenamefont {Torquato}}]{Jiao:2010aa}%
  \BibitemOpen
  \bibfield  {author} {\bibinfo {author} {\bibnamefont {Jiao}, \bibfnamefont
  {Y.}}, \bibinfo {author} {\bibfnamefont {F.~H.}\ \bibnamefont {Stillinger}},
  \ and\ \bibinfo {author} {\bibfnamefont {S.}~\bibnamefont {Torquato}}}
  (\bibinfo {year} {2010}),\ \href {\doibase 10.1103/PhysRevE.81.041304}
  {\bibfield  {journal} {\bibinfo  {journal} {Phys. Rev. E}\ }\textbf {\bibinfo
  {volume} {81}},\ \bibinfo {pages} {041304}}\BibitemShut {NoStop}%
\bibitem [{\citenamefont {Jiao}\ and\ \citenamefont
  {Torquato}(2011)}]{Jiao:2011aa}%
  \BibitemOpen
  \bibfield  {author} {\bibinfo {author} {\bibnamefont {Jiao}, \bibfnamefont
  {Y.}}, \ and\ \bibinfo {author} {\bibfnamefont {S.}~\bibnamefont {Torquato}}}
  (\bibinfo {year} {2011}),\ \href {\doibase 10.1103/PhysRevE.84.041309}
  {\bibfield  {journal} {\bibinfo  {journal} {Phys. Rev. E}\ }\textbf {\bibinfo
  {volume} {84}},\ \bibinfo {pages} {041309}}\BibitemShut {NoStop}%
\bibitem [{\citenamefont {Jin}\ \emph {et~al.}(2010)\citenamefont {Jin},
  \citenamefont {Charbonneau}, \citenamefont {Meyer}, \citenamefont {Song},\
  and\ \citenamefont {Zamponi}}]{Jin:2010ab}%
  \BibitemOpen
  \bibfield  {author} {\bibinfo {author} {\bibnamefont {Jin}, \bibfnamefont
  {Y.}}, \bibinfo {author} {\bibfnamefont {P.}~\bibnamefont {Charbonneau}},
  \bibinfo {author} {\bibfnamefont {S.}~\bibnamefont {Meyer}}, \bibinfo
  {author} {\bibfnamefont {C.}~\bibnamefont {Song}}, \ and\ \bibinfo {author}
  {\bibfnamefont {F.}~\bibnamefont {Zamponi}}} (\bibinfo {year} {2010}),\
  \href@noop {} {\bibfield  {journal} {\bibinfo  {journal} {Phys. Rev. E}\
  }\textbf {\bibinfo {volume} {82}}}\BibitemShut {NoStop}%
\bibitem [{\citenamefont {Jin}\ and\ \citenamefont {Makse}(2010)}]{Jin:2010aa}%
  \BibitemOpen
  \bibfield  {author} {\bibinfo {author} {\bibnamefont {Jin}, \bibfnamefont
  {Y.}}, \ and\ \bibinfo {author} {\bibfnamefont {H.~A.}\ \bibnamefont
  {Makse}}} (\bibinfo {year} {2010}),\ \href {\doibase
  10.1016/j.physa.2010.08.010} {\bibfield  {journal} {\bibinfo  {journal}
  {Physica A}\ }\textbf {\bibinfo {volume} {389}},\ \bibinfo {pages}
  {5362}}\BibitemShut {NoStop}%
\bibitem [{\citenamefont {Jin}\ \emph {et~al.}(2014)\citenamefont {Jin},
  \citenamefont {Puckett},\ and\ \citenamefont {Makse}}]{Jin:2014aa}%
  \BibitemOpen
  \bibfield  {author} {\bibinfo {author} {\bibnamefont {Jin}, \bibfnamefont
  {Y.}}, \bibinfo {author} {\bibfnamefont {J.~G.}\ \bibnamefont {Puckett}}, \
  and\ \bibinfo {author} {\bibfnamefont {H.~A.}\ \bibnamefont {Makse}}}
  (\bibinfo {year} {2014}),\ \href {\doibase 10.1103/PhysRevE.89.052207}
  {\bibfield  {journal} {\bibinfo  {journal} {Phys. Rev. E}\ }\textbf {\bibinfo
  {volume} {89}},\ \bibinfo {pages} {052207}}\BibitemShut {NoStop}%
\bibitem [{\citenamefont {Johnson}\ \emph {et~al.}(2015)\citenamefont
  {Johnson}, \citenamefont {Hu},\ and\ \citenamefont {Makse}}]{Johnson:2015aa}%
  \BibitemOpen
  \bibfield  {author} {\bibinfo {author} {\bibnamefont {Johnson}, \bibfnamefont
  {D.~L.}}, \bibinfo {author} {\bibfnamefont {Y.}~\bibnamefont {Hu}}, \ and\
  \bibinfo {author} {\bibfnamefont {H.}~\bibnamefont {Makse}}} (\bibinfo {year}
  {2015}),\ \href {\doibase 10.1103/PhysRevE.91.062208} {\bibfield  {journal}
  {\bibinfo  {journal} {Phys. Rev. E}\ }\textbf {\bibinfo {volume} {91}},\
  \bibinfo {pages} {062208}}\BibitemShut {NoStop}%
\bibitem [{\citenamefont {Johnson}(1985)}]{Johnson:1985aa}%
  \BibitemOpen
  \bibfield  {author} {\bibinfo {author} {\bibnamefont {Johnson}, \bibfnamefont
  {K.~L.}}} (\bibinfo {year} {1985}),\ \href@noop {} {\emph {\bibinfo {title}
  {Contact Mechanics}}}\ (\bibinfo  {publisher} {Cambrdige University
  Press})\BibitemShut {NoStop}%
\bibitem [{\citenamefont {Jorjadze}\ \emph {et~al.}(2011)\citenamefont
  {Jorjadze}, \citenamefont {Pontani}, \citenamefont {Newhall},\ and\
  \citenamefont {Bruji\'c}}]{Jorjadze:2011aa}%
  \BibitemOpen
  \bibfield  {author} {\bibinfo {author} {\bibnamefont {Jorjadze},
  \bibfnamefont {I.}}, \bibinfo {author} {\bibfnamefont {L.-L.}\ \bibnamefont
  {Pontani}}, \bibinfo {author} {\bibfnamefont {K.~A.}\ \bibnamefont
  {Newhall}}, \ and\ \bibinfo {author} {\bibfnamefont {J.}~\bibnamefont
  {Bruji\'c}}} (\bibinfo {year} {2011}),\ \href {\doibase
  10.1073/pnas.1017716108} {\bibfield  {journal} {\bibinfo  {journal} {Proc.
  Nat. Acad. Sci.}\ }\textbf {\bibinfo {volume} {108}},\ \bibinfo {pages}
  {4286}}\BibitemShut {NoStop}%
\bibitem [{\citenamefont {Kadanoff}(1999)}]{Kadanoff:1999aa}%
  \BibitemOpen
  \bibfield  {author} {\bibinfo {author} {\bibnamefont {Kadanoff},
  \bibfnamefont {L.~P.}}} (\bibinfo {year} {1999}),\ \href {\doibase
  10.1103/RevModPhys.71.435} {\bibfield  {journal} {\bibinfo  {journal} {Rev.
  Mod. Phys.}\ }\textbf {\bibinfo {volume} {71}},\ \bibinfo {pages}
  {435}}\BibitemShut {NoStop}%
\bibitem [{\citenamefont {Kadau}\ and\ \citenamefont
  {Herrmann}(2011)}]{Kadau:2011aa}%
  \BibitemOpen
  \bibfield  {author} {\bibinfo {author} {\bibnamefont {Kadau}, \bibfnamefont
  {D.}}, \ and\ \bibinfo {author} {\bibfnamefont {H.~J.}\ \bibnamefont
  {Herrmann}}} (\bibinfo {year} {2011}),\ \href {\doibase
  10.1103/PhysRevE.83.031301} {\bibfield  {journal} {\bibinfo  {journal} {Phys.
  Rev. E}\ }\textbf {\bibinfo {volume} {83}},\ \bibinfo {pages}
  {031301}}\BibitemShut {NoStop}%
\bibitem [{\citenamefont {Kallus}(2016)}]{Kallus:2016aa}%
  \BibitemOpen
  \bibfield  {author} {\bibinfo {author} {\bibnamefont {Kallus}, \bibfnamefont
  {Y.}}} (\bibinfo {year} {2016}),\ \href {\doibase 10.1039/C6SM00213G}
  {\bibfield  {journal} {\bibinfo  {journal} {Soft Matter}\ }\textbf {\bibinfo
  {volume} {12}},\ \bibinfo {pages} {4123}}\BibitemShut {NoStop}%
\bibitem [{\citenamefont {Kallus}\ and\ \citenamefont
  {Elser}(2011)}]{Kallus:2011aa}%
  \BibitemOpen
  \bibfield  {author} {\bibinfo {author} {\bibnamefont {Kallus}, \bibfnamefont
  {Y.}}, \ and\ \bibinfo {author} {\bibfnamefont {V.}~\bibnamefont {Elser}}}
  (\bibinfo {year} {2011}),\ \href {\doibase 10.1103/PhysRevE.83.036703}
  {\bibfield  {journal} {\bibinfo  {journal} {Phys. Rev. E}\ }\textbf {\bibinfo
  {volume} {83}},\ \bibinfo {pages} {036703}}\BibitemShut {NoStop}%
\bibitem [{\citenamefont {Kamien}\ and\ \citenamefont
  {Liu}(2007)}]{Kamien:2007aa}%
  \BibitemOpen
  \bibfield  {author} {\bibinfo {author} {\bibnamefont {Kamien}, \bibfnamefont
  {R.~D.}}, \ and\ \bibinfo {author} {\bibfnamefont {A.~J.}\ \bibnamefont
  {Liu}}} (\bibinfo {year} {2007}),\ \href {\doibase
  10.1103/PhysRevLett.99.155501} {\bibfield  {journal} {\bibinfo  {journal}
  {Phys. Rev. Lett.}\ }\textbf {\bibinfo {volume} {99}},\ \bibinfo {pages}
  {155501}}\BibitemShut {NoStop}%
\bibitem [{\citenamefont {Kapfer}\ \emph {et~al.}(2012)\citenamefont {Kapfer},
  \citenamefont {Mickel}, \citenamefont {Mecke},\ and\ \citenamefont
  {Schr\"oder-Turk}}]{Kapfer:2012aa}%
  \BibitemOpen
  \bibfield  {author} {\bibinfo {author} {\bibnamefont {Kapfer}, \bibfnamefont
  {S.~C.}}, \bibinfo {author} {\bibfnamefont {W.}~\bibnamefont {Mickel}},
  \bibinfo {author} {\bibfnamefont {K.}~\bibnamefont {Mecke}}, \ and\ \bibinfo
  {author} {\bibfnamefont {G.~E.}\ \bibnamefont {Schr\"oder-Turk}}} (\bibinfo
  {year} {2012}),\ \href {\doibase 10.1103/PhysRevE.85.030301} {\bibfield
  {journal} {\bibinfo  {journal} {Phys. Rev. E}\ }\textbf {\bibinfo {volume}
  {85}},\ \bibinfo {pages} {030301}}\BibitemShut {NoStop}%
\bibitem [{\citenamefont {Kasahara}\ and\ \citenamefont
  {Nakanishi}(2004)}]{Kasahara:2004aa}%
  \BibitemOpen
  \bibfield  {author} {\bibinfo {author} {\bibnamefont {Kasahara},
  \bibfnamefont {A.}}, \ and\ \bibinfo {author} {\bibfnamefont
  {H.}~\bibnamefont {Nakanishi}}} (\bibinfo {year} {2004}),\ \href {\doibase
  10.1103/PhysRevE.70.051309} {\bibfield  {journal} {\bibinfo  {journal} {Phys.
  Rev. E}\ }\textbf {\bibinfo {volume} {70}},\ \bibinfo {pages}
  {051309}}\BibitemShut {NoStop}%
\bibitem [{\citenamefont {Kepler}(1611)}]{Kepler:1611aa}%
  \BibitemOpen
  \bibfield  {author} {\bibinfo {author} {\bibnamefont {Kepler}, \bibfnamefont
  {J.}}} (\bibinfo {year} {1611}),\ \href@noop {} {\emph {\bibinfo {title}
  {Strena seu de nive sexangula (The six-cornered snowflake)}}}\ (\bibinfo
  {publisher} {http://www.thelatinlibrary.com/kepler/strena.html})\BibitemShut
  {NoStop}%
\bibitem [{\citenamefont {Kirkpatrick}\ and\ \citenamefont
  {Selman}(1994)}]{Kirkpatrick:1994aa}%
  \BibitemOpen
  \bibfield  {author} {\bibinfo {author} {\bibnamefont {Kirkpatrick},
  \bibfnamefont {S.}}, \ and\ \bibinfo {author} {\bibfnamefont
  {B.}~\bibnamefont {Selman}}} (\bibinfo {year} {1994}),\ \href@noop {}
  {\bibfield  {journal} {\bibinfo  {journal} {Science}\ }\textbf {\bibinfo
  {volume} {264}},\ \bibinfo {pages} {1297}}\BibitemShut {NoStop}%
\bibitem [{\citenamefont {Kirkpatrick}\ and\ \citenamefont
  {Wolynes}(1987)}]{Kirkpatrick:1987aa}%
  \BibitemOpen
  \bibfield  {author} {\bibinfo {author} {\bibnamefont {Kirkpatrick},
  \bibfnamefont {T.~R.}}, \ and\ \bibinfo {author} {\bibfnamefont {P.~G.}\
  \bibnamefont {Wolynes}}} (\bibinfo {year} {1987}),\ \href {\doibase
  10.1103/PhysRevA.35.3072} {\bibfield  {journal} {\bibinfo  {journal} {Phys.
  Rev. A}\ }\textbf {\bibinfo {volume} {35}},\ \bibinfo {pages}
  {3072}}\BibitemShut {NoStop}%
\bibitem [{\citenamefont {Kirkwood}(1935)}]{Kirkwood:1935aa}%
  \BibitemOpen
  \bibfield  {author} {\bibinfo {author} {\bibnamefont {Kirkwood},
  \bibfnamefont {J.~G.}}} (\bibinfo {year} {1935}),\ \href {\doibase
  http://dx.doi.org/10.1063/1.1749657} {\bibfield  {journal} {\bibinfo
  {journal} {J. Chem. Phys.}\ }\textbf {\bibinfo {volume} {3}},\ \bibinfo
  {pages} {300}}\BibitemShut {NoStop}%
\bibitem [{\citenamefont {Klumov}\ \emph {et~al.}(2014)\citenamefont {Klumov},
  \citenamefont {Jin},\ and\ \citenamefont {Makse}}]{Klumov:2014aa}%
  \BibitemOpen
  \bibfield  {author} {\bibinfo {author} {\bibnamefont {Klumov}, \bibfnamefont
  {B.~A.}}, \bibinfo {author} {\bibfnamefont {Y.}~\bibnamefont {Jin}}, \ and\
  \bibinfo {author} {\bibfnamefont {H.~A.}\ \bibnamefont {Makse}}} (\bibinfo
  {year} {2014}),\ \href {\doibase 10.1021/jp504537n} {\bibfield  {journal}
  {\bibinfo  {journal} {The J. Phys. Chem. B}\ }\textbf {\bibinfo {volume}
  {118}},\ \bibinfo {pages} {10761}}\BibitemShut {NoStop}%
\bibitem [{\citenamefont {Klumov}\ \emph {et~al.}(2011)\citenamefont {Klumov},
  \citenamefont {Khrapak},\ and\ \citenamefont {Morfill}}]{Klumov:2011aa}%
  \BibitemOpen
  \bibfield  {author} {\bibinfo {author} {\bibnamefont {Klumov}, \bibfnamefont
  {B.~A.}}, \bibinfo {author} {\bibfnamefont {S.~A.}\ \bibnamefont {Khrapak}},
  \ and\ \bibinfo {author} {\bibfnamefont {G.~E.}\ \bibnamefont {Morfill}}}
  (\bibinfo {year} {2011}),\ \href {\doibase 10.1103/PhysRevB.83.184105}
  {\bibfield  {journal} {\bibinfo  {journal} {Phys. Rev. B}\ }\textbf {\bibinfo
  {volume} {83}},\ \bibinfo {pages} {184105}}\BibitemShut {NoStop}%
\bibitem [{\citenamefont {Knight}\ \emph {et~al.}(1995)\citenamefont {Knight},
  \citenamefont {Fandrich}, \citenamefont {Lau}, \citenamefont {Jaeger},\ and\
  \citenamefont {Nagel}}]{Knight:1995aa}%
  \BibitemOpen
  \bibfield  {author} {\bibinfo {author} {\bibnamefont {Knight}, \bibfnamefont
  {J.~B.}}, \bibinfo {author} {\bibfnamefont {C.~G.}\ \bibnamefont {Fandrich}},
  \bibinfo {author} {\bibfnamefont {C.~N.}\ \bibnamefont {Lau}}, \bibinfo
  {author} {\bibfnamefont {H.~M.}\ \bibnamefont {Jaeger}}, \ and\ \bibinfo
  {author} {\bibfnamefont {S.~R.}\ \bibnamefont {Nagel}}} (\bibinfo {year}
  {1995}),\ \href {\doibase 10.1103/PhysRevE.51.3957} {\bibfield  {journal}
  {\bibinfo  {journal} {Phys. Rev. E}\ }\textbf {\bibinfo {volume} {51}},\
  \bibinfo {pages} {3957}}\BibitemShut {NoStop}%
\bibitem [{\citenamefont {Krapivsky}\ and\ \citenamefont
  {Ben‐Naim}(1994)}]{Krapivsky:1994aa}%
  \BibitemOpen
  \bibfield  {author} {\bibinfo {author} {\bibnamefont {Krapivsky},
  \bibfnamefont {P.~L.}}, \ and\ \bibinfo {author} {\bibfnamefont
  {E.}~\bibnamefont {Ben‐Naim}}} (\bibinfo {year} {1994}),\ \href {\doibase
  http://dx.doi.org/10.1063/1.467037} {\bibfield  {journal} {\bibinfo
  {journal} {The Journal of Chemical Physics}\ }\textbf {\bibinfo {volume}
  {100}},\ \bibinfo {pages} {6778}}\BibitemShut {NoStop}%
\bibitem [{\citenamefont {Kruyt}\ and\ \citenamefont
  {Rothenburg}(2002)}]{Kruyt:2002aa}%
  \BibitemOpen
  \bibfield  {author} {\bibinfo {author} {\bibnamefont {Kruyt}, \bibfnamefont
  {N.}}, \ and\ \bibinfo {author} {\bibfnamefont {L.}~\bibnamefont
  {Rothenburg}}} (\bibinfo {year} {2002}),\ \href {\doibase
  10.1016/S0020-7683(01)00190-1} {\bibfield  {journal} {\bibinfo  {journal}
  {Int. J. Solids. Struct.}\ }\textbf {\bibinfo {volume} {39}},\ \bibinfo
  {pages} {571}}\BibitemShut {NoStop}%
\bibitem [{\citenamefont {Krzakala}\ and\ \citenamefont
  {Kurchan}(2007)}]{Krzakala:2007aa}%
  \BibitemOpen
  \bibfield  {author} {\bibinfo {author} {\bibnamefont {Krzakala},
  \bibfnamefont {F.}}, \ and\ \bibinfo {author} {\bibfnamefont
  {J.}~\bibnamefont {Kurchan}}} (\bibinfo {year} {2007}),\ \href {\doibase
  10.1103/PhysRevE.76.021122} {\bibfield  {journal} {\bibinfo  {journal} {Phys.
  Rev. E}\ }\textbf {\bibinfo {volume} {76}},\ \bibinfo {pages}
  {021122}}\BibitemShut {NoStop}%
\bibitem [{\citenamefont {Kumar}\ \emph {et~al.}(1992)\citenamefont {Kumar},
  \citenamefont {Kurtz}, \citenamefont {Banavar},\ and\ \citenamefont
  {Sharma}}]{Kumar:1992aa}%
  \BibitemOpen
  \bibfield  {author} {\bibinfo {author} {\bibnamefont {Kumar}, \bibfnamefont
  {S.}}, \bibinfo {author} {\bibfnamefont {S.~K.}\ \bibnamefont {Kurtz}},
  \bibinfo {author} {\bibfnamefont {J.~R.}\ \bibnamefont {Banavar}}, \ and\
  \bibinfo {author} {\bibfnamefont {M.~G.}\ \bibnamefont {Sharma}}} (\bibinfo
  {year} {1992}),\ \href {\doibase 10.1007/BF01049719} {\bibfield  {journal}
  {\bibinfo  {journal} {Journal of Statistical Physics}\ }\textbf {\bibinfo
  {volume} {67}}~(\bibinfo {number} {3}),\ \bibinfo {pages} {523}}\BibitemShut
  {NoStop}%
\bibitem [{\citenamefont {Kurchan}(2000)}]{Kurchan:2000aa}%
  \BibitemOpen
  \bibfield  {author} {\bibinfo {author} {\bibnamefont {Kurchan}, \bibfnamefont
  {J.}}} (\bibinfo {year} {2000}),\ \href@noop {} {\bibfield  {journal}
  {\bibinfo  {journal} {Journal of Physics: Condensed Matter}\ }\textbf
  {\bibinfo {volume} {12}},\ \bibinfo {pages} {6611}}\BibitemShut {NoStop}%
\bibitem [{\citenamefont {Kurchan}(2001)}]{Kurchan:2001aa}%
  \BibitemOpen
  \bibfield  {author} {\bibinfo {author} {\bibnamefont {Kurchan}, \bibfnamefont
  {J.}}} (\bibinfo {year} {2001}),\ \enquote {\bibinfo {title} {Rheology and
  how to stop aging},}\ in\ \href@noop {} {\emph {\bibinfo {booktitle} {Jamming
  and Rheology: Constrained Dynamics on Microscopic and Macroscopic Scales}}},\
  \bibinfo {editor} {edited by\ \bibinfo {editor} {\bibfnamefont
  {A.}~\bibnamefont {Liu}}\ and\ \bibinfo {editor} {\bibfnamefont {S.~R.}\
  \bibnamefont {Nagel}}}\ (\bibinfo  {publisher} {Taylor \& Francis},\ \bibinfo
  {address} {London})\BibitemShut {NoStop}%
\bibitem [{\citenamefont {Kurchan}\ \emph {et~al.}(2016)\citenamefont
  {Kurchan}, \citenamefont {Maimbourg},\ and\ \citenamefont
  {Zamponi}}]{Kurchan:2016aa}%
  \BibitemOpen
  \bibfield  {author} {\bibinfo {author} {\bibnamefont {Kurchan}, \bibfnamefont
  {J.}}, \bibinfo {author} {\bibfnamefont {T.}~\bibnamefont {Maimbourg}}, \
  and\ \bibinfo {author} {\bibfnamefont {F.}~\bibnamefont {Zamponi}}} (\bibinfo
  {year} {2016}),\ \href {http://stacks.iop.org/1742-5468/2016/i=3/a=033210}
  {\bibfield  {journal} {\bibinfo  {journal} {Journal of Statistical Mechanics:
  Theory and Experiment}\ }\textbf {\bibinfo {volume} {2016}}~(\bibinfo
  {number} {3}),\ \bibinfo {pages} {033210}}\BibitemShut {NoStop}%
\bibitem [{\citenamefont {Kyeyune-Nyombi}\ \emph {et~al.}(2018)\citenamefont
  {Kyeyune-Nyombi}, \citenamefont {Morone}, \citenamefont {Liu}, \citenamefont
  {Li}, \citenamefont {Gilchrist},\ and\ \citenamefont
  {Makse}}]{Kyeyune-Nyombi:2018aa}%
  \BibitemOpen
  \bibfield  {author} {\bibinfo {author} {\bibnamefont {Kyeyune-Nyombi},
  \bibfnamefont {E.}}, \bibinfo {author} {\bibfnamefont {F.}~\bibnamefont
  {Morone}}, \bibinfo {author} {\bibfnamefont {W.}~\bibnamefont {Liu}},
  \bibinfo {author} {\bibfnamefont {S.}~\bibnamefont {Li}}, \bibinfo {author}
  {\bibfnamefont {M.~L.}\ \bibnamefont {Gilchrist}}, \ and\ \bibinfo {author}
  {\bibfnamefont {H.~A.}\ \bibnamefont {Makse}}} (\bibinfo {year} {2018}),\
  \href {\doibase https://doi.org/10.1016/j.physa.2017.08.029} {\bibfield
  {journal} {\bibinfo  {journal} {Physica A}\ }\textbf {\bibinfo {volume}
  {490}},\ \bibinfo {pages} {1387 }}\BibitemShut {NoStop}%
\bibitem [{\citenamefont {Kyrylyuk}\ \emph {et~al.}(2011)\citenamefont
  {Kyrylyuk}, \citenamefont {van~de Haar}, \citenamefont {Rossi}, \citenamefont
  {Wouterse},\ and\ \citenamefont {Philipse}}]{Kyrylyuk:2011aa}%
  \BibitemOpen
  \bibfield  {author} {\bibinfo {author} {\bibnamefont {Kyrylyuk},
  \bibfnamefont {A.~V.}}, \bibinfo {author} {\bibfnamefont {M.~A.}\
  \bibnamefont {van~de Haar}}, \bibinfo {author} {\bibfnamefont
  {L.}~\bibnamefont {Rossi}}, \bibinfo {author} {\bibfnamefont
  {A.}~\bibnamefont {Wouterse}}, \ and\ \bibinfo {author} {\bibfnamefont
  {A.~P.}\ \bibnamefont {Philipse}}} (\bibinfo {year} {2011}),\ \href@noop {}
  {\bibfield  {journal} {\bibinfo  {journal} {Soft Matter}\ }\textbf {\bibinfo
  {volume} {7}},\ \bibinfo {pages} {1671}}\BibitemShut {NoStop}%
\bibitem [{\citenamefont {Landau}\ and\ \citenamefont
  {Lifshitz}(1980)}]{Landau:1980aa}%
  \BibitemOpen
  \bibfield  {author} {\bibinfo {author} {\bibnamefont {Landau}, \bibfnamefont
  {L.~D.}}, \ and\ \bibinfo {author} {\bibfnamefont {E.~M.}\ \bibnamefont
  {Lifshitz}}} (\bibinfo {year} {1980}),\ \href@noop {} {\emph {\bibinfo
  {title} {Statistical Physics}}}\ (\bibinfo  {publisher}
  {Butterworth-Heinemann})\BibitemShut {NoStop}%
\bibitem [{\citenamefont {Landau}\ \emph {et~al.}(1986)\citenamefont {Landau},
  \citenamefont {Pitaevskii}, \citenamefont {Kosevich},\ and\ \citenamefont
  {Lifshitz}}]{Landau:1959aa}%
  \BibitemOpen
  \bibfield  {author} {\bibinfo {author} {\bibnamefont {Landau}, \bibfnamefont
  {L.~D.}}, \bibinfo {author} {\bibfnamefont {L.~P.}\ \bibnamefont
  {Pitaevskii}}, \bibinfo {author} {\bibfnamefont {A.~M.}\ \bibnamefont
  {Kosevich}}, \ and\ \bibinfo {author} {\bibfnamefont {E.~M.}\ \bibnamefont
  {Lifshitz}}} (\bibinfo {year} {1986}),\ \href@noop {} {\emph {\bibinfo
  {title} {Theory of Elasticity}}}\ (\bibinfo  {publisher}
  {Butterworth-Heinemann})\BibitemShut {NoStop}%
\bibitem [{\citenamefont {de~Lange~Kristiansen}\ \emph
  {et~al.}(2005)\citenamefont {de~Lange~Kristiansen}, \citenamefont
  {Wouterse},\ and\ \citenamefont {Philipse}}]{Lange-Kristiansen:2005aa}%
  \BibitemOpen
  \bibfield  {author} {\bibinfo {author} {\bibnamefont {de~Lange~Kristiansen},
  \bibfnamefont {K.}}, \bibinfo {author} {\bibfnamefont {A.}~\bibnamefont
  {Wouterse}}, \ and\ \bibinfo {author} {\bibfnamefont {A.}~\bibnamefont
  {Philipse}}} (\bibinfo {year} {2005}),\ \href {\doibase
  http://dx.doi.org/10.1016/j.physa.2005.03.057} {\bibfield  {journal}
  {\bibinfo  {journal} {Physica A: Statistical Mechanics and its Applications}\
  }\textbf {\bibinfo {volume} {358}}~(\bibinfo {number} {2--4}),\ \bibinfo
  {pages} {249 }}\BibitemShut {NoStop}%
\bibitem [{\citenamefont {de~Larrard}(1999)}]{Larrard:1999aa}%
  \BibitemOpen
  \bibfield  {author} {\bibinfo {author} {\bibnamefont {de~Larrard},
  \bibfnamefont {F.}}} (\bibinfo {year} {1999}),\ \enquote {\bibinfo {title}
  {Concrete mixture proportioning: A scientific approach},}\ \ (\bibinfo
  {publisher} {CRC Press})\BibitemShut {NoStop}%
\bibitem [{\citenamefont {Lazar}\ \emph {et~al.}(2013)\citenamefont {Lazar},
  \citenamefont {Mason}, \citenamefont {MacPherson},\ and\ \citenamefont
  {Srolovitz}}]{Lazar:2013aa}%
  \BibitemOpen
  \bibfield  {author} {\bibinfo {author} {\bibnamefont {Lazar}, \bibfnamefont
  {E.~A.}}, \bibinfo {author} {\bibfnamefont {J.~K.}\ \bibnamefont {Mason}},
  \bibinfo {author} {\bibfnamefont {R.~D.}\ \bibnamefont {MacPherson}}, \ and\
  \bibinfo {author} {\bibfnamefont {D.~J.}\ \bibnamefont {Srolovitz}}}
  (\bibinfo {year} {2013}),\ \href {\doibase 10.1103/PhysRevE.88.063309}
  {\bibfield  {journal} {\bibinfo  {journal} {Phys. Rev. E}\ }\textbf {\bibinfo
  {volume} {88}},\ \bibinfo {pages} {063309}}\BibitemShut {NoStop}%
\bibitem [{\citenamefont {Lechenault}\ \emph {et~al.}(2006)\citenamefont
  {Lechenault}, \citenamefont {da~Cruz}, \citenamefont {Dauchot},\ and\
  \citenamefont {Bertin}}]{Lechenault:2006aa}%
  \BibitemOpen
  \bibfield  {author} {\bibinfo {author} {\bibnamefont {Lechenault},
  \bibfnamefont {F.}}, \bibinfo {author} {\bibfnamefont {F.}~\bibnamefont
  {da~Cruz}}, \bibinfo {author} {\bibfnamefont {O.}~\bibnamefont {Dauchot}}, \
  and\ \bibinfo {author} {\bibfnamefont {E.}~\bibnamefont {Bertin}}} (\bibinfo
  {year} {2006}),\ \href {\doibase 10.1088/1742-5468/2006/07/P07009} {\bibinfo
  {journal} {J. Stat. Mech.}\ ,\ \bibinfo {pages} {P07009}}\BibitemShut
  {NoStop}%
\bibitem [{\citenamefont {Lefevre}(2002)}]{Lefevre:2002ab}%
  \BibitemOpen
\bibfield  {journal} {  }\bibfield  {author} {\bibinfo {author} {\bibnamefont
  {Lefevre}, \bibfnamefont {A.}}} (\bibinfo {year} {2002}),\ \href@noop {}
  {\bibfield  {journal} {\bibinfo  {journal} {J. Phys. A}\ }\textbf {\bibinfo
  {volume} {35}},\ \bibinfo {pages} {9037}}\BibitemShut {NoStop}%
\bibitem [{\citenamefont {Lef\`evre}\ and\ \citenamefont
  {Dean}(2002)}]{Lefevre:2002aa}%
  \BibitemOpen
  \bibfield  {author} {\bibinfo {author} {\bibnamefont {Lef\`evre},
  \bibfnamefont {A.}}, \ and\ \bibinfo {author} {\bibfnamefont {D.~S.}\
  \bibnamefont {Dean}}} (\bibinfo {year} {2002}),\ \href {\doibase
  10.1103/PhysRevB.65.220403} {\bibfield  {journal} {\bibinfo  {journal} {Phys.
  Rev. B}\ }\textbf {\bibinfo {volume} {65}},\ \bibinfo {pages}
  {220403}}\BibitemShut {NoStop}%
\bibitem [{\citenamefont {Lerner}\ \emph {et~al.}(2014)\citenamefont {Lerner},
  \citenamefont {DeGiuli}, \citenamefont {During},\ and\ \citenamefont
  {Wyart}}]{Lerner:2014aa}%
  \BibitemOpen
  \bibfield  {author} {\bibinfo {author} {\bibnamefont {Lerner}, \bibfnamefont
  {E.}}, \bibinfo {author} {\bibfnamefont {E.}~\bibnamefont {DeGiuli}},
  \bibinfo {author} {\bibfnamefont {G.}~\bibnamefont {During}}, \ and\ \bibinfo
  {author} {\bibfnamefont {M.}~\bibnamefont {Wyart}}} (\bibinfo {year}
  {2014}),\ \href {\doibase 10.1039/C4SM00311J} {\bibfield  {journal} {\bibinfo
   {journal} {Soft Matter}\ }\textbf {\bibinfo {volume} {10}},\ \bibinfo
  {pages} {5085}}\BibitemShut {NoStop}%
\bibitem [{\citenamefont {Lerner}\ \emph {et~al.}(2016)\citenamefont {Lerner},
  \citenamefont {D\"uring},\ and\ \citenamefont {Bouchbinder}}]{Lerner:2016aa}%
  \BibitemOpen
  \bibfield  {author} {\bibinfo {author} {\bibnamefont {Lerner}, \bibfnamefont
  {E.}}, \bibinfo {author} {\bibfnamefont {G.}~\bibnamefont {D\"uring}}, \ and\
  \bibinfo {author} {\bibfnamefont {E.}~\bibnamefont {Bouchbinder}}} (\bibinfo
  {year} {2016}),\ \href {\doibase 10.1103/PhysRevLett.117.035501} {\bibfield
  {journal} {\bibinfo  {journal} {Phys. Rev. Lett.}\ }\textbf {\bibinfo
  {volume} {117}},\ \bibinfo {pages} {035501}}\BibitemShut {NoStop}%
\bibitem [{\citenamefont {Lerner}\ \emph {et~al.}(2013)\citenamefont {Lerner},
  \citenamefont {During},\ and\ \citenamefont {Wyart}}]{Lerner:2013aa}%
  \BibitemOpen
  \bibfield  {author} {\bibinfo {author} {\bibnamefont {Lerner}, \bibfnamefont
  {E.}}, \bibinfo {author} {\bibfnamefont {G.}~\bibnamefont {During}}, \ and\
  \bibinfo {author} {\bibfnamefont {M.}~\bibnamefont {Wyart}}} (\bibinfo {year}
  {2013}),\ \href {\doibase 10.1039/C3SM50515D} {\bibfield  {journal} {\bibinfo
   {journal} {Soft Matter}\ }\textbf {\bibinfo {volume} {9}},\ \bibinfo {pages}
  {8252}}\BibitemShut {NoStop}%
\bibitem [{\citenamefont {Leuzzi}(2009)}]{Leuzzi:2009aa}%
  \BibitemOpen
  \bibfield  {author} {\bibinfo {author} {\bibnamefont {Leuzzi}, \bibfnamefont
  {L.}}} (\bibinfo {year} {2009}),\ \href {\doibase
  http://dx.doi.org/10.1016/j.jnoncrysol.2009.01.035} {\bibfield  {journal}
  {\bibinfo  {journal} {J. Non-Crystall. Solids}\ }\textbf {\bibinfo {volume}
  {355}},\ \bibinfo {pages} {686 }}\BibitemShut {NoStop}%
\bibitem [{\citenamefont {Li}\ and\ \citenamefont
  {Marshall}(2007)}]{Li:2007aa}%
  \BibitemOpen
  \bibfield  {author} {\bibinfo {author} {\bibnamefont {Li}, \bibfnamefont
  {S.-Q.}}, \ and\ \bibinfo {author} {\bibfnamefont {J.}~\bibnamefont
  {Marshall}}} (\bibinfo {year} {2007}),\ \href {\doibase
  http://dx.doi.org/10.1016/j.jaerosci.2007.08.004} {\bibfield  {journal}
  {\bibinfo  {journal} {J. Aeros. Sci.}\ }\textbf {\bibinfo {volume} {38}},\
  \bibinfo {pages} {1031 }}\BibitemShut {NoStop}%
\bibitem [{\citenamefont {Lieou}\ and\ \citenamefont
  {Langer}(2012)}]{Lieou:2012aa}%
  \BibitemOpen
  \bibfield  {author} {\bibinfo {author} {\bibnamefont {Lieou}, \bibfnamefont
  {C.~K.~C.}}, \ and\ \bibinfo {author} {\bibfnamefont {J.~S.}\ \bibnamefont
  {Langer}}} (\bibinfo {year} {2012}),\ \href {\doibase
  10.1103/PhysRevE.85.061308} {\bibfield  {journal} {\bibinfo  {journal} {Phys.
  Rev. E}\ }\textbf {\bibinfo {volume} {85}},\ \bibinfo {pages}
  {061308}}\BibitemShut {NoStop}%
\bibitem [{\citenamefont {Lin}\ \emph {et~al.}(2016)\citenamefont {Lin},
  \citenamefont {Jorjadze}, \citenamefont {Pontani}, \citenamefont {Wyart},\
  and\ \citenamefont {Brujic}}]{Lin:2016aa}%
  \BibitemOpen
  \bibfield  {author} {\bibinfo {author} {\bibnamefont {Lin}, \bibfnamefont
  {J.}}, \bibinfo {author} {\bibfnamefont {I.}~\bibnamefont {Jorjadze}},
  \bibinfo {author} {\bibfnamefont {L.-L.}\ \bibnamefont {Pontani}}, \bibinfo
  {author} {\bibfnamefont {M.}~\bibnamefont {Wyart}}, \ and\ \bibinfo {author}
  {\bibfnamefont {J.}~\bibnamefont {Brujic}}} (\bibinfo {year} {2016}),\ \href
  {\doibase 10.1103/PhysRevLett.117.208001} {\bibfield  {journal} {\bibinfo
  {journal} {Phys. Rev. Lett.}\ }\textbf {\bibinfo {volume} {117}},\ \bibinfo
  {pages} {208001}}\BibitemShut {NoStop}%
\bibitem [{\citenamefont {Liu}\ and\ \citenamefont {Nagel}(2010)}]{Liu:2010aa}%
  \BibitemOpen
  \bibfield  {author} {\bibinfo {author} {\bibnamefont {Liu}, \bibfnamefont
  {A.~J.}}, \ and\ \bibinfo {author} {\bibfnamefont {S.~R.}\ \bibnamefont
  {Nagel}}} (\bibinfo {year} {2010}),\ \href {\doibase
  10.1146/annurev-conmatphys-070909-104045} {\bibfield  {journal} {\bibinfo
  {journal} {Annu. Rev. Cond. Matt. Phys.}\ }\textbf {\bibinfo {volume} {1}},\
  \bibinfo {pages} {347}}\BibitemShut {NoStop}%
\bibitem [{\citenamefont {Liu}\ \emph {et~al.}(1995)\citenamefont {Liu},
  \citenamefont {Nagel}, \citenamefont {Schecter}, \citenamefont {Coppersmith},
  \citenamefont {Majumdar}, \citenamefont {Narayan},\ and\ \citenamefont
  {Witten}}]{Liu:1995aa}%
  \BibitemOpen
  \bibfield  {author} {\bibinfo {author} {\bibnamefont {Liu}, \bibfnamefont
  {C.~h.}}, \bibinfo {author} {\bibfnamefont {S.~R.}\ \bibnamefont {Nagel}},
  \bibinfo {author} {\bibfnamefont {D.~A.}\ \bibnamefont {Schecter}}, \bibinfo
  {author} {\bibfnamefont {S.~N.}\ \bibnamefont {Coppersmith}}, \bibinfo
  {author} {\bibfnamefont {S.}~\bibnamefont {Majumdar}}, \bibinfo {author}
  {\bibfnamefont {O.}~\bibnamefont {Narayan}}, \ and\ \bibinfo {author}
  {\bibfnamefont {T.~A.}\ \bibnamefont {Witten}}} (\bibinfo {year} {1995}),\
  \href {\doibase 10.1126/science.269.5223.513} {\bibfield  {journal} {\bibinfo
   {journal} {Science}\ }\textbf {\bibinfo {volume} {269}},\ \bibinfo {pages}
  {513}}\BibitemShut {NoStop}%
\bibitem [{\citenamefont {Liu}\ \emph {et~al.}(2017)\citenamefont {Liu},
  \citenamefont {Jin}, \citenamefont {Chen}, \citenamefont {Makse},\ and\
  \citenamefont {Li}}]{Liu:2017aa}%
  \BibitemOpen
  \bibfield  {author} {\bibinfo {author} {\bibnamefont {Liu}, \bibfnamefont
  {W.}}, \bibinfo {author} {\bibfnamefont {Y.}~\bibnamefont {Jin}}, \bibinfo
  {author} {\bibfnamefont {S.}~\bibnamefont {Chen}}, \bibinfo {author}
  {\bibfnamefont {H.~A.}\ \bibnamefont {Makse}}, \ and\ \bibinfo {author}
  {\bibfnamefont {S.}~\bibnamefont {Li}}} (\bibinfo {year} {2017}),\ \href
  {\doibase 10.1039/C6SM02216B} {\bibfield  {journal} {\bibinfo  {journal}
  {Soft Matter}\ }\textbf {\bibinfo {volume} {13}},\ \bibinfo {pages}
  {421}}\BibitemShut {NoStop}%
\bibitem [{\citenamefont {Liu}\ \emph {et~al.}(2015)\citenamefont {Liu},
  \citenamefont {Li}, \citenamefont {Baule},\ and\ \citenamefont
  {Makse}}]{Liu:2015aa}%
  \BibitemOpen
  \bibfield  {author} {\bibinfo {author} {\bibnamefont {Liu}, \bibfnamefont
  {W.}}, \bibinfo {author} {\bibfnamefont {S.}~\bibnamefont {Li}}, \bibinfo
  {author} {\bibfnamefont {A.}~\bibnamefont {Baule}}, \ and\ \bibinfo {author}
  {\bibfnamefont {H.~A.}\ \bibnamefont {Makse}}} (\bibinfo {year} {2015}),\
  \href {\doibase 10.1039/C5SM01169H} {\bibfield  {journal} {\bibinfo
  {journal} {Soft Matter}\ }\textbf {\bibinfo {volume} {11}},\ \bibinfo {pages}
  {6492}}\BibitemShut {NoStop}%
\bibitem [{\citenamefont {Loi}\ \emph {et~al.}(2008)\citenamefont {Loi},
  \citenamefont {Mossa},\ and\ \citenamefont {Cugliandolo}}]{Loi:2008aa}%
  \BibitemOpen
  \bibfield  {author} {\bibinfo {author} {\bibnamefont {Loi}, \bibfnamefont
  {D.}}, \bibinfo {author} {\bibfnamefont {S.}~\bibnamefont {Mossa}}, \ and\
  \bibinfo {author} {\bibfnamefont {L.~F.}\ \bibnamefont {Cugliandolo}}}
  (\bibinfo {year} {2008}),\ \href {\doibase 10.1103/PhysRevE.77.051111}
  {\bibfield  {journal} {\bibinfo  {journal} {Phys. Rev. E}\ }\textbf {\bibinfo
  {volume} {77}},\ \bibinfo {pages} {051111}}\BibitemShut {NoStop}%
\bibitem [{\citenamefont {Lois}\ \emph {et~al.}(2008)\citenamefont {Lois},
  \citenamefont {Blawzdziewicz},\ and\ \citenamefont {O'Hern}}]{Lois:2008aa}%
  \BibitemOpen
  \bibfield  {author} {\bibinfo {author} {\bibnamefont {Lois}, \bibfnamefont
  {G.}}, \bibinfo {author} {\bibfnamefont {J.}~\bibnamefont {Blawzdziewicz}}, \
  and\ \bibinfo {author} {\bibfnamefont {C.~S.}\ \bibnamefont {O'Hern}}}
  (\bibinfo {year} {2008}),\ \href {\doibase 10.1103/PhysRevLett.100.028001}
  {\bibfield  {journal} {\bibinfo  {journal} {Phys. Rev. Lett.}\ }\textbf
  {\bibinfo {volume} {100}},\ \bibinfo {pages} {028001}}\BibitemShut {NoStop}%
\bibitem [{\citenamefont {L\o{}voll}\ \emph {et~al.}(1999)\citenamefont
  {L\o{}voll}, \citenamefont {M\aa{}l\o{}y},\ and\ \citenamefont
  {Flekk\o{}y}}]{Lovoll:1999aa}%
  \BibitemOpen
  \bibfield  {author} {\bibinfo {author} {\bibnamefont {L\o{}voll},
  \bibfnamefont {G.}}, \bibinfo {author} {\bibfnamefont {K.~J.}\ \bibnamefont
  {M\aa{}l\o{}y}}, \ and\ \bibinfo {author} {\bibfnamefont {E.~G.}\
  \bibnamefont {Flekk\o{}y}}} (\bibinfo {year} {1999}),\ \href {\doibase
  10.1103/PhysRevE.60.5872} {\bibfield  {journal} {\bibinfo  {journal} {Phys.
  Rev. E}\ }\textbf {\bibinfo {volume} {60}},\ \bibinfo {pages}
  {5872}}\BibitemShut {NoStop}%
\bibitem [{\citenamefont {Lu}\ \emph {et~al.}(2008{\natexlab{a}})\citenamefont
  {Lu}, \citenamefont {Brodsky},\ and\ \citenamefont {Kavehpour}}]{Lu:2008aa}%
  \BibitemOpen
  \bibfield  {author} {\bibinfo {author} {\bibnamefont {Lu}, \bibfnamefont
  {K.}}, \bibinfo {author} {\bibfnamefont {E.~E.}\ \bibnamefont {Brodsky}}, \
  and\ \bibinfo {author} {\bibfnamefont {H.~P.}\ \bibnamefont {Kavehpour}}}
  (\bibinfo {year} {2008}{\natexlab{a}}),\ \href {\doibase 10.1038/nphys934}
  {\bibfield  {journal} {\bibinfo  {journal} {Nature Phys.}\ }\textbf {\bibinfo
  {volume} {4}},\ \bibinfo {pages} {404}}\BibitemShut {NoStop}%
\bibitem [{\citenamefont {Lu}\ \emph {et~al.}(2010)\citenamefont {Lu},
  \citenamefont {Li}, \citenamefont {Zhao},\ and\ \citenamefont
  {Meng}}]{Lu:2010aa}%
  \BibitemOpen
  \bibfield  {author} {\bibinfo {author} {\bibnamefont {Lu}, \bibfnamefont
  {P.}}, \bibinfo {author} {\bibfnamefont {S.}~\bibnamefont {Li}}, \bibinfo
  {author} {\bibfnamefont {J.}~\bibnamefont {Zhao}}, \ and\ \bibinfo {author}
  {\bibfnamefont {L.}~\bibnamefont {Meng}}} (\bibinfo {year} {2010}),\
  \href@noop {} {\bibfield  {journal} {\bibinfo  {journal} {Science China}\
  }\textbf {\bibinfo {volume} {53}},\ \bibinfo {pages} {2284}}\BibitemShut
  {NoStop}%
\bibitem [{\citenamefont {Lu}\ \emph {et~al.}(2008{\natexlab{b}})\citenamefont
  {Lu}, \citenamefont {Zaccarelli}, \citenamefont {Ciulla}, \citenamefont
  {Schofield}, \citenamefont {Sciortino},\ and\ \citenamefont
  {Weitz}}]{Lu:2008ab}%
  \BibitemOpen
  \bibfield  {author} {\bibinfo {author} {\bibnamefont {Lu}, \bibfnamefont
  {P.~J.}}, \bibinfo {author} {\bibfnamefont {E.}~\bibnamefont {Zaccarelli}},
  \bibinfo {author} {\bibfnamefont {F.}~\bibnamefont {Ciulla}}, \bibinfo
  {author} {\bibfnamefont {A.~B.}\ \bibnamefont {Schofield}}, \bibinfo {author}
  {\bibfnamefont {F.}~\bibnamefont {Sciortino}}, \ and\ \bibinfo {author}
  {\bibfnamefont {D.~A.}\ \bibnamefont {Weitz}}} (\bibinfo {year}
  {2008}{\natexlab{b}}),\ \href {http://dx.doi.org/10.1038/nature06931}
  {\bibfield  {journal} {\bibinfo  {journal} {Nature}\ }\textbf {\bibinfo
  {volume} {453}},\ \bibinfo {pages} {499}}\BibitemShut {NoStop}%
\bibitem [{\citenamefont {Lubachevsky}\ and\ \citenamefont
  {Stillinger}(1990)}]{Lubachevsky:1990aa}%
  \BibitemOpen
  \bibfield  {author} {\bibinfo {author} {\bibnamefont {Lubachevsky},
  \bibfnamefont {B.~D.}}, \ and\ \bibinfo {author} {\bibfnamefont {F.~H.}\
  \bibnamefont {Stillinger}}} (\bibinfo {year} {1990}),\ \href
  {http://dx.doi.org/10.1007/BF01025983} {\bibfield  {journal} {\bibinfo
  {journal} {J. Stat. Phys.}\ }\textbf {\bibinfo {volume} {60}},\ \bibinfo
  {pages} {561}}\BibitemShut {NoStop}%
\bibitem [{\citenamefont {Luchnikov}\ \emph {et~al.}(1999)\citenamefont
  {Luchnikov}, \citenamefont {Medvedev}, \citenamefont {Oger},\ and\
  \citenamefont {Troadec}}]{Luchnikov:1999aa}%
  \BibitemOpen
  \bibfield  {author} {\bibinfo {author} {\bibnamefont {Luchnikov},
  \bibfnamefont {V.~A.}}, \bibinfo {author} {\bibfnamefont {N.~N.}\
  \bibnamefont {Medvedev}}, \bibinfo {author} {\bibfnamefont {L.}~\bibnamefont
  {Oger}}, \ and\ \bibinfo {author} {\bibfnamefont {J.-P.}\ \bibnamefont
  {Troadec}}} (\bibinfo {year} {1999}),\ \href {\doibase
  10.1103/PhysRevE.59.7205} {\bibfield  {journal} {\bibinfo  {journal} {Phys.
  Rev. E}\ }\textbf {\bibinfo {volume} {59}},\ \bibinfo {pages}
  {7205}}\BibitemShut {NoStop}%
\bibitem [{\citenamefont {Magnanimo}\ \emph {et~al.}(2008)\citenamefont
  {Magnanimo}, \citenamefont {Ragione}, \citenamefont {Jenkins}, \citenamefont
  {Wang},\ and\ \citenamefont {Makse}}]{Magnanimo:2008aa}%
  \BibitemOpen
  \bibfield  {author} {\bibinfo {author} {\bibnamefont {Magnanimo},
  \bibfnamefont {V.}}, \bibinfo {author} {\bibfnamefont {L.~L.}\ \bibnamefont
  {Ragione}}, \bibinfo {author} {\bibfnamefont {J.~T.}\ \bibnamefont
  {Jenkins}}, \bibinfo {author} {\bibfnamefont {P.}~\bibnamefont {Wang}}, \
  and\ \bibinfo {author} {\bibfnamefont {H.~A.}\ \bibnamefont {Makse}}}
  (\bibinfo {year} {2008}),\ \href
  {http://stacks.iop.org/0295-5075/81/i=3/a=34006} {\bibfield  {journal}
  {\bibinfo  {journal} {EPL}\ }\textbf {\bibinfo {volume} {81}},\ \bibinfo
  {pages} {34006}}\BibitemShut {NoStop}%
\bibitem [{\citenamefont {Mailman}\ and\ \citenamefont
  {Chakraborty}(2011)}]{Mailman:2011aa}%
  \BibitemOpen
  \bibfield  {author} {\bibinfo {author} {\bibnamefont {Mailman}, \bibfnamefont
  {M.}}, \ and\ \bibinfo {author} {\bibfnamefont {B.}~\bibnamefont
  {Chakraborty}}} (\bibinfo {year} {2011}),\ \href
  {http://stacks.iop.org/1742-5468/2011/i=07/a=L07002} {\bibfield  {journal}
  {\bibinfo  {journal} {J. Stat. Mech.}\ }\textbf {\bibinfo {volume} {2011}},\
  \bibinfo {pages} {L07002}}\BibitemShut {NoStop}%
\bibitem [{\citenamefont {Mailman}\ and\ \citenamefont
  {Chakraborty}(2012)}]{Mailman:2012aa}%
  \BibitemOpen
  \bibfield  {author} {\bibinfo {author} {\bibnamefont {Mailman}, \bibfnamefont
  {M.}}, \ and\ \bibinfo {author} {\bibfnamefont {B.}~\bibnamefont
  {Chakraborty}}} (\bibinfo {year} {2012}),\ \href
  {http://stacks.iop.org/1742-5468/2012/i=05/a=P05001} {\bibfield  {journal}
  {\bibinfo  {journal} {J. Stat. Mech.}\ }\textbf {\bibinfo {volume} {2012}},\
  \bibinfo {pages} {P05001}}\BibitemShut {NoStop}%
\bibitem [{\citenamefont {Maimbourg}\ \emph {et~al.}(2016)\citenamefont
  {Maimbourg}, \citenamefont {Kurchan},\ and\ \citenamefont
  {Zamponi}}]{Maimbourg:2016aa}%
  \BibitemOpen
  \bibfield  {author} {\bibinfo {author} {\bibnamefont {Maimbourg},
  \bibfnamefont {T.}}, \bibinfo {author} {\bibfnamefont {J.}~\bibnamefont
  {Kurchan}}, \ and\ \bibinfo {author} {\bibfnamefont {F.}~\bibnamefont
  {Zamponi}}} (\bibinfo {year} {2016}),\ \href {\doibase
  10.1103/PhysRevLett.116.015902} {\bibfield  {journal} {\bibinfo  {journal}
  {Phys. Rev. Lett.}\ }\textbf {\bibinfo {volume} {116}},\ \bibinfo {pages}
  {015902}}\BibitemShut {NoStop}%
\bibitem [{\citenamefont {Majmudar}\ and\ \citenamefont
  {Behringer}(2005)}]{Majmudar:2005aa}%
  \BibitemOpen
  \bibfield  {author} {\bibinfo {author} {\bibnamefont {Majmudar},
  \bibfnamefont {T.~S.}}, \ and\ \bibinfo {author} {\bibfnamefont {R.~P.}\
  \bibnamefont {Behringer}}} (\bibinfo {year} {2005}),\ \href
  {http://dx.doi.org/10.1038/nature03805} {\bibfield  {journal} {\bibinfo
  {journal} {Nature}\ }\textbf {\bibinfo {volume} {435}},\ \bibinfo {pages}
  {1079}}\BibitemShut {NoStop}%
\bibitem [{\citenamefont {Majmudar}\ \emph {et~al.}(2007)\citenamefont
  {Majmudar}, \citenamefont {Sperl}, \citenamefont {Luding},\ and\
  \citenamefont {Behringer}}]{Majmudar:2007aa}%
  \BibitemOpen
  \bibfield  {author} {\bibinfo {author} {\bibnamefont {Majmudar},
  \bibfnamefont {T.~S.}}, \bibinfo {author} {\bibfnamefont {M.}~\bibnamefont
  {Sperl}}, \bibinfo {author} {\bibfnamefont {S.}~\bibnamefont {Luding}}, \
  and\ \bibinfo {author} {\bibfnamefont {R.~P.}\ \bibnamefont {Behringer}}}
  (\bibinfo {year} {2007}),\ \href {\doibase 10.1103/PhysRevLett.98.058001}
  {\bibfield  {journal} {\bibinfo  {journal} {Phys. Rev. Lett.}\ }\textbf
  {\bibinfo {volume} {98}},\ 10.1103/PhysRevLett.98.058001}\BibitemShut
  {NoStop}%
\bibitem [{\citenamefont {Makse}\ and\ \citenamefont
  {Kurchan}(2002)}]{Makse:2002aa}%
  \BibitemOpen
  \bibfield  {author} {\bibinfo {author} {\bibnamefont {Makse}, \bibfnamefont
  {H.}}, \ and\ \bibinfo {author} {\bibfnamefont {J.}~\bibnamefont {Kurchan}}}
  (\bibinfo {year} {2002}),\ \href {\doibase 10.1038/415614a} {\bibfield
  {journal} {\bibinfo  {journal} {Nature}\ }\textbf {\bibinfo {volume} {415}},\
  \bibinfo {pages} {614}}\BibitemShut {NoStop}%
\bibitem [{\citenamefont {Makse}\ \emph {et~al.}(2005)\citenamefont {Makse},
  \citenamefont {Bruji\'c},\ and\ \citenamefont {Edwards}}]{Makse:2005aa}%
  \BibitemOpen
  \bibfield  {author} {\bibinfo {author} {\bibnamefont {Makse}, \bibfnamefont
  {H.~A.}}, \bibinfo {author} {\bibfnamefont {J.}~\bibnamefont {Bruji\'c}}, \
  and\ \bibinfo {author} {\bibfnamefont {S.~F.}\ \bibnamefont {Edwards}}}
  (\bibinfo {year} {2005}),\ \enquote {\bibinfo {title} {Statistical mechanics
  of jammed matter},}\ in\ \href@noop {} {\emph {\bibinfo {booktitle} {The
  Physics of Granular Media}}},\ \bibinfo {editor} {edited by\ \bibinfo
  {editor} {\bibfnamefont {H.}~\bibnamefont {Hinrichsen}}\ and\ \bibinfo
  {editor} {\bibfnamefont {D.~E.}\ \bibnamefont {Wolf}}}\ (\bibinfo
  {publisher} {Wiley-VCH})\BibitemShut {NoStop}%
\bibitem [{\citenamefont {Makse}\ \emph {et~al.}(2004)\citenamefont {Makse},
  \citenamefont {Gland}, \citenamefont {Johnson},\ and\ \citenamefont
  {Schwartz}}]{Makse:2004aa}%
  \BibitemOpen
  \bibfield  {author} {\bibinfo {author} {\bibnamefont {Makse}, \bibfnamefont
  {H.~A.}}, \bibinfo {author} {\bibfnamefont {N.}~\bibnamefont {Gland}},
  \bibinfo {author} {\bibfnamefont {D.~L.}\ \bibnamefont {Johnson}}, \ and\
  \bibinfo {author} {\bibfnamefont {L.}~\bibnamefont {Schwartz}}} (\bibinfo
  {year} {2004}),\ \href {\doibase 10.1103/PhysRevE.70.061302} {\bibfield
  {journal} {\bibinfo  {journal} {Phys. Rev. E}\ }\textbf {\bibinfo {volume}
  {70}},\ \bibinfo {pages} {061302}}\BibitemShut {NoStop}%
\bibitem [{\citenamefont {Makse}\ \emph {et~al.}(1999)\citenamefont {Makse},
  \citenamefont {Gland}, \citenamefont {Johnson},\ and\ \citenamefont
  {Schwartz}}]{Makse:1999aa}%
  \BibitemOpen
  \bibfield  {author} {\bibinfo {author} {\bibnamefont {Makse}, \bibfnamefont
  {H.~A.}}, \bibinfo {author} {\bibfnamefont {N.}~\bibnamefont {Gland}},
  \bibinfo {author} {\bibfnamefont {D.~L.}\ \bibnamefont {Johnson}}, \ and\
  \bibinfo {author} {\bibfnamefont {L.~M.}\ \bibnamefont {Schwartz}}} (\bibinfo
  {year} {1999}),\ \href {\doibase 10.1103/PhysRevLett.83.5070} {\bibfield
  {journal} {\bibinfo  {journal} {Phys. Rev. Lett.}\ }\textbf {\bibinfo
  {volume} {83}},\ \bibinfo {pages} {5070}}\BibitemShut {NoStop}%
\bibitem [{\citenamefont {Makse}\ \emph {et~al.}(2000)\citenamefont {Makse},
  \citenamefont {Johnson},\ and\ \citenamefont {Schwartz}}]{Makse:2000aa}%
  \BibitemOpen
  \bibfield  {author} {\bibinfo {author} {\bibnamefont {Makse}, \bibfnamefont
  {H.~A.}}, \bibinfo {author} {\bibfnamefont {D.~L.}\ \bibnamefont {Johnson}},
  \ and\ \bibinfo {author} {\bibfnamefont {L.~M.}\ \bibnamefont {Schwartz}}}
  (\bibinfo {year} {2000}),\ \href {\doibase 10.1103/PhysRevLett.84.4160}
  {\bibfield  {journal} {\bibinfo  {journal} {Phys. Rev. Lett.}\ }\textbf
  {\bibinfo {volume} {84}},\ \bibinfo {pages} {4160}}\BibitemShut {NoStop}%
\bibitem [{\citenamefont {Man}\ \emph {et~al.}(2005)\citenamefont {Man},
  \citenamefont {Donev}, \citenamefont {Stillinger}, \citenamefont {Sullivan},
  \citenamefont {Russel}, \citenamefont {Heeger}, \citenamefont {Inati},
  \citenamefont {Torquato},\ and\ \citenamefont {Chaikin}}]{Man:2005aa}%
  \BibitemOpen
  \bibfield  {author} {\bibinfo {author} {\bibnamefont {Man}, \bibfnamefont
  {W.}}, \bibinfo {author} {\bibfnamefont {A.}~\bibnamefont {Donev}}, \bibinfo
  {author} {\bibfnamefont {F.~H.}\ \bibnamefont {Stillinger}}, \bibinfo
  {author} {\bibfnamefont {M.~T.}\ \bibnamefont {Sullivan}}, \bibinfo {author}
  {\bibfnamefont {W.~B.}\ \bibnamefont {Russel}}, \bibinfo {author}
  {\bibfnamefont {D.}~\bibnamefont {Heeger}}, \bibinfo {author} {\bibfnamefont
  {S.}~\bibnamefont {Inati}}, \bibinfo {author} {\bibfnamefont
  {S.}~\bibnamefont {Torquato}}, \ and\ \bibinfo {author} {\bibfnamefont
  {P.~M.}\ \bibnamefont {Chaikin}}} (\bibinfo {year} {2005}),\ \href {\doibase
  10.1103/PhysRevLett.94.198001} {\bibfield  {journal} {\bibinfo  {journal}
  {Phys. Rev. Lett.}\ }\textbf {\bibinfo {volume} {94}},\ \bibinfo {pages}
  {198001}}\BibitemShut {NoStop}%
\bibitem [{\citenamefont {Mangeat}\ and\ \citenamefont
  {Zamponi}(2016)}]{Mangeat:2016aa}%
  \BibitemOpen
  \bibfield  {author} {\bibinfo {author} {\bibnamefont {Mangeat}, \bibfnamefont
  {M.}}, \ and\ \bibinfo {author} {\bibfnamefont {F.}~\bibnamefont {Zamponi}}}
  (\bibinfo {year} {2016}),\ \href {\doibase 10.1103/PhysRevE.93.012609}
  {\bibfield  {journal} {\bibinfo  {journal} {Phys. Rev. E}\ }\textbf {\bibinfo
  {volume} {93}},\ \bibinfo {pages} {012609}}\BibitemShut {NoStop}%
\bibitem [{\citenamefont {Marconi}\ \emph {et~al.}(2008)\citenamefont
  {Marconi}, \citenamefont {Puglisi}, \citenamefont {Rondoni},\ and\
  \citenamefont {Vulpiani}}]{Marconi:2008aa}%
  \BibitemOpen
  \bibfield  {author} {\bibinfo {author} {\bibnamefont {Marconi}, \bibfnamefont
  {U.~M.~B.}}, \bibinfo {author} {\bibfnamefont {A.}~\bibnamefont {Puglisi}},
  \bibinfo {author} {\bibfnamefont {L.}~\bibnamefont {Rondoni}}, \ and\
  \bibinfo {author} {\bibfnamefont {A.}~\bibnamefont {Vulpiani}}} (\bibinfo
  {year} {2008}),\ \href {\doibase
  http://dx.doi.org/10.1016/j.physrep.2008.02.002} {\bibfield  {journal}
  {\bibinfo  {journal} {Phys. Rep.}\ }\textbf {\bibinfo {volume} {461}},\
  \bibinfo {pages} {111 }}\BibitemShut {NoStop}%
\bibitem [{\citenamefont {Marechal}\ and\ \citenamefont
  {L\"owen}(2013)}]{Marechal:2013aa}%
  \BibitemOpen
  \bibfield  {author} {\bibinfo {author} {\bibnamefont {Marechal},
  \bibfnamefont {M.}}, \ and\ \bibinfo {author} {\bibfnamefont
  {H.}~\bibnamefont {L\"owen}}} (\bibinfo {year} {2013}),\ \href {\doibase
  10.1103/PhysRevLett.110.137801} {\bibfield  {journal} {\bibinfo  {journal}
  {Phys. Rev. Lett.}\ }\textbf {\bibinfo {volume} {110}},\ \bibinfo {pages}
  {137801}}\BibitemShut {NoStop}%
\bibitem [{\citenamefont {Mari}\ \emph {et~al.}(2009)\citenamefont {Mari},
  \citenamefont {Krzakala},\ and\ \citenamefont {Kurchan}}]{Mari:2009aa}%
  \BibitemOpen
  \bibfield  {author} {\bibinfo {author} {\bibnamefont {Mari}, \bibfnamefont
  {R.}}, \bibinfo {author} {\bibfnamefont {F.}~\bibnamefont {Krzakala}}, \ and\
  \bibinfo {author} {\bibfnamefont {J.}~\bibnamefont {Kurchan}}} (\bibinfo
  {year} {2009}),\ \href {\doibase 10.1103/PhysRevLett.103.025701} {\bibfield
  {journal} {\bibinfo  {journal} {Phys. Rev. Lett.}\ }\textbf {\bibinfo
  {volume} {103}},\ \bibinfo {pages} {025701}}\BibitemShut {NoStop}%
\bibitem [{\citenamefont {Marshall}\ and\ \citenamefont
  {Li}(2014)}]{Marshall:2014aa}%
  \BibitemOpen
  \bibfield  {author} {\bibinfo {author} {\bibnamefont {Marshall},
  \bibfnamefont {J.~S.}}, \ and\ \bibinfo {author} {\bibfnamefont
  {S.}~\bibnamefont {Li}}} (\bibinfo {year} {2014}),\ \href@noop {} {\emph
  {\bibinfo {title} {Adhesive Particle Flow}}}\ (\bibinfo  {publisher}
  {Cambridge University Press})\BibitemShut {NoStop}%
\bibitem [{\citenamefont {Martin}\ and\ \citenamefont
  {Bordia}(2008)}]{Martin:2008aa}%
  \BibitemOpen
  \bibfield  {author} {\bibinfo {author} {\bibnamefont {Martin}, \bibfnamefont
  {C.~L.}}, \ and\ \bibinfo {author} {\bibfnamefont {R.~K.}\ \bibnamefont
  {Bordia}}} (\bibinfo {year} {2008}),\ \href {\doibase
  10.1103/PhysRevE.77.031307} {\bibfield  {journal} {\bibinfo  {journal} {Phys.
  Rev. E}\ }\textbf {\bibinfo {volume} {77}},\ \bibinfo {pages}
  {031307}}\BibitemShut {NoStop}%
\bibitem [{\citenamefont {Martiniani}\ \emph {et~al.}(2017)\citenamefont
  {Martiniani}, \citenamefont {Schrenk}, \citenamefont {Ramola}, \citenamefont
  {Chakraborty},\ and\ \citenamefont {Frenkel}}]{Martiniani:2017aa}%
  \BibitemOpen
  \bibfield  {author} {\bibinfo {author} {\bibnamefont {Martiniani},
  \bibfnamefont {S.}}, \bibinfo {author} {\bibfnamefont {K.~J.}\ \bibnamefont
  {Schrenk}}, \bibinfo {author} {\bibfnamefont {K.}~\bibnamefont {Ramola}},
  \bibinfo {author} {\bibfnamefont {B.}~\bibnamefont {Chakraborty}}, \ and\
  \bibinfo {author} {\bibfnamefont {D.}~\bibnamefont {Frenkel}}} (\bibinfo
  {year} {2017}),\ \href {http://dx.doi.org/10.1038/nphys4168} {\bibfield
  {journal} {\bibinfo  {journal} {Nat Phys}\ }\textbf {\bibinfo {volume}
  {13}}~(\bibinfo {number} {9}),\ \bibinfo {pages} {848}}\BibitemShut {NoStop}%
\bibitem [{\citenamefont {Martiniani}\ \emph
  {et~al.}(2016{\natexlab{a}})\citenamefont {Martiniani}, \citenamefont
  {Schrenk}, \citenamefont {Stevenson}, \citenamefont {Wales},\ and\
  \citenamefont {Frenkel}}]{Martiniani:2016aa}%
  \BibitemOpen
  \bibfield  {author} {\bibinfo {author} {\bibnamefont {Martiniani},
  \bibfnamefont {S.}}, \bibinfo {author} {\bibfnamefont {K.~J.}\ \bibnamefont
  {Schrenk}}, \bibinfo {author} {\bibfnamefont {J.~D.}\ \bibnamefont
  {Stevenson}}, \bibinfo {author} {\bibfnamefont {D.~J.}\ \bibnamefont
  {Wales}}, \ and\ \bibinfo {author} {\bibfnamefont {D.}~\bibnamefont
  {Frenkel}}} (\bibinfo {year} {2016}{\natexlab{a}}),\ \href {\doibase
  10.1103/PhysRevE.94.031301} {\bibfield  {journal} {\bibinfo  {journal} {Phys.
  Rev. E}\ }\textbf {\bibinfo {volume} {94}},\ \bibinfo {pages}
  {031301}}\BibitemShut {NoStop}%
\bibitem [{\citenamefont {Martiniani}\ \emph
  {et~al.}(2016{\natexlab{b}})\citenamefont {Martiniani}, \citenamefont
  {Schrenk}, \citenamefont {Stevenson}, \citenamefont {Wales},\ and\
  \citenamefont {Frenkel}}]{Martiniani:2016ad}%
  \BibitemOpen
  \bibfield  {author} {\bibinfo {author} {\bibnamefont {Martiniani},
  \bibfnamefont {S.}}, \bibinfo {author} {\bibfnamefont {K.~J.}\ \bibnamefont
  {Schrenk}}, \bibinfo {author} {\bibfnamefont {J.~D.}\ \bibnamefont
  {Stevenson}}, \bibinfo {author} {\bibfnamefont {D.~J.}\ \bibnamefont
  {Wales}}, \ and\ \bibinfo {author} {\bibfnamefont {D.}~\bibnamefont
  {Frenkel}}} (\bibinfo {year} {2016}{\natexlab{b}}),\ \href {\doibase
  10.1103/PhysRevE.94.031301} {\bibfield  {journal} {\bibinfo  {journal} {Phys.
  Rev. E}\ }\textbf {\bibinfo {volume} {94}},\ \bibinfo {pages}
  {031301}}\BibitemShut {NoStop}%
\bibitem [{\citenamefont {Martiniani}\ \emph
  {et~al.}(2016{\natexlab{c}})\citenamefont {Martiniani}, \citenamefont
  {Schrenk}, \citenamefont {Stevenson}, \citenamefont {Wales},\ and\
  \citenamefont {Frenkel}}]{Martiniani:2016ab}%
  \BibitemOpen
  \bibfield  {author} {\bibinfo {author} {\bibnamefont {Martiniani},
  \bibfnamefont {S.}}, \bibinfo {author} {\bibfnamefont {K.~J.}\ \bibnamefont
  {Schrenk}}, \bibinfo {author} {\bibfnamefont {J.~D.}\ \bibnamefont
  {Stevenson}}, \bibinfo {author} {\bibfnamefont {D.~J.}\ \bibnamefont
  {Wales}}, \ and\ \bibinfo {author} {\bibfnamefont {D.}~\bibnamefont
  {Frenkel}}} (\bibinfo {year} {2016}{\natexlab{c}}),\ \href {\doibase
  10.1103/PhysRevE.93.012906} {\bibfield  {journal} {\bibinfo  {journal} {Phys.
  Rev. E}\ }\textbf {\bibinfo {volume} {93}},\ \bibinfo {pages}
  {012906}}\BibitemShut {NoStop}%
\bibitem [{\citenamefont {Matsushima}\ and\ \citenamefont
  {Blumenfeld}(2014)}]{Matsushima:2014aa}%
  \BibitemOpen
  \bibfield  {author} {\bibinfo {author} {\bibnamefont {Matsushima},
  \bibfnamefont {T.}}, \ and\ \bibinfo {author} {\bibfnamefont
  {R.}~\bibnamefont {Blumenfeld}}} (\bibinfo {year} {2014}),\ \href {\doibase
  10.1103/PhysRevLett.112.098003} {\bibfield  {journal} {\bibinfo  {journal}
  {Phys. Rev. Lett.}\ }\textbf {\bibinfo {volume} {112}},\ \bibinfo {pages}
  {098003}}\BibitemShut {NoStop}%
\bibitem [{\citenamefont {Maxwell}(1864)}]{Maxwell:1864aa}%
  \BibitemOpen
  \bibfield  {author} {\bibinfo {author} {\bibnamefont {Maxwell}, \bibfnamefont
  {J.~C.}}} (\bibinfo {year} {1864}),\ \href@noop {} {\bibfield  {journal}
  {\bibinfo  {journal} {Philosophical Magazine}\ }\textbf {\bibinfo {volume}
  {26}},\ \bibinfo {pages} {250}}\BibitemShut {NoStop}%
\bibitem [{\citenamefont {Maxwell}(1870)}]{Maxwell:1870aa}%
  \BibitemOpen
  \bibfield  {author} {\bibinfo {author} {\bibnamefont {Maxwell}, \bibfnamefont
  {J.~C.}}} (\bibinfo {year} {1870}),\ \href@noop {} {\bibfield  {journal}
  {\bibinfo  {journal} {Transactions of the Royal Society of Edinburgh}\
  }\textbf {\bibinfo {volume} {26}},\ \bibinfo {pages} {1}}\BibitemShut
  {NoStop}%
\bibitem [{\citenamefont {McNamara}\ and\ \citenamefont
  {Herrmann}(2004)}]{McNamara:2004aa}%
  \BibitemOpen
  \bibfield  {author} {\bibinfo {author} {\bibnamefont {McNamara},
  \bibfnamefont {S.}}, \ and\ \bibinfo {author} {\bibfnamefont
  {H.}~\bibnamefont {Herrmann}}} (\bibinfo {year} {2004}),\ \href {\doibase
  10.1103/PhysRevE.70.061303} {\bibfield  {journal} {\bibinfo  {journal} {Phys.
  Rev. E}\ }\textbf {\bibinfo {volume} {70}},\ \bibinfo {pages}
  {061303}}\BibitemShut {NoStop}%
\bibitem [{\citenamefont {McNamara}\ \emph
  {et~al.}(2009{\natexlab{a}})\citenamefont {McNamara}, \citenamefont
  {Richard}, \citenamefont {de~Richter}, \citenamefont {Le~Caer},\ and\
  \citenamefont {Delannay}}]{McNamara:2009aa}%
  \BibitemOpen
  \bibfield  {author} {\bibinfo {author} {\bibnamefont {McNamara},
  \bibfnamefont {S.}}, \bibinfo {author} {\bibfnamefont {P.}~\bibnamefont
  {Richard}}, \bibinfo {author} {\bibfnamefont {S.~K.}\ \bibnamefont
  {de~Richter}}, \bibinfo {author} {\bibfnamefont {G.}~\bibnamefont {Le~Caer}},
  \ and\ \bibinfo {author} {\bibfnamefont {R.}~\bibnamefont {Delannay}}}
  (\bibinfo {year} {2009}{\natexlab{a}}),\ \href {\doibase
  10.1103/PhysRevE.80.031301} {\bibfield  {journal} {\bibinfo  {journal} {Phys.
  Rev. E}\ }\textbf {\bibinfo {volume} {80}},\ \bibinfo {pages}
  {031301}}\BibitemShut {NoStop}%
\bibitem [{\citenamefont {McNamara}\ \emph
  {et~al.}(2009{\natexlab{b}})\citenamefont {McNamara}, \citenamefont
  {Richard}, \citenamefont {de~Richter}, \citenamefont {Le~Caer},\ and\
  \citenamefont {Delannay}}]{McNamara:2009ab}%
  \BibitemOpen
  \bibfield  {author} {\bibinfo {author} {\bibnamefont {McNamara},
  \bibfnamefont {S.}}, \bibinfo {author} {\bibfnamefont {P.}~\bibnamefont
  {Richard}}, \bibinfo {author} {\bibfnamefont {S.~K.}\ \bibnamefont
  {de~Richter}}, \bibinfo {author} {\bibfnamefont {G.}~\bibnamefont {Le~Caer}},
  \ and\ \bibinfo {author} {\bibfnamefont {R.}~\bibnamefont {Delannay}}}
  (\bibinfo {year} {2009}{\natexlab{b}}),\ \enquote {\bibinfo {title}
  {Overlapping histogram method for testing edward's statistical mechanics of
  powders},}\ in\ \href@noop {} {\emph {\bibinfo {booktitle} {Powders and
  Grains 2009}}},\ \bibinfo {series} {AIP Conference Proceedings}, Vol.\
  \bibinfo {volume} {1145},\ \bibinfo {editor} {edited by\ \bibinfo {editor}
  {\bibfnamefont {M.}~\bibnamefont {Nakagawa}}\ and\ \bibinfo {editor}
  {\bibfnamefont {S.}~\bibnamefont {Luding}}}\ (\bibinfo  {publisher} {AIP})\
  pp.\ \bibinfo {pages} {465--468}\BibitemShut {NoStop}%
\bibitem [{\citenamefont {Medvedev}\ and\ \citenamefont
  {Naberukhin}(1987)}]{Medvedev:1987aa}%
  \BibitemOpen
  \bibfield  {author} {\bibinfo {author} {\bibnamefont {Medvedev},
  \bibfnamefont {N.}}, \ and\ \bibinfo {author} {\bibfnamefont
  {Y.}~\bibnamefont {Naberukhin}}} (\bibinfo {year} {1987}),\ \href {\doibase
  http://dx.doi.org/10.1016/S0022-3093(87)80074-1} {\bibfield  {journal}
  {\bibinfo  {journal} {J. Non-Crystall. Solids}\ }\textbf {\bibinfo {volume}
  {94}},\ \bibinfo {pages} {402 }}\BibitemShut {NoStop}%
\bibitem [{\citenamefont {van Meel}\ \emph
  {et~al.}(2009{\natexlab{a}})\citenamefont {van Meel}, \citenamefont
  {Charbonneau}, \citenamefont {Fortini},\ and\ \citenamefont
  {Charbonneau}}]{Meel:2009ab}%
  \BibitemOpen
  \bibfield  {author} {\bibinfo {author} {\bibnamefont {van Meel},
  \bibfnamefont {J.~A.}}, \bibinfo {author} {\bibfnamefont {B.}~\bibnamefont
  {Charbonneau}}, \bibinfo {author} {\bibfnamefont {A.}~\bibnamefont
  {Fortini}}, \ and\ \bibinfo {author} {\bibfnamefont {P.}~\bibnamefont
  {Charbonneau}}} (\bibinfo {year} {2009}{\natexlab{a}}),\ \href {\doibase
  10.1103/PhysRevE.80.061110} {\bibfield  {journal} {\bibinfo  {journal} {Phys.
  Rev. E}\ }\textbf {\bibinfo {volume} {80}},\ \bibinfo {pages}
  {061110}}\BibitemShut {NoStop}%
\bibitem [{\citenamefont {van Meel}\ \emph
  {et~al.}(2009{\natexlab{b}})\citenamefont {van Meel}, \citenamefont
  {Frenkel},\ and\ \citenamefont {Charbonneau}}]{Meel:2009aa}%
  \BibitemOpen
  \bibfield  {author} {\bibinfo {author} {\bibnamefont {van Meel},
  \bibfnamefont {J.~A.}}, \bibinfo {author} {\bibfnamefont {D.}~\bibnamefont
  {Frenkel}}, \ and\ \bibinfo {author} {\bibfnamefont {P.}~\bibnamefont
  {Charbonneau}}} (\bibinfo {year} {2009}{\natexlab{b}}),\ \href {\doibase
  10.1103/PhysRevE.79.030201} {\bibfield  {journal} {\bibinfo  {journal} {Phys.
  Rev. E}\ }\textbf {\bibinfo {volume} {79}},\ \bibinfo {pages}
  {030201}}\BibitemShut {NoStop}%
\bibitem [{\citenamefont {Mehta}\ and\ \citenamefont
  {Edwards}(1990)}]{Mehta:1990aa}%
  \BibitemOpen
  \bibfield  {author} {\bibinfo {author} {\bibnamefont {Mehta}, \bibfnamefont
  {A.}}, \ and\ \bibinfo {author} {\bibfnamefont {S.}~\bibnamefont {Edwards}}}
  (\bibinfo {year} {1990}),\ \href {\doibase 10.1016/0378-4371(90)90026-O}
  {\bibfield  {journal} {\bibinfo  {journal} {Physica A}\ }\textbf {\bibinfo
  {volume} {168}},\ \bibinfo {pages} {714}}\BibitemShut {NoStop}%
\bibitem [{\citenamefont {Metzger}(2004)}]{Metzger:2004aa}%
  \BibitemOpen
  \bibfield  {author} {\bibinfo {author} {\bibnamefont {Metzger}, \bibfnamefont
  {P.}}} (\bibinfo {year} {2004}),\ \href {\doibase 10.1103/PhysRevE.70.051303}
  {\bibfield  {journal} {\bibinfo  {journal} {Phys. Rev. E}\ }\textbf {\bibinfo
  {volume} {70}},\ \bibinfo {pages} {051303}}\BibitemShut {NoStop}%
\bibitem [{\citenamefont {Metzger}\ and\ \citenamefont
  {Donahue}(2005)}]{Metzger:2005aa}%
  \BibitemOpen
  \bibfield  {author} {\bibinfo {author} {\bibnamefont {Metzger}, \bibfnamefont
  {P.}}, \ and\ \bibinfo {author} {\bibfnamefont {C.}~\bibnamefont {Donahue}}}
  (\bibinfo {year} {2005}),\ \href {\doibase 10.1103/PhysRevLett.94.148001}
  {\bibfield  {journal} {\bibinfo  {journal} {Phys. Rev. Lett.}\ }\textbf
  {\bibinfo {volume} {94}},\ \bibinfo {pages} {148001}}\BibitemShut {NoStop}%
\bibitem [{\citenamefont {Meyer}\ \emph {et~al.}(2010)\citenamefont {Meyer},
  \citenamefont {Song}, \citenamefont {Jin}, \citenamefont {Wang},\ and\
  \citenamefont {Makse}}]{Meyer:2010aa}%
  \BibitemOpen
  \bibfield  {author} {\bibinfo {author} {\bibnamefont {Meyer}, \bibfnamefont
  {S.}}, \bibinfo {author} {\bibfnamefont {C.}~\bibnamefont {Song}}, \bibinfo
  {author} {\bibfnamefont {Y.}~\bibnamefont {Jin}}, \bibinfo {author}
  {\bibfnamefont {K.}~\bibnamefont {Wang}}, \ and\ \bibinfo {author}
  {\bibfnamefont {H.~A.}\ \bibnamefont {Makse}}} (\bibinfo {year} {2010}),\
  \href {\doibase 10.1016/j.physa.2010.07.030} {\bibfield  {journal} {\bibinfo
  {journal} {Physica A}\ }\textbf {\bibinfo {volume} {389}},\ \bibinfo {pages}
  {5137}}\BibitemShut {NoStop}%
\bibitem [{\citenamefont {M\'ezard}\ and\ \citenamefont
  {Montanari}(2009)}]{Mezard:2009aa}%
  \BibitemOpen
  \bibfield  {author} {\bibinfo {author} {\bibnamefont {M\'ezard},
  \bibfnamefont {M.}}, \ and\ \bibinfo {author} {\bibfnamefont
  {A.}~\bibnamefont {Montanari}}} (\bibinfo {year} {2009}),\ \href@noop {}
  {\emph {\bibinfo {title} {Information, Physics, and Computation}}}\ (\bibinfo
   {publisher} {Oxford University Press})\BibitemShut {NoStop}%
\bibitem [{\citenamefont {M\'ezard}\ and\ \citenamefont
  {Parisi}(2001)}]{Mezard:2001aa}%
  \BibitemOpen
  \bibfield  {author} {\bibinfo {author} {\bibnamefont {M\'ezard},
  \bibfnamefont {M.}}, \ and\ \bibinfo {author} {\bibfnamefont
  {G.}~\bibnamefont {Parisi}}} (\bibinfo {year} {2001}),\ \href
  {http://dx.doi.org/10.1007/PL00011099} {\bibfield  {journal} {\bibinfo
  {journal} {Eur. Phys. J. B}\ }\textbf {\bibinfo {volume} {20}},\ \bibinfo
  {pages} {217}}\BibitemShut {NoStop}%
\bibitem [{\citenamefont {M{\'e}zard}\ and\ \citenamefont
  {Parisi}(2003)}]{Mezard:2003aa}%
  \BibitemOpen
  \bibfield  {author} {\bibinfo {author} {\bibnamefont {M{\'e}zard},
  \bibfnamefont {M.}}, \ and\ \bibinfo {author} {\bibfnamefont
  {G.}~\bibnamefont {Parisi}}} (\bibinfo {year} {2003}),\ \href {\doibase
  10.1023/A:1022221005097} {\bibfield  {journal} {\bibinfo  {journal} {Journal
  of Statistical Physics}\ }\textbf {\bibinfo {volume} {111}}~(\bibinfo
  {number} {1}),\ \bibinfo {pages} {1}}\BibitemShut {NoStop}%
\bibitem [{\citenamefont {Mindlin}(1949)}]{Mindlin:1949aa}%
  \BibitemOpen
  \bibfield  {author} {\bibinfo {author} {\bibnamefont {Mindlin}, \bibfnamefont
  {R.~D.}}} (\bibinfo {year} {1949}),\ \href@noop {} {\bibfield  {journal}
  {\bibinfo  {journal} {Journal of Applied Mechanics (ASME)}\ }\textbf
  {\bibinfo {volume} {71}},\ \bibinfo {pages} {259}}\BibitemShut {NoStop}%
\bibitem [{\citenamefont {Miskin}\ and\ \citenamefont
  {Jaeger}(2013)}]{Miskin:2013aa}%
  \BibitemOpen
  \bibfield  {author} {\bibinfo {author} {\bibnamefont {Miskin}, \bibfnamefont
  {M.~Z.}}, \ and\ \bibinfo {author} {\bibfnamefont {H.~M.}\ \bibnamefont
  {Jaeger}}} (\bibinfo {year} {2013}),\ \href
  {http://dx.doi.org/10.1038/nmat3543} {\bibfield  {journal} {\bibinfo
  {journal} {Nature Mater.}\ }\textbf {\bibinfo {volume} {12}},\ \bibinfo
  {pages} {326}}\BibitemShut {NoStop}%
\bibitem [{\citenamefont {Miskin}\ and\ \citenamefont
  {Jaeger}(2014)}]{Miskin:2014aa}%
  \BibitemOpen
  \bibfield  {author} {\bibinfo {author} {\bibnamefont {Miskin}, \bibfnamefont
  {M.~Z.}}, \ and\ \bibinfo {author} {\bibfnamefont {H.~M.}\ \bibnamefont
  {Jaeger}}} (\bibinfo {year} {2014}),\ \href {\doibase 10.1039/C4SM00539B}
  {\bibfield  {journal} {\bibinfo  {journal} {Soft Matter}\ }\textbf {\bibinfo
  {volume} {10}},\ \bibinfo {pages} {3708}}\BibitemShut {NoStop}%
\bibitem [{\citenamefont {{Mizuno}}\ \emph {et~al.}(2017)\citenamefont
  {{Mizuno}}, \citenamefont {{Shiba}},\ and\ \citenamefont
  {{Ikeda}}}]{Mizuno:2017aa}%
  \BibitemOpen
  \bibfield  {author} {\bibinfo {author} {\bibnamefont {{Mizuno}},
  \bibfnamefont {H.}}, \bibinfo {author} {\bibfnamefont {H.}~\bibnamefont
  {{Shiba}}}, \ and\ \bibinfo {author} {\bibfnamefont {A.}~\bibnamefont
  {{Ikeda}}}} (\bibinfo {year} {2017}),\ \href@noop {} {\bibfield  {journal}
  {\bibinfo  {journal} {ArXiv e-prints}\ }}\Eprint
  {http://arxiv.org/abs/1703.10004} {arXiv:1703.10004 [cond-mat.soft]}
  \BibitemShut {NoStop}%
\bibitem [{\citenamefont {Monasson}(1995)}]{Monasson:1995aa}%
  \BibitemOpen
  \bibfield  {author} {\bibinfo {author} {\bibnamefont {Monasson},
  \bibfnamefont {R.}}} (\bibinfo {year} {1995}),\ \href {\doibase
  10.1103/PhysRevLett.75.2847} {\bibfield  {journal} {\bibinfo  {journal}
  {Phys. Rev. Lett.}\ }\textbf {\bibinfo {volume} {75}},\ \bibinfo {pages}
  {2847}}\BibitemShut {NoStop}%
\bibitem [{\citenamefont {Moukarzel}(1998)}]{Moukarzel:1998ab}%
  \BibitemOpen
  \bibfield  {author} {\bibinfo {author} {\bibnamefont {Moukarzel},
  \bibfnamefont {C.~F.}}} (\bibinfo {year} {1998}),\ \href {\doibase
  10.1103/PhysRevLett.81.1634} {\bibfield  {journal} {\bibinfo  {journal}
  {Phys. Rev. Lett.}\ }\textbf {\bibinfo {volume} {81}},\ \bibinfo {pages}
  {1634}}\BibitemShut {NoStop}%
\bibitem [{\citenamefont {Mounfield}\ and\ \citenamefont
  {Edwards}(1994)}]{Mounfield:1994aa}%
  \BibitemOpen
  \bibfield  {author} {\bibinfo {author} {\bibnamefont {Mounfield},
  \bibfnamefont {C.}}, \ and\ \bibinfo {author} {\bibfnamefont
  {S.}~\bibnamefont {Edwards}}} (\bibinfo {year} {1994}),\ \href {\doibase
  10.1016/0378-4371(94)90076-0} {\bibfield  {journal} {\bibinfo  {journal}
  {Physica A}\ }\textbf {\bibinfo {volume} {210}},\ \bibinfo {pages}
  {279}}\BibitemShut {NoStop}%
\bibitem [{\citenamefont {Mueth}\ \emph {et~al.}(1998)\citenamefont {Mueth},
  \citenamefont {Jaeger},\ and\ \citenamefont {Nagel}}]{Mueth:1998aa}%
  \BibitemOpen
  \bibfield  {author} {\bibinfo {author} {\bibnamefont {Mueth}, \bibfnamefont
  {D.~M.}}, \bibinfo {author} {\bibfnamefont {H.~M.}\ \bibnamefont {Jaeger}}, \
  and\ \bibinfo {author} {\bibfnamefont {S.~R.}\ \bibnamefont {Nagel}}}
  (\bibinfo {year} {1998}),\ \href {\doibase 10.1103/PhysRevE.57.3164}
  {\bibfield  {journal} {\bibinfo  {journal} {Phys. Rev. E}\ }\textbf {\bibinfo
  {volume} {57}},\ \bibinfo {pages} {3164}}\BibitemShut {NoStop}%
\bibitem [{\citenamefont {M{\"u}ller}\ and\ \citenamefont
  {Wyart}(2015)}]{Muller:2015aa}%
  \BibitemOpen
  \bibfield  {author} {\bibinfo {author} {\bibnamefont {M{\"u}ller},
  \bibfnamefont {M.}}, \ and\ \bibinfo {author} {\bibfnamefont
  {M.}~\bibnamefont {Wyart}}} (\bibinfo {year} {2015}),\ \href {\doibase
  10.1146/annurev-conmatphys-031214-014614} {\bibfield  {journal} {\bibinfo
  {journal} {Annual Review of Condensed Matter Physics}\ }\textbf {\bibinfo
  {volume} {6}}~(\bibinfo {number} {1}),\ \bibinfo {pages} {177}}\BibitemShut
  {NoStop}%
\bibitem [{\citenamefont {Neudecker}\ \emph {et~al.}(2013)\citenamefont
  {Neudecker}, \citenamefont {Ulrich}, \citenamefont {Herminghaus},\ and\
  \citenamefont {Schr\"oter}}]{Neudecker:2013aa}%
  \BibitemOpen
  \bibfield  {author} {\bibinfo {author} {\bibnamefont {Neudecker},
  \bibfnamefont {M.}}, \bibinfo {author} {\bibfnamefont {S.}~\bibnamefont
  {Ulrich}}, \bibinfo {author} {\bibfnamefont {S.}~\bibnamefont {Herminghaus}},
  \ and\ \bibinfo {author} {\bibfnamefont {M.}~\bibnamefont {Schr\"oter}}}
  (\bibinfo {year} {2013}),\ \href {\doibase 10.1103/PhysRevLett.111.028001}
  {\bibfield  {journal} {\bibinfo  {journal} {Phys. Rev. Lett.}\ }\textbf
  {\bibinfo {volume} {111}},\ \bibinfo {pages} {028001}}\BibitemShut {NoStop}%
\bibitem [{\citenamefont {Newhall}\ \emph {et~al.}(2011)\citenamefont
  {Newhall}, \citenamefont {Jorjadze}, \citenamefont {Vanden-Eijnden},\ and\
  \citenamefont {Bruji\'c}}]{Newhall:2011aa}%
  \BibitemOpen
  \bibfield  {author} {\bibinfo {author} {\bibnamefont {Newhall}, \bibfnamefont
  {K.~A.}}, \bibinfo {author} {\bibfnamefont {I.}~\bibnamefont {Jorjadze}},
  \bibinfo {author} {\bibfnamefont {E.}~\bibnamefont {Vanden-Eijnden}}, \ and\
  \bibinfo {author} {\bibfnamefont {J.}~\bibnamefont {Bruji\'c}}} (\bibinfo
  {year} {2011}),\ \href {\doibase 10.1039/c1sm06243c} {\bibfield  {journal}
  {\bibinfo  {journal} {Soft Matter}\ }\textbf {\bibinfo {volume} {7}},\
  \bibinfo {pages} {11518}}\BibitemShut {NoStop}%
\bibitem [{\citenamefont {Newman}\ and\ \citenamefont
  {Stein}(1999)}]{Newman:1999aa}%
  \BibitemOpen
  \bibfield  {author} {\bibinfo {author} {\bibnamefont {Newman}, \bibfnamefont
  {C.~M.}}, \ and\ \bibinfo {author} {\bibfnamefont {D.~L.}\ \bibnamefont
  {Stein}}} (\bibinfo {year} {1999}),\ \href {\doibase
  10.1103/PhysRevE.60.5244} {\bibfield  {journal} {\bibinfo  {journal} {Phys.
  Rev. E}\ }\textbf {\bibinfo {volume} {60}},\ \bibinfo {pages}
  {5244}}\BibitemShut {NoStop}%
\bibitem [{\citenamefont {Ngan}(2004)}]{Ngan:2004aa}%
  \BibitemOpen
  \bibfield  {author} {\bibinfo {author} {\bibnamefont {Ngan}, \bibfnamefont
  {A.}}} (\bibinfo {year} {2004}),\ \href {\doibase
  http://dx.doi.org/10.1016/j.physa.2004.03.068} {\bibfield  {journal}
  {\bibinfo  {journal} {Physica A}\ }\textbf {\bibinfo {volume} {339}},\
  \bibinfo {pages} {207 }}\BibitemShut {NoStop}%
\bibitem [{\citenamefont {Ngan}(2003)}]{Ngan:2003aa}%
  \BibitemOpen
  \bibfield  {author} {\bibinfo {author} {\bibnamefont {Ngan}, \bibfnamefont
  {A.~H.~W.}}} (\bibinfo {year} {2003}),\ \href {\doibase
  10.1103/PhysRevE.68.011301} {\bibfield  {journal} {\bibinfo  {journal} {Phys.
  Rev. E}\ }\textbf {\bibinfo {volume} {68}},\ \bibinfo {pages}
  {011301}}\BibitemShut {NoStop}%
\bibitem [{\citenamefont {Ni}\ \emph {et~al.}(2013)\citenamefont {Ni},
  \citenamefont {Stuart},\ and\ \citenamefont {Dijkstra}}]{Ni:2013aa}%
  \BibitemOpen
  \bibfield  {author} {\bibinfo {author} {\bibnamefont {Ni}, \bibfnamefont
  {R.}}, \bibinfo {author} {\bibfnamefont {M.~A.~C.}\ \bibnamefont {Stuart}}, \
  and\ \bibinfo {author} {\bibfnamefont {M.}~\bibnamefont {Dijkstra}}}
  (\bibinfo {year} {2013}),\ \href {http://dx.doi.org/10.1038/ncomms3704}
  {\bibfield  {journal} {\bibinfo  {journal} {Nature Communications}\ }\textbf
  {\bibinfo {volume} {4}},\ \bibinfo {pages} {2704}}\BibitemShut {NoStop}%
\bibitem [{\citenamefont {Nicodemi}(1999)}]{Nicodemi:1999aa}%
  \BibitemOpen
  \bibfield  {author} {\bibinfo {author} {\bibnamefont {Nicodemi},
  \bibfnamefont {M.}}} (\bibinfo {year} {1999}),\ \href {\doibase
  10.1103/PhysRevLett.82.3734} {\bibfield  {journal} {\bibinfo  {journal}
  {Phys. Rev. Lett.}\ }\textbf {\bibinfo {volume} {82}},\ \bibinfo {pages}
  {3734}}\BibitemShut {NoStop}%
\bibitem [{\citenamefont {Nicodemi}\ \emph {et~al.}(2004)\citenamefont
  {Nicodemi}, \citenamefont {Coniglio}, \citenamefont {de~Candia},
  \citenamefont {Fierro}, \citenamefont {Ciamarra},\ and\ \citenamefont
  {Tarzia}}]{Nicodemi:2004aa}%
  \BibitemOpen
  \bibfield  {author} {\bibinfo {author} {\bibnamefont {Nicodemi},
  \bibfnamefont {M.}}, \bibinfo {author} {\bibfnamefont {A.}~\bibnamefont
  {Coniglio}}, \bibinfo {author} {\bibfnamefont {A.}~\bibnamefont {de~Candia}},
  \bibinfo {author} {\bibfnamefont {A.}~\bibnamefont {Fierro}}, \bibinfo
  {author} {\bibfnamefont {M.}~\bibnamefont {Ciamarra}}, \ and\ \bibinfo
  {author} {\bibfnamefont {M.}~\bibnamefont {Tarzia}}} (\bibinfo {year}
  {2004}),\ \enquote {\bibinfo {title} {Statistical mechanics of jamming and
  segregation in granular media},}\ in\ \href {\doibase
  10.1016/B978-044451607-7/50005-4} {\emph {\bibinfo {booktitle} {Unifying
  Concepts in Granular Media and Glasses}}},\ \bibinfo {editor} {edited by\
  \bibinfo {editor} {\bibfnamefont {A.}~\bibnamefont {Coniglio}}, \bibinfo
  {editor} {\bibfnamefont {A.}~\bibnamefont {Fierro}}, \bibinfo {editor}
  {\bibfnamefont {H.}~\bibnamefont {Herrmann}}, \ and\ \bibinfo {editor}
  {\bibfnamefont {M.}~\bibnamefont {Nicodemi}}}\ (\bibinfo  {publisher}
  {Elsevier})\ pp.\ \bibinfo {pages} {47--61}\BibitemShut {NoStop}%
\bibitem [{\citenamefont {Nicodemi}\ \emph
  {et~al.}(1997{\natexlab{a}})\citenamefont {Nicodemi}, \citenamefont
  {Coniglio},\ and\ \citenamefont {Herrmann}}]{Nicodemi:1997ac}%
  \BibitemOpen
  \bibfield  {author} {\bibinfo {author} {\bibnamefont {Nicodemi},
  \bibfnamefont {M.}}, \bibinfo {author} {\bibfnamefont {A.}~\bibnamefont
  {Coniglio}}, \ and\ \bibinfo {author} {\bibfnamefont {H.}~\bibnamefont
  {Herrmann}}} (\bibinfo {year} {1997}{\natexlab{a}}),\ \href {\doibase
  10.1016/S0378-4371(97)00162-3} {\bibfield  {journal} {\bibinfo  {journal}
  {Physica A}\ }\textbf {\bibinfo {volume} {240}},\ \bibinfo {pages}
  {405}}\BibitemShut {NoStop}%
\bibitem [{\citenamefont {Nicodemi}\ \emph
  {et~al.}(1997{\natexlab{b}})\citenamefont {Nicodemi}, \citenamefont
  {Coniglio},\ and\ \citenamefont {Herrmann}}]{Nicodemi:1997ab}%
  \BibitemOpen
  \bibfield  {author} {\bibinfo {author} {\bibnamefont {Nicodemi},
  \bibfnamefont {M.}}, \bibinfo {author} {\bibfnamefont {A.}~\bibnamefont
  {Coniglio}}, \ and\ \bibinfo {author} {\bibfnamefont {H.~J.}\ \bibnamefont
  {Herrmann}}} (\bibinfo {year} {1997}{\natexlab{b}}),\ \href
  {http://stacks.iop.org/0305-4470/30/i=11/a=006} {\bibfield  {journal}
  {\bibinfo  {journal} {J. Phys. A}\ }\textbf {\bibinfo {volume} {30}},\
  \bibinfo {pages} {L379}}\BibitemShut {NoStop}%
\bibitem [{\citenamefont {Nicodemi}\ \emph
  {et~al.}(1997{\natexlab{c}})\citenamefont {Nicodemi}, \citenamefont
  {Coniglio},\ and\ \citenamefont {Herrmann}}]{Nicodemi:1997aa}%
  \BibitemOpen
  \bibfield  {author} {\bibinfo {author} {\bibnamefont {Nicodemi},
  \bibfnamefont {M.}}, \bibinfo {author} {\bibfnamefont {A.}~\bibnamefont
  {Coniglio}}, \ and\ \bibinfo {author} {\bibfnamefont {H.~J.}\ \bibnamefont
  {Herrmann}}} (\bibinfo {year} {1997}{\natexlab{c}}),\ \href {\doibase
  10.1103/PhysRevE.55.3962} {\bibfield  {journal} {\bibinfo  {journal} {Phys.
  Rev. E}\ }\textbf {\bibinfo {volume} {55}},\ \bibinfo {pages}
  {3962}}\BibitemShut {NoStop}%
\bibitem [{\citenamefont {Nicodemi}\ \emph {et~al.}(1999)\citenamefont
  {Nicodemi}, \citenamefont {Coniglio},\ and\ \citenamefont
  {Herrmann}}]{Nicodemi:1999ab}%
  \BibitemOpen
  \bibfield  {author} {\bibinfo {author} {\bibnamefont {Nicodemi},
  \bibfnamefont {M.}}, \bibinfo {author} {\bibfnamefont {A.}~\bibnamefont
  {Coniglio}}, \ and\ \bibinfo {author} {\bibfnamefont {H.~J.}\ \bibnamefont
  {Herrmann}}} (\bibinfo {year} {1999}),\ \href {\doibase
  10.1103/PhysRevE.59.6830} {\bibfield  {journal} {\bibinfo  {journal} {Phys.
  Rev. E}\ }\textbf {\bibinfo {volume} {59}},\ \bibinfo {pages}
  {6830}}\BibitemShut {NoStop}%
\bibitem [{\citenamefont {Norris}\ and\ \citenamefont
  {Johnson}(1997)}]{Norris:1997aa}%
  \BibitemOpen
  \bibfield  {author} {\bibinfo {author} {\bibnamefont {Norris}, \bibfnamefont
  {A.~N.}}, \ and\ \bibinfo {author} {\bibfnamefont {D.~L.}\ \bibnamefont
  {Johnson}}} (\bibinfo {year} {1997}),\ \href@noop {} {\bibfield  {journal}
  {\bibinfo  {journal} {J. App. Mech.}\ }\textbf {\bibinfo {volume} {64}},\
  \bibinfo {pages} {39}}\BibitemShut {NoStop}%
\bibitem [{\citenamefont {Nowak}\ \emph {et~al.}(1998)\citenamefont {Nowak},
  \citenamefont {Knight}, \citenamefont {Ben-Naim}, \citenamefont {Jaeger},\
  and\ \citenamefont {Nagel}}]{Nowak:1998aa}%
  \BibitemOpen
  \bibfield  {author} {\bibinfo {author} {\bibnamefont {Nowak}, \bibfnamefont
  {E.}}, \bibinfo {author} {\bibfnamefont {J.}~\bibnamefont {Knight}}, \bibinfo
  {author} {\bibfnamefont {E.}~\bibnamefont {Ben-Naim}}, \bibinfo {author}
  {\bibfnamefont {H.}~\bibnamefont {Jaeger}}, \ and\ \bibinfo {author}
  {\bibfnamefont {S.}~\bibnamefont {Nagel}}} (\bibinfo {year} {1998}),\ \href
  {\doibase 10.1103/PhysRevE.57.1971} {\bibfield  {journal} {\bibinfo
  {journal} {Phys. Rev. E}\ }\textbf {\bibinfo {volume} {57}},\ \bibinfo
  {pages} {1971}}\BibitemShut {NoStop}%
\bibitem [{\citenamefont {Nowak}\ \emph {et~al.}(1997)\citenamefont {Nowak},
  \citenamefont {Knight}, \citenamefont {Povinelli}, \citenamefont {Jaeger},\
  and\ \citenamefont {Nagel}}]{Nowak:1997aa}%
  \BibitemOpen
  \bibfield  {author} {\bibinfo {author} {\bibnamefont {Nowak}, \bibfnamefont
  {E.}}, \bibinfo {author} {\bibfnamefont {J.}~\bibnamefont {Knight}}, \bibinfo
  {author} {\bibfnamefont {M.}~\bibnamefont {Povinelli}}, \bibinfo {author}
  {\bibfnamefont {H.}~\bibnamefont {Jaeger}}, \ and\ \bibinfo {author}
  {\bibfnamefont {S.}~\bibnamefont {Nagel}}} (\bibinfo {year} {1997}),\ \href
  {\doibase 10.1016/S0032-5910(97)03291-9} {\bibfield  {journal} {\bibinfo
  {journal} {Powd. Tech.}\ }\textbf {\bibinfo {volume} {94}},\ \bibinfo {pages}
  {79}}\BibitemShut {NoStop}%
\bibitem [{\citenamefont {O'Hern}\ \emph {et~al.}(2004)\citenamefont {O'Hern},
  \citenamefont {Liu},\ and\ \citenamefont {Nagel}}]{OHern:2004aa}%
  \BibitemOpen
  \bibfield  {author} {\bibinfo {author} {\bibnamefont {O'Hern}, \bibfnamefont
  {C.}}, \bibinfo {author} {\bibfnamefont {A.}~\bibnamefont {Liu}}, \ and\
  \bibinfo {author} {\bibfnamefont {S.}~\bibnamefont {Nagel}}} (\bibinfo {year}
  {2004}),\ \href {\doibase 10.1103/PhysRevLett.93.165702} {\bibfield
  {journal} {\bibinfo  {journal} {Phys. Rev. Lett.}\ }\textbf {\bibinfo
  {volume} {93}},\ \bibinfo {pages} {165702}}\BibitemShut {NoStop}%
\bibitem [{\citenamefont {O'Hern}\ \emph {et~al.}(2003)\citenamefont {O'Hern},
  \citenamefont {Silbert}, \citenamefont {Liu},\ and\ \citenamefont
  {Nagel}}]{OHern:2003aa}%
  \BibitemOpen
  \bibfield  {author} {\bibinfo {author} {\bibnamefont {O'Hern}, \bibfnamefont
  {C.}}, \bibinfo {author} {\bibfnamefont {L.}~\bibnamefont {Silbert}},
  \bibinfo {author} {\bibfnamefont {A.}~\bibnamefont {Liu}}, \ and\ \bibinfo
  {author} {\bibfnamefont {S.}~\bibnamefont {Nagel}}} (\bibinfo {year}
  {2003}),\ \href {\doibase 10.1103/PhysRevE.68.011306} {\bibfield  {journal}
  {\bibinfo  {journal} {Phys. Rev. E}\ }\textbf {\bibinfo {volume} {68}},\
  \bibinfo {pages} {011306}}\BibitemShut {NoStop}%
\bibitem [{\citenamefont {O'Hern}\ \emph {et~al.}(2001)\citenamefont {O'Hern},
  \citenamefont {Langer}, \citenamefont {Liu},\ and\ \citenamefont
  {Nagel}}]{OHern:2001aa}%
  \BibitemOpen
  \bibfield  {author} {\bibinfo {author} {\bibnamefont {O'Hern}, \bibfnamefont
  {C.~S.}}, \bibinfo {author} {\bibfnamefont {S.~A.}\ \bibnamefont {Langer}},
  \bibinfo {author} {\bibfnamefont {A.~J.}\ \bibnamefont {Liu}}, \ and\
  \bibinfo {author} {\bibfnamefont {S.~R.}\ \bibnamefont {Nagel}}} (\bibinfo
  {year} {2001}),\ \href {\doibase 10.1103/PhysRevLett.86.111} {\bibfield
  {journal} {\bibinfo  {journal} {Phys. Rev. Lett.}\ }\textbf {\bibinfo
  {volume} {86}},\ \bibinfo {pages} {111}}\BibitemShut {NoStop}%
\bibitem [{\citenamefont {O'Hern}\ \emph {et~al.}(2002)\citenamefont {O'Hern},
  \citenamefont {Langer}, \citenamefont {Liu},\ and\ \citenamefont
  {Nagel}}]{OHern:2002aa}%
  \BibitemOpen
  \bibfield  {author} {\bibinfo {author} {\bibnamefont {O'Hern}, \bibfnamefont
  {C.~S.}}, \bibinfo {author} {\bibfnamefont {S.~A.}\ \bibnamefont {Langer}},
  \bibinfo {author} {\bibfnamefont {A.~J.}\ \bibnamefont {Liu}}, \ and\
  \bibinfo {author} {\bibfnamefont {S.~R.}\ \bibnamefont {Nagel}}} (\bibinfo
  {year} {2002}),\ \href {\doibase 10.1103/PhysRevLett.88.075507} {\bibfield
  {journal} {\bibinfo  {journal} {Phys. Rev. Lett.}\ }\textbf {\bibinfo
  {volume} {88}},\ \bibinfo {pages} {075507}}\BibitemShut {NoStop}%
\bibitem [{\citenamefont {Okabe}\ \emph {et~al.}(2000)\citenamefont {Okabe},
  \citenamefont {Boots}, \citenamefont {Sugihara},\ and\ \citenamefont
  {Nok~Chiu}}]{Okabe:2000aa}%
  \BibitemOpen
  \bibfield  {author} {\bibinfo {author} {\bibnamefont {Okabe}, \bibfnamefont
  {A.}}, \bibinfo {author} {\bibfnamefont {B.}~\bibnamefont {Boots}}, \bibinfo
  {author} {\bibfnamefont {K.}~\bibnamefont {Sugihara}}, \ and\ \bibinfo
  {author} {\bibfnamefont {S.}~\bibnamefont {Nok~Chiu}}} (\bibinfo {year}
  {2000}),\ \href@noop {} {\emph {\bibinfo {title} {Spatial Tessellations:
  Concepts and Applications of Voronoi Diagrams}}}\ (\bibinfo  {publisher}
  {Wiley-Blackwell})\BibitemShut {NoStop}%
\bibitem [{\citenamefont {Olsson}\ and\ \citenamefont
  {Teitel}(2007)}]{Olsson:2007aa}%
  \BibitemOpen
  \bibfield  {author} {\bibinfo {author} {\bibnamefont {Olsson}, \bibfnamefont
  {P.}}, \ and\ \bibinfo {author} {\bibfnamefont {S.}~\bibnamefont {Teitel}}}
  (\bibinfo {year} {2007}),\ \href {\doibase 10.1103/PhysRevLett.99.178001}
  {\bibfield  {journal} {\bibinfo  {journal} {Phys. Rev. Lett.}\ }\textbf
  {\bibinfo {volume} {99}},\ \bibinfo {pages} {178001}}\BibitemShut {NoStop}%
\bibitem [{\citenamefont {Ono}\ \emph {et~al.}(2002)\citenamefont {Ono},
  \citenamefont {O'Hern}, \citenamefont {Durian}, \citenamefont {Langer},
  \citenamefont {Liu},\ and\ \citenamefont {Nagel}}]{Ono:2002aa}%
  \BibitemOpen
  \bibfield  {author} {\bibinfo {author} {\bibnamefont {Ono}, \bibfnamefont
  {I.}}, \bibinfo {author} {\bibfnamefont {C.}~\bibnamefont {O'Hern}}, \bibinfo
  {author} {\bibfnamefont {D.}~\bibnamefont {Durian}}, \bibinfo {author}
  {\bibfnamefont {S.}~\bibnamefont {Langer}}, \bibinfo {author} {\bibfnamefont
  {A.}~\bibnamefont {Liu}}, \ and\ \bibinfo {author} {\bibfnamefont
  {S.}~\bibnamefont {Nagel}}} (\bibinfo {year} {2002}),\ \href {\doibase
  10.1103/PhysRevLett.89.095703} {\bibfield  {journal} {\bibinfo  {journal}
  {Phys. Rev. Lett.}\ }\textbf {\bibinfo {volume} {89}},\ \bibinfo {pages}
  {095703}}\BibitemShut {NoStop}%
\bibitem [{\citenamefont {Onoda}\ and\ \citenamefont
  {Liniger}(1990)}]{Onoda:1990aa}%
  \BibitemOpen
  \bibfield  {author} {\bibinfo {author} {\bibnamefont {Onoda}, \bibfnamefont
  {G.~Y.}}, \ and\ \bibinfo {author} {\bibfnamefont {E.~G.}\ \bibnamefont
  {Liniger}}} (\bibinfo {year} {1990}),\ \href {\doibase
  10.1103/PhysRevLett.64.2727} {\bibfield  {journal} {\bibinfo  {journal}
  {Phys. Rev. Lett.}\ }\textbf {\bibinfo {volume} {64}},\ \bibinfo {pages}
  {2727}}\BibitemShut {NoStop}%
\bibitem [{\citenamefont {Onsager}(1949)}]{Onsager:1949aa}%
  \BibitemOpen
  \bibfield  {author} {\bibinfo {author} {\bibnamefont {Onsager}, \bibfnamefont
  {L.}}} (\bibinfo {year} {1949}),\ \href@noop {} {\bibfield  {journal}
  {\bibinfo  {journal} {Ann. N. Y. Acad. Sci.}\ }\textbf {\bibinfo {volume}
  {51}},\ \bibinfo {pages} {627}}\BibitemShut {NoStop}%
\bibitem [{\citenamefont {Oquendo}\ \emph {et~al.}(2016)\citenamefont
  {Oquendo}, \citenamefont {Mu{\~n}oz},\ and\ \citenamefont
  {Radjai}}]{Oquendo:2016aa}%
  \BibitemOpen
  \bibfield  {author} {\bibinfo {author} {\bibnamefont {Oquendo}, \bibfnamefont
  {W.~F.}}, \bibinfo {author} {\bibfnamefont {J.~D.}\ \bibnamefont
  {Mu{\~n}oz}}, \ and\ \bibinfo {author} {\bibfnamefont {F.}~\bibnamefont
  {Radjai}}} (\bibinfo {year} {2016}),\ \href
  {http://stacks.iop.org/0295-5075/114/i=1/a=14004} {\bibfield  {journal}
  {\bibinfo  {journal} {EPL (Europhysics Letters)}\ }\textbf {\bibinfo {volume}
  {114}}~(\bibinfo {number} {1}),\ \bibinfo {pages} {14004}}\BibitemShut
  {NoStop}%
\bibitem [{\citenamefont {Oron}\ and\ \citenamefont
  {Herrmann}(1999)}]{Oron:1999aa}%
  \BibitemOpen
  \bibfield  {author} {\bibinfo {author} {\bibnamefont {Oron}, \bibfnamefont
  {G.}}, \ and\ \bibinfo {author} {\bibfnamefont {H.}~\bibnamefont {Herrmann}}}
  (\bibinfo {year} {1999}),\ \href {\doibase
  http://dx.doi.org/10.1016/S0378-4371(98)00648-7} {\bibfield  {journal}
  {\bibinfo  {journal} {Physica A}\ }\textbf {\bibinfo {volume} {265}},\
  \bibinfo {pages} {455 }}\BibitemShut {NoStop}%
\bibitem [{\citenamefont {Oron}\ and\ \citenamefont
  {Herrmann}(1998)}]{Oron:1998aa}%
  \BibitemOpen
  \bibfield  {author} {\bibinfo {author} {\bibnamefont {Oron}, \bibfnamefont
  {G.}}, \ and\ \bibinfo {author} {\bibfnamefont {H.~J.}\ \bibnamefont
  {Herrmann}}} (\bibinfo {year} {1998}),\ \href {\doibase
  10.1103/PhysRevE.58.2079} {\bibfield  {journal} {\bibinfo  {journal} {Phys.
  Rev. E}\ }\textbf {\bibinfo {volume} {58}},\ \bibinfo {pages}
  {2079}}\BibitemShut {NoStop}%
\bibitem [{\citenamefont {{Ozawa}}\ \emph {et~al.}(2017)\citenamefont
  {{Ozawa}}, \citenamefont {{Berthier}},\ and\ \citenamefont
  {{Coslovich}}}]{Ozawa:2017aa}%
  \BibitemOpen
  \bibfield  {author} {\bibinfo {author} {\bibnamefont {{Ozawa}}, \bibfnamefont
  {M.}}, \bibinfo {author} {\bibfnamefont {L.}~\bibnamefont {{Berthier}}}, \
  and\ \bibinfo {author} {\bibfnamefont {D.}~\bibnamefont {{Coslovich}}}}
  (\bibinfo {year} {2017}),\ \href@noop {} {\bibfield  {journal} {\bibinfo
  {journal} {ArXiv e-prints}\ }}\Eprint {http://arxiv.org/abs/1705.10156}
  {arXiv:1705.10156 [cond-mat.stat-mech]} \BibitemShut {NoStop}%
\bibitem [{\citenamefont {Ozawa}\ \emph {et~al.}(2012)\citenamefont {Ozawa},
  \citenamefont {Kuroiwa}, \citenamefont {Ikeda},\ and\ \citenamefont
  {Miyazaki}}]{Ozawa:2012aa}%
  \BibitemOpen
  \bibfield  {author} {\bibinfo {author} {\bibnamefont {Ozawa}, \bibfnamefont
  {M.}}, \bibinfo {author} {\bibfnamefont {T.}~\bibnamefont {Kuroiwa}},
  \bibinfo {author} {\bibfnamefont {A.}~\bibnamefont {Ikeda}}, \ and\ \bibinfo
  {author} {\bibfnamefont {K.}~\bibnamefont {Miyazaki}}} (\bibinfo {year}
  {2012}),\ \href {\doibase 10.1103/PhysRevLett.109.205701} {\bibfield
  {journal} {\bibinfo  {journal} {Phys. Rev. Lett.}\ }\textbf {\bibinfo
  {volume} {109}},\ \bibinfo {pages} {205701}}\BibitemShut {NoStop}%
\bibitem [{\citenamefont {Paillusson}(2015)}]{Paillusson:2015aa}%
  \BibitemOpen
  \bibfield  {author} {\bibinfo {author} {\bibnamefont {Paillusson},
  \bibfnamefont {F.}}} (\bibinfo {year} {2015}),\ \href {\doibase
  10.1103/PhysRevE.91.012204} {\bibfield  {journal} {\bibinfo  {journal} {Phys.
  Rev. E}\ }\textbf {\bibinfo {volume} {91}},\ \bibinfo {pages}
  {012204}}\BibitemShut {NoStop}%
\bibitem [{\citenamefont {Paillusson}\ and\ \citenamefont
  {Frenkel}(2012)}]{Paillusson:2012aa}%
  \BibitemOpen
  \bibfield  {author} {\bibinfo {author} {\bibnamefont {Paillusson},
  \bibfnamefont {F.}}, \ and\ \bibinfo {author} {\bibfnamefont
  {D.}~\bibnamefont {Frenkel}}} (\bibinfo {year} {2012}),\ \href {\doibase
  10.1103/PhysRevLett.109.208001} {\bibfield  {journal} {\bibinfo  {journal}
  {Phys. Rev. Lett.}\ }\textbf {\bibinfo {volume} {109}},\ \bibinfo {pages}
  {208001}}\BibitemShut {NoStop}%
\bibitem [{\citenamefont {Pal\'asti}(1960)}]{Palasti:1960aa}%
  \BibitemOpen
  \bibfield  {author} {\bibinfo {author} {\bibnamefont {Pal\'asti},
  \bibfnamefont {I.}}} (\bibinfo {year} {1960}),\ \href@noop {} {\bibfield
  {journal} {\bibinfo  {journal} {Publ. Math. Res. Inst. Hung. Acad. Sci.}\
  }\textbf {\bibinfo {volume} {5}},\ \bibinfo {pages} {353}}\BibitemShut
  {NoStop}%
\bibitem [{\citenamefont {Panaitescu}\ and\ \citenamefont
  {Kudrolli}(2014)}]{Panaitescu:2014aa}%
  \BibitemOpen
  \bibfield  {author} {\bibinfo {author} {\bibnamefont {Panaitescu},
  \bibfnamefont {A.}}, \ and\ \bibinfo {author} {\bibfnamefont
  {A.}~\bibnamefont {Kudrolli}}} (\bibinfo {year} {2014}),\ \href {\doibase
  10.1103/PhysRevE.90.032203} {\bibfield  {journal} {\bibinfo  {journal} {Phys.
  Rev. E}\ }\textbf {\bibinfo {volume} {90}},\ \bibinfo {pages}
  {032203}}\BibitemShut {NoStop}%
\bibitem [{\citenamefont {Papanikolaou}\ \emph {et~al.}(2013)\citenamefont
  {Papanikolaou}, \citenamefont {O'Hern},\ and\ \citenamefont
  {Shattuck}}]{Papanikolaou:2013aa}%
  \BibitemOpen
  \bibfield  {author} {\bibinfo {author} {\bibnamefont {Papanikolaou},
  \bibfnamefont {S.}}, \bibinfo {author} {\bibfnamefont {C.~S.}\ \bibnamefont
  {O'Hern}}, \ and\ \bibinfo {author} {\bibfnamefont {M.~D.}\ \bibnamefont
  {Shattuck}}} (\bibinfo {year} {2013}),\ \href {\doibase
  10.1103/PhysRevLett.110.198002} {\bibfield  {journal} {\bibinfo  {journal}
  {Phys. Rev. Lett.}\ }\textbf {\bibinfo {volume} {110}},\ \bibinfo {pages}
  {198002}}\BibitemShut {NoStop}%
\bibitem [{\citenamefont {Parisi}(1988)}]{Parisi:1988aa}%
  \BibitemOpen
  \bibfield  {author} {\bibinfo {author} {\bibnamefont {Parisi}, \bibfnamefont
  {G.}}} (\bibinfo {year} {1988}),\ \href@noop {} {\emph {\bibinfo {title}
  {Statistical Field Theory}}}\ (\bibinfo  {publisher}
  {Addison-Wesley})\BibitemShut {NoStop}%
\bibitem [{\citenamefont {Parisi}(2017)}]{Parisi:2017aa}%
  \BibitemOpen
  \bibfield  {author} {\bibinfo {author} {\bibnamefont {Parisi}, \bibfnamefont
  {G.}}} (\bibinfo {year} {2017}),\ \href {\doibase 10.1007/s10955-017-1724-z}
  {\bibfield  {journal} {\bibinfo  {journal} {Journal of Statistical Physics}\
  }\textbf {\bibinfo {volume} {167}}~(\bibinfo {number} {3}),\ \bibinfo {pages}
  {515}}\BibitemShut {NoStop}%
\bibitem [{\citenamefont {Parisi}\ and\ \citenamefont
  {Zamponi}(2005)}]{Parisi:2005aa}%
  \BibitemOpen
  \bibfield  {author} {\bibinfo {author} {\bibnamefont {Parisi}, \bibfnamefont
  {G.}}, \ and\ \bibinfo {author} {\bibfnamefont {F.}~\bibnamefont {Zamponi}}}
  (\bibinfo {year} {2005}),\ \href@noop {} {\bibfield  {journal} {\bibinfo
  {journal} {J. Chem. Phys.}\ }\textbf {\bibinfo {volume} {123}},\ \bibinfo
  {pages} {144501}}\BibitemShut {NoStop}%
\bibitem [{\citenamefont {Parisi}\ and\ \citenamefont
  {Zamponi}(2010)}]{Parisi:2010aa}%
  \BibitemOpen
  \bibfield  {author} {\bibinfo {author} {\bibnamefont {Parisi}, \bibfnamefont
  {G.}}, \ and\ \bibinfo {author} {\bibfnamefont {F.}~\bibnamefont {Zamponi}}}
  (\bibinfo {year} {2010}),\ \href {\doibase 10.1103/RevModPhys.82.789}
  {\bibfield  {journal} {\bibinfo  {journal} {Rev. Mod. Phys.}\ }\textbf
  {\bibinfo {volume} {82}},\ \bibinfo {pages} {789}}\BibitemShut {NoStop}%
\bibitem [{\citenamefont {Parteli}\ \emph {et~al.}(2014)\citenamefont
  {Parteli}, \citenamefont {Schmidt}, \citenamefont {Blumel}, \citenamefont
  {Wirth}, \citenamefont {Peukert},\ and\ \citenamefont
  {Poschel}}]{Parteli:2014aa}%
  \BibitemOpen
  \bibfield  {author} {\bibinfo {author} {\bibnamefont {Parteli}, \bibfnamefont
  {E.~J.~R.}}, \bibinfo {author} {\bibfnamefont {J.}~\bibnamefont {Schmidt}},
  \bibinfo {author} {\bibfnamefont {C.}~\bibnamefont {Blumel}}, \bibinfo
  {author} {\bibfnamefont {K.-E.}\ \bibnamefont {Wirth}}, \bibinfo {author}
  {\bibfnamefont {W.}~\bibnamefont {Peukert}}, \ and\ \bibinfo {author}
  {\bibfnamefont {T.}~\bibnamefont {Poschel}}} (\bibinfo {year} {2014}),\ \href
  {http://dx.doi.org/10.1038/srep06227} {\bibfield  {journal} {\bibinfo
  {journal} {Sci. Rep.}\ }\textbf {\bibinfo {volume} {4}}}\BibitemShut
  {NoStop}%
\bibitem [{\citenamefont {Pathria}\ and\ \citenamefont
  {Beale}(2011)}]{Pathria:2011aa}%
  \BibitemOpen
  \bibfield  {author} {\bibinfo {author} {\bibnamefont {Pathria}, \bibfnamefont
  {R.~K.}}, \ and\ \bibinfo {author} {\bibfnamefont {P.~D.}\ \bibnamefont
  {Beale}}} (\bibinfo {year} {2011}),\ \href@noop {} {\emph {\bibinfo {title}
  {Statistical Mechanics}}}\ (\bibinfo  {publisher} {Elsevier})\BibitemShut
  {NoStop}%
\bibitem [{\citenamefont {Philippe}\ and\ \citenamefont
  {Bideau}(2002)}]{Philippe:2002aa}%
  \BibitemOpen
  \bibfield  {author} {\bibinfo {author} {\bibnamefont {Philippe},
  \bibfnamefont {P.}}, \ and\ \bibinfo {author} {\bibfnamefont
  {D.}~\bibnamefont {Bideau}}} (\bibinfo {year} {2002}),\ \href {\doibase
  10.1209/epl/i2002-00362-7} {\bibfield  {journal} {\bibinfo  {journal}
  {Europhys. Lett.}\ }\textbf {\bibinfo {volume} {60}},\ \bibinfo {pages}
  {677}}\BibitemShut {NoStop}%
\bibitem [{\citenamefont {Philipse}(1996)}]{Philipse:1996aa}%
  \BibitemOpen
  \bibfield  {author} {\bibinfo {author} {\bibnamefont {Philipse},
  \bibfnamefont {A.}}} (\bibinfo {year} {1996}),\ \href@noop {} {\bibfield
  {journal} {\bibinfo  {journal} {Langmuir}\ }\textbf {\bibinfo {volume}
  {12}},\ \bibinfo {pages} {1127}}\BibitemShut {NoStop}%
\bibitem [{\citenamefont {Phillips}\ \emph {et~al.}(2012)\citenamefont
  {Phillips}, \citenamefont {Anderson}, \citenamefont {Huber},\ and\
  \citenamefont {Glotzer}}]{Phillips:2012aa}%
  \BibitemOpen
  \bibfield  {author} {\bibinfo {author} {\bibnamefont {Phillips},
  \bibfnamefont {C.~L.}}, \bibinfo {author} {\bibfnamefont {J.~A.}\
  \bibnamefont {Anderson}}, \bibinfo {author} {\bibfnamefont {G.}~\bibnamefont
  {Huber}}, \ and\ \bibinfo {author} {\bibfnamefont {S.~C.}\ \bibnamefont
  {Glotzer}}} (\bibinfo {year} {2012}),\ \href@noop {} {\bibfield  {journal}
  {\bibinfo  {journal} {Phys. Rev. Lett.}\ }\textbf {\bibinfo {volume} {108}},\
  \bibinfo {pages} {198304}}\BibitemShut {NoStop}%
\bibitem [{\citenamefont {Portal}\ \emph {et~al.}(2013)\citenamefont {Portal},
  \citenamefont {Danisch}, \citenamefont {Baule}, \citenamefont {Mari},\ and\
  \citenamefont {Makse}}]{Portal:2013aa}%
  \BibitemOpen
  \bibfield  {author} {\bibinfo {author} {\bibnamefont {Portal}, \bibfnamefont
  {L.}}, \bibinfo {author} {\bibfnamefont {M.}~\bibnamefont {Danisch}},
  \bibinfo {author} {\bibfnamefont {A.}~\bibnamefont {Baule}}, \bibinfo
  {author} {\bibfnamefont {R.}~\bibnamefont {Mari}}, \ and\ \bibinfo {author}
  {\bibfnamefont {H.~A.}\ \bibnamefont {Makse}}} (\bibinfo {year} {2013}),\
  \href@noop {} {\bibinfo  {journal} {J. Stat. Mech.}\ ,\ \bibinfo {pages}
  {P11009}}\BibitemShut {NoStop}%
\bibitem [{\citenamefont {Potiguar}\ and\ \citenamefont
  {Makse}(2006)}]{Potiguar:2006aa}%
  \BibitemOpen
\bibfield  {journal} {  }\bibfield  {author} {\bibinfo {author} {\bibnamefont
  {Potiguar}, \bibfnamefont {F.}}, \ and\ \bibinfo {author} {\bibfnamefont
  {H.}~\bibnamefont {Makse}}} (\bibinfo {year} {2006}),\ \href {\doibase
  10.1140/epje/e2006-00017-4} {\bibfield  {journal} {\bibinfo  {journal} {Eur.
  Phys. J. E}\ }\textbf {\bibinfo {volume} {19}},\ \bibinfo {pages}
  {171}}\BibitemShut {NoStop}%
\bibitem [{\citenamefont {Pouliquen}\ \emph {et~al.}(1997)\citenamefont
  {Pouliquen}, \citenamefont {Nicolas},\ and\ \citenamefont
  {Weidman}}]{Pouliquen:1997aa}%
  \BibitemOpen
  \bibfield  {author} {\bibinfo {author} {\bibnamefont {Pouliquen},
  \bibfnamefont {O.}}, \bibinfo {author} {\bibfnamefont {M.}~\bibnamefont
  {Nicolas}}, \ and\ \bibinfo {author} {\bibfnamefont {P.~D.}\ \bibnamefont
  {Weidman}}} (\bibinfo {year} {1997}),\ \href {\doibase
  10.1103/PhysRevLett.79.3640} {\bibfield  {journal} {\bibinfo  {journal}
  {Phys. Rev. Lett.}\ }\textbf {\bibinfo {volume} {79}},\ \bibinfo {pages}
  {3640}}\BibitemShut {NoStop}%
\bibitem [{\citenamefont {Prados}\ \emph {et~al.}(2000)\citenamefont {Prados},
  \citenamefont {Brey},\ and\ \citenamefont {Sanchez-Rey}}]{Prados:2000aa}%
  \BibitemOpen
  \bibfield  {author} {\bibinfo {author} {\bibnamefont {Prados}, \bibfnamefont
  {A.}}, \bibinfo {author} {\bibfnamefont {J.}~\bibnamefont {Brey}}, \ and\
  \bibinfo {author} {\bibfnamefont {B.}~\bibnamefont {Sanchez-Rey}}} (\bibinfo
  {year} {2000}),\ \href {\doibase 10.1016/S0378-4371(00)00115-1} {\bibfield
  {journal} {\bibinfo  {journal} {Physica A}\ }\textbf {\bibinfo {volume}
  {284}},\ \bibinfo {pages} {277}}\BibitemShut {NoStop}%
\bibitem [{\citenamefont {Prados}\ and\ \citenamefont
  {Brey}(2002)}]{Prados:2002aa}%
  \BibitemOpen
  \bibfield  {author} {\bibinfo {author} {\bibnamefont {Prados}, \bibfnamefont
  {A.}}, \ and\ \bibinfo {author} {\bibfnamefont {J.~J.}\ \bibnamefont {Brey}}}
  (\bibinfo {year} {2002}),\ \href {\doibase 10.1103/PhysRevE.66.041308}
  {\bibfield  {journal} {\bibinfo  {journal} {Phys. Rev. E}\ }\textbf {\bibinfo
  {volume} {66}},\ \bibinfo {pages} {041308}}\BibitemShut {NoStop}%
\bibitem [{\citenamefont {Puckett}\ and\ \citenamefont
  {Daniels}(2013)}]{Puckett:2013aa}%
  \BibitemOpen
  \bibfield  {author} {\bibinfo {author} {\bibnamefont {Puckett}, \bibfnamefont
  {J.~G.}}, \ and\ \bibinfo {author} {\bibfnamefont {K.~E.}\ \bibnamefont
  {Daniels}}} (\bibinfo {year} {2013}),\ \href {\doibase
  10.1103/PhysRevLett.110.058001} {\bibfield  {journal} {\bibinfo  {journal}
  {Phys. Rev. Lett.}\ }\textbf {\bibinfo {volume} {110}},\ \bibinfo {pages}
  {058001}}\BibitemShut {NoStop}%
\bibitem [{\citenamefont {Puckett}\ \emph {et~al.}(2011)\citenamefont
  {Puckett}, \citenamefont {Lechenault},\ and\ \citenamefont
  {Daniels}}]{Puckett:2011aa}%
  \BibitemOpen
  \bibfield  {author} {\bibinfo {author} {\bibnamefont {Puckett}, \bibfnamefont
  {J.~G.}}, \bibinfo {author} {\bibfnamefont {F.}~\bibnamefont {Lechenault}}, \
  and\ \bibinfo {author} {\bibfnamefont {K.~E.}\ \bibnamefont {Daniels}}}
  (\bibinfo {year} {2011}),\ \href {\doibase 10.1103/PhysRevE.83.041301}
  {\bibfield  {journal} {\bibinfo  {journal} {Phys. Rev. E}\ }\textbf {\bibinfo
  {volume} {83}},\ \bibinfo {pages} {041301}}\BibitemShut {NoStop}%
\bibitem [{\citenamefont {Pugnaloni}\ \emph {et~al.}(2010)\citenamefont
  {Pugnaloni}, \citenamefont {Sanchez}, \citenamefont {Gago}, \citenamefont
  {Damas}, \citenamefont {Zuriguel},\ and\ \citenamefont
  {Maza}}]{Pugnaloni:2010aa}%
  \BibitemOpen
  \bibfield  {author} {\bibinfo {author} {\bibnamefont {Pugnaloni},
  \bibfnamefont {L.~A.}}, \bibinfo {author} {\bibfnamefont {I.}~\bibnamefont
  {Sanchez}}, \bibinfo {author} {\bibfnamefont {P.~A.}\ \bibnamefont {Gago}},
  \bibinfo {author} {\bibfnamefont {J.}~\bibnamefont {Damas}}, \bibinfo
  {author} {\bibfnamefont {I.}~\bibnamefont {Zuriguel}}, \ and\ \bibinfo
  {author} {\bibfnamefont {D.}~\bibnamefont {Maza}}} (\bibinfo {year} {2010}),\
  \href {\doibase 10.1103/PhysRevE.82.050301} {\bibfield  {journal} {\bibinfo
  {journal} {Phys. Rev. E}\ }\textbf {\bibinfo {volume} {82}},\
  10.1103/PhysRevE.82.050301}\BibitemShut {NoStop}%
\bibitem [{\citenamefont {Qiong}\ and\ \citenamefont
  {Mei-Ying}(2014)}]{Qiong:2014aa}%
  \BibitemOpen
  \bibfield  {author} {\bibinfo {author} {\bibnamefont {Qiong}, \bibfnamefont
  {C.}}, \ and\ \bibinfo {author} {\bibfnamefont {H.}~\bibnamefont {Mei-Ying}}}
  (\bibinfo {year} {2014}),\ \href
  {http://stacks.iop.org/1674-1056/23/i=7/a=074501} {\bibfield  {journal}
  {\bibinfo  {journal} {Chin. Phys. B}\ }\textbf {\bibinfo {volume} {23}},\
  \bibinfo {pages} {074501}}\BibitemShut {NoStop}%
\bibitem [{\citenamefont {Radeke}\ \emph {et~al.}(2004)\citenamefont {Radeke},
  \citenamefont {Bagi}, \citenamefont {Pal{\'a}ncz},\ and\ \citenamefont
  {Stoyan}}]{Radeke:2004aa}%
  \BibitemOpen
  \bibfield  {author} {\bibinfo {author} {\bibnamefont {Radeke}, \bibfnamefont
  {C.}}, \bibinfo {author} {\bibfnamefont {K.}~\bibnamefont {Bagi}}, \bibinfo
  {author} {\bibfnamefont {B.}~\bibnamefont {Pal{\'a}ncz}}, \ and\ \bibinfo
  {author} {\bibfnamefont {D.}~\bibnamefont {Stoyan}}} (\bibinfo {year}
  {2004}),\ \href {\doibase 10.1007/s10035-004-0154-1} {\bibfield  {journal}
  {\bibinfo  {journal} {Granular Matter}\ }\textbf {\bibinfo {volume} {6}},\
  \bibinfo {pages} {17}}\BibitemShut {NoStop}%
\bibitem [{\citenamefont {Radin}(2008)}]{Radin:2008aa}%
  \BibitemOpen
  \bibfield  {author} {\bibinfo {author} {\bibnamefont {Radin}, \bibfnamefont
  {C.}}} (\bibinfo {year} {2008}),\ \href {\doibase 10.1007/s10955-008-9523-1}
  {\bibfield  {journal} {\bibinfo  {journal} {J. Stat. Phys.}\ }\textbf
  {\bibinfo {volume} {131}},\ \bibinfo {pages} {567}}\BibitemShut {NoStop}%
\bibitem [{\citenamefont {Radjai}\ \emph {et~al.}(1996)\citenamefont {Radjai},
  \citenamefont {Jean}, \citenamefont {Moreau},\ and\ \citenamefont
  {Roux}}]{Radjai:1996aa}%
  \BibitemOpen
  \bibfield  {author} {\bibinfo {author} {\bibnamefont {Radjai}, \bibfnamefont
  {F.}}, \bibinfo {author} {\bibfnamefont {M.}~\bibnamefont {Jean}}, \bibinfo
  {author} {\bibfnamefont {J.-J.}\ \bibnamefont {Moreau}}, \ and\ \bibinfo
  {author} {\bibfnamefont {S.}~\bibnamefont {Roux}}} (\bibinfo {year} {1996}),\
  \href {\doibase 10.1103/PhysRevLett.77.274} {\bibfield  {journal} {\bibinfo
  {journal} {Phys. Rev. Lett.}\ }\textbf {\bibinfo {volume} {77}},\ \bibinfo
  {pages} {274}}\BibitemShut {NoStop}%
\bibitem [{\citenamefont {Radjai}\ and\ \citenamefont
  {Richefeu}(2009)}]{Radjai:2009aa}%
  \BibitemOpen
  \bibfield  {author} {\bibinfo {author} {\bibnamefont {Radjai}, \bibfnamefont
  {F.}}, \ and\ \bibinfo {author} {\bibfnamefont {V.}~\bibnamefont {Richefeu}}}
  (\bibinfo {year} {2009}),\ \href {\doibase
  http://dx.doi.org/10.1016/j.mechmat.2009.01.028} {\bibfield  {journal}
  {\bibinfo  {journal} {Mech. Mater.}\ }\textbf {\bibinfo {volume} {41}},\
  \bibinfo {pages} {715 }}\BibitemShut {NoStop}%
\bibitem [{\citenamefont {Rainone}\ and\ \citenamefont
  {Urbani}(2016)}]{Rainone:2016aa}%
  \BibitemOpen
  \bibfield  {author} {\bibinfo {author} {\bibnamefont {Rainone}, \bibfnamefont
  {C.}}, \ and\ \bibinfo {author} {\bibfnamefont {P.}~\bibnamefont {Urbani}}}
  (\bibinfo {year} {2016}),\ \href
  {http://stacks.iop.org/1742-5468/2016/i=5/a=053302} {\bibfield  {journal}
  {\bibinfo  {journal} {Journal of Statistical Mechanics: Theory and
  Experiment}\ }\textbf {\bibinfo {volume} {2016}}~(\bibinfo {number} {5}),\
  \bibinfo {pages} {053302}}\BibitemShut {NoStop}%
\bibitem [{\citenamefont {R\'enyi}(1958)}]{Renyi:1958aa}%
  \BibitemOpen
  \bibfield  {author} {\bibinfo {author} {\bibnamefont {R\'enyi}, \bibfnamefont
  {A.}}} (\bibinfo {year} {1958}),\ \href@noop {} {\bibfield  {journal}
  {\bibinfo  {journal} {Publ. Math. Res. Inst. Hung. Acad. Sci.}\ }\textbf
  {\bibinfo {volume} {3}},\ \bibinfo {pages} {109}}\BibitemShut {NoStop}%
\bibitem [{\citenamefont {Ribiere}\ \emph {et~al.}(2007)\citenamefont
  {Ribiere}, \citenamefont {Richard}, \citenamefont {Philippe}, \citenamefont
  {Bideau},\ and\ \citenamefont {Delannay}}]{Ribiere:2007aa}%
  \BibitemOpen
  \bibfield  {author} {\bibinfo {author} {\bibnamefont {Ribiere}, \bibfnamefont
  {P.}}, \bibinfo {author} {\bibfnamefont {P.}~\bibnamefont {Richard}},
  \bibinfo {author} {\bibfnamefont {P.}~\bibnamefont {Philippe}}, \bibinfo
  {author} {\bibfnamefont {D.}~\bibnamefont {Bideau}}, \ and\ \bibinfo {author}
  {\bibfnamefont {R.}~\bibnamefont {Delannay}}} (\bibinfo {year} {2007}),\
  \href {\doibase 10.1140/epje/e2007-00017-x} {\bibfield  {journal} {\bibinfo
  {journal} {Eur. Phys. J. E}\ }\textbf {\bibinfo {volume} {22}},\ \bibinfo
  {pages} {249}}\BibitemShut {NoStop}%
\bibitem [{\citenamefont {Richard}\ \emph {et~al.}(2005)\citenamefont
  {Richard}, \citenamefont {Nicodemi}, \citenamefont {Delannay}, \citenamefont
  {Ribiere},\ and\ \citenamefont {Bideau}}]{Richard:2005aa}%
  \BibitemOpen
  \bibfield  {author} {\bibinfo {author} {\bibnamefont {Richard}, \bibfnamefont
  {P.}}, \bibinfo {author} {\bibfnamefont {M.}~\bibnamefont {Nicodemi}},
  \bibinfo {author} {\bibfnamefont {R.}~\bibnamefont {Delannay}}, \bibinfo
  {author} {\bibfnamefont {P.}~\bibnamefont {Ribiere}}, \ and\ \bibinfo
  {author} {\bibfnamefont {D.}~\bibnamefont {Bideau}}} (\bibinfo {year}
  {2005}),\ \href {\doibase 10.1038/nmat1300} {\bibfield  {journal} {\bibinfo
  {journal} {Nature Mater.}\ }\textbf {\bibinfo {volume} {4}},\ \bibinfo
  {pages} {121}}\BibitemShut {NoStop}%
\bibitem [{\citenamefont {Richard}\ \emph {et~al.}(2003)\citenamefont
  {Richard}, \citenamefont {Philippe}, \citenamefont {Barbe}, \citenamefont
  {Bourles}, \citenamefont {Thibault},\ and\ \citenamefont
  {Bideau}}]{Richard:2003aa}%
  \BibitemOpen
  \bibfield  {author} {\bibinfo {author} {\bibnamefont {Richard}, \bibfnamefont
  {P.}}, \bibinfo {author} {\bibfnamefont {P.}~\bibnamefont {Philippe}},
  \bibinfo {author} {\bibfnamefont {F.}~\bibnamefont {Barbe}}, \bibinfo
  {author} {\bibfnamefont {S.}~\bibnamefont {Bourles}}, \bibinfo {author}
  {\bibfnamefont {X.}~\bibnamefont {Thibault}}, \ and\ \bibinfo {author}
  {\bibfnamefont {D.}~\bibnamefont {Bideau}}} (\bibinfo {year} {2003}),\ \href
  {\doibase 10.1103/PhysRevE.68.020301} {\bibfield  {journal} {\bibinfo
  {journal} {Phys. Rev. E}\ }\textbf {\bibinfo {volume} {68}},\ \bibinfo
  {pages} {020301}}\BibitemShut {NoStop}%
\bibitem [{\citenamefont {Roberts}(1981)}]{Roberts:1981aa}%
  \BibitemOpen
  \bibfield  {author} {\bibinfo {author} {\bibnamefont {Roberts}, \bibfnamefont
  {S.~A.}}} (\bibinfo {year} {1981}),\ \href
  {http://stacks.iop.org/0022-3719/14/i=21/a=018} {\bibfield  {journal}
  {\bibinfo  {journal} {Journal of Physics C: Solid State Physics}\ }\textbf
  {\bibinfo {volume} {14}}~(\bibinfo {number} {21}),\ \bibinfo {pages}
  {3015}}\BibitemShut {NoStop}%
\bibitem [{\citenamefont {Roth}\ and\ \citenamefont
  {Jaeger}(2016)}]{Roth:2016aa}%
  \BibitemOpen
  \bibfield  {author} {\bibinfo {author} {\bibnamefont {Roth}, \bibfnamefont
  {L.~K.}}, \ and\ \bibinfo {author} {\bibfnamefont {H.~M.}\ \bibnamefont
  {Jaeger}}} (\bibinfo {year} {2016}),\ \href {\doibase 10.1039/C5SM02335A}
  {\bibfield  {journal} {\bibinfo  {journal} {Soft Matter}\ }\textbf {\bibinfo
  {volume} {12}},\ \bibinfo {pages} {1107}}\BibitemShut {NoStop}%
\bibitem [{\citenamefont {Rothenburg}\ and\ \citenamefont
  {Kruyt}(2009)}]{Rothenburg:2009aa}%
  \BibitemOpen
  \bibfield  {author} {\bibinfo {author} {\bibnamefont {Rothenburg},
  \bibfnamefont {L.}}, \ and\ \bibinfo {author} {\bibfnamefont {N.~P.}\
  \bibnamefont {Kruyt}}} (\bibinfo {year} {2009}),\ \href {\doibase
  10.1016/j.jmps.2008.09.018} {\bibfield  {journal} {\bibinfo  {journal} {J.
  Mech. Phys. Solids}\ }\textbf {\bibinfo {volume} {57}},\ \bibinfo {pages}
  {634}}\BibitemShut {NoStop}%
\bibitem [{\citenamefont {Roux}(2000)}]{Roux:2000aa}%
  \BibitemOpen
  \bibfield  {author} {\bibinfo {author} {\bibnamefont {Roux}, \bibfnamefont
  {J.-N.}}} (\bibinfo {year} {2000}),\ \href {\doibase
  10.1103/PhysRevE.61.6802} {\bibfield  {journal} {\bibinfo  {journal} {Phys.
  Rev. E}\ }\textbf {\bibinfo {volume} {61}},\ \bibinfo {pages}
  {6802}}\BibitemShut {NoStop}%
\bibitem [{\citenamefont {Saadatfar}\ \emph {et~al.}(2012)\citenamefont
  {Saadatfar}, \citenamefont {Sheppard}, \citenamefont {Senden},\ and\
  \citenamefont {Kabla}}]{Saadatfar:2012aa}%
  \BibitemOpen
  \bibfield  {author} {\bibinfo {author} {\bibnamefont {Saadatfar},
  \bibfnamefont {M.}}, \bibinfo {author} {\bibfnamefont {A.~P.}\ \bibnamefont
  {Sheppard}}, \bibinfo {author} {\bibfnamefont {T.~J.}\ \bibnamefont
  {Senden}}, \ and\ \bibinfo {author} {\bibfnamefont {A.~J.}\ \bibnamefont
  {Kabla}}} (\bibinfo {year} {2012}),\ \href {\doibase
  http://dx.doi.org/10.1016/j.jmps.2011.10.001} {\bibfield  {journal} {\bibinfo
   {journal} {Journal of the Mechanics and Physics of Solids}\ }\textbf
  {\bibinfo {volume} {60}}~(\bibinfo {number} {1}),\ \bibinfo {pages} {55
  }}\BibitemShut {NoStop}%
\bibitem [{\citenamefont {Santiso}\ and\ \citenamefont
  {M\"uller}(2002)}]{Santiso:2002aa}%
  \BibitemOpen
  \bibfield  {author} {\bibinfo {author} {\bibnamefont {Santiso}, \bibfnamefont
  {E.}}, \ and\ \bibinfo {author} {\bibfnamefont {E.~A.}\ \bibnamefont
  {M\"uller}}} (\bibinfo {year} {2002}),\ \href {\doibase
  10.1080/00268970210125313} {\bibfield  {journal} {\bibinfo  {journal}
  {Molecular Physics}\ }\textbf {\bibinfo {volume} {100}}~(\bibinfo {number}
  {15}),\ \bibinfo {pages} {2461}}\BibitemShut {NoStop}%
\bibitem [{\citenamefont {Satake}(1993)}]{Satake:1993aa}%
  \BibitemOpen
  \bibfield  {author} {\bibinfo {author} {\bibnamefont {Satake}, \bibfnamefont
  {M.}}} (\bibinfo {year} {1993}),\ \href {\doibase
  http://dx.doi.org/10.1016/0167-6636(93)90028-P} {\bibfield  {journal}
  {\bibinfo  {journal} {Mech. Mater.}\ }\textbf {\bibinfo {volume} {16}},\
  \bibinfo {pages} {65 }}\BibitemShut {NoStop}%
\bibitem [{\citenamefont {Schaller}\ \emph {et~al.}(2013)\citenamefont
  {Schaller}, \citenamefont {Kapfer}, \citenamefont {Evans}, \citenamefont
  {Hoffmann}, \citenamefont {Aste}, \citenamefont {Saadatfar}, \citenamefont
  {Mecke}, \citenamefont {Delaney},\ and\ \citenamefont
  {Schr\"oder-Turk}}]{Schaller:2013ab}%
  \BibitemOpen
  \bibfield  {author} {\bibinfo {author} {\bibnamefont {Schaller},
  \bibfnamefont {F.~M.}}, \bibinfo {author} {\bibfnamefont {S.~C.}\
  \bibnamefont {Kapfer}}, \bibinfo {author} {\bibfnamefont {M.~E.}\
  \bibnamefont {Evans}}, \bibinfo {author} {\bibfnamefont {M.~J.~F.}\
  \bibnamefont {Hoffmann}}, \bibinfo {author} {\bibfnamefont {T.}~\bibnamefont
  {Aste}}, \bibinfo {author} {\bibfnamefont {M.}~\bibnamefont {Saadatfar}},
  \bibinfo {author} {\bibfnamefont {K.}~\bibnamefont {Mecke}}, \bibinfo
  {author} {\bibfnamefont {G.~W.}\ \bibnamefont {Delaney}}, \ and\ \bibinfo
  {author} {\bibfnamefont {G.~E.}\ \bibnamefont {Schr\"oder-Turk}}} (\bibinfo
  {year} {2013}),\ \href {\doibase 10.1080/14786435.2013.834389} {\bibfield
  {journal} {\bibinfo  {journal} {Philos. Mag.}\ }\textbf {\bibinfo {volume}
  {93}},\ \bibinfo {pages} {3993}}\BibitemShut {NoStop}%
\bibitem [{\citenamefont {Schaller}\ \emph
  {et~al.}(2015{\natexlab{a}})\citenamefont {Schaller}, \citenamefont {Kapfer},
  \citenamefont {Hilton}, \citenamefont {Cleary}, \citenamefont {Mecke},
  \citenamefont {Michele}, \citenamefont {Schilling}, \citenamefont
  {Saadatfar}, \citenamefont {Schr\"oter}, \citenamefont {Delaney},\ and\
  \citenamefont {Schr\"oder-Turk}}]{Schaller:2015aa}%
  \BibitemOpen
  \bibfield  {author} {\bibinfo {author} {\bibnamefont {Schaller},
  \bibfnamefont {F.~M.}}, \bibinfo {author} {\bibfnamefont {S.~C.}\
  \bibnamefont {Kapfer}}, \bibinfo {author} {\bibfnamefont {J.~E.}\
  \bibnamefont {Hilton}}, \bibinfo {author} {\bibfnamefont {P.~W.}\
  \bibnamefont {Cleary}}, \bibinfo {author} {\bibfnamefont {K.}~\bibnamefont
  {Mecke}}, \bibinfo {author} {\bibfnamefont {C.~D.}\ \bibnamefont {Michele}},
  \bibinfo {author} {\bibfnamefont {T.}~\bibnamefont {Schilling}}, \bibinfo
  {author} {\bibfnamefont {M.}~\bibnamefont {Saadatfar}}, \bibinfo {author}
  {\bibfnamefont {M.}~\bibnamefont {Schr\"oter}}, \bibinfo {author}
  {\bibfnamefont {G.~W.}\ \bibnamefont {Delaney}}, \ and\ \bibinfo {author}
  {\bibfnamefont {G.~E.}\ \bibnamefont {Schr\"oder-Turk}}} (\bibinfo {year}
  {2015}{\natexlab{a}}),\ \href
  {http://stacks.iop.org/0295-5075/111/i=2/a=24002} {\bibfield  {journal}
  {\bibinfo  {journal} {EPL}\ }\textbf {\bibinfo {volume} {111}},\ \bibinfo
  {pages} {24002}}\BibitemShut {NoStop}%
\bibitem [{\citenamefont {Schaller}\ \emph
  {et~al.}(2015{\natexlab{b}})\citenamefont {Schaller}, \citenamefont
  {Neudecker}, \citenamefont {Saadatfar}, \citenamefont {Delaney},
  \citenamefont {Schr\"oder-Turk},\ and\ \citenamefont
  {Schr\"oter}}]{Schaller:2015ab}%
  \BibitemOpen
  \bibfield  {author} {\bibinfo {author} {\bibnamefont {Schaller},
  \bibfnamefont {F.~M.}}, \bibinfo {author} {\bibfnamefont {M.}~\bibnamefont
  {Neudecker}}, \bibinfo {author} {\bibfnamefont {M.}~\bibnamefont
  {Saadatfar}}, \bibinfo {author} {\bibfnamefont {G.~W.}\ \bibnamefont
  {Delaney}}, \bibinfo {author} {\bibfnamefont {G.~E.}\ \bibnamefont
  {Schr\"oder-Turk}}, \ and\ \bibinfo {author} {\bibfnamefont {M.}~\bibnamefont
  {Schr\"oter}}} (\bibinfo {year} {2015}{\natexlab{b}}),\ \href {\doibase
  10.1103/PhysRevLett.114.158001} {\bibfield  {journal} {\bibinfo  {journal}
  {Phys. Rev. Lett.}\ }\textbf {\bibinfo {volume} {114}},\ \bibinfo {pages}
  {158001}}\BibitemShut {NoStop}%
\bibitem [{\citenamefont {Schopenhauer}(1974)}]{Schopenhauer:1974aa}%
  \BibitemOpen
  \bibfield  {author} {\bibinfo {author} {\bibnamefont {Schopenhauer},
  \bibfnamefont {A.}}} (\bibinfo {year} {1974}),\ \enquote {\bibinfo {title}
  {Transcendent speculation on the apparent deliberateness in the fate of the
  individual},}\ in\ \href@noop {} {\emph {\bibinfo {booktitle} {Parerga and
  Paralipomena: Short Philosophical Essays}}},\ Vol.~\bibinfo {volume} {I}\
  (\bibinfo  {publisher} {Oxford University Press})\ pp.\ \bibinfo {pages}
  {199--225}\BibitemShut {NoStop}%
\bibitem [{\citenamefont {Schreck}\ \emph {et~al.}(2012)\citenamefont
  {Schreck}, \citenamefont {Mailman}, \citenamefont {Chakraborty},\ and\
  \citenamefont {O'Hern}}]{Schreck:2012aa}%
  \BibitemOpen
  \bibfield  {author} {\bibinfo {author} {\bibnamefont {Schreck}, \bibfnamefont
  {C.~F.}}, \bibinfo {author} {\bibfnamefont {M.}~\bibnamefont {Mailman}},
  \bibinfo {author} {\bibfnamefont {B.}~\bibnamefont {Chakraborty}}, \ and\
  \bibinfo {author} {\bibfnamefont {C.~S.}\ \bibnamefont {O'Hern}}} (\bibinfo
  {year} {2012}),\ \href {\doibase 10.1103/PhysRevE.85.061305} {\bibfield
  {journal} {\bibinfo  {journal} {Phys. Rev. E}\ }\textbf {\bibinfo {volume}
  {85}},\ \bibinfo {pages} {061305}}\BibitemShut {NoStop}%
\bibitem [{\citenamefont {Schreck}\ and\ \citenamefont
  {O'Hern}(2011)}]{Schreck:2011aa}%
  \BibitemOpen
  \bibfield  {author} {\bibinfo {author} {\bibnamefont {Schreck}, \bibfnamefont
  {C.~F.}}, \ and\ \bibinfo {author} {\bibfnamefont {C.~S.}\ \bibnamefont
  {O'Hern}}} (\bibinfo {year} {2011}),\ \href@noop {} {\ }\BibitemShut
  {NoStop}%
\bibitem [{\citenamefont {Schreck}\ \emph {et~al.}(2010)\citenamefont
  {Schreck}, \citenamefont {Xu},\ and\ \citenamefont
  {O{'}Hern}}]{Schreck:2010aa}%
  \BibitemOpen
  \bibfield  {author} {\bibinfo {author} {\bibnamefont {Schreck}, \bibfnamefont
  {C.~F.}}, \bibinfo {author} {\bibfnamefont {N.}~\bibnamefont {Xu}}, \ and\
  \bibinfo {author} {\bibfnamefont {C.~S.}\ \bibnamefont {O{'}Hern}}} (\bibinfo
  {year} {2010}),\ \href {\doibase 10.1039/C001085E} {\bibfield  {journal}
  {\bibinfo  {journal} {Soft Matter}\ }\textbf {\bibinfo {volume} {6}},\
  \bibinfo {pages} {2960}}\BibitemShut {NoStop}%
\bibitem [{\citenamefont {Schr\"oder-Turk}\ \emph {et~al.}(2010)\citenamefont
  {Schr\"oder-Turk}, \citenamefont {Mickel}, \citenamefont {Schr\"oter},
  \citenamefont {Delaney}, \citenamefont {Saadatfar}, \citenamefont {Senden},
  \citenamefont {Mecke},\ and\ \citenamefont {Aste}}]{Schroeder-Turk:2010aa}%
  \BibitemOpen
  \bibfield  {author} {\bibinfo {author} {\bibnamefont {Schr\"oder-Turk},
  \bibfnamefont {G.~E.}}, \bibinfo {author} {\bibfnamefont {W.}~\bibnamefont
  {Mickel}}, \bibinfo {author} {\bibfnamefont {M.}~\bibnamefont {Schr\"oter}},
  \bibinfo {author} {\bibfnamefont {G.~W.}\ \bibnamefont {Delaney}}, \bibinfo
  {author} {\bibfnamefont {M.}~\bibnamefont {Saadatfar}}, \bibinfo {author}
  {\bibfnamefont {T.~J.}\ \bibnamefont {Senden}}, \bibinfo {author}
  {\bibfnamefont {K.}~\bibnamefont {Mecke}}, \ and\ \bibinfo {author}
  {\bibfnamefont {T.}~\bibnamefont {Aste}}} (\bibinfo {year} {2010}),\ \href
  {\doibase 10.1209/0295-5075/90/34001} {\bibfield  {journal} {\bibinfo
  {journal} {EPL}\ }\textbf {\bibinfo {volume} {90}},\ \bibinfo {pages}
  {34001}}\BibitemShut {NoStop}%
\bibitem [{\citenamefont {Schr\"oter}\ \emph {et~al.}(2005)\citenamefont
  {Schr\"oter}, \citenamefont {Goldman},\ and\ \citenamefont
  {Swinney}}]{Schroter:2005aa}%
  \BibitemOpen
  \bibfield  {author} {\bibinfo {author} {\bibnamefont {Schr\"oter},
  \bibfnamefont {M.}}, \bibinfo {author} {\bibfnamefont {D.}~\bibnamefont
  {Goldman}}, \ and\ \bibinfo {author} {\bibfnamefont {H.}~\bibnamefont
  {Swinney}}} (\bibinfo {year} {2005}),\ \href {\doibase
  10.1103/PhysRevE.71.030301} {\bibfield  {journal} {\bibinfo  {journal} {Phys.
  Rev. E}\ }\textbf {\bibinfo {volume} {71}},\ \bibinfo {pages}
  {030301}}\BibitemShut {NoStop}%
\bibitem [{\citenamefont {Schwarz}\ \emph {et~al.}(2006)\citenamefont
  {Schwarz}, \citenamefont {Liu},\ and\ \citenamefont
  {Chayes}}]{Schwarz:2006aa}%
  \BibitemOpen
  \bibfield  {author} {\bibinfo {author} {\bibnamefont {Schwarz}, \bibfnamefont
  {J.~M.}}, \bibinfo {author} {\bibfnamefont {A.~J.}\ \bibnamefont {Liu}}, \
  and\ \bibinfo {author} {\bibfnamefont {L.~Q.}\ \bibnamefont {Chayes}}}
  (\bibinfo {year} {2006}),\ \href
  {http://stacks.iop.org/0295-5075/73/i=4/a=560} {\bibfield  {journal}
  {\bibinfo  {journal} {EPL}\ }\textbf {\bibinfo {volume} {73}},\ \bibinfo
  {pages} {560}}\BibitemShut {NoStop}%
\bibitem [{\citenamefont {Scott}(1960)}]{Scott:1960aa}%
  \BibitemOpen
  \bibfield  {author} {\bibinfo {author} {\bibnamefont {Scott}, \bibfnamefont
  {G.~D.}}} (\bibinfo {year} {1960}),\ \href
  {http://dx.doi.org/10.1038/188908a0} {\bibfield  {journal} {\bibinfo
  {journal} {Nature}\ }\textbf {\bibinfo {volume} {188}},\ \bibinfo {pages}
  {908}}\BibitemShut {NoStop}%
\bibitem [{\citenamefont {Scott}(1962)}]{Scott:1962aa}%
  \BibitemOpen
  \bibfield  {author} {\bibinfo {author} {\bibnamefont {Scott}, \bibfnamefont
  {G.~D.}}} (\bibinfo {year} {1962}),\ \href
  {http://dx.doi.org/10.1038/194956a0} {\bibfield  {journal} {\bibinfo
  {journal} {Nature}\ }\textbf {\bibinfo {volume} {194}},\ \bibinfo {pages}
  {956}}\BibitemShut {NoStop}%
\bibitem [{\citenamefont {Sharma}\ \emph {et~al.}(2016)\citenamefont {Sharma},
  \citenamefont {Yeo},\ and\ \citenamefont {Moore}}]{Sharma:2016aa}%
  \BibitemOpen
  \bibfield  {author} {\bibinfo {author} {\bibnamefont {Sharma}, \bibfnamefont
  {A.}}, \bibinfo {author} {\bibfnamefont {J.}~\bibnamefont {Yeo}}, \ and\
  \bibinfo {author} {\bibfnamefont {M.~A.}\ \bibnamefont {Moore}}} (\bibinfo
  {year} {2016}),\ \href {\doibase 10.1103/PhysRevE.94.052143} {\bibfield
  {journal} {\bibinfo  {journal} {Phys. Rev. E}\ }\textbf {\bibinfo {volume}
  {94}},\ \bibinfo {pages} {052143}}\BibitemShut {NoStop}%
\bibitem [{\citenamefont {Shen}\ \emph {et~al.}(2014)\citenamefont {Shen},
  \citenamefont {Papanikolaou}, \citenamefont {O'Hern},\ and\ \citenamefont
  {Shattuck}}]{Shen:2014aa}%
  \BibitemOpen
  \bibfield  {author} {\bibinfo {author} {\bibnamefont {Shen}, \bibfnamefont
  {T.}}, \bibinfo {author} {\bibfnamefont {S.}~\bibnamefont {Papanikolaou}},
  \bibinfo {author} {\bibfnamefont {C.~S.}\ \bibnamefont {O'Hern}}, \ and\
  \bibinfo {author} {\bibfnamefont {M.~D.}\ \bibnamefont {Shattuck}}} (\bibinfo
  {year} {2014}),\ \href {\doibase 10.1103/PhysRevLett.113.128302} {\bibfield
  {journal} {\bibinfo  {journal} {Phys. Rev. Lett.}\ }\textbf {\bibinfo
  {volume} {113}},\ \bibinfo {pages} {128302}}\BibitemShut {NoStop}%
\bibitem [{\citenamefont {Sherrington}\ and\ \citenamefont
  {Kirkpatrick}(1975)}]{Sherrington:1975aa}%
  \BibitemOpen
  \bibfield  {author} {\bibinfo {author} {\bibnamefont {Sherrington},
  \bibfnamefont {D.}}, \ and\ \bibinfo {author} {\bibfnamefont
  {S.}~\bibnamefont {Kirkpatrick}}} (\bibinfo {year} {1975}),\ \href {\doibase
  10.1103/PhysRevLett.35.1792} {\bibfield  {journal} {\bibinfo  {journal}
  {Phys. Rev. Lett.}\ }\textbf {\bibinfo {volume} {35}},\ \bibinfo {pages}
  {1792}}\BibitemShut {NoStop}%
\bibitem [{\citenamefont {Shundyak}\ \emph {et~al.}(2007)\citenamefont
  {Shundyak}, \citenamefont {van Hecke},\ and\ \citenamefont {van
  Saarloos}}]{Shundyak:2007aa}%
  \BibitemOpen
  \bibfield  {author} {\bibinfo {author} {\bibnamefont {Shundyak},
  \bibfnamefont {K.}}, \bibinfo {author} {\bibfnamefont {M.}~\bibnamefont {van
  Hecke}}, \ and\ \bibinfo {author} {\bibfnamefont {W.}~\bibnamefont {van
  Saarloos}}} (\bibinfo {year} {2007}),\ \href {\doibase
  10.1103/PhysRevE.75.010301} {\bibfield  {journal} {\bibinfo  {journal} {Phys.
  Rev. E}\ }\textbf {\bibinfo {volume} {75}},\ \bibinfo {pages}
  {010301}}\BibitemShut {NoStop}%
\bibitem [{\citenamefont {Silbert}\ \emph {et~al.}(2005)\citenamefont
  {Silbert}, \citenamefont {Liu},\ and\ \citenamefont
  {Nagel}}]{Silbert:2005aa}%
  \BibitemOpen
  \bibfield  {author} {\bibinfo {author} {\bibnamefont {Silbert}, \bibfnamefont
  {L.}}, \bibinfo {author} {\bibfnamefont {A.}~\bibnamefont {Liu}}, \ and\
  \bibinfo {author} {\bibfnamefont {S.}~\bibnamefont {Nagel}}} (\bibinfo {year}
  {2005}),\ \href {\doibase 10.1103/PhysRevLett.95.098301} {\bibfield
  {journal} {\bibinfo  {journal} {Phys. Rev. Lett.}\ }\textbf {\bibinfo
  {volume} {95}},\ \bibinfo {pages} {098301}}\BibitemShut {NoStop}%
\bibitem [{\citenamefont {Silbert}(2010)}]{Silbert:2010aa}%
  \BibitemOpen
  \bibfield  {author} {\bibinfo {author} {\bibnamefont {Silbert}, \bibfnamefont
  {L.~E.}}} (\bibinfo {year} {2010}),\ \href {\doibase 10.1039/C001973A}
  {\bibfield  {journal} {\bibinfo  {journal} {Soft Matter}\ }\textbf {\bibinfo
  {volume} {6}},\ \bibinfo {pages} {2918}}\BibitemShut {NoStop}%
\bibitem [{\citenamefont {Silbert}\ \emph
  {et~al.}(2002{\natexlab{a}})\citenamefont {Silbert}, \citenamefont {Ertas},
  \citenamefont {Grest}, \citenamefont {Halsey},\ and\ \citenamefont
  {Levine}}]{Silbert:2002ab}%
  \BibitemOpen
  \bibfield  {author} {\bibinfo {author} {\bibnamefont {Silbert}, \bibfnamefont
  {L.~E.}}, \bibinfo {author} {\bibfnamefont {D.}~\bibnamefont {Ertas}},
  \bibinfo {author} {\bibfnamefont {G.~S.}\ \bibnamefont {Grest}}, \bibinfo
  {author} {\bibfnamefont {T.~C.}\ \bibnamefont {Halsey}}, \ and\ \bibinfo
  {author} {\bibfnamefont {D.}~\bibnamefont {Levine}}} (\bibinfo {year}
  {2002}{\natexlab{a}}),\ \href {\doibase 10.1103/PhysRevE.65.031304}
  {\bibfield  {journal} {\bibinfo  {journal} {Phys. Rev. E}\ }\textbf {\bibinfo
  {volume} {65}},\ \bibinfo {pages} {031304}}\BibitemShut {NoStop}%
\bibitem [{\citenamefont {Silbert}\ \emph
  {et~al.}(2002{\natexlab{b}})\citenamefont {Silbert}, \citenamefont {Grest},\
  and\ \citenamefont {Landry}}]{Silbert:2002aa}%
  \BibitemOpen
  \bibfield  {author} {\bibinfo {author} {\bibnamefont {Silbert}, \bibfnamefont
  {L.~E.}}, \bibinfo {author} {\bibfnamefont {G.~S.}\ \bibnamefont {Grest}}, \
  and\ \bibinfo {author} {\bibfnamefont {J.~W.}\ \bibnamefont {Landry}}}
  (\bibinfo {year} {2002}{\natexlab{b}}),\ \href {\doibase
  10.1103/PhysRevE.66.061303} {\bibfield  {journal} {\bibinfo  {journal} {Phys.
  Rev. E}\ }\textbf {\bibinfo {volume} {66}},\ \bibinfo {pages}
  {061303}}\BibitemShut {NoStop}%
\bibitem [{\citenamefont {Silbert}\ \emph {et~al.}(2006)\citenamefont
  {Silbert}, \citenamefont {Liu},\ and\ \citenamefont
  {Nagel}}]{Silbert:2006aa}%
  \BibitemOpen
  \bibfield  {author} {\bibinfo {author} {\bibnamefont {Silbert}, \bibfnamefont
  {L.~E.}}, \bibinfo {author} {\bibfnamefont {A.~J.}\ \bibnamefont {Liu}}, \
  and\ \bibinfo {author} {\bibfnamefont {S.~R.}\ \bibnamefont {Nagel}}}
  (\bibinfo {year} {2006}),\ \href {\doibase 10.1103/PhysRevE.73.041304}
  {\bibfield  {journal} {\bibinfo  {journal} {Phys. Rev. E}\ }\textbf {\bibinfo
  {volume} {73}},\ \bibinfo {pages} {041304}}\BibitemShut {NoStop}%
\bibitem [{\citenamefont {Silbert}\ \emph {et~al.}(2009)\citenamefont
  {Silbert}, \citenamefont {Liu},\ and\ \citenamefont
  {Nagel}}]{Silbert:2009aa}%
  \BibitemOpen
  \bibfield  {author} {\bibinfo {author} {\bibnamefont {Silbert}, \bibfnamefont
  {L.~E.}}, \bibinfo {author} {\bibfnamefont {A.~J.}\ \bibnamefont {Liu}}, \
  and\ \bibinfo {author} {\bibfnamefont {S.~R.}\ \bibnamefont {Nagel}}}
  (\bibinfo {year} {2009}),\ \href {\doibase 10.1103/PhysRevE.79.021308}
  {\bibfield  {journal} {\bibinfo  {journal} {Phys. Rev. E}\ }\textbf {\bibinfo
  {volume} {79}},\ \bibinfo {pages} {021308}}\BibitemShut {NoStop}%
\bibitem [{\citenamefont {Skoge}\ \emph {et~al.}(2006)\citenamefont {Skoge},
  \citenamefont {Donev}, \citenamefont {Stillinger},\ and\ \citenamefont
  {Torquato}}]{Skoge:2006aa}%
  \BibitemOpen
  \bibfield  {author} {\bibinfo {author} {\bibnamefont {Skoge}, \bibfnamefont
  {M.}}, \bibinfo {author} {\bibfnamefont {A.}~\bibnamefont {Donev}}, \bibinfo
  {author} {\bibfnamefont {F.~H.}\ \bibnamefont {Stillinger}}, \ and\ \bibinfo
  {author} {\bibfnamefont {S.}~\bibnamefont {Torquato}}} (\bibinfo {year}
  {2006}),\ \href {\doibase 10.1103/PhysRevE.74.041127} {\bibfield  {journal}
  {\bibinfo  {journal} {Phys. Rev. E}\ }\textbf {\bibinfo {volume} {74}},\
  \bibinfo {pages} {041127}}\BibitemShut {NoStop}%
\bibitem [{\citenamefont {Slobinsky}\ and\ \citenamefont
  {Pugnaloni}(2015{\natexlab{a}})}]{Slobinsky:2015ab}%
  \BibitemOpen
  \bibfield  {author} {\bibinfo {author} {\bibnamefont {Slobinsky},
  \bibfnamefont {D.}}, \ and\ \bibinfo {author} {\bibfnamefont
  {L.}~\bibnamefont {Pugnaloni}}} (\bibinfo {year} {2015}{\natexlab{a}}),\
  \href {http://www.papersinphysics.org/papersinphysics/article/view/239}
  {\bibfield  {journal} {\bibinfo  {journal} {Papers in Physics}\ }\textbf
  {\bibinfo {volume} {7}}~(\bibinfo {number} {0})}\BibitemShut {NoStop}%
\bibitem [{\citenamefont {Slobinsky}\ and\ \citenamefont
  {Pugnaloni}(2015{\natexlab{b}})}]{Slobinsky:2015aa}%
  \BibitemOpen
  \bibfield  {author} {\bibinfo {author} {\bibnamefont {Slobinsky},
  \bibfnamefont {D.}}, \ and\ \bibinfo {author} {\bibfnamefont {L.~A.}\
  \bibnamefont {Pugnaloni}}} (\bibinfo {year} {2015}{\natexlab{b}}),\ \href
  {http://stacks.iop.org/1742-5468/2015/i=2/a=P02005} {\bibfield  {journal}
  {\bibinfo  {journal} {Journal of Statistical Mechanics: Theory and
  Experiment}\ }\textbf {\bibinfo {volume} {2015}}~(\bibinfo {number} {2}),\
  \bibinfo {pages} {P02005}}\BibitemShut {NoStop}%
\bibitem [{\citenamefont {Smith}\ \emph {et~al.}(2011)\citenamefont {Smith},
  \citenamefont {Fisher},\ and\ \citenamefont {Alam}}]{Smith:2011aa}%
  \BibitemOpen
  \bibfield  {author} {\bibinfo {author} {\bibnamefont {Smith}, \bibfnamefont
  {K.~C.}}, \bibinfo {author} {\bibfnamefont {T.~S.}\ \bibnamefont {Fisher}}, \
  and\ \bibinfo {author} {\bibfnamefont {M.}~\bibnamefont {Alam}}} (\bibinfo
  {year} {2011}),\ \href {\doibase 10.1103/PhysRevE.84.030301} {\bibfield
  {journal} {\bibinfo  {journal} {Phys. Rev. E}\ }\textbf {\bibinfo {volume}
  {84}},\ \bibinfo {pages} {030301}}\BibitemShut {NoStop}%
\bibitem [{\citenamefont {Snoeijer}\ \emph {et~al.}(2004)\citenamefont
  {Snoeijer}, \citenamefont {Vlugt}, \citenamefont {Ellenbroek}, \citenamefont
  {van Hecke},\ and\ \citenamefont {van Leeuwen}}]{Snoeijer:2004ab}%
  \BibitemOpen
  \bibfield  {author} {\bibinfo {author} {\bibnamefont {Snoeijer},
  \bibfnamefont {J.}}, \bibinfo {author} {\bibfnamefont {T.}~\bibnamefont
  {Vlugt}}, \bibinfo {author} {\bibfnamefont {W.}~\bibnamefont {Ellenbroek}},
  \bibinfo {author} {\bibfnamefont {M.}~\bibnamefont {van Hecke}}, \ and\
  \bibinfo {author} {\bibfnamefont {J.}~\bibnamefont {van Leeuwen}}} (\bibinfo
  {year} {2004}),\ \href {\doibase 10.1103/PhysRevE.70.061306} {\bibfield
  {journal} {\bibinfo  {journal} {Phys. Rev. E}\ }\textbf {\bibinfo {volume}
  {70}},\ \bibinfo {pages} {061306}}\BibitemShut {NoStop}%
\bibitem [{\citenamefont {Sohn}\ and\ \citenamefont
  {Moreland}(1968)}]{Sohn:1968aa}%
  \BibitemOpen
  \bibfield  {author} {\bibinfo {author} {\bibnamefont {Sohn}, \bibfnamefont
  {H.~Y.}}, \ and\ \bibinfo {author} {\bibfnamefont {C.}~\bibnamefont
  {Moreland}}} (\bibinfo {year} {1968}),\ \href {\doibase
  10.1002/cjce.5450460305} {\bibfield  {journal} {\bibinfo  {journal} {Canad.
  J. Chem. Eng.}\ }\textbf {\bibinfo {volume} {46}},\ \bibinfo {pages}
  {162}}\BibitemShut {NoStop}%
\bibitem [{\citenamefont {Somfai}\ \emph {et~al.}(2007)\citenamefont {Somfai},
  \citenamefont {van Hecke}, \citenamefont {Ellenbroek}, \citenamefont
  {Shundyak},\ and\ \citenamefont {van Saarloos}}]{Somfai:2007aa}%
  \BibitemOpen
  \bibfield  {author} {\bibinfo {author} {\bibnamefont {Somfai}, \bibfnamefont
  {E.}}, \bibinfo {author} {\bibfnamefont {M.}~\bibnamefont {van Hecke}},
  \bibinfo {author} {\bibfnamefont {W.~G.}\ \bibnamefont {Ellenbroek}},
  \bibinfo {author} {\bibfnamefont {K.}~\bibnamefont {Shundyak}}, \ and\
  \bibinfo {author} {\bibfnamefont {W.}~\bibnamefont {van Saarloos}}} (\bibinfo
  {year} {2007}),\ \href {\doibase 10.1103/PhysRevE.75.020301} {\bibfield
  {journal} {\bibinfo  {journal} {Phys. Rev. E}\ }\textbf {\bibinfo {volume}
  {75}},\ \bibinfo {pages} {020301}}\BibitemShut {NoStop}%
\bibitem [{\citenamefont {Song}\ \emph {et~al.}(2010)\citenamefont {Song},
  \citenamefont {Wang}, \citenamefont {Jin},\ and\ \citenamefont
  {Makse}}]{Song:2010aa}%
  \BibitemOpen
  \bibfield  {author} {\bibinfo {author} {\bibnamefont {Song}, \bibfnamefont
  {C.}}, \bibinfo {author} {\bibfnamefont {P.}~\bibnamefont {Wang}}, \bibinfo
  {author} {\bibfnamefont {Y.}~\bibnamefont {Jin}}, \ and\ \bibinfo {author}
  {\bibfnamefont {H.~A.}\ \bibnamefont {Makse}}} (\bibinfo {year} {2010}),\
  \href {\doibase http://dx.doi.org/10.1016/j.physa.2010.06.043} {\bibfield
  {journal} {\bibinfo  {journal} {Physica A}\ }\textbf {\bibinfo {volume}
  {389}},\ \bibinfo {pages} {4497 }}\BibitemShut {NoStop}%
\bibitem [{\citenamefont {Song}\ \emph {et~al.}(2005)\citenamefont {Song},
  \citenamefont {Wang},\ and\ \citenamefont {Makse}}]{Song:2005aa}%
  \BibitemOpen
  \bibfield  {author} {\bibinfo {author} {\bibnamefont {Song}, \bibfnamefont
  {C.}}, \bibinfo {author} {\bibfnamefont {P.}~\bibnamefont {Wang}}, \ and\
  \bibinfo {author} {\bibfnamefont {H.}~\bibnamefont {Makse}}} (\bibinfo {year}
  {2005}),\ \href {\doibase 10.1073/pnas.0409911102} {\bibfield  {journal}
  {\bibinfo  {journal} {Proc. Nat. Acad. Sci.}\ }\textbf {\bibinfo {volume}
  {102}},\ \bibinfo {pages} {2299}}\BibitemShut {NoStop}%
\bibitem [{\citenamefont {Song}\ \emph {et~al.}(2008)\citenamefont {Song},
  \citenamefont {Wang},\ and\ \citenamefont {Makse}}]{Song:2008aa}%
  \BibitemOpen
  \bibfield  {author} {\bibinfo {author} {\bibnamefont {Song}, \bibfnamefont
  {C.}}, \bibinfo {author} {\bibfnamefont {P.}~\bibnamefont {Wang}}, \ and\
  \bibinfo {author} {\bibfnamefont {H.~A.}\ \bibnamefont {Makse}}} (\bibinfo
  {year} {2008}),\ \href {\doibase 10.1038/nature06981} {\bibfield  {journal}
  {\bibinfo  {journal} {Nature}\ }\textbf {\bibinfo {volume} {453}},\ \bibinfo
  {pages} {629}}\BibitemShut {NoStop}%
\bibitem [{\citenamefont {Steinhardt}\ \emph {et~al.}(1983)\citenamefont
  {Steinhardt}, \citenamefont {Nelson},\ and\ \citenamefont
  {Ronchetti}}]{Steinhardt:1983aa}%
  \BibitemOpen
  \bibfield  {author} {\bibinfo {author} {\bibnamefont {Steinhardt},
  \bibfnamefont {P.~J.}}, \bibinfo {author} {\bibfnamefont {D.~R.}\
  \bibnamefont {Nelson}}, \ and\ \bibinfo {author} {\bibfnamefont
  {M.}~\bibnamefont {Ronchetti}}} (\bibinfo {year} {1983}),\ \href {\doibase
  10.1103/PhysRevB.28.784} {\bibfield  {journal} {\bibinfo  {journal} {Phys.
  Rev. B}\ }\textbf {\bibinfo {volume} {28}},\ \bibinfo {pages}
  {784}}\BibitemShut {NoStop}%
\bibitem [{\citenamefont {Swendsen}(2006)}]{Swendsen:2006aa}%
  \BibitemOpen
  \bibfield  {author} {\bibinfo {author} {\bibnamefont {Swendsen},
  \bibfnamefont {R.~H.}}} (\bibinfo {year} {2006}),\ \href {\doibase
  10.1119/1.2174962} {\bibfield  {journal} {\bibinfo  {journal} {American
  Journal of Physics}\ }\textbf {\bibinfo {volume} {74}}~(\bibinfo {number}
  {3}),\ \bibinfo {pages} {187}}\BibitemShut {NoStop}%
\bibitem [{\citenamefont {Tarjus}\ and\ \citenamefont
  {Viot}(2004)}]{Tarjus:2004aa}%
  \BibitemOpen
  \bibfield  {author} {\bibinfo {author} {\bibnamefont {Tarjus}, \bibfnamefont
  {G.}}, \ and\ \bibinfo {author} {\bibfnamefont {P.}~\bibnamefont {Viot}}}
  (\bibinfo {year} {2004}),\ \href {\doibase 10.1103/PhysRevE.69.011307}
  {\bibfield  {journal} {\bibinfo  {journal} {Phys. Rev. E}\ }\textbf {\bibinfo
  {volume} {69}},\ \bibinfo {pages} {011307}}\BibitemShut {NoStop}%
\bibitem [{\citenamefont {Thomas}(1941)}]{Thomas:1941aa}%
  \BibitemOpen
  \bibfield  {author} {\bibinfo {author} {\bibnamefont {Thomas}, \bibfnamefont
  {I.}}} (\bibinfo {year} {1941}),\ \enquote {\bibinfo {title} {Greek
  mathematical works, volume ii: Aristarchus to pappus},}\ \ (\bibinfo
  {publisher} {Harvard University Press})\BibitemShut {NoStop}%
\bibitem [{\citenamefont {Thue}(1892)}]{Thue:1892aa}%
  \BibitemOpen
  \bibfield  {author} {\bibinfo {author} {\bibnamefont {Thue}, \bibfnamefont
  {A.}}} (\bibinfo {year} {1892}),\ \href@noop {} {\bibfield  {journal}
  {\bibinfo  {journal} {Forand. Skand. Natur.}\ }\textbf {\bibinfo {volume}
  {14}},\ \bibinfo {pages} {352}}\BibitemShut {NoStop}%
\bibitem [{\citenamefont {Tian}\ \emph {et~al.}(2015)\citenamefont {Tian},
  \citenamefont {Xu}, \citenamefont {Jiao},\ and\ \citenamefont
  {Torquato}}]{Tian:2015aa}%
  \BibitemOpen
  \bibfield  {author} {\bibinfo {author} {\bibnamefont {Tian}, \bibfnamefont
  {J.}}, \bibinfo {author} {\bibfnamefont {Y.}~\bibnamefont {Xu}}, \bibinfo
  {author} {\bibfnamefont {Y.}~\bibnamefont {Jiao}}, \ and\ \bibinfo {author}
  {\bibfnamefont {S.}~\bibnamefont {Torquato}}} (\bibinfo {year} {2015}),\
  \href {http://dx.doi.org/10.1038/srep16722} {\bibfield  {journal} {\bibinfo
  {journal} {Sci. Rep.}\ }\textbf {\bibinfo {volume} {5}},\ \bibinfo {pages}
  {16722 EP }}\BibitemShut {NoStop}%
\bibitem [{\citenamefont {Tighe}\ \emph {et~al.}(2008)\citenamefont {Tighe},
  \citenamefont {van Eerd},\ and\ \citenamefont {Vlugt}}]{Tighe:2008aa}%
  \BibitemOpen
  \bibfield  {author} {\bibinfo {author} {\bibnamefont {Tighe}, \bibfnamefont
  {B.~P.}}, \bibinfo {author} {\bibfnamefont {A.~R.~T.}\ \bibnamefont {van
  Eerd}}, \ and\ \bibinfo {author} {\bibfnamefont {T.~J.~H.}\ \bibnamefont
  {Vlugt}}} (\bibinfo {year} {2008}),\ \href {\doibase
  10.1103/PhysRevLett.100.238001} {\bibfield  {journal} {\bibinfo  {journal}
  {Phys. Rev. Lett.}\ }\textbf {\bibinfo {volume} {100}},\ \bibinfo {pages}
  {238001}}\BibitemShut {NoStop}%
\bibitem [{\citenamefont {Tighe}\ \emph {et~al.}(2010)\citenamefont {Tighe},
  \citenamefont {Snoeijer}, \citenamefont {Vlugt},\ and\ \citenamefont {van
  Hecke}}]{Tighe:2010aa}%
  \BibitemOpen
  \bibfield  {author} {\bibinfo {author} {\bibnamefont {Tighe}, \bibfnamefont
  {B.~P.}}, \bibinfo {author} {\bibfnamefont {J.~H.}\ \bibnamefont {Snoeijer}},
  \bibinfo {author} {\bibfnamefont {T.~J.~H.}\ \bibnamefont {Vlugt}}, \ and\
  \bibinfo {author} {\bibfnamefont {M.}~\bibnamefont {van Hecke}}} (\bibinfo
  {year} {2010}),\ \href {\doibase 10.1039/b926592a} {\bibfield  {journal}
  {\bibinfo  {journal} {Soft Matter}\ }\textbf {\bibinfo {volume} {6}},\
  \bibinfo {pages} {2908}}\BibitemShut {NoStop}%
\bibitem [{\citenamefont {Tighe}\ and\ \citenamefont
  {Vlugt}(2010)}]{Tighe:2010ab}%
  \BibitemOpen
  \bibfield  {author} {\bibinfo {author} {\bibnamefont {Tighe}, \bibfnamefont
  {B.~P.}}, \ and\ \bibinfo {author} {\bibfnamefont {T.~J.~H.}\ \bibnamefont
  {Vlugt}}} (\bibinfo {year} {2010}),\ \href {\doibase
  10.1088/1742-5468/2010/01/P01015} {\bibinfo  {journal} {J. Stat. Mech.}\ ,\
  \bibinfo {pages} {P01015}}\BibitemShut {NoStop}%
\bibitem [{\citenamefont {Tighe}\ and\ \citenamefont
  {Vlugt}(2011)}]{Tighe:2011aa}%
  \BibitemOpen
\bibfield  {journal} {  }\bibfield  {author} {\bibinfo {author} {\bibnamefont
  {Tighe}, \bibfnamefont {B.~P.}}, \ and\ \bibinfo {author} {\bibfnamefont
  {T.~J.~H.}\ \bibnamefont {Vlugt}}} (\bibinfo {year} {2011}),\ \href {\doibase
  10.1088/1742-5468/2011/04/P04002} {\bibinfo  {journal} {J. Stat. Mech.}\ ,\
  \bibinfo {pages} {P04002}}\BibitemShut {NoStop}%
\bibitem [{\citenamefont {Tkachenko}\ and\ \citenamefont
  {Witten}(2000)}]{Tkachenko:2000aa}%
  \BibitemOpen
\bibfield  {journal} {  }\bibfield  {author} {\bibinfo {author} {\bibnamefont
  {Tkachenko}, \bibfnamefont {A.~V.}}, \ and\ \bibinfo {author} {\bibfnamefont
  {T.~A.}\ \bibnamefont {Witten}}} (\bibinfo {year} {2000}),\ \href {\doibase
  10.1103/PhysRevE.62.2510} {\bibfield  {journal} {\bibinfo  {journal} {Phys.
  Rev. E}\ }\textbf {\bibinfo {volume} {62}},\ \bibinfo {pages}
  {2510}}\BibitemShut {NoStop}%
\bibitem [{\citenamefont {Toninelli}\ \emph {et~al.}(2006)\citenamefont
  {Toninelli}, \citenamefont {Biroli},\ and\ \citenamefont
  {Fisher}}]{Toninelli:2006aa}%
  \BibitemOpen
  \bibfield  {author} {\bibinfo {author} {\bibnamefont {Toninelli},
  \bibfnamefont {C.}}, \bibinfo {author} {\bibfnamefont {G.}~\bibnamefont
  {Biroli}}, \ and\ \bibinfo {author} {\bibfnamefont {D.~S.}\ \bibnamefont
  {Fisher}}} (\bibinfo {year} {2006}),\ \href {\doibase
  10.1103/PhysRevLett.96.035702} {\bibfield  {journal} {\bibinfo  {journal}
  {Phys. Rev. Lett.}\ }\textbf {\bibinfo {volume} {96}},\ \bibinfo {pages}
  {035702}}\BibitemShut {NoStop}%
\bibitem [{\citenamefont {Tonks}(1936)}]{Tonks:1936aa}%
  \BibitemOpen
  \bibfield  {author} {\bibinfo {author} {\bibnamefont {Tonks}, \bibfnamefont
  {L.}}} (\bibinfo {year} {1936}),\ \href {\doibase 10.1103/PhysRev.50.955}
  {\bibfield  {journal} {\bibinfo  {journal} {Phys. Rev.}\ }\textbf {\bibinfo
  {volume} {50}},\ \bibinfo {pages} {955}}\BibitemShut {NoStop}%
\bibitem [{\citenamefont {Torquato}\ and\ \citenamefont
  {Jiao}(2009)}]{Torquato:2009aa}%
  \BibitemOpen
  \bibfield  {author} {\bibinfo {author} {\bibnamefont {Torquato},
  \bibfnamefont {S.}}, \ and\ \bibinfo {author} {\bibfnamefont
  {Y.}~\bibnamefont {Jiao}}} (\bibinfo {year} {2009}),\ \href@noop {}
  {\bibfield  {journal} {\bibinfo  {journal} {Nature}\ }\textbf {\bibinfo
  {volume} {460}},\ \bibinfo {pages} {876}}\BibitemShut {NoStop}%
\bibitem [{\citenamefont {Torquato}\ and\ \citenamefont
  {Jiao}(2010)}]{Torquato:2010ac}%
  \BibitemOpen
  \bibfield  {author} {\bibinfo {author} {\bibnamefont {Torquato},
  \bibfnamefont {S.}}, \ and\ \bibinfo {author} {\bibfnamefont
  {Y.}~\bibnamefont {Jiao}}} (\bibinfo {year} {2010}),\ \href {\doibase
  10.1103/PhysRevE.82.061302} {\bibfield  {journal} {\bibinfo  {journal} {Phys.
  Rev. E}\ }\textbf {\bibinfo {volume} {82}},\ \bibinfo {pages}
  {061302}}\BibitemShut {NoStop}%
\bibitem [{\citenamefont {Torquato}\ and\ \citenamefont
  {Jiao}(2012)}]{Torquato:2012aa}%
  \BibitemOpen
  \bibfield  {author} {\bibinfo {author} {\bibnamefont {Torquato},
  \bibfnamefont {S.}}, \ and\ \bibinfo {author} {\bibfnamefont
  {Y.}~\bibnamefont {Jiao}}} (\bibinfo {year} {2012}),\ \href {\doibase
  10.1103/PhysRevE.86.011102} {\bibfield  {journal} {\bibinfo  {journal} {Phys.
  Rev. E}\ }\textbf {\bibinfo {volume} {86}},\ \bibinfo {pages}
  {011102}}\BibitemShut {NoStop}%
\bibitem [{\citenamefont {Torquato}\ and\ \citenamefont
  {Stillinger}(2010)}]{Torquato:2010aa}%
  \BibitemOpen
  \bibfield  {author} {\bibinfo {author} {\bibnamefont {Torquato},
  \bibfnamefont {S.}}, \ and\ \bibinfo {author} {\bibfnamefont
  {F.}~\bibnamefont {Stillinger}}} (\bibinfo {year} {2010}),\ \href {\doibase
  10.1103/RevModPhys.82.2633} {\bibfield  {journal} {\bibinfo  {journal} {Rev.
  Mod. Phys.}\ }\textbf {\bibinfo {volume} {82}},\ \bibinfo {pages}
  {2633}}\BibitemShut {NoStop}%
\bibitem [{\citenamefont {Torquato}\ and\ \citenamefont
  {Stillinger}(2001)}]{Torquato:2001aa}%
  \BibitemOpen
  \bibfield  {author} {\bibinfo {author} {\bibnamefont {Torquato},
  \bibfnamefont {S.}}, \ and\ \bibinfo {author} {\bibfnamefont {F.~H.}\
  \bibnamefont {Stillinger}}} (\bibinfo {year} {2001}),\ \href {\doibase
  10.1021/jp011960q} {\bibfield  {journal} {\bibinfo  {journal} {J. Phys. Chem.
  B}\ }\textbf {\bibinfo {volume} {105}},\ \bibinfo {pages}
  {11849}}\BibitemShut {NoStop}%
\bibitem [{\citenamefont {Torquato}\ and\ \citenamefont
  {Stillinger}(2003)}]{Torquato:2003aa}%
  \BibitemOpen
  \bibfield  {author} {\bibinfo {author} {\bibnamefont {Torquato},
  \bibfnamefont {S.}}, \ and\ \bibinfo {author} {\bibfnamefont {F.~H.}\
  \bibnamefont {Stillinger}}} (\bibinfo {year} {2003}),\ \href {\doibase
  10.1103/PhysRevE.68.041113} {\bibfield  {journal} {\bibinfo  {journal} {Phys.
  Rev. E}\ }\textbf {\bibinfo {volume} {68}},\ \bibinfo {pages}
  {041113}}\BibitemShut {NoStop}%
\bibitem [{\citenamefont {Torquato}\ and\ \citenamefont
  {Stillinger}(2006)}]{Torquato:2006aa}%
  \BibitemOpen
  \bibfield  {author} {\bibinfo {author} {\bibnamefont {Torquato},
  \bibfnamefont {S.}}, \ and\ \bibinfo {author} {\bibfnamefont {F.~H.}\
  \bibnamefont {Stillinger}}} (\bibinfo {year} {2006}),\ \href {\doibase
  10.1103/PhysRevE.73.031106} {\bibfield  {journal} {\bibinfo  {journal} {Phys.
  Rev. E}\ }\textbf {\bibinfo {volume} {73}},\ \bibinfo {pages}
  {031106}}\BibitemShut {NoStop}%
\bibitem [{\citenamefont {Torquato}\ \emph {et~al.}(2000)\citenamefont
  {Torquato}, \citenamefont {Truskett},\ and\ \citenamefont
  {Debenedetti}}]{Torquato:2000aa}%
  \BibitemOpen
  \bibfield  {author} {\bibinfo {author} {\bibnamefont {Torquato},
  \bibfnamefont {S.}}, \bibinfo {author} {\bibfnamefont {T.~M.}\ \bibnamefont
  {Truskett}}, \ and\ \bibinfo {author} {\bibfnamefont {P.~G.}\ \bibnamefont
  {Debenedetti}}} (\bibinfo {year} {2000}),\ \href {\doibase
  10.1103/PhysRevLett.84.2064} {\bibfield  {journal} {\bibinfo  {journal}
  {Phys. Rev. Lett.}\ }\textbf {\bibinfo {volume} {84}},\ \bibinfo {pages}
  {2064}}\BibitemShut {NoStop}%
\bibitem [{\citenamefont {Trappe}\ \emph {et~al.}(2001)\citenamefont {Trappe},
  \citenamefont {Prasad}, \citenamefont {Cipelletti}, \citenamefont {Segre},\
  and\ \citenamefont {Weitz}}]{Trappe:2001aa}%
  \BibitemOpen
  \bibfield  {author} {\bibinfo {author} {\bibnamefont {Trappe}, \bibfnamefont
  {V.}}, \bibinfo {author} {\bibfnamefont {V.}~\bibnamefont {Prasad}}, \bibinfo
  {author} {\bibfnamefont {L.}~\bibnamefont {Cipelletti}}, \bibinfo {author}
  {\bibfnamefont {P.~N.}\ \bibnamefont {Segre}}, \ and\ \bibinfo {author}
  {\bibfnamefont {D.~A.}\ \bibnamefont {Weitz}}} (\bibinfo {year} {2001}),\
  \href {http://dx.doi.org/10.1038/35081021} {\bibfield  {journal} {\bibinfo
  {journal} {Nature}\ }\textbf {\bibinfo {volume} {411}},\ \bibinfo {pages}
  {772}}\BibitemShut {NoStop}%
\bibitem [{\citenamefont {Unger}\ \emph {et~al.}(2005)\citenamefont {Unger},
  \citenamefont {Kertesz},\ and\ \citenamefont {Wolf}}]{Unger:2005aa}%
  \BibitemOpen
  \bibfield  {author} {\bibinfo {author} {\bibnamefont {Unger}, \bibfnamefont
  {T.}}, \bibinfo {author} {\bibfnamefont {J.}~\bibnamefont {Kertesz}}, \ and\
  \bibinfo {author} {\bibfnamefont {D.}~\bibnamefont {Wolf}}} (\bibinfo {year}
  {2005}),\ \href {\doibase 10.1103/PhysRevLett.94.178001} {\bibfield
  {journal} {\bibinfo  {journal} {Phys. Rev. Lett.}\ }\textbf {\bibinfo
  {volume} {94}},\ \bibinfo {pages} {178001}}\BibitemShut {NoStop}%
\bibitem [{\citenamefont {Valverde}\ \emph {et~al.}(2004)\citenamefont
  {Valverde}, \citenamefont {Quintanilla},\ and\ \citenamefont
  {Castellanos}}]{Valverde:2004aa}%
  \BibitemOpen
  \bibfield  {author} {\bibinfo {author} {\bibnamefont {Valverde},
  \bibfnamefont {J.~M.}}, \bibinfo {author} {\bibfnamefont {M.~A.~S.}\
  \bibnamefont {Quintanilla}}, \ and\ \bibinfo {author} {\bibfnamefont
  {A.}~\bibnamefont {Castellanos}}} (\bibinfo {year} {2004}),\ \href {\doibase
  10.1103/PhysRevLett.92.258303} {\bibfield  {journal} {\bibinfo  {journal}
  {Phys. Rev. Lett.}\ }\textbf {\bibinfo {volume} {92}},\ \bibinfo {pages}
  {258303}}\BibitemShut {NoStop}%
\bibitem [{\citenamefont {Walton}(1987)}]{Walton:1987aa}%
  \BibitemOpen
  \bibfield  {author} {\bibinfo {author} {\bibnamefont {Walton}, \bibfnamefont
  {K.}}} (\bibinfo {year} {1987}),\ \href {\doibase
  http://dx.doi.org/10.1016/0022-5096(87)90036-6} {\bibfield  {journal}
  {\bibinfo  {journal} {J. Mech. Phys. Solids}\ }\textbf {\bibinfo {volume}
  {35}},\ \bibinfo {pages} {213 }}\BibitemShut {NoStop}%
\bibitem [{\citenamefont {Wang}\ \emph {et~al.}(2015)\citenamefont {Wang},
  \citenamefont {Dong},\ and\ \citenamefont {Yu}}]{Wang:2015aa}%
  \BibitemOpen
  \bibfield  {author} {\bibinfo {author} {\bibnamefont {Wang}, \bibfnamefont
  {C.}}, \bibinfo {author} {\bibfnamefont {K.}~\bibnamefont {Dong}}, \ and\
  \bibinfo {author} {\bibfnamefont {A.}~\bibnamefont {Yu}}} (\bibinfo {year}
  {2015}),\ \href {\doibase 10.1103/PhysRevE.92.062203} {\bibfield  {journal}
  {\bibinfo  {journal} {Phys. Rev. E}\ }\textbf {\bibinfo {volume} {92}},\
  \bibinfo {pages} {062203}}\BibitemShut {NoStop}%
\bibitem [{\citenamefont {Wang}\ \emph
  {et~al.}(2010{\natexlab{a}})\citenamefont {Wang}, \citenamefont {Song},
  \citenamefont {Wang},\ and\ \citenamefont {Makse}}]{Wang:2010aa}%
  \BibitemOpen
  \bibfield  {author} {\bibinfo {author} {\bibnamefont {Wang}, \bibfnamefont
  {K.}}, \bibinfo {author} {\bibfnamefont {C.}~\bibnamefont {Song}}, \bibinfo
  {author} {\bibfnamefont {P.}~\bibnamefont {Wang}}, \ and\ \bibinfo {author}
  {\bibfnamefont {H.~A.}\ \bibnamefont {Makse}}} (\bibinfo {year}
  {2010}{\natexlab{a}}),\ \href {\doibase 10.1209/0295-5075/91/68001}
  {\bibfield  {journal} {\bibinfo  {journal} {EPL}\ }\textbf {\bibinfo {volume}
  {91}},\ \bibinfo {pages} {68001}}\BibitemShut {NoStop}%
\bibitem [{\citenamefont {Wang}\ \emph {et~al.}(2012)\citenamefont {Wang},
  \citenamefont {Song}, \citenamefont {Wang},\ and\ \citenamefont
  {Makse}}]{Wang:2012aa}%
  \BibitemOpen
  \bibfield  {author} {\bibinfo {author} {\bibnamefont {Wang}, \bibfnamefont
  {K.}}, \bibinfo {author} {\bibfnamefont {C.}~\bibnamefont {Song}}, \bibinfo
  {author} {\bibfnamefont {P.}~\bibnamefont {Wang}}, \ and\ \bibinfo {author}
  {\bibfnamefont {H.~A.}\ \bibnamefont {Makse}}} (\bibinfo {year} {2012}),\
  \href {\doibase 10.1103/PhysRevE.86.011305} {\bibfield  {journal} {\bibinfo
  {journal} {Phys. Rev. E}\ }\textbf {\bibinfo {volume} {86}},\
  10.1103/PhysRevE.86.011305}\BibitemShut {NoStop}%
\bibitem [{\citenamefont {Wang}\ \emph {et~al.}(2008)\citenamefont {Wang},
  \citenamefont {Song}, \citenamefont {Briscoe},\ and\ \citenamefont
  {Makse}}]{Wang:2008aa}%
  \BibitemOpen
  \bibfield  {author} {\bibinfo {author} {\bibnamefont {Wang}, \bibfnamefont
  {P.}}, \bibinfo {author} {\bibfnamefont {C.}~\bibnamefont {Song}}, \bibinfo
  {author} {\bibfnamefont {C.}~\bibnamefont {Briscoe}}, \ and\ \bibinfo
  {author} {\bibfnamefont {H.~A.}\ \bibnamefont {Makse}}} (\bibinfo {year}
  {2008}),\ \href {\doibase 10.1103/PhysRevE.77.061309} {\bibfield  {journal}
  {\bibinfo  {journal} {Phys. Rev. E}\ }\textbf {\bibinfo {volume} {77}},\
  10.1103/PhysRevE.77.061309}\BibitemShut {NoStop}%
\bibitem [{\citenamefont {Wang}\ \emph
  {et~al.}(2010{\natexlab{b}})\citenamefont {Wang}, \citenamefont {Song},
  \citenamefont {Briscoe}, \citenamefont {Wang},\ and\ \citenamefont
  {Makse}}]{Wang:2010ab}%
  \BibitemOpen
  \bibfield  {author} {\bibinfo {author} {\bibnamefont {Wang}, \bibfnamefont
  {P.}}, \bibinfo {author} {\bibfnamefont {C.}~\bibnamefont {Song}}, \bibinfo
  {author} {\bibfnamefont {C.}~\bibnamefont {Briscoe}}, \bibinfo {author}
  {\bibfnamefont {K.}~\bibnamefont {Wang}}, \ and\ \bibinfo {author}
  {\bibfnamefont {H.~A.}\ \bibnamefont {Makse}}} (\bibinfo {year}
  {2010}{\natexlab{b}}),\ \href {\doibase 10.1016/j.physa.2010.05.053}
  {\bibfield  {journal} {\bibinfo  {journal} {Physica A}\ }\textbf {\bibinfo
  {volume} {389}},\ \bibinfo {pages} {3972}}\BibitemShut {NoStop}%
\bibitem [{\citenamefont {Wang}\ \emph {et~al.}(2011)\citenamefont {Wang},
  \citenamefont {Song}, \citenamefont {Jin},\ and\ \citenamefont
  {Makse}}]{Wang:2011aa}%
  \BibitemOpen
  \bibfield  {author} {\bibinfo {author} {\bibnamefont {Wang}, \bibfnamefont
  {P.}}, \bibinfo {author} {\bibfnamefont {C.}~\bibnamefont {Song}}, \bibinfo
  {author} {\bibfnamefont {Y.}~\bibnamefont {Jin}}, \ and\ \bibinfo {author}
  {\bibfnamefont {H.~A.}\ \bibnamefont {Makse}}} (\bibinfo {year} {2011}),\
  \href {\doibase 10.1016/j.physa.2010.10.017} {\bibfield  {journal} {\bibinfo
  {journal} {Physica A}\ }\textbf {\bibinfo {volume} {390}},\ \bibinfo {pages}
  {427}}\BibitemShut {NoStop}%
\bibitem [{\citenamefont {Wang}\ \emph
  {et~al.}(2010{\natexlab{c}})\citenamefont {Wang}, \citenamefont {Song},
  \citenamefont {Jin}, \citenamefont {Wang},\ and\ \citenamefont
  {Makse}}]{Wang:2010ac}%
  \BibitemOpen
  \bibfield  {author} {\bibinfo {author} {\bibnamefont {Wang}, \bibfnamefont
  {P.}}, \bibinfo {author} {\bibfnamefont {C.}~\bibnamefont {Song}}, \bibinfo
  {author} {\bibfnamefont {Y.}~\bibnamefont {Jin}}, \bibinfo {author}
  {\bibfnamefont {K.}~\bibnamefont {Wang}}, \ and\ \bibinfo {author}
  {\bibfnamefont {H.~A.}\ \bibnamefont {Makse}}} (\bibinfo {year}
  {2010}{\natexlab{c}}),\ \href {\doibase 10.1088/1742-5468/2010/12/P12005}
  {\bibfield  {journal} {\bibinfo  {journal} {J. Stat. Mech.}\
  }10.1088/1742-5468/2010/12/P12005}\BibitemShut {NoStop}%
\bibitem [{\citenamefont {Wang}\ \emph {et~al.}(2006)\citenamefont {Wang},
  \citenamefont {Song},\ and\ \citenamefont {Makse}}]{Wang:2006aa}%
  \BibitemOpen
  \bibfield  {author} {\bibinfo {author} {\bibnamefont {Wang}, \bibfnamefont
  {P.}}, \bibinfo {author} {\bibfnamefont {C.}~\bibnamefont {Song}}, \ and\
  \bibinfo {author} {\bibfnamefont {H.~A.}\ \bibnamefont {Makse}}} (\bibinfo
  {year} {2006}),\ \href {\doibase 10.1038/nphys366} {\bibfield  {journal}
  {\bibinfo  {journal} {Nature Phys.}\ }\textbf {\bibinfo {volume} {2}},\
  \bibinfo {pages} {526}}\BibitemShut {NoStop}%
\bibitem [{\citenamefont {Warner}(2017)}]{Warner:2017aa}%
  \BibitemOpen
  \bibfield  {author} {\bibinfo {author} {\bibnamefont {Warner}, \bibfnamefont
  {M.}}} (\bibinfo {year} {2017}),\ \href {\doibase 10.1098/rsbm.2016.0028}
  {\bibfield  {journal} {\bibinfo  {journal} {Biographical Memoirs of Fellows
  of the Royal Society}\ }10.1098/rsbm.2016.0028}\BibitemShut {NoStop}%
\bibitem [{\citenamefont {Weaire}\ and\ \citenamefont
  {Aste}(2008)}]{Weaire:2008aa}%
  \BibitemOpen
  \bibfield  {author} {\bibinfo {author} {\bibnamefont {Weaire}, \bibfnamefont
  {D.}}, \ and\ \bibinfo {author} {\bibfnamefont {T.}~\bibnamefont {Aste}}}
  (\bibinfo {year} {2008}),\ \href@noop {} {\emph {\bibinfo {title} {The
  pursuit of perfect packing}}}\ (\bibinfo  {publisher} {CRC
  Press})\BibitemShut {NoStop}%
\bibitem [{\citenamefont {Williams}\ and\ \citenamefont
  {Philipse}(2003)}]{Williams:2003aa}%
  \BibitemOpen
  \bibfield  {author} {\bibinfo {author} {\bibnamefont {Williams},
  \bibfnamefont {S.~R.}}, \ and\ \bibinfo {author} {\bibfnamefont {A.~P.}\
  \bibnamefont {Philipse}}} (\bibinfo {year} {2003}),\ \href {\doibase
  10.1103/PhysRevE.67.051301} {\bibfield  {journal} {\bibinfo  {journal} {Phys.
  Rev. E}\ }\textbf {\bibinfo {volume} {67}},\ \bibinfo {pages}
  {051301}}\BibitemShut {NoStop}%
\bibitem [{\citenamefont {Wouterse}\ \emph {et~al.}(2009)\citenamefont
  {Wouterse}, \citenamefont {Luding},\ and\ \citenamefont
  {Philipse}}]{Wouterse:2009aa}%
  \BibitemOpen
  \bibfield  {author} {\bibinfo {author} {\bibnamefont {Wouterse},
  \bibfnamefont {A.}}, \bibinfo {author} {\bibfnamefont {S.}~\bibnamefont
  {Luding}}, \ and\ \bibinfo {author} {\bibfnamefont {A.~P.}\ \bibnamefont
  {Philipse}}} (\bibinfo {year} {2009}),\ \href@noop {} {\bibfield  {journal}
  {\bibinfo  {journal} {Granular Matter}\ }\textbf {\bibinfo {volume} {11}},\
  \bibinfo {pages} {169}}\BibitemShut {NoStop}%
\bibitem [{\citenamefont {Wu}\ \emph {et~al.}(2015)\citenamefont {Wu},
  \citenamefont {Olsson},\ and\ \citenamefont {Teitel}}]{Wu:2015aa}%
  \BibitemOpen
  \bibfield  {author} {\bibinfo {author} {\bibnamefont {Wu}, \bibfnamefont
  {Y.}}, \bibinfo {author} {\bibfnamefont {P.}~\bibnamefont {Olsson}}, \ and\
  \bibinfo {author} {\bibfnamefont {S.}~\bibnamefont {Teitel}}} (\bibinfo
  {year} {2015}),\ \href {\doibase 10.1103/PhysRevE.92.052206} {\bibfield
  {journal} {\bibinfo  {journal} {Phys. Rev. E}\ }\textbf {\bibinfo {volume}
  {92}},\ \bibinfo {pages} {052206}}\BibitemShut {NoStop}%
\bibitem [{\citenamefont {Wyart}(2005)}]{Wyart:2005ac}%
  \BibitemOpen
  \bibfield  {author} {\bibinfo {author} {\bibnamefont {Wyart}, \bibfnamefont
  {M.}}} (\bibinfo {year} {2005}),\ \href {\doibase 10.1051/anphys:2006003}
  {\bibfield  {journal} {\bibinfo  {journal} {Annales de Physique}\ }\textbf
  {\bibinfo {volume} {30}},\ \bibinfo {pages} {1}}\BibitemShut {NoStop}%
\bibitem [{\citenamefont {Wyart}(2010)}]{Wyart:2010aa}%
  \BibitemOpen
  \bibfield  {author} {\bibinfo {author} {\bibnamefont {Wyart}, \bibfnamefont
  {M.}}} (\bibinfo {year} {2010}),\ \href
  {http://stacks.iop.org/0295-5075/89/i=6/a=64001} {\bibfield  {journal}
  {\bibinfo  {journal} {EPL}\ }\textbf {\bibinfo {volume} {89}},\ \bibinfo
  {pages} {64001}}\BibitemShut {NoStop}%
\bibitem [{\citenamefont {Wyart}(2012)}]{Wyart:2012aa}%
  \BibitemOpen
  \bibfield  {author} {\bibinfo {author} {\bibnamefont {Wyart}, \bibfnamefont
  {M.}}} (\bibinfo {year} {2012}),\ \href {\doibase
  10.1103/PhysRevLett.109.125502} {\bibfield  {journal} {\bibinfo  {journal}
  {Phys. Rev. Lett.}\ }\textbf {\bibinfo {volume} {109}},\ \bibinfo {pages}
  {125502}}\BibitemShut {NoStop}%
\bibitem [{\citenamefont {Wyart}\ \emph
  {et~al.}(2005{\natexlab{a}})\citenamefont {Wyart}, \citenamefont {Nagel},\
  and\ \citenamefont {Witten}}]{Wyart:2005aa}%
  \BibitemOpen
  \bibfield  {author} {\bibinfo {author} {\bibnamefont {Wyart}, \bibfnamefont
  {M.}}, \bibinfo {author} {\bibfnamefont {S.}~\bibnamefont {Nagel}}, \ and\
  \bibinfo {author} {\bibfnamefont {T.}~\bibnamefont {Witten}}} (\bibinfo
  {year} {2005}{\natexlab{a}}),\ \href {\doibase 10.1209/epl/i2005-10245-5}
  {\bibfield  {journal} {\bibinfo  {journal} {EPL}\ }\textbf {\bibinfo {volume}
  {72}},\ \bibinfo {pages} {486}}\BibitemShut {NoStop}%
\bibitem [{\citenamefont {Wyart}\ \emph
  {et~al.}(2005{\natexlab{b}})\citenamefont {Wyart}, \citenamefont {Silbert},
  \citenamefont {Nagel},\ and\ \citenamefont {Witten}}]{Wyart:2005ab}%
  \BibitemOpen
  \bibfield  {author} {\bibinfo {author} {\bibnamefont {Wyart}, \bibfnamefont
  {M.}}, \bibinfo {author} {\bibfnamefont {L.}~\bibnamefont {Silbert}},
  \bibinfo {author} {\bibfnamefont {S.}~\bibnamefont {Nagel}}, \ and\ \bibinfo
  {author} {\bibfnamefont {T.}~\bibnamefont {Witten}}} (\bibinfo {year}
  {2005}{\natexlab{b}}),\ \href {\doibase 10.1103/PhysRevE.72.051306}
  {\bibfield  {journal} {\bibinfo  {journal} {Phys. Rev. E}\ }\textbf {\bibinfo
  {volume} {72}},\ \bibinfo {pages} {051306}}\BibitemShut {NoStop}%
\bibitem [{\citenamefont {Xu}\ \emph {et~al.}(2005)\citenamefont {Xu},
  \citenamefont {Blawzdziewicz},\ and\ \citenamefont {O'Hern}}]{Xu:2005aa}%
  \BibitemOpen
  \bibfield  {author} {\bibinfo {author} {\bibnamefont {Xu}, \bibfnamefont
  {N.}}, \bibinfo {author} {\bibfnamefont {J.}~\bibnamefont {Blawzdziewicz}}, \
  and\ \bibinfo {author} {\bibfnamefont {C.}~\bibnamefont {O'Hern}}} (\bibinfo
  {year} {2005}),\ \href {\doibase 10.1103/PhysRevE.71.061306} {\bibfield
  {journal} {\bibinfo  {journal} {Phys. Rev. E}\ }\textbf {\bibinfo {volume}
  {71}},\ \bibinfo {pages} {061306}}\BibitemShut {NoStop}%
\bibitem [{\citenamefont {Xu}\ \emph {et~al.}(2011)\citenamefont {Xu},
  \citenamefont {Frenkel},\ and\ \citenamefont {Liu}}]{Xu:2011aa}%
  \BibitemOpen
  \bibfield  {author} {\bibinfo {author} {\bibnamefont {Xu}, \bibfnamefont
  {N.}}, \bibinfo {author} {\bibfnamefont {D.}~\bibnamefont {Frenkel}}, \ and\
  \bibinfo {author} {\bibfnamefont {A.~J.}\ \bibnamefont {Liu}}} (\bibinfo
  {year} {2011}),\ \href {\doibase 10.1103/PhysRevLett.106.245502} {\bibfield
  {journal} {\bibinfo  {journal} {Phys. Rev. Lett.}\ }\textbf {\bibinfo
  {volume} {106}},\ \bibinfo {pages} {245502}}\BibitemShut {NoStop}%
\bibitem [{\citenamefont {Yadav}\ \emph {et~al.}(2013)\citenamefont {Yadav},
  \citenamefont {Chastaing},\ and\ \citenamefont {Kudrolli}}]{Yadav:2013aa}%
  \BibitemOpen
  \bibfield  {author} {\bibinfo {author} {\bibnamefont {Yadav}, \bibfnamefont
  {V.}}, \bibinfo {author} {\bibfnamefont {J.-Y.}\ \bibnamefont {Chastaing}}, \
  and\ \bibinfo {author} {\bibfnamefont {A.}~\bibnamefont {Kudrolli}}}
  (\bibinfo {year} {2013}),\ \href {\doibase 10.1103/PhysRevE.88.052203}
  {\bibfield  {journal} {\bibinfo  {journal} {Phys. Rev. E}\ }\textbf {\bibinfo
  {volume} {88}},\ \bibinfo {pages} {052203}}\BibitemShut {NoStop}%
\bibitem [{\citenamefont {Yang}\ \emph {et~al.}(2000)\citenamefont {Yang},
  \citenamefont {Zou},\ and\ \citenamefont {Yu}}]{Yang:2000aa}%
  \BibitemOpen
  \bibfield  {author} {\bibinfo {author} {\bibnamefont {Yang}, \bibfnamefont
  {R.}}, \bibinfo {author} {\bibfnamefont {R.}~\bibnamefont {Zou}}, \ and\
  \bibinfo {author} {\bibfnamefont {A.}~\bibnamefont {Yu}}} (\bibinfo {year}
  {2000}),\ \href {\doibase 10.1103/PhysRevE.62.3900} {\bibfield  {journal}
  {\bibinfo  {journal} {Phys. Rev. E}\ }\textbf {\bibinfo {volume} {62}},\
  \bibinfo {pages} {3900}}\BibitemShut {NoStop}%
\bibitem [{\citenamefont {Zaccarelli}(2007)}]{Zaccarelli:2007aa}%
  \BibitemOpen
  \bibfield  {author} {\bibinfo {author} {\bibnamefont {Zaccarelli},
  \bibfnamefont {E.}}} (\bibinfo {year} {2007}),\ \href
  {http://stacks.iop.org/0953-8984/19/i=32/a=323101} {\bibfield  {journal}
  {\bibinfo  {journal} {J. Phys. Cond. Mat.}\ }\textbf {\bibinfo {volume}
  {19}},\ \bibinfo {pages} {323101}}\BibitemShut {NoStop}%
\bibitem [{\citenamefont {Zaccone}\ and\ \citenamefont
  {Scossa-Romano}(2011)}]{Zaccone:2011aa}%
  \BibitemOpen
  \bibfield  {author} {\bibinfo {author} {\bibnamefont {Zaccone}, \bibfnamefont
  {A.}}, \ and\ \bibinfo {author} {\bibfnamefont {E.}~\bibnamefont
  {Scossa-Romano}}} (\bibinfo {year} {2011}),\ \href {\doibase
  10.1103/PhysRevB.83.184205} {\bibfield  {journal} {\bibinfo  {journal} {Phys.
  Rev. B}\ }\textbf {\bibinfo {volume} {83}},\ \bibinfo {pages}
  {184205}}\BibitemShut {NoStop}%
\bibitem [{\citenamefont {Zhang}\ and\ \citenamefont
  {Makse}(2005)}]{Zhang:2005aa}%
  \BibitemOpen
  \bibfield  {author} {\bibinfo {author} {\bibnamefont {Zhang}, \bibfnamefont
  {H.}}, \ and\ \bibinfo {author} {\bibfnamefont {H.}~\bibnamefont {Makse}}}
  (\bibinfo {year} {2005}),\ \href {\doibase 10.1103/PhysRevE.72.011301}
  {\bibfield  {journal} {\bibinfo  {journal} {Phys. Rev. E}\ }\textbf {\bibinfo
  {volume} {72}},\ \bibinfo {pages} {011301}}\BibitemShut {NoStop}%
\bibitem [{\citenamefont {Zhao}\ \emph {et~al.}(2012)\citenamefont {Zhao},
  \citenamefont {Li}, \citenamefont {Zou},\ and\ \citenamefont
  {Yu}}]{Zhao:2012aa}%
  \BibitemOpen
  \bibfield  {author} {\bibinfo {author} {\bibnamefont {Zhao}, \bibfnamefont
  {J.}}, \bibinfo {author} {\bibfnamefont {S.}~\bibnamefont {Li}}, \bibinfo
  {author} {\bibfnamefont {R.}~\bibnamefont {Zou}}, \ and\ \bibinfo {author}
  {\bibfnamefont {A.}~\bibnamefont {Yu}}} (\bibinfo {year} {2012}),\ \href@noop
  {} {\bibfield  {journal} {\bibinfo  {journal} {Soft Matter}\ }\textbf
  {\bibinfo {volume} {8}},\ \bibinfo {pages} {1003}}\BibitemShut {NoStop}%
\bibitem [{\citenamefont {Zhao}\ and\ \citenamefont
  {Schr\"oter}(2014)}]{Zhao:2014aa}%
  \BibitemOpen
  \bibfield  {author} {\bibinfo {author} {\bibnamefont {Zhao}, \bibfnamefont
  {S.-C.}}, \ and\ \bibinfo {author} {\bibfnamefont {M.}~\bibnamefont
  {Schr\"oter}}} (\bibinfo {year} {2014}),\ \href {\doibase 10.1039/C3SM53176G}
  {\bibfield  {journal} {\bibinfo  {journal} {Soft Matter}\ }\textbf {\bibinfo
  {volume} {10}},\ \bibinfo {pages} {4208}}\BibitemShut {NoStop}%
\bibitem [{\citenamefont {Zhou}\ \emph {et~al.}(2006)\citenamefont {Zhou},
  \citenamefont {Long}, \citenamefont {Wang},\ and\ \citenamefont
  {Dinsmore}}]{Zhou:2006aa}%
  \BibitemOpen
  \bibfield  {author} {\bibinfo {author} {\bibnamefont {Zhou}, \bibfnamefont
  {J.}}, \bibinfo {author} {\bibfnamefont {S.}~\bibnamefont {Long}}, \bibinfo
  {author} {\bibfnamefont {Q.}~\bibnamefont {Wang}}, \ and\ \bibinfo {author}
  {\bibfnamefont {A.~D.}\ \bibnamefont {Dinsmore}}} (\bibinfo {year} {2006}),\
  \href {\doibase 10.1126/science.1125151} {\bibfield  {journal} {\bibinfo
  {journal} {Science}\ }\textbf {\bibinfo {volume} {312}},\ \bibinfo {pages}
  {1631}}\BibitemShut {NoStop}%
\end{thebibliography}

%

\end{document}